\documentclass[aps,twocolumn,superscriptaddress,preprintnumbers,nofootinbib,nolongbibliography,floatfix,reprint]{revtex4-2}

\usepackage{epsfig,amssymb,amsmath,rotating,graphicx}
\usepackage{lineno}
\usepackage{longtable}
\usepackage{rotating}
\usepackage{multirow}
\usepackage{graphicx}
\usepackage{times,natbib,graphicx,amsmath,multirow,xspace}
\usepackage{bm}
\usepackage{orcidlink}
\usepackage[graphicx]{realboxes}
\usepackage{adjustbox}

\newcommand\xmm{{\it XMM-Newton}}
\newcommand\nustar{{\it NuSTAR}}
\newcommand\ixpe{{\it IXPE}}

\newcommand\colibri{\textit {Colibr\'i}}
\newcommand\athena{\textit {NewAthena}}
\newcommand\heroix{\textit {HEROIX}}
\newcommand\jwst{\textit {JWST}}

\newcommand\keV{{\rm~keV}}
\newcommand\ev{{\rm~eV}}

\begin{document}

\title{Constraining disk-to-corona power transfer fraction, soft X-ray excess origin, and black hole spin population of Type-1 AGN across mass scales}

\author{Labani Mallick\,\orcidlink{0000-0001-8624-9162}}\thanks{CITA National Fellow}
\email{Corresponding author (lmallick@cita.utoronto.ca)}
\affiliation{Department of Physics \& Astronomy, University of Manitoba, Winnipeg, Manitoba R3T 2N2, Canada}
\affiliation{CITA, University of Toronto, 60 St. George Street, Toronto, Ontario M5S 3H8, Canada}
\affiliation{Cahill Center for Astronomy and Astrophysics, California Institute of Technology, Pasadena, CA 91125, USA}

\author{Ciro Pinto\,\orcidlink{0000-0003-2532-7379}}
\affiliation{INAF/IASF Palermo, via Ugo La Malfa 153, I-90146 Palermo, Italy}

\author{John A. Tomsick\,\orcidlink{0000-0001-5506-9855}}
\affiliation{Space Sciences Laboratory, 7 Gauss Way, University of California, Berkeley, CA 94720-7450, USA}

\author{Alex G. Markowitz\,\orcidlink{0000-0002-2173-0673}}
\affiliation{Nicolaus Copernicus Astronomical Center, Polish Academy of Sciences, Bartycka 18, 00-716, Warsaw, Poland}

\author{Andrew C. Fabian\,\orcidlink{0000-0002-9378-4072}}
\affiliation{Institute of Astronomy, University of Cambridge, Madingley Road, Cambridge CB3 0HA, UK}

\author{Samar Safi-Harb\,\orcidlink{0000-0001-6189-7665}}
\affiliation{Department of Physics \& Astronomy, University of Manitoba, Winnipeg, Manitoba R3T 2N2, Canada}

\author{James F. Steiner\,\orcidlink{0000-0002-5872-6061}}
\affiliation{Center for Astrophysics $\vert$ Harvard \& Smithsonian, 60 Garden Street, Cambridge, MA 02138, USA}

\author{Fabio Pacucci\,\orcidlink{0000-0001-9879-7780}}
\affiliation{Center for Astrophysics $\vert$ Harvard \& Smithsonian, 60 Garden Street, Cambridge, MA 02138, USA}
\affiliation{Black Hole Initiative, Harvard University, 20 Garden Street, Cambridge, MA 02138, USA}

\author{William N. Alston\,\orcidlink{0000-0003-2658-6559}}
\affiliation{Centre for Astrophysics Research, University of Hertfordshire, College Lane, Hatfield AL10 9AB, UK}

\begin{abstract}
Understanding the nature of the accretion disk, its interplay with the X-ray corona, and assessing black hole spin demographics remain open challenges in astrophysics. In this paper, we examine the predictions of the standard $\alpha$-disk model, origin of the puzzling soft X-ray excess, and measure the black hole spin parameter by applying an updated high-density disk reflection model to the {\textit{XMM-Newton}}/{\textit{NuSTAR}} broadband (0.3--78~keV) X-ray spectra of a sample of 11 Type-1 AGN. Our Bayesian analysis confirms that a variable-density relativistic disk reflection model with a broken power-law emissivity profile can simultaneously fit the soft X-ray excess, broad iron~K line emission, and Compton hump in 3 out of 11 AGN. For the remaining sources, a distinct warm Comptonization component is still required, which supports a hybrid origin for the soft X-ray excess. The measured temperature and optical depth of the warm corona span nearly the entire theoretically allowed range, with median values of $0.43_{-0.18}^{+0.40}$~keV and $12.5_{-3.9}^{+3.1}$, respectively. Our first systematic calculation of the disk-to-corona power transfer fraction reveals that the fraction of power released from the accretion disk into the hot corona spans a wide range, with a sample median of $0.68_{-0.25}^{+0.25}$. The sample median values for the hot coronal plasma temperature and optical depth are $54_{-12}^{+11}$~keV and $0.98_{-0.28}^{+0.22}$, respectively. Finally, through both hard X-ray (3--78~keV) and broadband (0.3--78~keV) relativistic reflection spectroscopy, we systematically constrain the black hole spin parameter across the mass scales of $\log(M_{\rm BH}/M_{\odot}) \sim 5.5-9.0$, thereby increasing or refining the available spin measurements in the AGN population by $\sim$20\%.
\end{abstract}

\maketitle

\section{Introduction}
\label{introduction}
Active galactic nuclei (AGN) are the most luminous ($10^{41}-10^{48}~{\rm erg~s^{-1}}$), compact regions at the center of galaxies, fed by mass accretion from their host galaxies onto supermassive black holes (SMBHs) of mass $M_{\rm BH} \sim 10^{5-10} M_{\odot}$ (e.g. \cite{Lynden_1969,Gurzadian_Ozernoi_1979,Rees_1984,Reines_2013}). The gradual loss of angular momentum causes the inflowing matter to move toward the center of gravity, forming an accretion disk around the SMBH, which is believed to be the engine that powers AGN (e.g. \cite{Frank_2002}). The observed correlation between SMBH mass and host galaxy--bulge velocity dispersion (e.g. \cite{Ferrarese_Merritt_2000,Gultekin_2009,Kormendy_Ho_2013}) implies that the growth of the host galaxy is coupled with the energy output from the central SMBH, and feedback from AGN can play a key role in the evolution of galaxies (e.g. \cite{DiMatteo2005,Hopkins_Elvis_2010,Fabian_2012,Heckman_Best_2014}). Indeed, the AGN accretion disk can influence the evolution of the host galaxy via a sustained release of gravitational energy in the form of radiation or outflows (e.g. \cite{Silk_Rees_1998,Tombesi_2011,Parker_2017,Pinto_2018}). However, understanding the nature and dynamics of the accretion disk remains an open question in both theoretical and observational astrophysics.

AGN are inherently multi-wavelength phenomena \citep{Elvis_1994}, and their spectral energy distributions (SEDs) at different energies are the outcome of various physical processes arising from distinct regions, dominating the observed emission for different AGN sub-classes. However, the most effective way to probe the immediate vicinity of the central SMBH or innermost regions of AGN is to study the X-ray emission from Type-1 AGN, the line of sight of which is not obscured by the molecular torus (e.g. \cite{Pier_Krolik_1993,Cappi_2006,Combes_2019}). In the unified view of AGN \citep{Antonucci_1993,Netzer_2015}, the primary X-ray emission is produced by Compton up-scattering of the optical/UV seed photons in an optically thin, hot plasma \citep{Shakura_Titarchuk_1980,Haardt_Maraschi_1993} called `corona' surrounding the SMBH, as also suggested by the global radiation magnetohydrodynamic (MHD) simulations \citep{Yan_Jiang_2019}, where the seed photons are thought to be supplied by the accretion disk \citep{Page_Thorne_1974}. However, it is not well known how the accretion disk and corona are coupled and what fraction of power from the accretion disk gets transported into the corona. When this coronal radiation illuminates the accretion disk, it produces an intrinsic reflection spectrum containing a forest of fluorescent emission lines below $\sim 2$\keV{}, narrow iron (Fe)~K emission lines (K$_\alpha$/K$_\beta$), and a Compton scattering hump above $10$\keV{} \citep{George_Fabian_1991,Ross_Fabian_2005,Garcia_Kallman_2010}. If the disk is close to the black hole, the intrinsic reflection spectrum gets smeared by the strong gravitational field and produces the blurred or relativistic reflection spectrum containing a soft X-ray excess below around 1$-$2\keV{}, a broad Fe~K emission line (6$-$7\keV{}), and the Compton hump (15$-$30\keV{}) (e.g. \cite{Fabian_2000,Garcia_2014,Matt_2014}). Additionally, the Compton shoulder arising from line broadening in the disk photosphere due to Compton scattering in ionized gas can significantly contribute to the apparent red wing of the broad Fe K emission line \citep{Rozanska_Madej_2008}. Although the {\tt RELXILL} family of reflection models incorporates Compton broadening, it still requires a more realistic treatment of the ionization structure of the accretion disk, both vertically and radially. Thus, measuring spin is not a straightforward diagnostic of general relativity; it relies on complex astrophysical modeling. Currently, relativistic reflection spectroscopy of the innermost accretion disk is the most widely adopted technique for measuring the spins of massive black holes in AGN (e.g. \cite{Brenneman_Reynolds_2006, Reynolds_2021,Bambi_2021,Mallick_2022}), which is a crucial step in studying the formation and growth channels of SMBHs (e.g. \cite{Volonteri_2005_spin,Berti_Volonteri_2008,Volonteri_2013_spin,Pacucci_Loeb_2020}). Therefore, it is of central importance to precisely model disk reflection signatures and systematically assess the black hole spin population in AGN.

The origin of the soft X-ray excess observed below $\sim$ 1$-$2\keV{} in AGN is still unknown despite its discovery nearly 40 years ago by \citep{Singh_1985} and \cite{Arnaud_1985}. At first, it was thought to be the high-energy tail of the standard disk emission. However, the soft excess temperature is much higher than the maximum disk temperature and remains constant within the range of $\sim$ 0.1$-$0.3\keV{}, irrespective of the black hole mass (see \cite{Gierlinski_Done_2004} for high-mass quasars and \cite{Mallick_2022} for low-mass dwarf AGN).

Currently, two models are used to describe the soft X-ray excess emission. One model requires low-temperature Comptonization of optical/UV disk photons in an optically thick, warm corona (e.g. \cite{Mallick_2017,Petrucci_2018,Chalise_2022}). On the other hand, the reflection model naturally produces soft X-ray excess as relativistically smeared, broadened fluorescent lines arising from the innermost accretion disk (e.g. \cite{Crummy_2006, Garcia_2014, Mallick_Dewangan_2018}). However, in some AGN, relativistic reflection alone cannot model the entire soft X-ray excess, and a warm Comptonization component is still required, particularly when the density of the accretion disk is fixed at the canonical value of $n_{\rm e}=10^{15}~\rm{cm}^{-3}$ (e.g. Ark~120: \cite{Mallick_2017,Porquet_2018}, Mrk~110: \cite{Porquet_2024}). This issue was tackled by employing a high-density disk reflection model \citep{Garcia_2016}, which can potentially fit the entire soft X-ray excess by boosting the strength of the excess emission below $\sim$1\keV{} in the model, where the density parameter can reach as high as $\log [n_{\rm e}/\rm{cm}^{-3}] = 20$, first incorporated by \cite{Mallick_2022}. For a geometrically thin and optically thick standard accretion disk (\cite{Shakura_Sunyaev_1973}; hereafter SS73), the electron density of the inner disk is likely to be higher than $10^{15}$~cm$^{-3}$. Moreover, the theoretical work of \cite{Svensson_Zdziarski_1994} (hereafter SZ94) proposed that the inner disk density would be even higher than that predicted by the SS73 model after including the fraction of power transferred from the accretion disk to the corona, i.e., the disk-to-corona power transfer fraction. While higher-density disk reflection has become more prevalent in explaining the origin of the soft X-ray excess in several AGN \citep{Mallick_2018,Garcia_2019,Mallick_2022}, the disk-to-corona power transfer fraction has yet to be explored.

In this paper, we study the broadband (0.3$-$78\keV{}) spectra of a sample of Type-1 AGN across the central SMBH mass scales of $M_{\rm BH}\approx 10^{5.5-9} M_{\rm \odot}$, employing both \xmm{} (0.3$-$10\keV{}; \cite{Jansen_2001}) and \nustar{} (3$-$78\keV{}; \cite{Harrison_2013}) data available in the public archive as of July 2024. Previously, \cite{Jiang_2019} (hereafter JJ19) fit the averaged 0.5$-$10\keV{} spectra extracted solely from the \xmm{} observations conducted before 2016 for the sample. They utilized a preliminary version of the high-density disk reflection model ({\tt relxillD} v.1.2.0), where gas density was variable in the range of $\log[n_{\rm e}/{\rm cm^{-3}}]=15-19$. The earlier versions of the model did not have coronal temperature as a parameter, as it assumed a simple power law without a cut-off energy for the incident continuum. Our work employs the updated high-density disk reflection model ({\tt relxillCp} v.2.3)\footnote{\url{https://www.sternwarte.uni-erlangen.de/~dauser/research}}, where the disk density can freely vary from $\log[n_{\rm e}/{\rm cm^{-3}}]=15-20$. This new model also incorporates the coronal temperature as a variable parameter that can only be measured via \nustar{} spectroscopy. We will verify whether high-density disk reflection is robust enough to explain both hard and soft X-ray excess self-consistently or if an extra warm Comptonization model is still required to fit the observed soft X-ray excess. We will calculate the disk-to-corona power transfer fraction and test the validity of the standard $\alpha$-disk model for the AGN sample. We will utilize the unique capability of \nustar{} to unambiguously probe the hard band reflection associated with the broad Fe~K emission together with \xmm{}'s lower-energy coverage to reveal the soft X-ray reflected continuum and measure the SMBH spin population spanning almost the complete SMBH mass range. Modeling the broadband reflection components through joint \xmm{}+\nustar{} spectroscopy is currently the most effective technique for constraining the SMBH spin parameter in AGN.

The paper is organized in the following way. In Section~II, we describe the sample selection criteria with source properties, details of the observations analyzed in this work, and spectral extraction methods from the raw \xmm{} and \nustar{} data. Section~III describes our comprehensive broadband (0.3$-$78\keV{}) X-ray spectral fitting methodology, where we employ a Bayesian framework to assess the significance of both the variable-density disk reflection and warm Comptonization models for the origin of the observed soft X-ray excess. In Section~IV, we discuss our results and their implications. We summarize our conclusions in Section~V. The future prospects of this study are outlined in Section~VI.

\begin{table*}
\caption{Details of the AGN sample employed. The columns show (1) source name, (2) right ascension, (3) declination, (4) logarithmic of the central black hole mass, (5) source redshift, (6) Galactic hydrogen column density along the source line of sight in units of $10^{20}\rm{cm}^{-2}$\citep{Willingale_2013}, (7) dimensionless mass accretion rate ($\dot{m}=\frac{\dot{M}}{\dot{M}_{\rm E}}$) taken from JJ19, and (8) source type based on optical classification. The RA, DEC, redshift, and source type are obtained from {\tt NED} \cite{ipac_ned_2019}. The BH masses are taken from the AGN Black Hole Mass Database \cite{Bentz_Katz_2015} and measured through optical reverberation mapping using the most updated scaling factor of 4.8 given by \cite{Batiste_2017}.}
\begin{center}
\scalebox{0.95}{%
\begin{tabular}{cccccccccccc}
\hline 

Source  & $\alpha_{2000}$ [Degree]   & $\delta_{2000}$ [Degree]  &  $\log [\frac{M_{\rm BH}}{M_{\rm \odot}} ]$  & Redshift ($z$) &  $N_{\rm H, Gal}$    & Dimensionless Mass  & Optical Type \\
  &   &   &   &  &  $\left[10^{20} \rm{cm}^{-2} \right]$  &  Accretion Rate ($\dot{m}$) &  \\
(1)    &   (2)   &   (3)   &   (4)  &  (5)   & (6)  & (7) & (8)  \\                                                      
\hline 

UGC 6728 & $176.317$ & $79.682$ & $5.91_{-0.42}^{+0.19}$ & $0.0065$ & $5.48$ & $0.58_{-0.21}^{+0.76}$ & Sy1.2 \\ [0.2cm]
         
Mrk 1310 & $180.31$ & $-3.678$ & $6.26_{-0.09}^{+0.07}$ & $0.0196$ & $2.66$ & $0.6_{-0.1}^{+0.14}$ & BLS1 \\ [0.2cm]
         
NGC 4748 & $193.052$ & $-13.415$ & $6.46_{-0.18}^{+0.11}$ & $0.0146$ & $4.07$ & $2.2_{-0.5}^{+1.2}$ & NLS1 \\ [0.2cm]
         
Mrk 110 & $141.304$ & $52.286$ & $7.34_{-0.1}^{+0.1}$ & $0.0353$ & $1.39$ & $0.9_{-0.19}^{+0.23}$ & NLS1 \\ [0.2cm]
         
Mrk 279 & $208.264$ & $69.308$ & $7.48_{-0.13}^{+0.1}$ & $0.0305$ & $1.72$ & $0.75_{-0.16}^{+0.27}$ & BLS1 \\ [0.2cm]
         
Mrk 590 & $33.64$ & $-0.767$ & $7.62_{-0.07}^{+0.06}$ & $0.0264$ & $2.92$ & $0.31_{-0.05}^{+0.06}$ & BLS1 \\ [0.2cm]
         
Mrk 79 & $115.637$ & $49.81$ & $7.66_{-0.14}^{+0.11}$ & $0.0222$ & $6.73$ & $0.13_{-0.07}^{+0.05}$ & Sy 1.2 \\ [0.2cm]
         
PG 1229+204 & $188.015$ & $20.158$ & $7.81_{-0.22}^{+0.18}$ & $0.0636$ & $2.92$ & $0.5_{-0.2}^{+0.4}$ & BLS1 \\ [0.2cm]
         
PG 0844+349 & $131.927$ & $34.751$ & $7.91_{-0.23}^{+0.15}$ & $0.064$ & $3.13$ & $1.2_{-0.4}^{+1.0}$ & BLS1 \\ [0.2cm]
         
PG 0804+761 & $122.744$ & $76.045$ & $8.78_{-0.05}^{+0.05}$ & $0.1$ & $3.31$ & $1.13_{-0.12}^{+0.15}$ & BLS1 \\ [0.2cm]
         
PG 1426+015 & $217.277$ & $1.285$ & $9.06_{-0.16}^{+0.11}$ & $0.0866$ & $2.88$ & $0.28_{-0.06}^{+0.13}$ & BLS1 \\ [0.2cm]

\hline

\end{tabular}}
\end{center} 
\label{source_log}           
\end{table*}

\begin{figure*}
\centering
\includegraphics[width=0.48\textwidth,height=9.2cm]{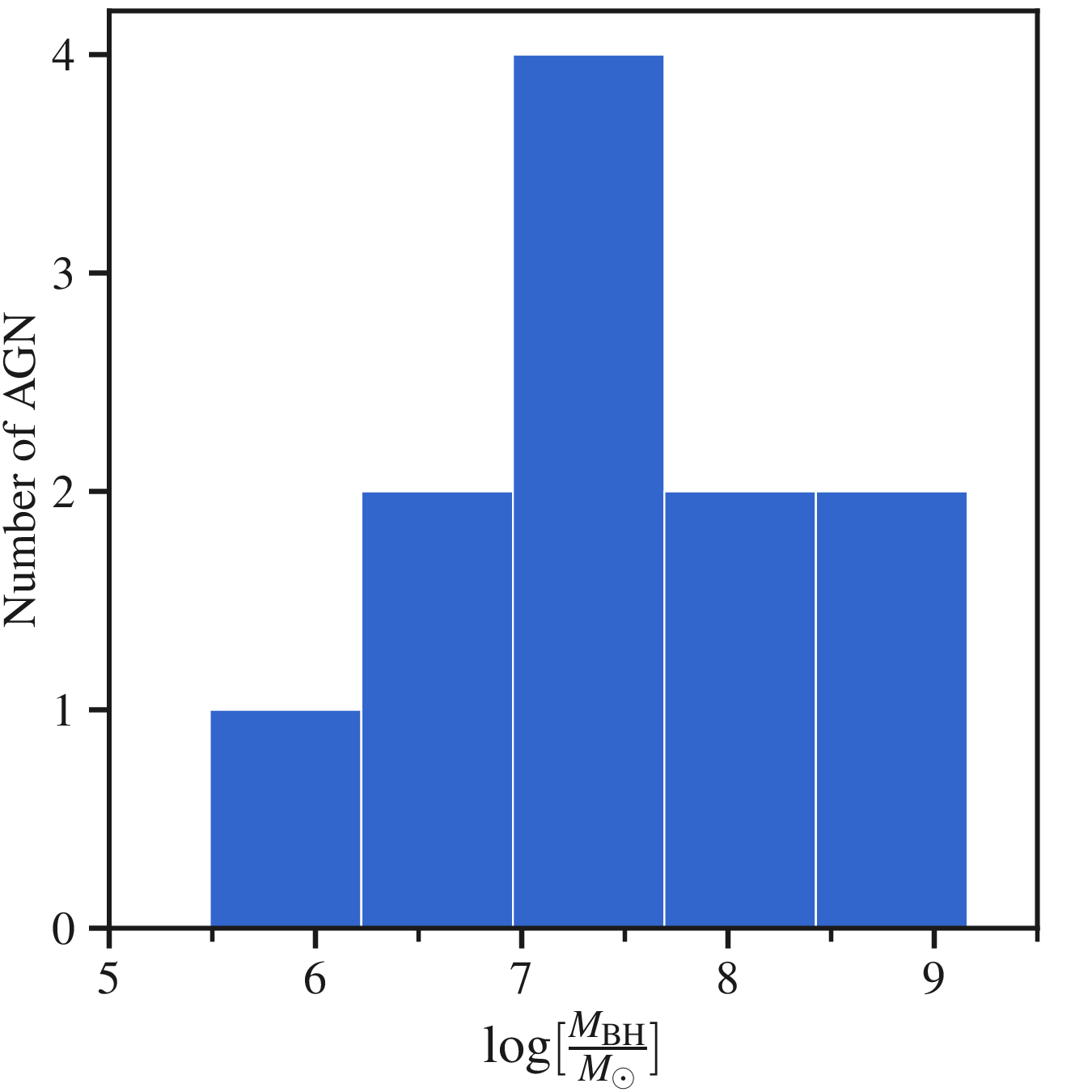}
\includegraphics[width=0.48\textwidth,height=9.2cm]{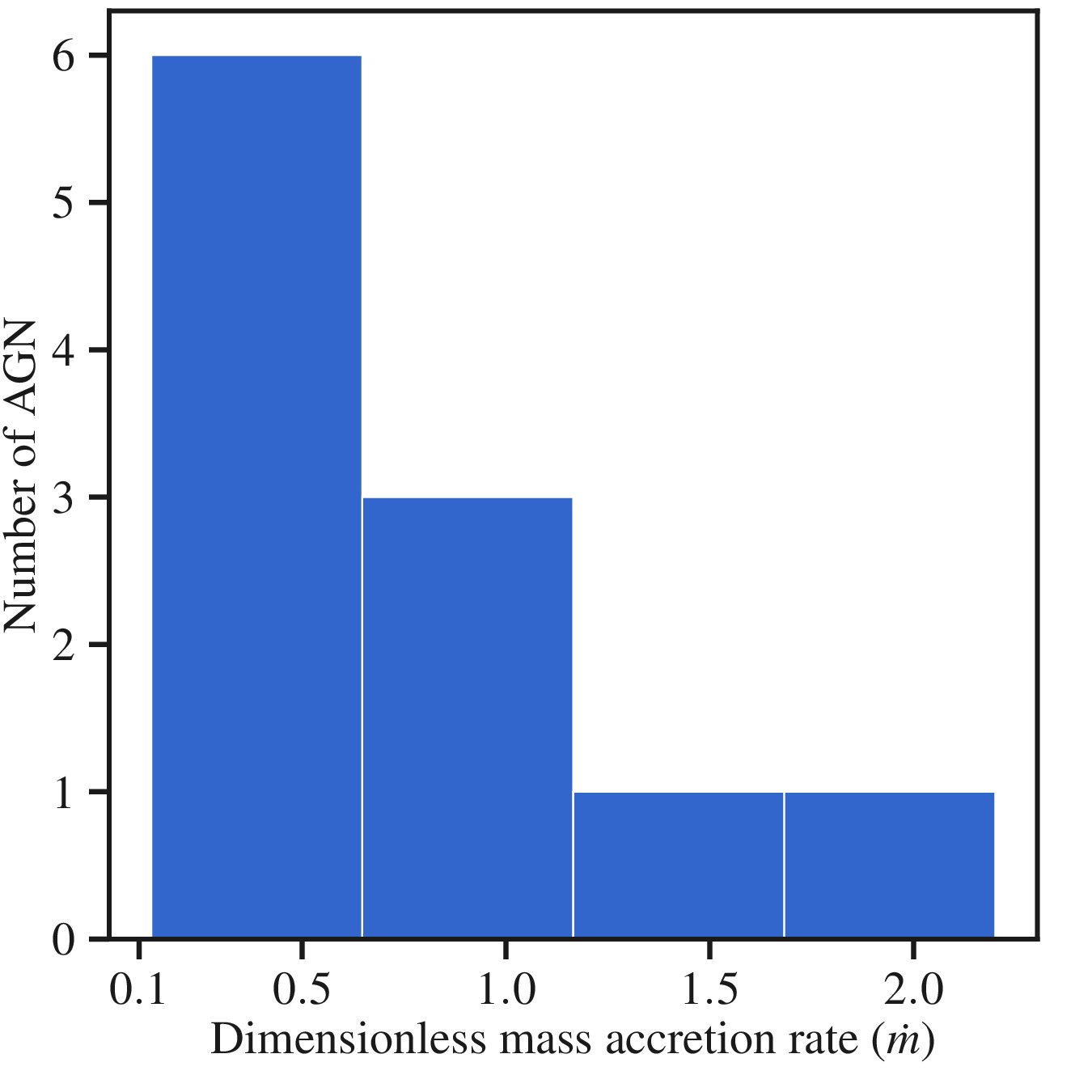}
\caption{Distribution of black hole mass (left panel) and dimensionless mass accretion rate (right panel) of the AGN sample employed in this work. We obtain black hole masses from the AGN Black Hole Mass Database \citep{Bentz_Katz_2015} and the dimensionless mass accretion rates from JJ19 \citep{Jiang_2019}.}
\label{logmbh_mdot}
\end{figure*}

\section{Sample Selection and Data Processing}
\label{sec:sample}
We selected our AGN sample from JJ19, who conducted high-density disk reflection modeling of 17 AGN using only \xmm{} data in the energy range of 0.5--10\keV{}. To better understand accretion disk/corona coupling and to unambiguously investigate the origin of the puzzling soft X-ray excess with the updated high-density disk reflection model, we selected 11 AGN from the JJ19 sample that do not show any absorption features in spectra extracted from the European Photon Imaging Camera (EPIC) on \xmm{}. The presence of absorption complicates spectral modeling and can lead to significant degeneracies. Therefore, we plan to study the 6 remaining AGN that exhibit absorption in the EPIC spectra in a follow-up paper. For the selected AGN sample, we utilize archival data from both the \xmm{} (0.3$-$10\keV{}) and \nustar{} (3$-$78\keV{}) observatories. This enables us to apply the updated variable-density relativistic disk reflection model to the broadband (0.3$-$78\keV{}) X-ray spectra of the sample for the first time in a systematic manner. We will investigate the origin of both the hard and soft X-ray excess emission by assessing whether broadband relativistic reflection from a higher-density inner disk alone can adequately explain all the excess emission, or if an additional warm Comptonization component is still required to account for the soft X-ray excess in these AGN. This approach will allow us to constrain the supermassive black hole (SMBH) spin population across mass scales using both hard X-ray (3$-$78\keV{}) and broadband (0.3$-$78\keV{}) relativistic reflection spectroscopy. Additionally, we will calculate the disk-to-corona power transfer fraction for the first time in any class of accreting objects. The observation details of the sample are provided in Table~A1.

The basic characteristics of the selected 11 AGN are presented in Table~I. Two key parameters for this study, black hole mass and dimensionless mass accretion rate, are listed in columns~(4) and (7), respectively. Black hole masses are determined using optical reverberation mapping and obtained from the AGN Black Hole Mass Database \citep{Bentz_Katz_2015}. The dimensionless mass accretion rate is calculated from the observed optical luminosity for each source (see JJ19 for details) and not from the bolometric luminosity due to large uncertainties in the bolometric conversion factor \citep{Vasudevan_2007} and radiative efficiency in AGN \citep{Raimundo_2012}. Figure~1 presents the distributions of black hole mass, ${\rm log}[M_{{\rm BH}}/M_{\odot}]$, and dimensionless mass accretion rate ($\dot{m}$) of the 11 AGN employed in this work.

\subsection{{\textit {XMM-Newton}} Data Extraction}
We start our work by collecting all the raw data of our sample from the \xmm{} observatory available in the {\tt HEASARC}\footnote{\url{https://heasarc.gsfc.nasa.gov/cgi-bin/W3Browse}} archive. We process the raw data from the European Photon Imaging Camera (EPIC) onboard \xmm{} in the Scientific Analysis System ({\tt SAS} v.21.0.0) with the most recent calibration files as of July 2024. First, we generated raw event files from EPIC-pn and MOS data with {\tt SAS} tasks {\tt epproc} and {\tt emproc}, respectively. To exclude the background flares, we created good time intervals ({\tt GTI}) above 10\keV{} for the full field using the technique detailed in \cite{Mallick_2021}. We extracted flare-filtered clean event files by applying the flare-corrected {\tt GTI} and unflagged events with {\tt PATTERN}$\leq 4$ for EPIC-pn and {\tt PATTERN}$\leq 12$ for EPIC-MOS. The EPIC-pn data have much better sensitivity above 6\keV{} and suffer less from pile-up effects compared to the EPIC-MOS data. Therefore, we concentrate on the EPIC-pn data of the sample. However, we notice that the EPIC-pn events of UGC~6728 are flaring background-dominated. Hence, we consider the EPIC-MOS data only for UGC~6728. The source and background events are extracted from a circular region of radius 35 arcsec centered on the point source and nearby source-free area, respectively. We checked for the presence of pile-up effects using the {\tt epatplot} task. Whenever pile-up was detected, we removed the central bright pixels by choosing an annular source region. We choose the inner radius of the annulus such that the distributions of the observed data match the model curves produced by {\tt epatplot}. We generate the redistribution matrix file ({\tt rmf}) and ancillary response file ({\tt arf}), source, and background spectra using the {\tt SAS} task {\tt especget}. We binned the source spectra including background with the {\tt FTOOL} task {\tt ftgrouppha}, where we set {\tt grouptype=optsnmin} and {\tt groupscale=5}. The {\tt grouptype=optsnmin} uses the optimal binning algorithm of \cite{Kaastra_2016} with an additional requirement of a minimum signal-to-noise ratio (SNR) per grouped bin set by {\tt groupscale}. We model the \xmm{}/EPIC spectra across the entire energy range of 0.3$-$10\keV{}. The EPIC camera, Obs. ID, start time of the observation, total elapsed time, net exposure, net count rate, and net counts in the 0.3$-$10\keV{} band for each source are listed in Table~\ref{table_obs_log}.

\subsection{{\textit {NuSTAR}} Data Extraction}
\nustar{} observed all the sources in our sample with its two co-aligned Focal Plane Modules, A (FPMA) and B (FPMB). We acquired all the available data from the {\tt HEASARC} archive and reduced the raw (level 1) data in the \nustar{} Data Analysis Software ({\tt NuSTARDAS} v.2.1.2). We produced the cleaned and calibrated event files with the {\tt nupipeline} task using the latest (as of 2024 July 24) calibration database ({\tt CALDB} version: 20240701). We employed conservative criteria, {\tt saamode=optimized} and {\tt tentacle=yes}, to treat the high background levels induced by the South Atlantic Anomaly (SAA) region. To maximize the SNR, we determine the optimal radius of a circular extraction region centered on the source using the \nustar{} tool {\tt nustar-gen-utils}\footnote{\url{https://github.com/NuSTAR/nustar-gen-utils}} for each observation. The corresponding background extraction region was selected from the same-sized circular off-source region. We produced the response matrices ({\tt rmf} and {\tt arf}), source, and background spectra for both FPMA and FPMB with the {\tt nuproducts} task. Finally, we generated background-subtracted binned spectra using the {\tt ftgrouppha} tool, where we employ the optimal binning algorithm of \cite{Kaastra_2016} and a minimum SNR of 5 per grouped bin for both FPMA and FPMB. We fit the calibrated energy range of 3$-$78\keV{} for \nustar{}/FPMA and FPMB spectra. The Obs. ID, observation start time, total elapsed time, net exposure time, net count rate, and net counts in the 3$-$78\keV{} range for both modules are shown in Table~\ref{table_obs_log}.

\section{Broadband X-ray Spectroscopy}
\label{sec:spec}

\subsection{Spectral Modeling Procedure}
\label{sec:fit_method}
We jointly fit the \xmm{} and \nustar{} spectral data for each source in our sample in the software package {\tt XSPEC} v.12.15.0 \citep{Arnaud_1996}. We model neutral photoelectric absorption ($N_{\rm H, Gal}$) along the line of sight (LOS) to the source using the Galactic absorption model {\tt TBabs}, adopting the cosmic elemental abundances of \cite{Wilms_2000} and photoionization cross-sections of \cite{Verner_1996}. The Galactic hydrogen column density ($N_{\rm H, Gal}$) incorporates both atomic ($N_{\rm HI}$) and molecular ($N_{\rm H_2}$) hydrogen. The $N_{\rm H, Gal}$ value for each source is obtained from \cite{Willingale_2013} (listed in Table~I) and kept fixed throughout the spectral fitting. We employ the Cash statistic ($C$-stat, \cite{cash_1979}) to find the best-fit model parameters, and the $\chi^{2}$ statistic to test the goodness of fit. Once the best-fit model parameters are found, we perform Markov Chain Monte Carlo (MCMC) analyses to determine the parameter uncertainties. The quoted uncertainties on best-fit parameters represent the 90\% confidence intervals.

We include a constant component ({\tt constant}) to consider cross-calibration uncertainties between FPMA and FPMB throughout our analysis. For simultaneous or quasi-simultaneous \xmm{} and \nustar{} observations, we multiply a constant component ({\tt constant}) to account for cross-calibration uncertainties between \xmm{}'s EPIC and \nustar{}'s FPM spectra. The {\tt constant} component is allowed to vary, while all other model parameters are tied between simultaneous or quasi-simultaneous EPIC and FPM spectra. For non-simultaneous \xmm{} and \nustar{} observations, we allow the photon index of the primary power-law continuum and the flux or normalization of each model component to vary independently between observations. The other two parameters likely to vary between non-simultaneous observations are the disk ionization parameter and the coronal temperature. However, when we allow them to vary independently across non-simultaneous observations, their values become unconstrained. This usually results in an upper limit for the ionization parameter and a lower limit for the coronal temperature. Moreover, we find that the fit statistics do not improve when these two parameters are allowed to vary independently between observations. Consequently, we have decided not to vary them independently across all observations of a source.

\subsection{Bayesian Analysis for Model Selection}
\label{sec:bayesian}
While we use the $\chi^{2}$-statistic to assess goodness of fit, comparing only the fit statistics between two models may not always be conclusive, as it is possible for a model to overfit the data due to having more free parameters, resulting in a lower $\chi^{2}$-statistic. Therefore, we implement a Bayesian model selection approach, where the posterior distributions of the models are computed from MCMC simulations. We employ the Deviance Information Criterion, DIC \citep{Spiegelhalter2002}, which is a Bayesian model selection metric and is defined by 

\begin{linenomath}
\[
{\rm DIC}=\overline{D(\theta)}+p_{D},
\]
\end{linenomath}
where $D(\theta)=-2\log\left(p(y|\theta)\right)$ and $p_{D}=\overline{D(\theta)}-D(\bar{\theta})$. Here $p(y|\theta)$ is the likelihood function, $D(\theta)$ is the deviance of a model parameter $\theta$, $y$ represents the data, $p_{D}$ is the effective number of parameters in the model, $\overline{D(\theta)}$ represents the average of $D(\theta)$ calculated over the samples of $\theta$, and $D(\bar{\theta})$ is the value of $D(\theta)$ evaluated at the average ($\bar{\theta}$) of the samples of $\theta$.

DIC considers both the goodness of fit evaluated by the likelihood function and the effective number of model parameters. DIC is a hierarchical modeling generalization of the Akaike information criterion, AIC \citep{Akaike_1974}. However, DIC does not penalize a model for parameters unconstrained by the data \citep{kass_1995} because it uses an effective number of parameters, unlike AIC. DIC considers a model parameter only if it affects the goodness of fit to the data and is altered by varying that parameter. The degree to which different parameters are constrained is reflected in the non-integer nature of the effective number of model parameters.

In order to confirm whether one model is preferred over another, we compare the DIC values calculated from the two models. Statistically, the model with a lower DIC is preferred by the data, and the difference in DIC values, $\Delta {\rm DIC}$, between the two models measures the strength of the preference. According to a scale proposed by \cite{Jeffreys_1961} and updated by \cite{kass_1995}, the $\Delta {\rm DIC}$ values between 0 and 2 hint only marginal evidence, $\Delta {\rm DIC}$ between 2 and 6 provides positive evidence, $\Delta {\rm DIC}$ between 6 and 10 suggests strong evidence, and $\Delta {\rm DIC}$ greater than 10 shows very strong evidence for one model over another.

\subsection{Basic Spectral Exploration}
To unambiguously detect any spectral features, we first consider the continuum bands, 3$-$5\keV{} and 7$-$10\keV{}, solely dominated by the primary coronal emission. We fit the 3$-$5\keV{} and 7$-$10\keV{} bands by the Galactic absorption corrected power-law model ({\tt{TBabs$*$zpowerlw}}). In search of various spectral features, we extrapolate the best-fit primary power-law continuum model over the whole 0.3$-$78\keV{} energy range. The \xmm{} EPIC-pn (MOS for UGC~6728) and \nustar{} FPMA/B spectral data, the Galactic absorption corrected power-law continuum model, and data-to-model ratio plots for the sample are shown in Fig.~A1. The ratio plots reveal a soft X-ray excess below $\sim$1--2\keV{} and Fe~K emission in the $\sim$5--7\keV{} band for all AGN, as well as a Compton hump in the $\sim$15--30\keV{} range for most sources in the sample through visual inspection. However, the detection of these features, especially the Compton hump and Fe~K emission, through visual inspection of ratio plots is insufficient, as the presence or strength of these residual features depends on the choice of bands presumed to be dominated by the primary power-law continuum.

\begin{figure*}
\centering
\includegraphics[width=0.75\textwidth,height=10cm]{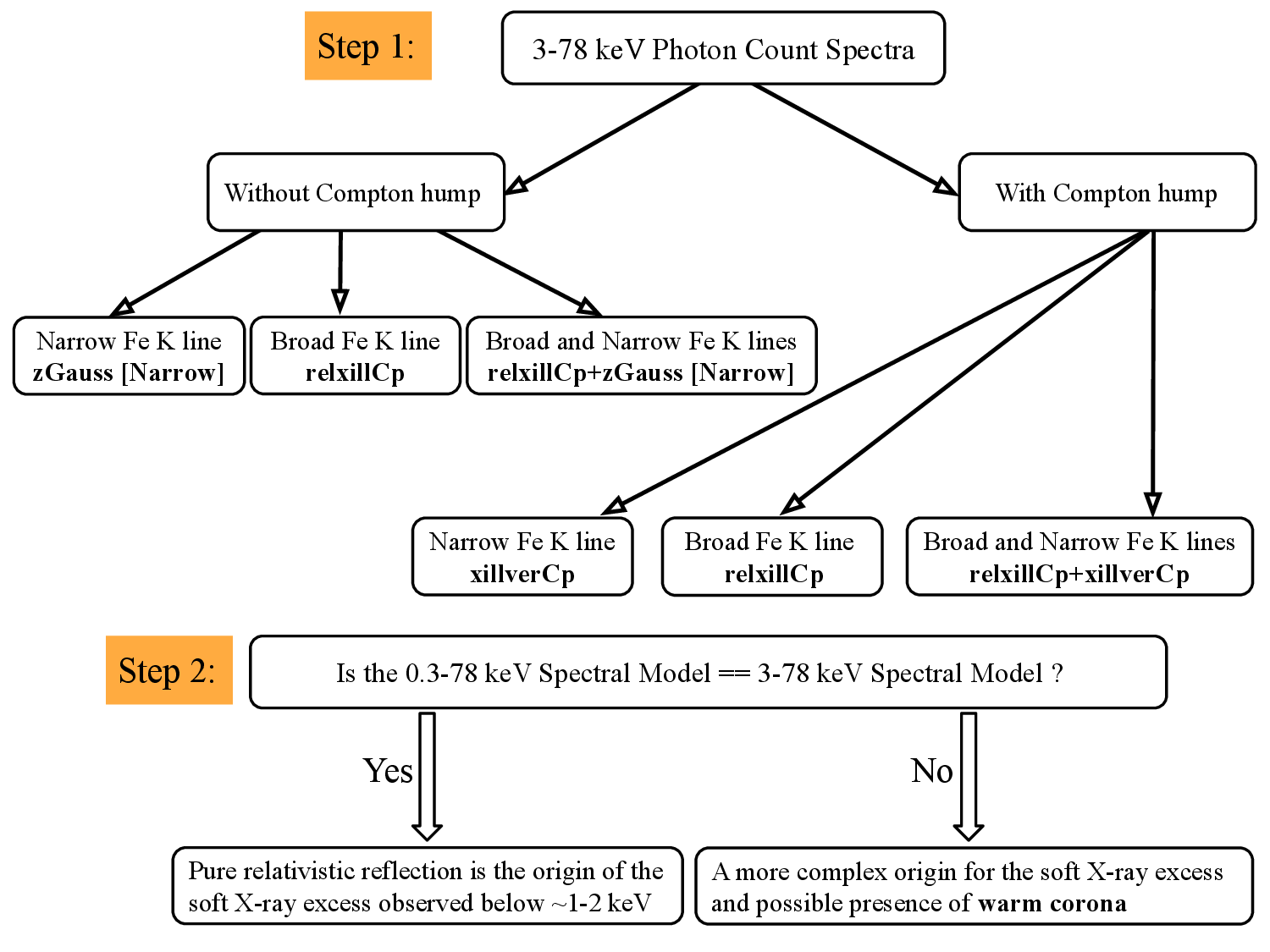}
\caption{The flowchart illustrates our systematic spectral fitting methodology, which unambiguously investigates the origin of soft X-ray excess in the AGN sample exhibiting diverse spectral features.}
\label{flowchart}
\end{figure*}

\subsection{Probing the Hard X-ray Excess Emission: 3$-$78 keV Spectral Modeling}
\label{sec:spec_3_78}
As evident from Fig.~A1, the X-ray photon count spectra of the sample are complex, with an excess emission in the soft X-ray band. Therefore, we start our spectral modeling first considering the hard X-ray (3$-$78\keV{}) photon count spectra and probe the Fe~K emission as well as the Compton hump. We describe the hard X-ray primary continuum by the physically motivated {\tt nthComp} model \cite{Zdziarski_1996}, which produces a power-law like continuum due to the thermal Comptonization of disk seed photons in a hot corona of electrons. The free parameters of the {\tt nthComp} model are photon index ($\Gamma$), electron temperature ($kT_{\rm e}$), and normalization.

In the 6$-$7\keV{} band, we can obtain either narrow or broad or both narrow and broad Fe~K emission features. However, the narrow Fe~K$_\alpha$/K$_\beta$ emission line cores are never resolved in low-resolution CCD data. Therefore, while performing progressive spectral fitting to assess the presence of Fe~K$_\alpha$/K$_\beta$ emission line core(s) from the torus or other distant material, we first add simple Gaussian line(s) [{\tt zGauss\_N} for Fe~K$_\alpha$ core or {\tt zGauss\_N+zGauss2\_N} for Fe K$_\alpha$ plus K$_\beta$ cores] with their width fixed at $10$\ev{} (see e.g. \cite{Mallick_2017}), which is smaller than the resolution of EPIC-pn at $\sim$6$-$7\keV{}. On the other hand, the broad emission feature in the Fe~K band is characteristic of relativistic reflection from the inner accretion disk \citep{Fabian_1989,Fabian_2002}. Therefore, to fit the broad Fe~K line emission, we apply the relativistic disk reflection model {\tt relxillCp} \citep{Garcia_2016,Garcia_2014,Dauser_2014}.

In the {\tt relxillCp} model, we set {\tt{refl$\_$frac}}$=-1$ to fit only the reprocessed emission from the accretion disk. The parameters (photon index $\Gamma$ and electron temperature $kT_{\rm e}$) that describe the properties of the corona are tied between {\tt relxillCp} and {\tt nthComp} models. The {\tt relxillCp} table model is calculated with the seed photon temperature of the accretion disk fixed at 50\ev{}. Accordingly, we set the input disk seed photon temperature at 50\ev{} in the {\tt nthComp} model for consistency. The relevant input parameters of the {\tt relxillCp} model are:

\begin{itemize}

\item Electron density ($n_{\rm e}$) of the accretion disk.

\noindent 
\item Iron abundance ($A_{\rm Fe}$) in the disk relative to solar.

\noindent 
\item The dimensionless spin ($a^{\ast}$) of the black hole, which is measured by setting the inner radius ($r_{\rm in}$) of the accretion disk at the innermost stable circular orbit ($r_{\rm isco}$). To meet the criterion, we need to set {\tt Rin=-1} in the model. 

\noindent 
\item Inclination angle ($\theta$) of the disk relative to the line of sight.

\noindent 
\item Disk ionization parameter, $\xi=\frac{4\pi F}{n_{\rm e}}$, where $F$ is the illuminating continuum flux. 

\noindent 
\item The emissivity profile of the accretion disk, which is a measure of the reflected flux as a function of disk radius and is parameterized by a broken power law: $\epsilon(r)\propto r^{-q_{\rm in}}$ for $r_{\rm in}\leq r\leq r_{\rm br}$, and $\epsilon(r)\propto r^{-q_{\rm out}}$ for $r_{\rm br}\leq r\leq r_{{\rm {\rm out}}}$. The inner emissivity index ($q_{\rm in}$) is a free parameter in the model. Over the outer disk, the emissivity profile falls as $r^{-3}$, as expected in flat spacetime. Therefore, we fix the outer emissivity index at $q_{\rm out}=3$. The break radius corresponds to the radial extent of the corona and is fixed at $r_{\rm br}=6 r_{\rm g}$, a typical value for the coronal radius in AGN \citep{Mallick_2021, Mallick_2022}.

\end{itemize}

To quantitatively assess the presence of the broad Fe~K emission feature, we evaluate the DIC values for models with and without the relativistic disk reflection ({\tt relxillCp}) component in the 3--10\keV{} band. Additionally, we estimated the signiﬁcance of the broad Fe K emission feature by employing the maximum likelihood ratio (MLR) test. The MLR test results agree with those obtained from the Bayesian model selection metric (Table~\ref{test_broad_FeK}). Both Bayesian analysis and MLR tests confirm that the relativistic disk reflection responsible for the broad Fe~K emission is evident in all 11 AGN with at least 90\% confidence.

We then determine the presence of Compton hump above 10\keV{} by fitting the 7--78\keV{} spectra with the Compton reflection model {\tt xillverCp} \citep{Garcia_2013}. Within {\tt xillverCp}, we set {\tt{refl$\_$frac}}$=-1$ and tied the coronal parameters (photon index $\Gamma$ and electron temperature $kT_{\rm e}$) to those in {\tt nthComp}. The density of the reflector in the distant reflection model ({\tt xillverCp}) is kept fixed at the canonical value of $n_{\rm e}=10^{15}~\rm{cm}^{-3}$ throughout the spectral fitting. The distant reflector is assumed to be near-neutral ($\log\xi=0$) and has a high inclination angle of $\theta=60$~degree (e.g. \cite{Mallick_2018,Zhao_2021}). For quantitative assessment of the Compton hump, we compute the DIC values for the primary power-law continuum ({\tt nthComp}) and the primary continuum plus Compton reflection ({\tt nthComp+xillverCp}) models fitted to the 7--78\keV{} spectra. The MLR test evaluates the detection significance of the Compton hump, which is in agreement with the result obtained from the Bayesian model selection metric, presented in Table~\ref{test_hump_xil}. To check the relativistic nature of the Compton hump, we then replace the distant reflection ({\tt xillverCp}) with the relativistic disk reflection model {\tt relxillCp} and refit the 7--78\keV{} spectra with {\tt nthComp+relxillCp} model. The fit statistics, along with the confidence levels and DIC values, are shown in Table~\ref{test_hump_rel}.

Our statistical analyses show that the relativistically broad Fe K emission feature is present in all 11 AGN. However, the Compton hump is either very weak or lacks statistical significance in some AGN, which can be entirely attributed to insufficient SNR in their \nustar{} observations. As shown in column~(9) of Table~\ref{table_obs_log}, all cases where the Compton reflection hump is weak or undetected correspond to low net source counts ($\lesssim 10^{4}$) in the \nustar{} data. When the Compton hump is weak or undetected, and the Fe~K band contains only a broad Fe~K emission feature, the model that best describes the hard X-ray (3--78\keV{}) spectra is {\tt nthComp+relxillCp}. In the absence of a Compton hump, when the Fe~K band contains both broad and narrow Fe~K emission features, the hard X-ray spectra are best described by models {\tt nthComp+zGauss\_N+relxillCp} or {\tt  nthComp+zGauss\_N+zGauss2\_N+relxillCp}. The second narrow Gaussian line {\tt zGauss2\_N} is included when the Fe~K$_{\beta}$ emission line is detected at around 7\keV{}. We employ the distant reflection model {\tt xillverCp} only when the Compton hump is statistically significant. Therefore, the model {\tt nthComp+relxillCp+xillverCp} best fits the hard X-ray spectra when both broad and narrow Fe~K emission features, as well as the Compton hump, are detected. The iron abundance ($A_{\rm Fe}$) parameter was tied between {\tt relxillCp} and {\tt xillverCp}. Figure~2 (Step~1) illustrates our methods of fitting various spectral features in the hard X-ray (3$-$78\keV{}) photon count spectra. The following four sets of models best fit the joint \xmm{}+\nustar{} hard X-ray spectra of the sample:  

\begin{itemize}

\item {\texttt {relxillCp+nthComp}} for UGC~6728, Mrk~1310, PG~1229+204 and PG~0844+349. 

\item {\tt relxillCp+zGauss\_N+zGauss2\_N+nthComp} for Mrk~590. 

\item {\tt relxillCp+zGauss\_N+nthComp} for PG~0804+761. 

\item {\tt relxillCp+xillverCp+nthComp} for NGC~4748, Mrk~110, Mrk~279, Mrk~79 and PG~1426+015.

\end{itemize}

We note that even when narrow Fe~K emission lines are detected, the Compton reflection hump remains undetected or weak in two AGN, Mrk~590 and PG~0804+761, indicating that a weak or undetected Compton hump is not exclusive to sources exhibiting only broad Fe~K emission lines. Some AGN lack a statistically significant Compton hump above 10\keV{}, even in the presence of Fe~K emission lines (broad or narrow), which can be fully explained by low net source counts ($\lesssim 10^{4}$) in their \nustar{} observations, reported in column~(9) of Table~\ref{table_obs_log}.

\subsection{Probing Both Soft and Hard X-ray Excess: 0.3--78 keV Spectral Modeling}
\label{sec:spec_p3_78}
Once the hard X-ray (3--78\keV{}) best-fit spectral models have been found for the sample, we extrapolate that model down to 0.3\keV{}, to see whether the same model can fit the whole 0.3$-$78\keV{} spectra or not. When the hard X-ray best-fit spectral model of an AGN can also explain the soft X-ray excess emission without the need for any extra blackbody or low-temperature Comptonization model, it will justify that the same physical mechanism is responsible for the origin of the soft X-ray excess emission, i.e., the relativistic reflection from an accretion disk with higher density. If the hard X-ray spectral model cannot fully describe the soft X-ray band and a warm Comptonization model is indeed required, we can conclude that the origin of the observed soft X-ray excess is relativistic disk reflection together with the warm coronal emission. Fig. 2 (Step 2) illustrates our methodology for fitting the hard-to-soft X-ray spectra.

For all sources in the sample, we quantitatively assess the possible presence of warm coronal emission by adding the warm Comptonization ({\tt compTT}) component to the hard X-ray (3--78\keV{}) best-fit model and refit the broadband (0.3--78\keV{}) spectra. The resulting fit statistics and DIC values, both without and with the warm Comptonization model, are presented in Table~\ref{test_wc}. We also estimated the confidence levels of the warm coronal emission through the MLR test, which agree with those obtained from the Bayesian model selection metric (Table~\ref{test_wc}). Both Bayesian analysis and MLR tests confirm that the observed soft X-ray excess requires an additional warm Comptonization ({\tt compTT}) for 8 out of 11 AGN in the sample. For the remaining 3 AGN, variable-density relativistic disk reflection ({\tt relxillCp}) is sufficient to fit the broad Fe~K emission feature and soft X-ray excess emission simultaneously. We find the following five sets of models best describe the joint \xmm{}+\nustar{} broadband X-ray spectra of the AGN sample: 

\begin{itemize}

\item {\tt relxillCp+nthComp} for UGC~6728 and PG~1229$+$204. 

\item {\tt relxillCp+zGauss\_N+zGauss2\_N+nthComp} for Mrk~590. 

\item {\tt compTT+relxillCp+nthComp} for Mrk~1310 and PG~0844$+$349. 

\item {\tt compTT+relxillCp+zGauss\_N+nthComp} for PG~0804$+$761. 

\item {\tt compTT+relxillCp+xillverCp+nthComp} for NGC~4748, Mrk~110, Mrk~279, Mrk~79 and PG~1426$+$015.

\end{itemize}

Figure~A2 shows the \xmm{}/EPIC, \nustar{}/FPMA, and FPMB photon count spectra, along with the best-fit count spectral models and their corresponding components. We present the best-fit flux spectral model with components in Figures~A3 and A4. The best-fit broadband (0.3$-$78\keV{}) spectral model parameters for each source are presented in Table~\ref{table_output_parameters}. In Appendix~\ref{sec:individual}, we discuss the hard-to-soft X-ray spectral fitting details for each source in the sample.

\subsection{Testing Pure Warm Coronal Origin of Soft X-ray Excess Without Relativistic Reflection} 
\label{sec:pure_wc}
We test the pure warm coronal origin of soft X-ray excess by completely removing the relativistic disk reflection ({\tt relxillCp}) model from the broadband (0.3--78\keV{}) spectra and adding the warm Comptonization ({\tt compTT}) component for all sources in the sample to reconfirm the presence of relativistic reflection in the broadband spectra and our conclusions on the origin of soft X-ray excess. We present the source model expression including {\tt compTT} and excluding {\tt relxillCp}, along with the resulting fit statistic and DIC value, in columns~(2), (4) and (7) of Table~\ref{test_rel}, respectively. The source model obtained in Section~III.E, along with the corresponding fit statistic and DIC value, are shown in columns~(3), (5), and (8) of Table~\ref{test_rel}, respectively. To quantitatively assess the significance of the source model obtained in Section~III.E (labelled as Model~2 in Table~\ref{test_rel}) against the pure warm corona model (labelled as Model~1 in Table~\ref{test_rel}), we evaluate the significance of Model~2 over Model~1 through the MLR test and calculate the difference between DIC values derived for Model~1 and Model~2. The confidence level computed by MLR and the Bayesian model selection metric agree with each other, reassuring the presence of relativistic disk reflection in the broadband spectra for all sources in the sample, as well as showing that a sole warm corona is insufficient to fit the entire soft X-ray excess without relativistic disk reflection.

\subsection{MCMC Analysis for Parameter Space Exploration}
\label{sec:mcmc}
To confirm that the parameters are not clustering at any local minima, we conduct an MCMC analysis on the best-fit model and explore the complete parameter space for each source. We draw the parameter distributions and determine confidence intervals for each free parameter from the converged MCMC chains. To run the MCMC chains, we use the algorithm of \cite{Goodman_Weare_2010} implemented in {\tt XSPEC}. We run MCMC chains with 100--300 walkers for $\sim 10^{6}-10^{7}$ iterations and burn the first $\sim [1-10]$\% iterations until the chains were converged. We notice that the number of walkers needs to be at least four times greater than the number of free parameters in the model for faster convergence of the chains. We ensured that the chains converged with Gelman--Rubin's MCMC convergence test, which resulted in the potential scale reduction factor being less than 1.1 for each parameter. The full posterior distributions of various model parameters and contour plots between each pair of parameters for the best-fit spectral models of all sources in the sample are shown in Figures~A5 and A6. The dark, medium, and light blue contours represent 68.3\%, 90\%, and 95\% confidence levels, respectively.

\begin{figure}
\centering
\begin{center}
\includegraphics[width=0.48\textwidth,height=8.7cm]{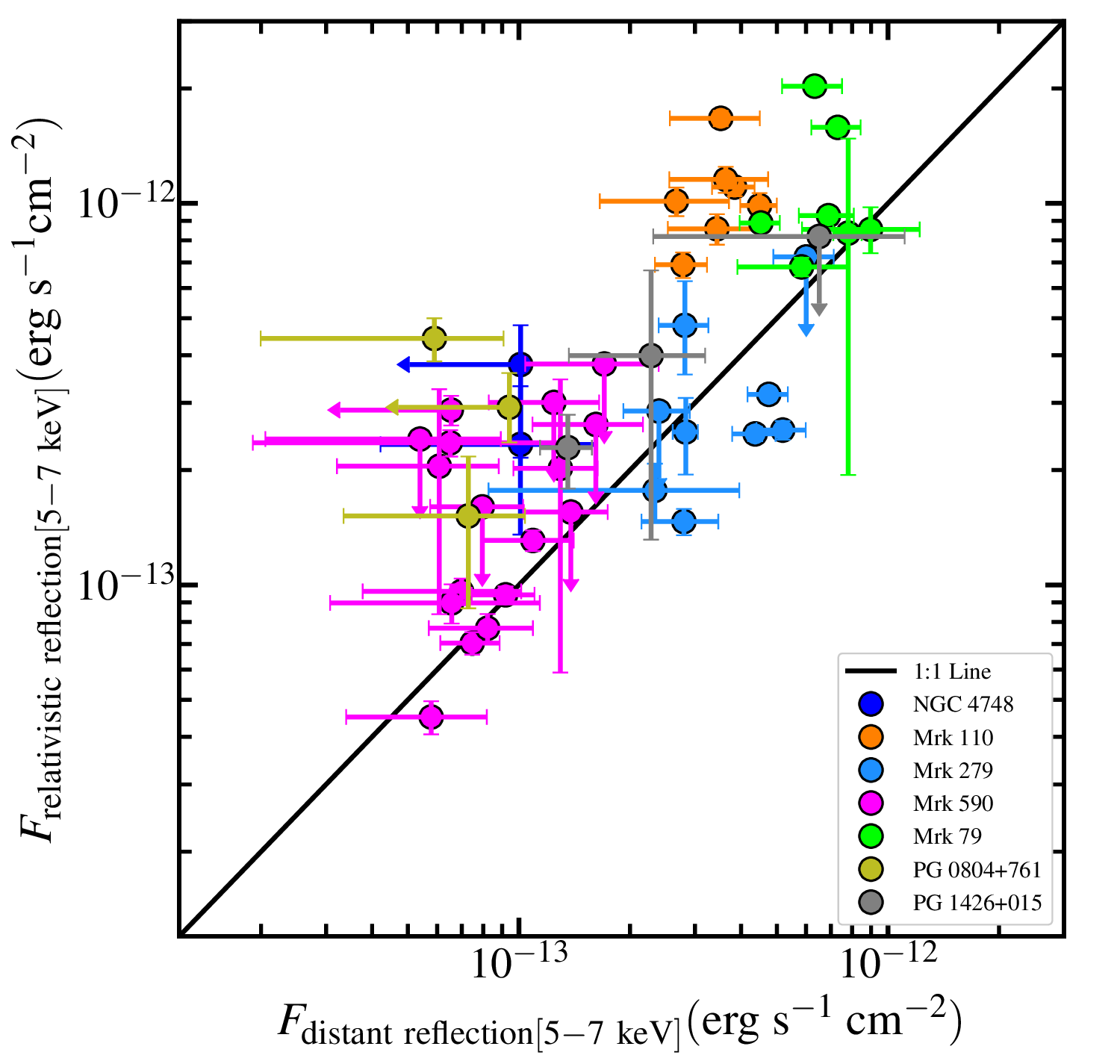}
\caption{The flux of relativistic reflection ({\tt relxillCp}) versus non-relativistic or distant reflection (either {\tt zGauss\_N} or {\tt zGauss\_N+zGauss2\_N} or {\tt xillverCp}) in the $5-7$\keV{} band, demonstrating the relative contributions of these components to the Fe~K band. The $1{:}1$ line marks equal contributions from relativistic disk reflection and distant reflection components to the Fe~K band.}
\end{center}
\label{Frel_Fdist}
\end{figure}

\begin{figure}
\centering
\begin{center}
\includegraphics[width=0.48\textwidth,height=8.8cm]{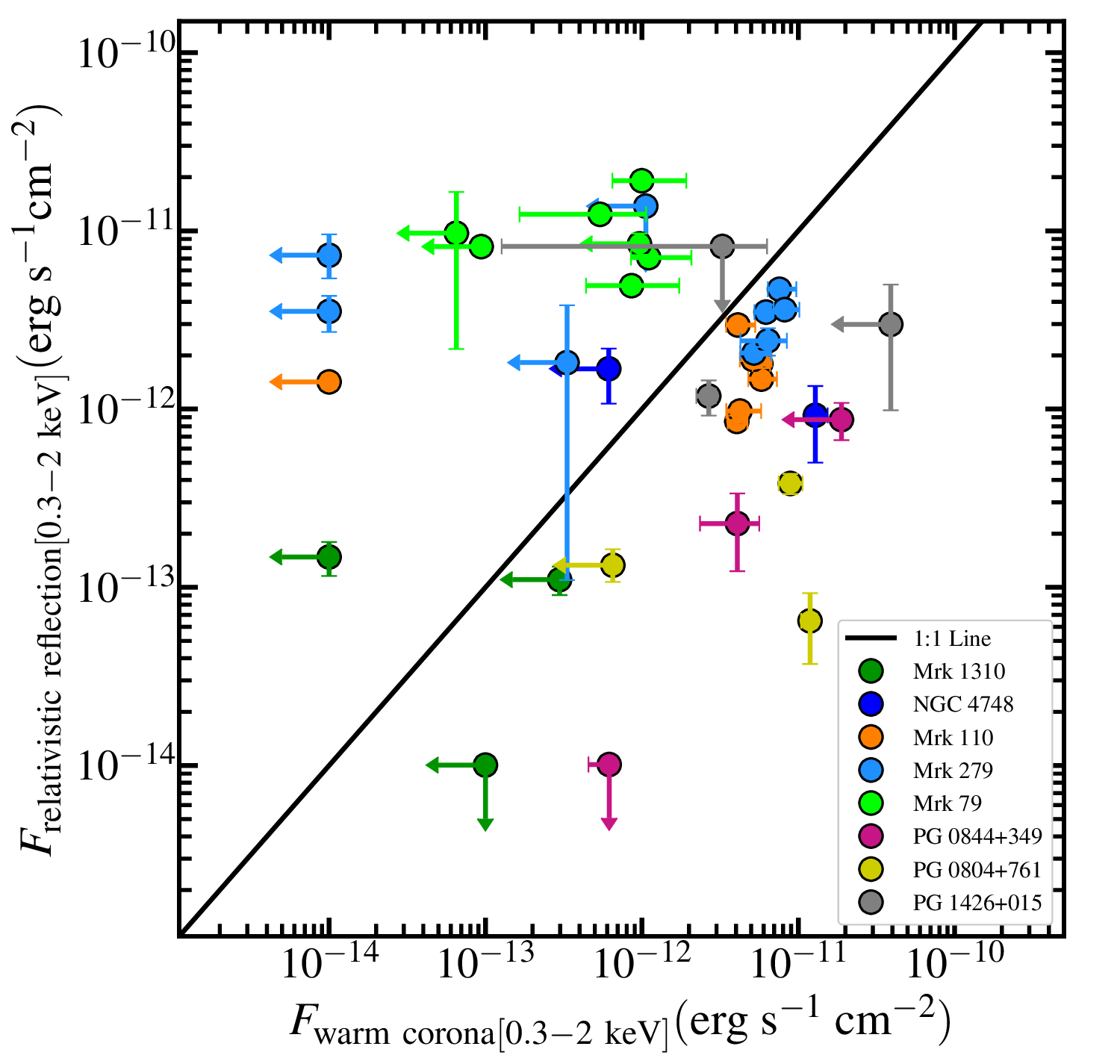}
\caption{The flux of relativistic reflection ({\tt relxillCp}) versus warm Comptonization ({\tt compTT}) in the $0.3-2$\keV{} band, showing the contributions of the relativistic reflection and warm Comptonization components to the soft X-ray band. The $1:1$ line indicates where relativistic disk reflection and warm coronal emission contribute equally to the observed soft X-ray excess.} 
\end{center}
\label{Frel_Fcomp}
\end{figure}

\subsection{Relative Contributions of Relativistic and Distant Reflection in the Fe~K Band}
We quantify the relative contributions of the relativistic reflection ({\tt relxillCp}) and distant reflection (either {\tt zGauss\_N} or {\tt zGauss\_N+zGauss2\_N} or {\tt xillverCp}) components in the $5-7$\keV{} Fe~K band to assess their respective strengths. Figure~3 demonstrates the relativistic disk reflected flux versus the non-relativistic or distant reflection flux within the Fe~K band. The 1:1 line denotes the point at which the relativistic and distant reflection components contribute equally to the Fe~K band emission. From Fig.~3, we can see that relativistic reflection contributes more than the distant reflection in the Fe~K band for all sources in the diagram except for some observations of Mrk~279. Additionally, we notice the variable nature of the relativistic disk reflected flux responsible for the broad Fe~K emission whenever we have multiple flux measurements of a source. However, as expected, the distant reflection flux characterizing the narrow Fe~K emission line(s) appears non-variable or constant within error bars. To confirm this, we attempt to quantify the flux variability of both relativistic and distant reflection components by calculating their fractional root mean square (rms) variability using the formula provided in \cite{Vaughan_2003}. However, for any descriptive statistics analysis, at least 10 data points are needed in order to obtain reliable results. The only source that satisfies this criterion is Mrk~590. The measured fractional rms variability of the relativistic disk reflection flux in the $5-7$\keV{} band is $F_{\rm var}=[28.6\pm 23.6]\%$. However, we are unable to measure the fractional rms variability of the distant reflection flux because the mean squared error in the measured distant reflection flux is greater than its variance, resulting in a negative excess variance\footnote{For detailed formulas, see equations (1) and (2) in \cite{Mallick_2016}.}. This confirms the non-variable nature of the observed distant reflection flux and is consistent with our visual inspection.

\begin{figure*}
\centering
\begin{center}
\includegraphics[scale=0.37,angle=-0]{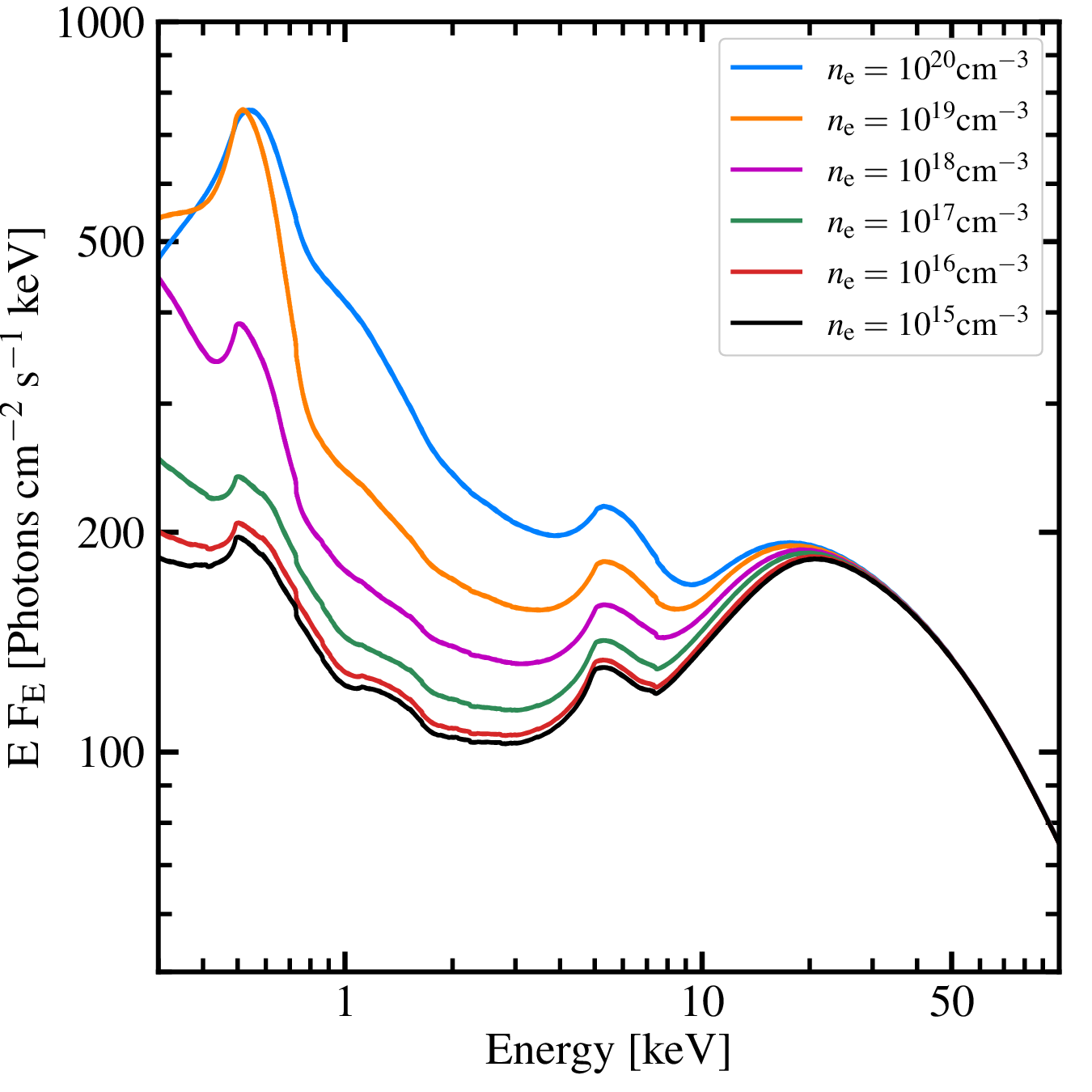}
\includegraphics[scale=0.37,angle=-0]{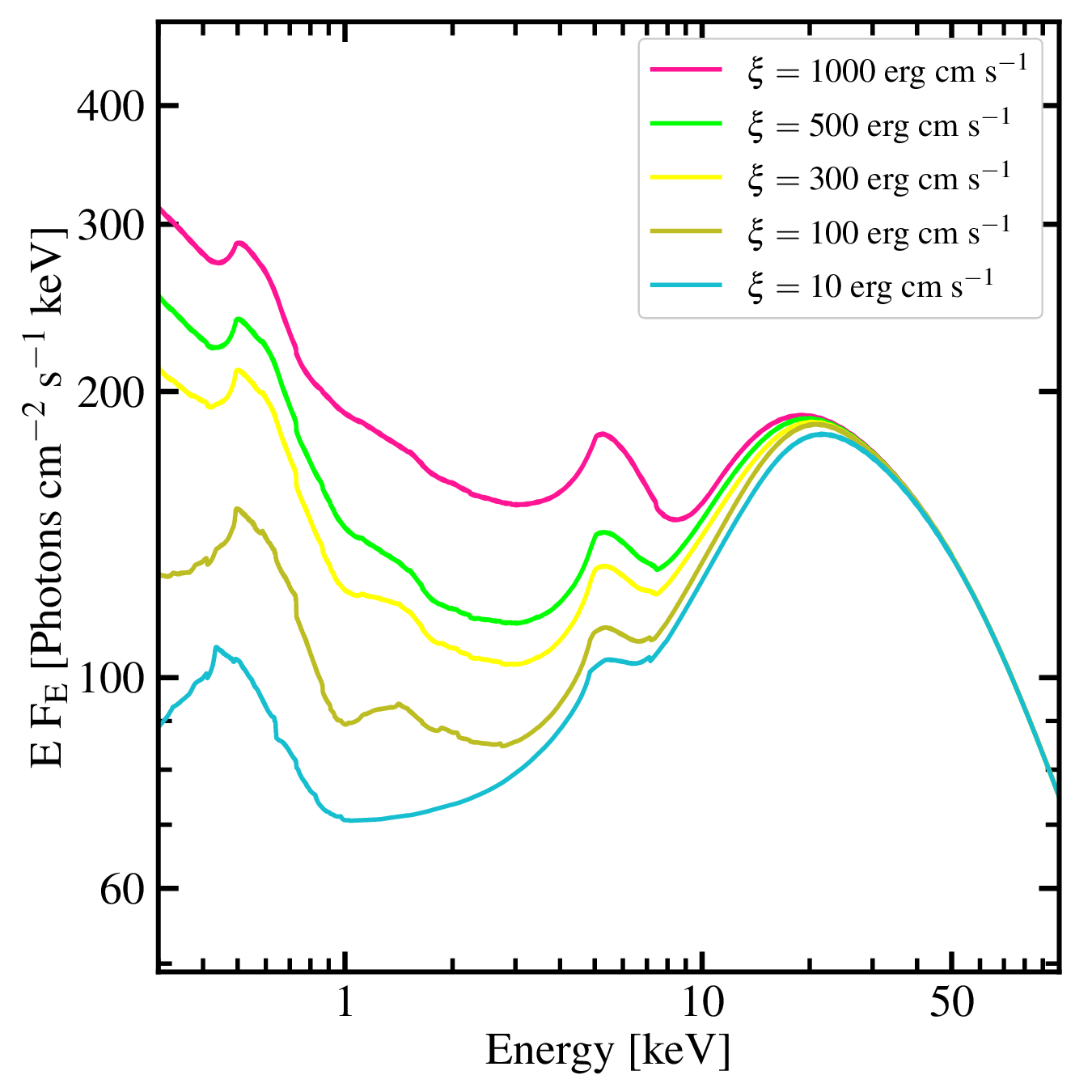}
\caption{The left panel shows the spectra calculated by the relativistic reflection model, {\tt relxillCp}, for a range of disk densities, $\log [n_{\rm e}/\rm{cm}^{-3}]=15$, $16$, $17$, $18$, $19$, and $20$. To make this plot, the model parameters are assumed to be $\Gamma=2$, $kT_{\rm e}=300$\keV{}, $\xi=500$~erg~cm~s$^{-1}$, $q_{\rm in}=8$, $a^{\ast}=0.9$, $\theta=45$~degree, $A_{\rm Fe}=1$, {\tt{refl$\_$frac}} $=3$, and {\tt{norm}} $=1$. We can see a significant boost in the strength of the soft X-ray component from the disk density of $n_{\rm e}=10^{17}\rm{cm}^{-3}$ onward. The right panel depicts the spectra for various ionization states ($\xi/{\rm erg~cm~s^{-1}}=10$, $100$, $300$, $500$, and $1000$) of the accretion disk in the {\tt relxillCp} model calculated for $\Gamma=2$, $kT_{\rm e}=300$\keV{}, $\log [n_{\rm e}/\rm{cm}^{-3}] =17$, $q_{\rm in}=8$, $a^{\ast}=0.9$, $\theta=45$~degree, $A_{\rm Fe}=1$, {\tt{refl$\_$frac}} $=3$, and {\tt{norm}} $=1$. Even for the same disk density, the soft X-ray excess flux is enhanced with a broader Fe~K emission line as the disk becomes more ionized upto $\log[\xi/$erg~cm~s$^{-1}]=3$.}
\end{center}
\label{eemo_ne}
\end{figure*}

\subsection{Relative Contributions of Relativistic Reflection and Warm Comptonization in the Soft X-ray Band}
For sources requiring an additional warm Comptonization component to model the observed soft X-ray excess, we examine the relative contributions of the relativistic disk reflection ({\tt relxillCp}) and warm Comptonization ({\tt compTT}) components in the $0.3-2$\keV{} soft X-ray band to evaluate their respective strengths. Figure 4 presents the relativistic disk reflected flux versus the warm coronal flux in this band. The 1:1 line indicates equal contributions from both relativistic disk reflection and warm Comptonization to the soft X-ray excess. As shown in Fig.~4, the warm corona contributes more to the soft X-ray excess in Mrk~110 (except for one observation), PG~0844$+$349, and PG~0804$+$761. The source for which relativistic disk reflection completely dominates over warm Comptonization is Mrk~79. For four AGN (Mrk~1310, NGC~4748, Mrk~279, and PG~1426$+$015), we find that relativistic disk reflected flux dominates the warm coronal flux in some observations, while in other observations of these four AGN, warm Comptonization dominates over relativistic disk reflection. Additionally, we notice that the warm coronal flux is unconstrained in several observations of some AGN (e.g. Mrk~1310, NGC~4748, Mrk~79), likely due to low SNR or insufficient net source counts in those specific observations. Our results suggest a strong degree of independence between the warm corona and relativistic reflection components.

\begin{figure*}
\centering
\begin{center}
\includegraphics[scale=0.355,angle=-0]{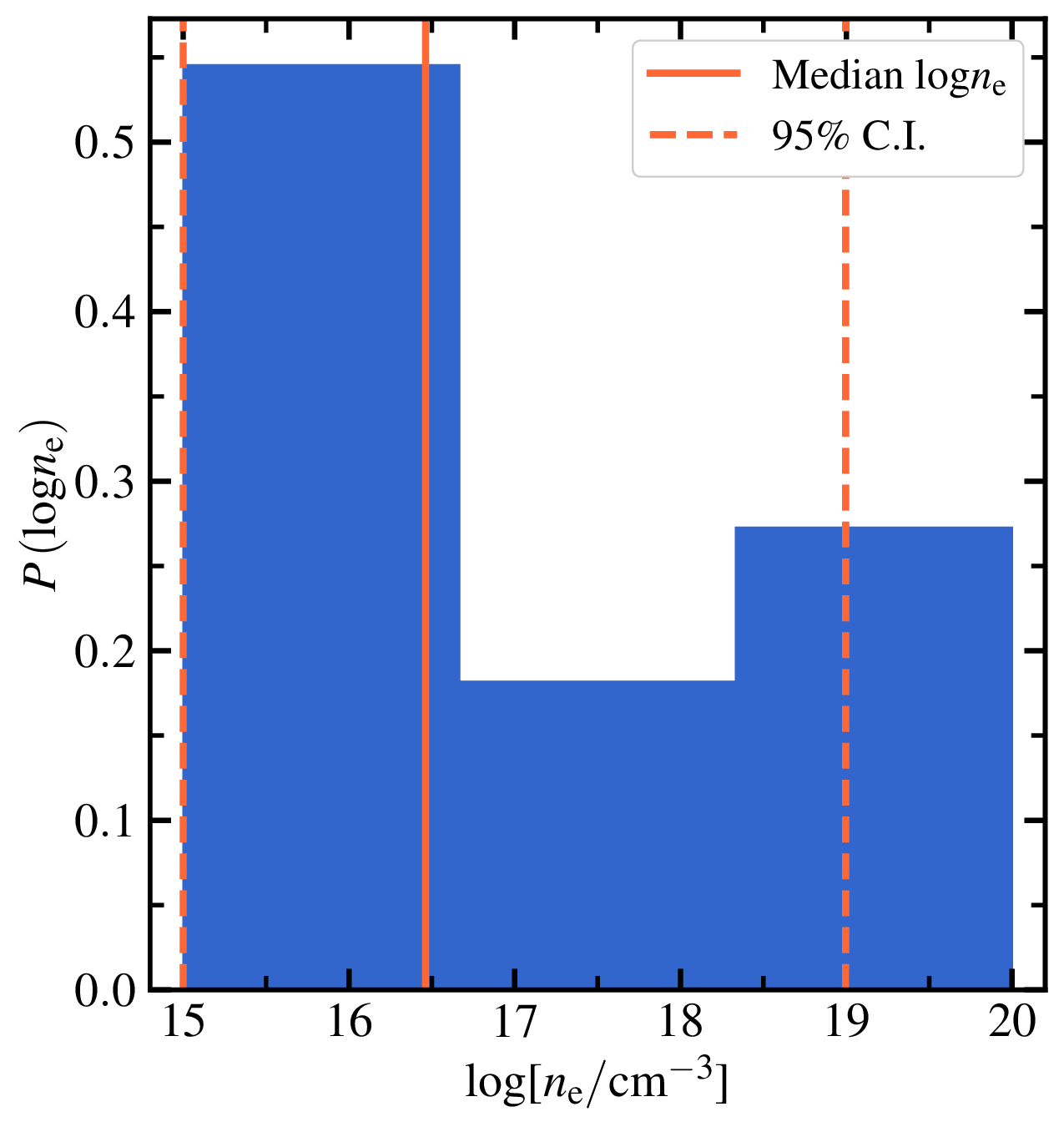}
\includegraphics[scale=0.545,angle=-0]{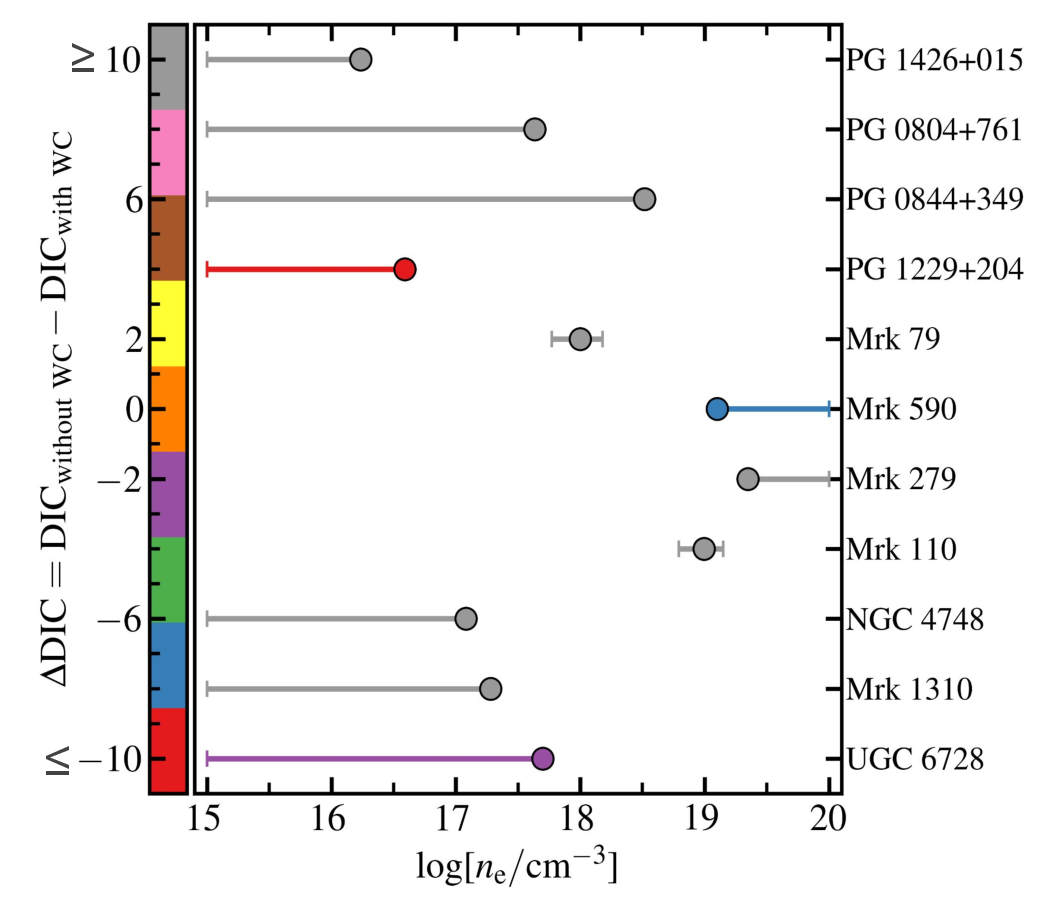}
\caption{The left panel shows the probability distribution of the measured electron density of the accretion disk for our sample. The colorbar in the right panel depicts the difference between the Deviance Information Criteria, ${\rm DIC}_{\rm without~WC}$ and ${\rm DIC}_{\rm with~WC}$, for the variable-density disk reflection model without and with warm Comptonization, respectively. The Bayesian model selection metric supports the relevance of additional warm Comptonization for the origin of soft X-ray excess in 8 AGN marked in grey color. As shown in column~(9) of Table~\ref{test_wc}, the $\Delta {\rm DIC}$ has an extensive range from $-13.7$ to $243.7$, so we cannot plot the full range in the colormap. Therefore, the upper and lower limits in the colormap are marked as $\geq 10$ and $\leq -10$, respectively, which signify `very strong' positive and negative evidence of additional warm Comptonization over the variable-density disk reflection.}
\end{center}
\label{dic_logne}
\end{figure*}

\section{Results and Discussion}
\label{sec:results}
In this section, we discuss all the results derived from our broadband X-ray spectral modeling, the implications of the physical reflection model for the origin of the soft X-ray excess, the validity of the standard SS73 accretion disk theory, the first-time calculation of the disk-to-corona power transfer fraction, coronal (hot as well as warm) properties, and black hole spin population across mass scales ($\log M_{\rm BH} \sim 5.5-9.0$). The details of broadband spectral modeling for each source in the sample are presented in Appendix~\ref{sec:individual}. In Table~\ref{table_output_parameters}, we report the best-fit source spectral model parameters and their 90\% confidence intervals determined through MCMC parameter space exploration of the best-fit model. Fig.~A2 shows the broadband \xmm{}/\nustar{} spectra, the best-fit count spectral model with components, and the corresponding residuals. The best-fit spectral energy flux models with components are presented in Figures~A3 and A4.

\subsection{Physical Origin of the Soft X-ray Excess Emission: High-density Disk Reflection or Warm Comptonization}
\label{sec:soft_excess}
Two models have been proposed to explain the observed soft X-ray excess. One model is the relativistic reflection or reprocessing of the incident hot coronal emission in the innermost part of the accretion disk \citep{George_Fabian_1991,Ross_Fabian_2005,Garcia_2014}. The other model considers Compton up-scattering of the optical/UV disk photons in a low-temperature ($kT\sim 0.1-2$\keV{}), optically thick ($\tau>1$) Comptonizing medium or warm corona \citep{Done_2012,Petrucci_2018}. However, the relativistic reflection as the origin of the soft X-ray excess is a more consistent explanation, as it is the only model that can explain the broad Fe~K emission line and Compton hump together with the soft X-ray excess. However, it was shown that the entire soft X-ray excess may not be well-fitted solely by relativistic reflection (e.g. Ark~120: \cite{Mallick_2017,Porquet_2018}), especially when the disk density is low and fixed at $\log [n_{\rm e}/\rm{cm}^{-3}] = 15$. Furthermore, fitting of soft X-ray excess with a fixed low-density relativistic disk reflection model resulted in unphysically high ($A_{\rm Fe}>10$) iron abundance in some sources (e.g. 1H~0707-495: \cite{Dauser_2012}). To resolve these issues, \cite{Garcia_2016} developed a new model where the density of the accretion disk is a free parameter varying in the range of $\log [n_{\rm e}/\rm{cm}^{-3}] = 15-20$, first implemented by \cite{Mallick_2022}. When the innermost part of the disk becomes radiation pressure dominated, extra heating from free-free absorption increases the disk density because of its quadratic dependence on density, thus boosting the strength of the observed soft X-ray excess below 1\keV{} \citep{Ross_Fabian_2007,Garcia_2016}. In Fig.~5 (left), we show the flux spectra calculated from the variable-density disk reflection model {\tt relxillCp} for the disk density of $\log [n_{\rm e}/\rm{cm}^{-3}] = 15$, $16$, $17$, $18$, $19$, and $20$. The parameters assumed for the model calculations shown in this plot are: $\Gamma=2$, $kT_{\rm e}=300$\keV{}, $\xi=500$~erg~cm~s$^{-1}$, $q_{\rm in}=8$, $a^{\ast}=0.9$, $\theta=45$~degree, $A_{\rm Fe}=1$, {\tt{refl$\_$frac}} $=3$, and {\tt{norm}} $=1$. Even with the solar iron abundance, model flux is noticeably enhanced at the soft X-ray band when the disk density is higher than $n_{\rm e}=10^{15}\rm{cm}^{-3}$, and the difference from the canonical disk reflection model becomes more prominent for $n_{\rm e} \geq 10^{17}\rm{cm}^{-3}$.

Disk ionization is also a crucial physical parameter in the relativistic disk reflection model ({\tt relxillCp}) and influences the strength of the observed soft X-ray excess. To illustrate how it affects the spectral components, especially the soft X-ray excess and broad Fe~K emission line, we show the flux spectra for the disk ionization of $\xi/{\rm erg~cm~s^{-1}}= 10$, $100$, $300$, $500$, and $1000$ in the right panel of Fig.~5. The other parameters considered for the model calculations shown in the plot are as follows: $\Gamma=2$, $kT_{\rm e}=300$\keV{}, $\log [n_{\rm e}/\rm{cm}^{-3}] = 17$, $q_{\rm in}=8$, $a^{\ast}=0.9$, $\theta=45$~degree, $A_{\rm Fe}=1$, {\tt{refl$\_$frac}} $=3$, and {\tt{norm}} $=1$. Evidently, as the disk becomes more ionized, the soft X-ray flux is enhanced with a broader Fe~K emission line, even for the same electron density of the accretion disk.

The probability distribution of the disk density parameter for the sample is shown in Fig.~6 (left panel). We obtained disk density measurements higher than the canonical value of $\log[n_{\rm e}/{\rm cm^{-3}]}=15$ with 90\% confidence for 4 AGN: Mrk~110, Mrk~279, Mrk~590, and Mrk~79. Of these 4 AGN, three sources (Mrk~110, Mrk~279, and Mrk~79) exhibited strong relativistic reflection features (broad Fe~K emission line, Compton hump), yet required a warm Comptonization component to account for the soft X-ray excess in addition to high-density relativistic disk reflection. For 7 out of 11 AGN, the 90\% lower limit of the density parameter has reached the canonical value of $n_{\rm e}=10^{15}$~cm$^{-3}$. Out of these 7 AGN, an additional warm Comptonization component is required to describe the observed soft X-ray excess in 5 AGN. 

Through joint \xmm{}+\nustar{} broadband spectroscopy, we find that variable-density relativistic reflection can self-consistently explain both the broad Fe~K emission line and soft X-ray excess in 3 out of 11 AGN without requiring any additional warm coronal emission. The inner accretion disk is found to be ionized and dense, with the median ionization of $\xi \sim 10^{2.0}$~erg~cm~s$^{-1}$, median density of $n_{\rm e} \sim 10^{16.5}\rm{cm}^{-3}$, and near-solar iron abundance of $A_{\rm Fe} \sim 2$ for the sample. The right panel in Fig.~6 shows $\Delta {\rm DIC} = {\rm DIC}_{\rm without~WC}-{\rm DIC}_{\rm with~WC}$ versus disk density for the sample, confirming the relevance of an additional warm coronal emission for the observed soft X-ray excess in 8 out of 11 AGN. We find that the temperature and optical depth of the warm coronae for these 8 AGN are in the range of $kT_{\rm wc}\sim 0.2-2$\keV{} and $\tau_{\rm wc}\sim 4-22$, respectively, which are consistent with the properties of warm coronae derived for a larger sample of AGN (see Fig.~5 of \cite{Petrucci_2018}). According to recent simulations, a warm corona can produce a featureless soft X-ray excess only if there is sufficient mechanical heating of the accretion disk \citep{Ballantyne_2020,Petrucci_2020,Kawanaka_Mineshige_2024}. The origin of such internal heating can be attributed to several physical processes, such as magnetic reconnection or viscous accretion within the corona \citep{Rozanska_2015,Gronkiewicz_2023}. The observed large optical depth of the warm corona supports the presence of intense magnetic pressure in the accretion disk \citep{Rozanska_2015}. 

A key simplification in all reflection models is the assumption that the density of the illuminated atmosphere remains constant. Therefore, our study investigates the effects of high density in X-ray illuminated disk atmospheres, assuming a constant-density vertical structure of the accretion disk (see section~2 of \cite{Garcia_2016} for more details), which may differ from reality. This simplification may explain why relativistic reflection alone cannot fully account for the soft X-ray excess without including additional warm Comptonized emission in most of the sources in the sample. However, developing a reflection model that incorporates variable density in the vertical direction of the accretion disk is beyond the scope of this paper.

\begin{figure*}
\centering
\begin{center}
\includegraphics[scale=0.35,angle=-0]{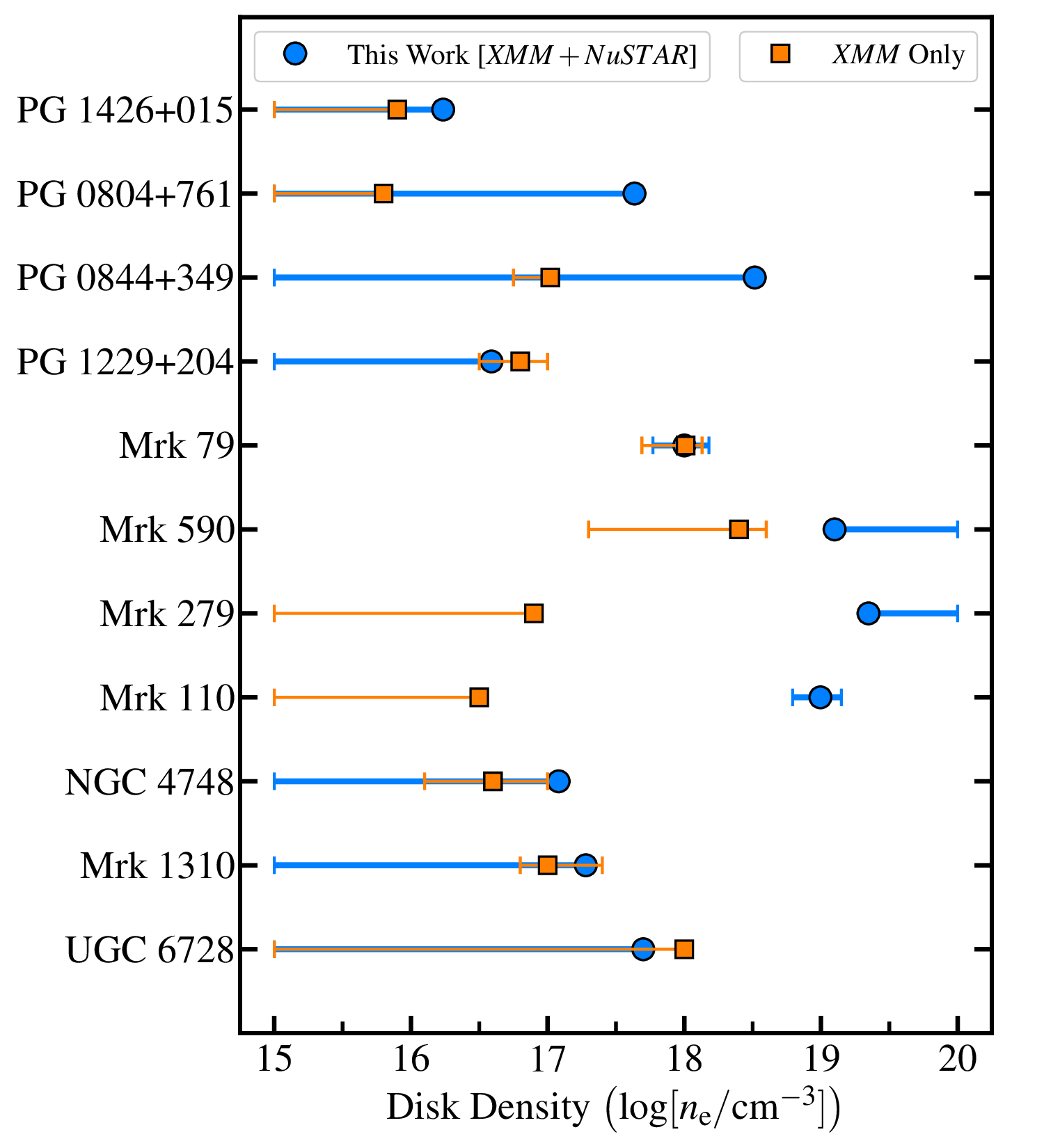}
\includegraphics[scale=0.35,angle=-0]{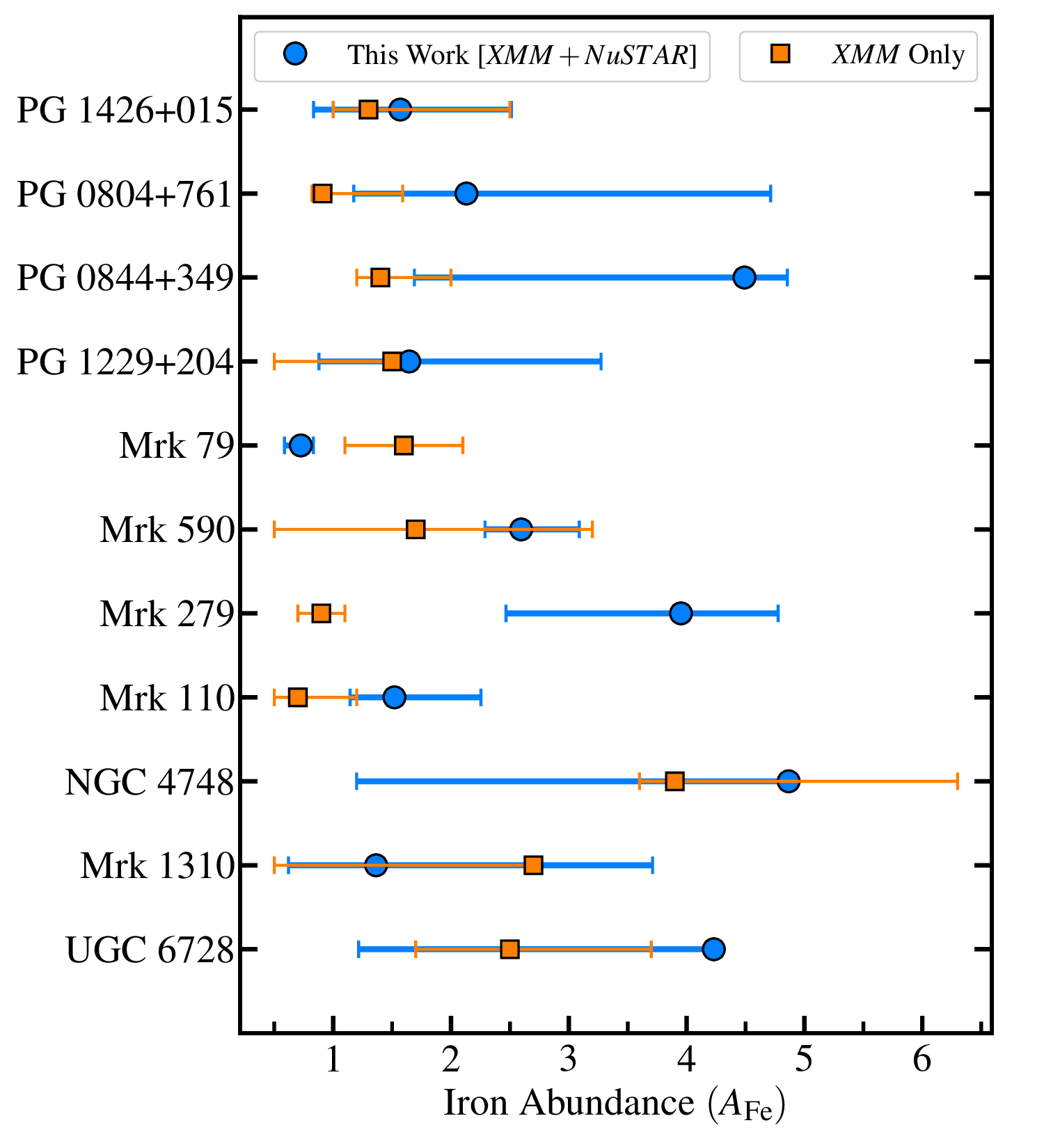}
\includegraphics[scale=0.35,angle=-0]{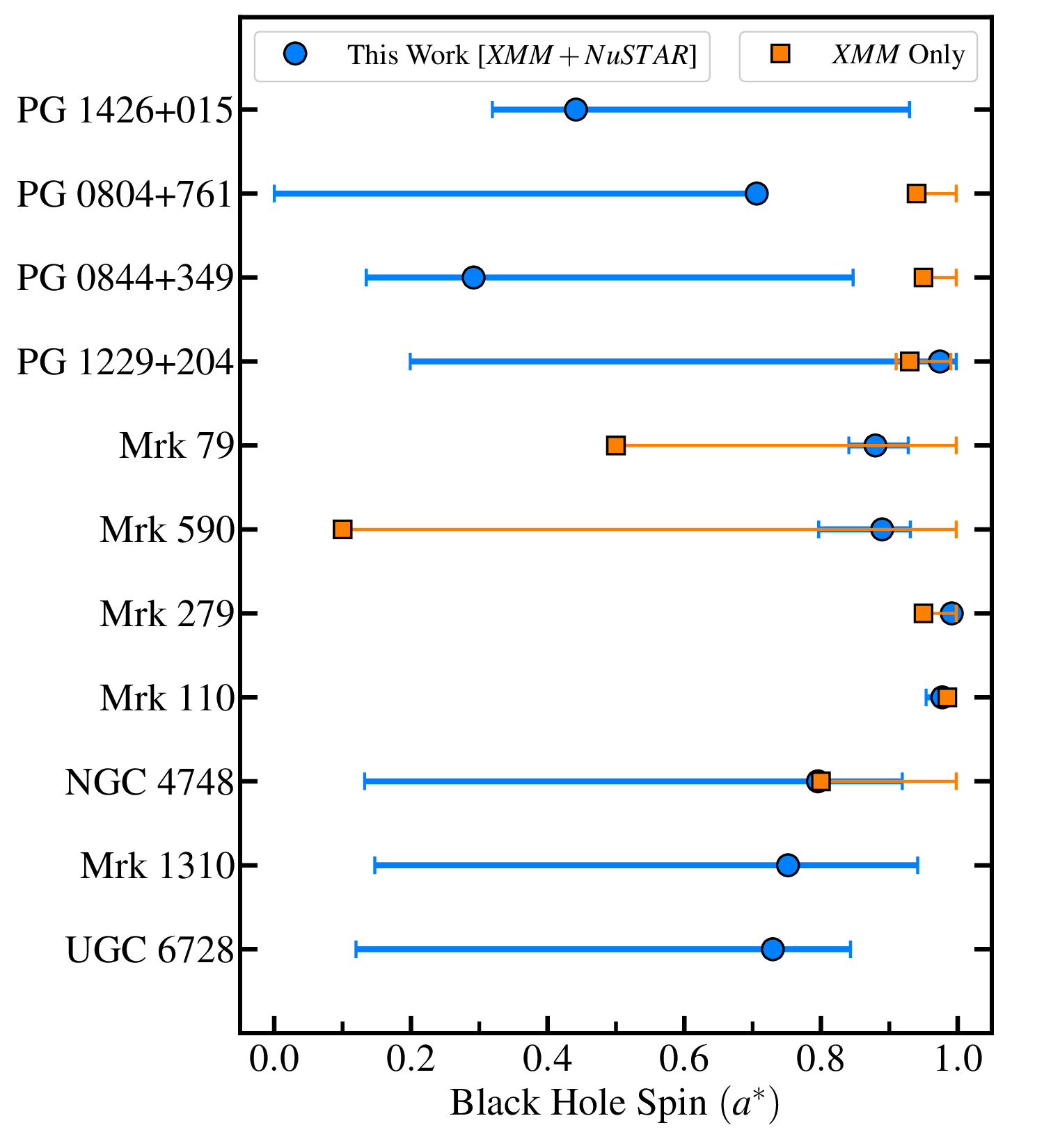}
\includegraphics[scale=0.35,angle=-0]{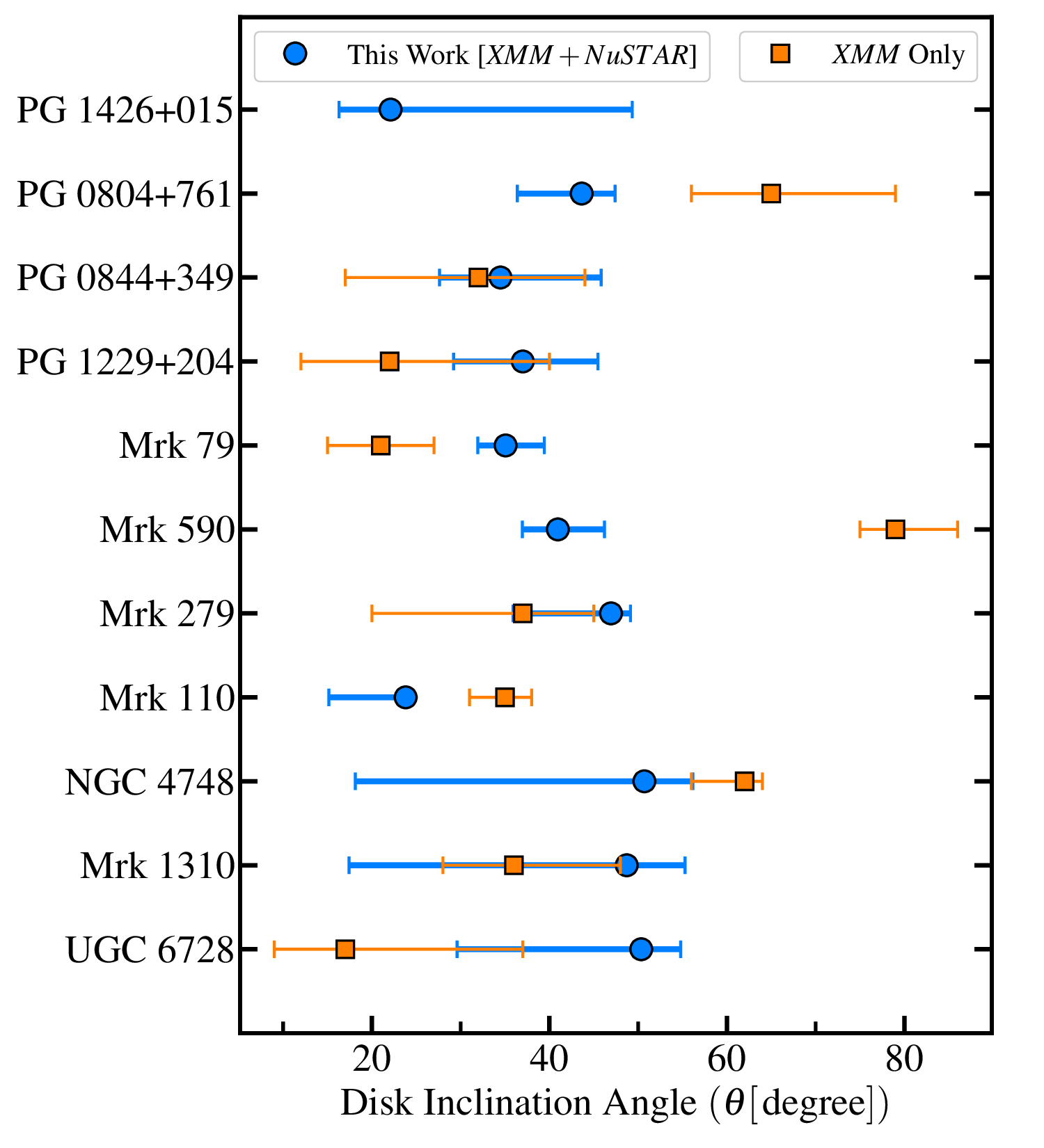}
\caption{Comparison of disk density, iron abundance, black hole spin, and disk inclination angle derived from our broadband (0.3--78\keV{}) joint \xmm{}+\nustar{} spectral modeling with those obtained from the previous 0.5--10\keV{} \xmm{} spectral fitting of the sample by JJ19. For UGC~6728, Mrk~1310, and PG~1426$+$015, JJ19 fixed the spin parameter at $a^{\ast}=0.998$, precluding a direct comparison of spin measurements. As a result, orange squares are absent for these three sources in the bottom left panel. Likewise, no comparison of disk inclination angle is shown for PG~1426$+$015 in the bottom right panel, since this parameter was also fixed in JJ19. The error bars in all four panels represent the 90\% confidence intervals.}
\end{center}
\label{compare_all_parms}
\end{figure*}

\subsection{Comparison with Previous Study of the Sample with {\textit {XMM-Newton}} Only}
 
\label{sec:comparison}
The comparison of four key parameters, i.e., disk density, iron abundance, black hole spin, and disk inclination angle of the AGN sample, as determined from our broadband joint \xmm{}+\nustar{} spectral modeling and from previous analyses by JJ19 using only 0.5--10\keV{} averaged \xmm{} spectra, are presented in Fig.~7. However, a direct comparison of parameter uncertainties should be treated with caution, since our uncertainties are derived using MCMC, while JJ19 did not employ MCMC for error estimation. Moreover, the spectral modeling statistics and residuals reported in JJ19 are high for most sources, raising further concerns about the robustness of their parameter estimation.

As evident from the top left panel in Fig.~7, the disk density we measured using joint \xmm{}/\nustar{} data is higher than that derived from \xmm{} spectral fitting alone. This is because the additional spectral features across the broad bandpass let us more accurately disentangle the reflection from the continuum. Moreover, some sources (e.g., Mrk~79) in the previous \xmm{} data did not reveal broad Fe~K line emission due to short exposure. In this work, with more data from both \xmm{} and \nustar{}, we prominently detected the broad Fe~K emission feature, the modeling of which resulted in an enhanced contribution from the disk reflection component.

The iron abundances measured from our joint \xmm{}+\nustar{} high-density reflection spectroscopy are consistent with those obtained from high-density reflection modeling of previous \xmm{} data for the sample (Fig.~7, top right). We measure solar or near-solar iron abundances for the sample, with a median value of $A_{\rm Fe}\sim 2$. This is expected since the higher-density reflection model can reduce the inferred iron abundance by enhancing the continuum in the reflection component.

Comparison of our black hole spin measurements with those inferred from previous high-density spectral fitting of \xmm{} data alone provides consistent results within the 90\% confidence limits (Fig.~7, bottom left). In the earlier \xmm{} spectral analysis by JJ19, the spin parameter was fixed at $a^{\ast}=0.998$ for three sources (UGC~6728, Mrk~1310, and PG~1426$+$015) due to either the non-detection of a broad Fe~K emission line in the \xmm{} data or the lack of hard X-ray data above 10\keV{}. In this work, the inclusion of \nustar{} data, and therefore the Compton reflection hump, constrains the continuum emission better, which can potentially impact the determination of the red wing of the broad Fe K emission line and thus affect black hole spin measurements. As a result, our broadband spectroscopy improves both the accuracy and precision of black hole spin measurements for all sources in the sample. Notably, joint \xmm{}/\nustar{} high-density relativistic reflection spectroscopy provides the first robust spin measurements for three AGN in our sample: UGC~6728, Mrk~1310, and PG~1426$+$015.

Previous measurements of the disk inclination angle using only \xmm{} data provided extreme values for the sample, ranging from as low as $5$~degree to as high as $90$~degree, as illustrated in Fig.~7 (bottom right). However, through joint \xmm{}/\nustar{} spectroscopy, we measure typical values for the disk inclination angle, clustering around $45$~degree. The median disk inclination for the sample is $\theta=41\pm 8$~degree at $2\sigma$ confidence. Accurate measurement of the inclination angle depends on the blue wing of the broad Fe~K emission line, which is better characterized when both the continuum and Compton hump are well constrained. The inclusion of \nustar{} data not only enables tighter constraints on the broadband continuum but also improves the modeling accuracy of both the broad Fe~K emission line and the Compton hump.

\subsection{Disk Density \& Disk-to-Corona Power Transfer Fraction}
\label{sec:disk_theory}
The electron density in the standard SS73 $\alpha$-disk model \citep{Shakura_Sunyaev_1973}, for a radiation pressure-dominated inner region and incorporating the transfer of a fraction of power from the disk to a corona, was derived by \cite{Svensson_Zdziarski_1994}: 
\begin{equation}
n_{\rm e}=\frac{1}{\sigma_{{\rm T}}r_{{\rm s}}}\frac{256\sqrt{2}}{27}\alpha^{-1}r^{3/2}\dot{m}^{-2}\Big[1-(r_{{\rm in}}/r)^{1/2}\Big]^{-2}(1-f)^{-3},
\label{eq1}
\end{equation}
where $f$ is the disk-to-corona power transfer fraction, representing the fraction of power released from the accretion disk into the hot corona. The range of the $f$-parameter is $f \in [0, 1)$. The $f=0$ solution can provide a non-zero value of the electron density according to SZ94. However, the $f=1$ solution is forbidden in the model. The viscosity parameter $\alpha$ is assumed to be 0.1 (see e.g. \cite{Salvesen_2016}). The Thomson scattering cross-section is $\sigma_{\rm T}=6.64\times10^{-25}$~cm$^2$. The radius ($r$) of the accretion disk is in units of the Schwarzschild radius $r_{\rm s}=\frac{2GM_{\rm BH}}{c^{2}}$. The inner disk radius ($r_{\rm in}$) is at the innermost stable circular orbit ($r_{\rm isco}$) in the relativistic disk reflection model, the median of which is estimated to be $\sim 2.8r_{\rm s}$ for the sample. $\dot{m}=\frac{\dot{M}}{\dot{M}_{\rm E}}$ denotes the dimensionless mass accretion rate.

\begin{table}
\centering
\caption{The disk-to-corona power transfer fraction ($f$-parameter), measured at $r=10 r_{\rm s}$ and expressed in percent for each source in the sample.}
\begin{center}
\scalebox{0.95}{%
\begin{tabular}{ccc}
\hline \\ [0.001cm]
Source Name  & $f$-parameter [percent] \\ [0.25cm]
\hline  \\ [0.001cm]
UGC~6728 & $\le 5.3$ \\ [0.25cm]
         
Mrk~1310 & $\le 2.5$ \\ [0.25cm]
         
NGC~4748 & $\le 58.9$ \\ [0.25cm]
         
Mrk~110 & $91.3_{-1.5}^{+1.0}$ \\ [0.25cm]
         
Mrk~279 & $\ge 93.3$ \\ [0.25cm]
         
Mrk~590 & $\ge 86.8$ \\ [0.25cm]
         
Mrk~79 & $46.9_{-10.2}^{+6.9}$ \\ [0.25cm]
         
PG~1229+204 & $\le 42.9$ \\ [0.25cm]
         
PG~0844+349 & $\le 93.3$ \\ [0.25cm]
         
PG~0804+761 & $\le 93.0$ \\ [0.25cm]
         
PG~1426+015 & $\le 57.8$ \\ [0.25cm]

\hline 
\end{tabular}}
\end{center} 
\label{source_f}           
\end{table}

\begin{figure*}
\centering
\begin{center}
\includegraphics[scale=0.33,angle=-0]{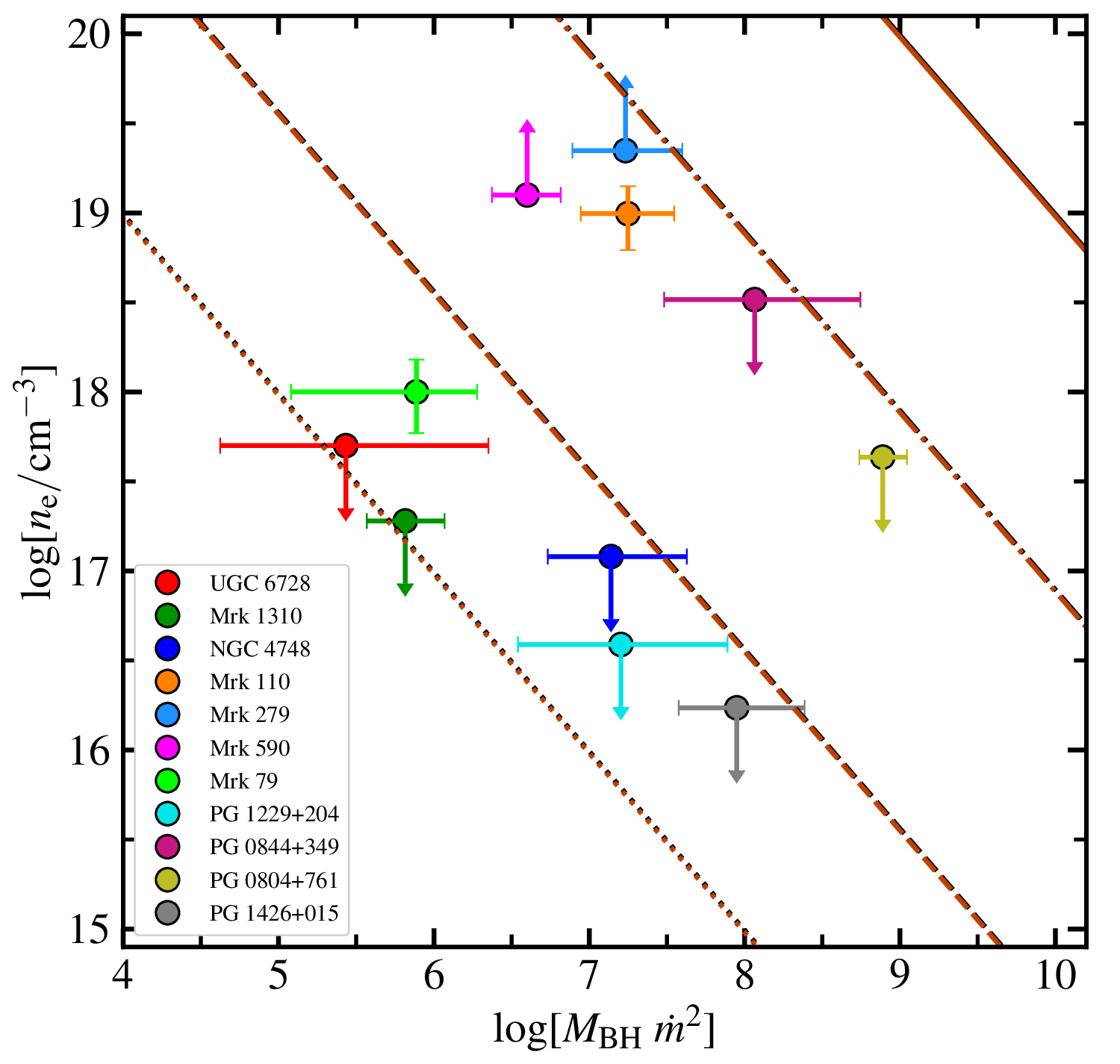}
\includegraphics[scale=0.33,angle=-0]{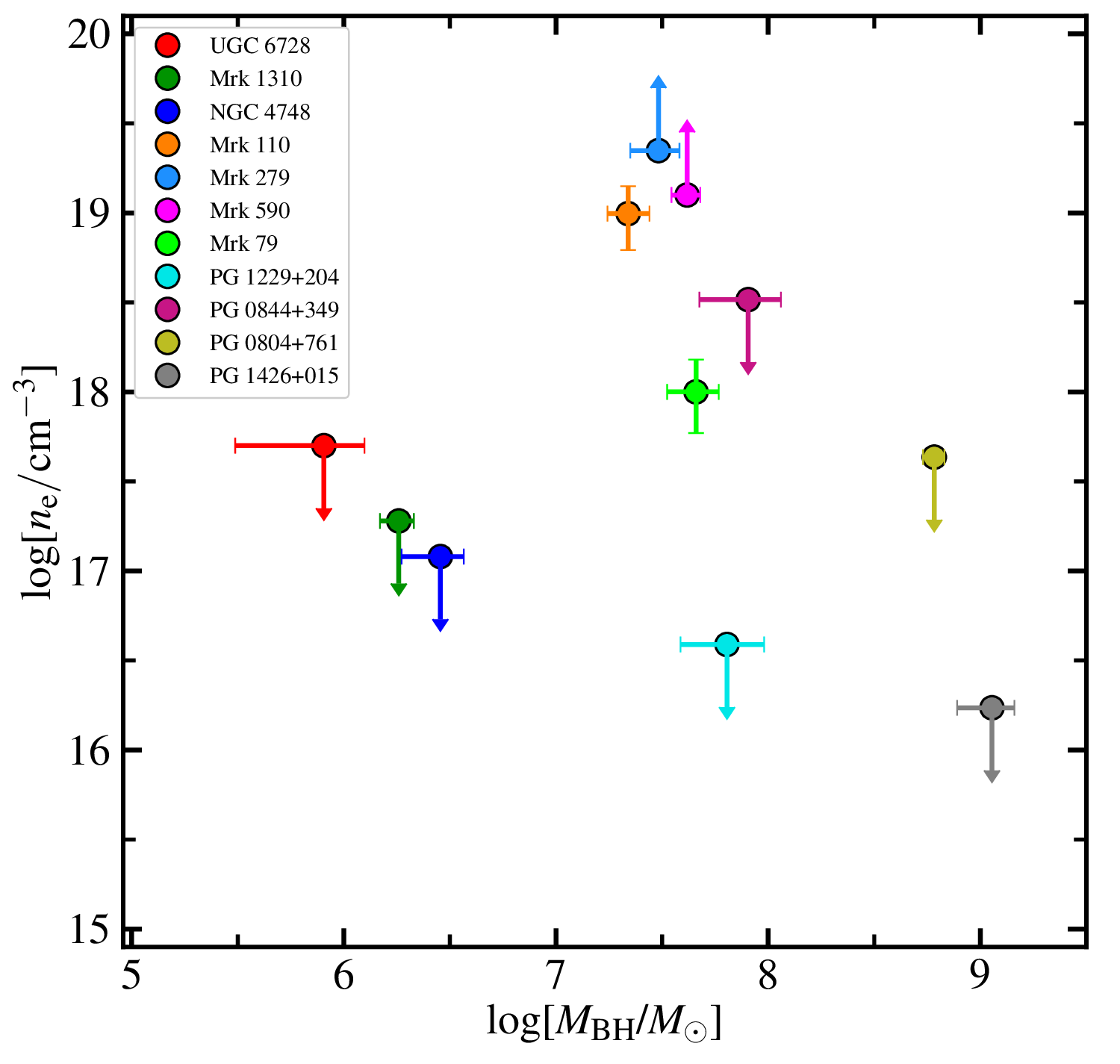}
\caption{Theoretically, the electron density ($n_{\rm e}$) of the accretion disk depends on five parameters: $M_{\rm BH}$, $\dot{m}$, $f$, $r$, and $r_{\rm in}$. The left panel shows how disk density varies as a function of the black hole mass times the accretion rate squared, $M_{\rm BH}\dot{m}^{2}$, plotted on a logarithmic scale. The density solutions for a radiation pressure-dominated disk at $r=6 r_{\rm s}$ (in black) and $r=10 r_{\rm s}$ (in brown) for $f=0$, $0.7$, $0.95$, and $0.99$ are depicted by the dotted, dashed, dash-dotted, and solid lines, respectively. The inner disk radius ($r_{\rm in}$) is set at the sample median value of $r_{\rm isco}=2.8 r_{\rm s}$. The right panel shows the dependence of $\log n_{\rm e}$ on $\log M_{\rm BH}$ for the sample. In both panels, no significant correlation is observed between the parameters.}
\end{center}
\label{logne_mbh_mdotsq}
\end{figure*}

\begin{figure*}
\centering
\begin{center}
\includegraphics[scale=0.33,angle=-0]{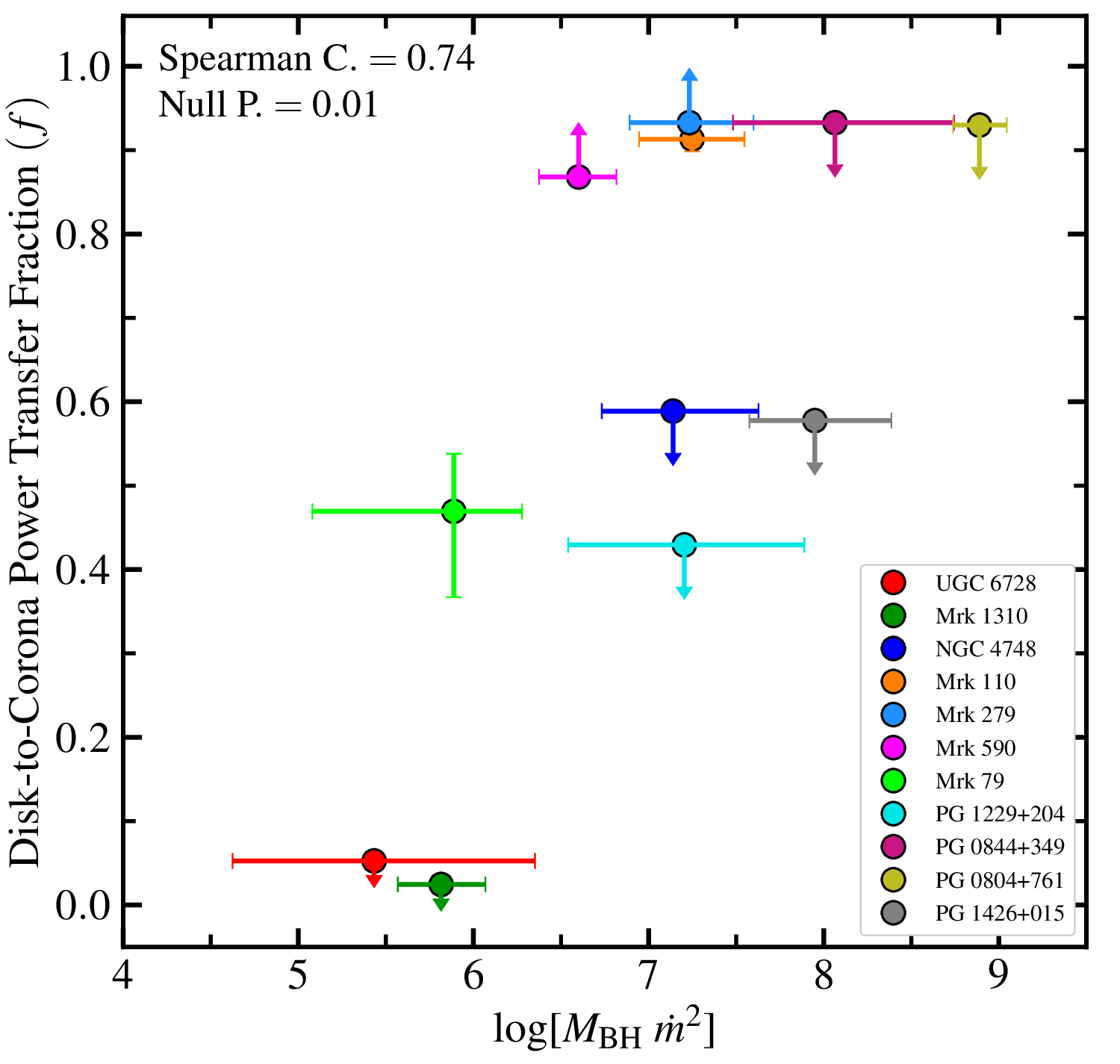}
\includegraphics[scale=0.33,angle=-0]{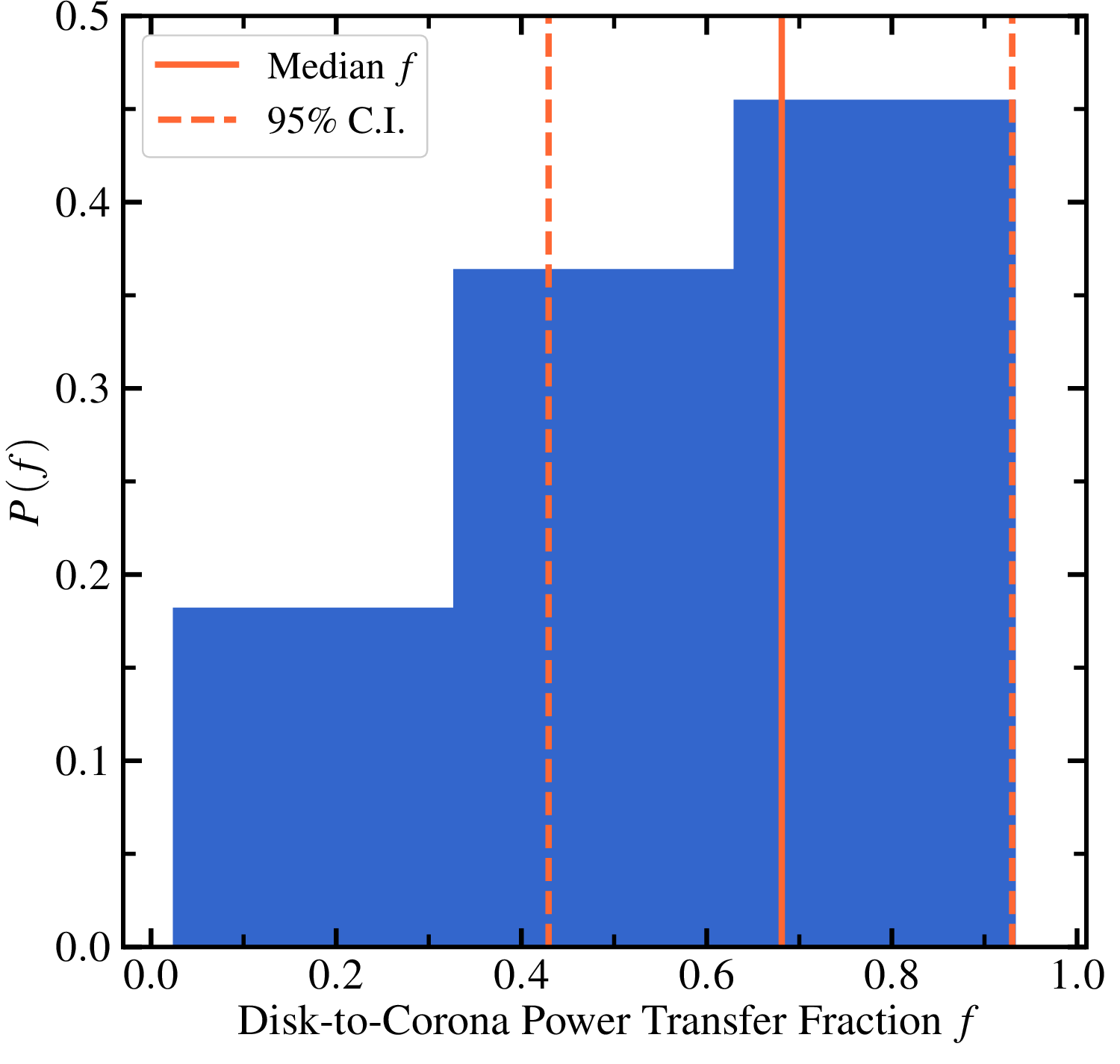}
\caption{The fraction ($f$) of power transferred from the disk into the corona, measured at $r=10 r_{\rm s}$, is plotted as a function of $\log [M_{\rm BH}$ $\dot{m}^{2}$] and $\log[M_{\rm BH}$/$M_{\odot}$] in the left panel. We find a strong positive correlation between $f$ and $\log [M_{\rm BH}$ $\dot{m}^{2}$] with a Spearman rank correlation coefficient of $0.74$ [p-value $=0.01$]. The right panel shows the distribution of the $f$-parameter, which has a median value of $f=0.68_{-0.25}^{+0.25}$.}
\end{center}
\label{f_mbh_mdotsq}
\end{figure*}

Theoretically, the disk density depends on five parameters: $M_{\rm BH}$, $\dot{m}$, $r_{\rm in}$, $r$, and $f$. The black hole mass ($M_{\rm BH}$) and dimensionless mass accretion rate ($\dot{m}$) for each source are obtained from the literature and listed in Table~I. In the relativistic reflection model, the inner disk radius ($r_{\rm in}$) is set to the innermost stable circular orbit ($r_{\rm isco}$), which can be directly estimated from the black hole spin parameter, reported in Table~\ref{table_output_parameters}. Thus, in the standard disk model, the two remaining unknown parameters for determining the disk density are the disk-to-corona power transfer fraction ($f$) and the disk radius ($r$).

To test the validity of the standard $\alpha$-disk model, we first examine the dependence of disk density on the black hole mass and accretion rate for specific disk radius and $f$-parameter values. Figure~8 shows the measured disk density as a function of black hole mass times the accretion rate squared ($M_{\rm BH}\dot{m}^{2}$) and black hole mass ($M_{\rm BH}$), both plotted on a logarithmic scale for our sample. The dotted, dashed, dash-dotted, and solid lines represent the theoretical density solutions for $f=0$, $0.7$, $0.95$, and $0.99$, respectively, each computed at $r=6 r_{\rm s}$ (black) and $r=10 r_{\rm s}$ (brown). As shown in Fig.~8 (left panel), the impact of disk radius ($r$) on density is much less than that of the disk-to-corona power transfer fraction, $f$. If the standard $\alpha$-disk theory holds and the intrinsic scatter in the $f$-parameter is negligible, we expect $\log n_{\rm e}$ to anti-correlate with both $\log[M_{\rm BH}\dot{m}^{2}]$ and $\log M_{\rm BH}$, following $\log n_{\rm e} \propto -\log[M_{\rm BH}\dot{m}^{2}]$. However, we do not observe any such correlation, suggesting that the intrinsic scatter associated with the $f$-parameter is substantial, and that $f$ varies significantly between different sources in our sample. In principle, disk density is most sensitive to variations in $f$, with $n_{\rm e} \propto (1-f)^{-3}$ or, $\log n_{\rm e} \propto -3\log(1-f)$. Next, we evaluate the model density as a function of disk radius for various values of the $f$-parameter and compare the results with the measured disk densities for the sample, as illustrated in Fig.~A7. The model density matches the measured density at different $f$-parameter values for individual sources. Notably, this agreement between the model and measured densities is observed at the common radius of $r=10 r_{\rm s}$ across all sources. Therefore, we determine the $f$-values from the measured disk densities at $r=10 r_{\rm s}$ for each source using equation~(\ref{eq1}) and report them in Table~II. As it stands, we have taken care of all the model intrinsic scatters involved, and if the standard $\alpha$-disk model is valid, we expect a positive correlation between $f$ and $\log[M_{\rm BH}\dot{m}^{2}]$. In Fig.~9 (left), we show the derived $f$-values for the sample as a function of $\log[M_{\rm BH}\dot{m}^{2}]$. To evaluate the correlation between these parameters, we perform Spearman's rank correlation test. The calculated Spearman coefficient for the $f-\log[M_{\rm BH}\dot{m}^{2}]$ correlation is $\rho_{\rm s}=0.74$ with a p-value of $0.01$, which indicates a strong positive correlation between these parameters, thereby validating the standard $\alpha$-disk model with a variable disk-to-corona power transfer fraction in the AGN sample.

\begin{figure}
\centering
\begin{center}
\includegraphics[width=0.48\textwidth]{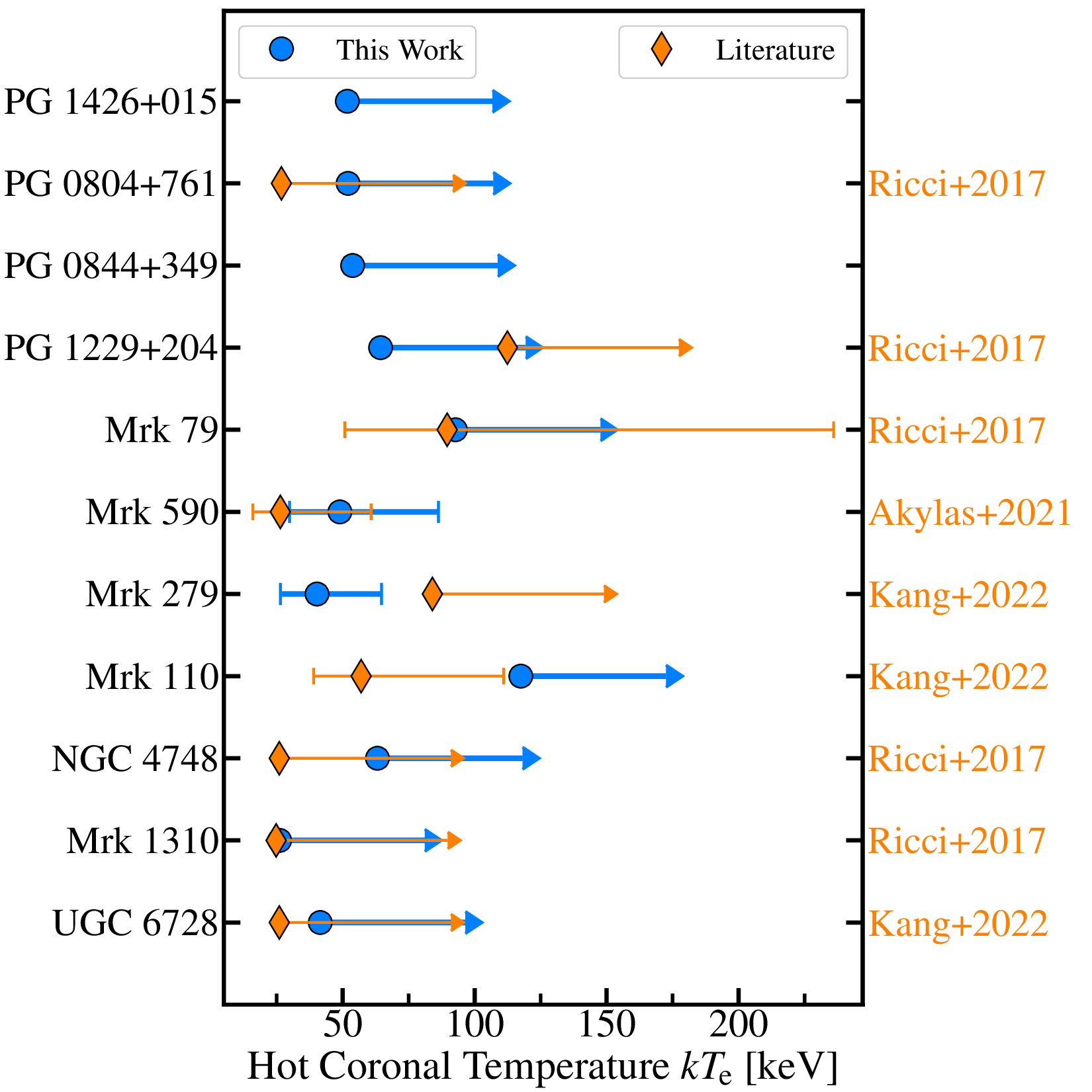}
\caption{Comparison of the electron temperature (blue circles) of hot corona obtained from our broadband \xmm{}+\nustar{} spectral modeling of the sample and their previous measurements (orange diamonds) from the literature \citep{Ricci_2017,Akylas_Georg_2021,Kang_Wang_2022}, labeled on the right-hand side of the Y-axis. The errors are reported at the 90\% confidence level. Orange diamonds are absent for PG~0844+349 and PG~1426+015 since no estimate of their coronal temperature is available in the literature.}
\end{center}
\label{compare_kTe}
\end{figure}

\begin{figure*}
\centering
\begin{center}
\includegraphics[scale=0.38,angle=-0]{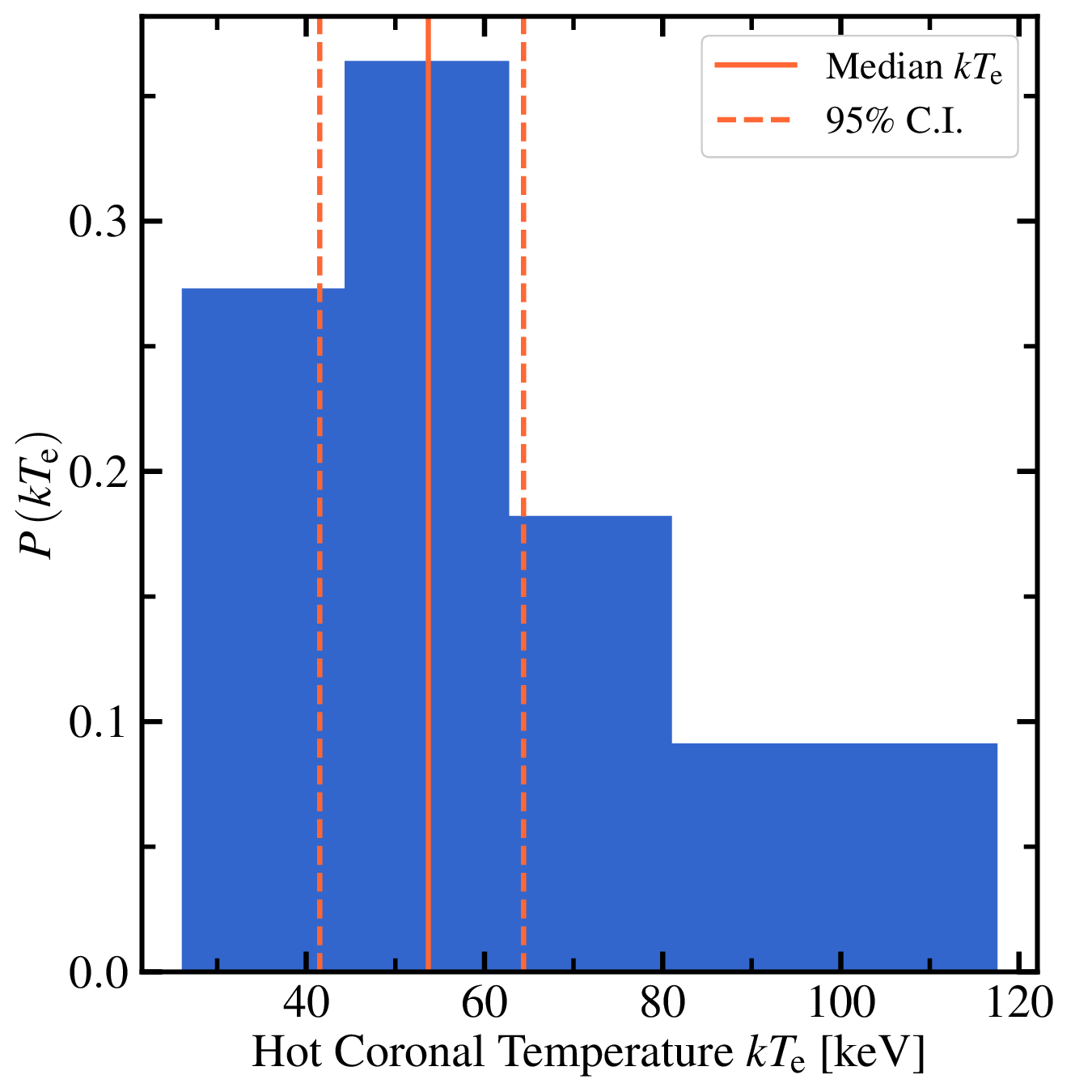}
\includegraphics[scale=0.38,angle=-0]{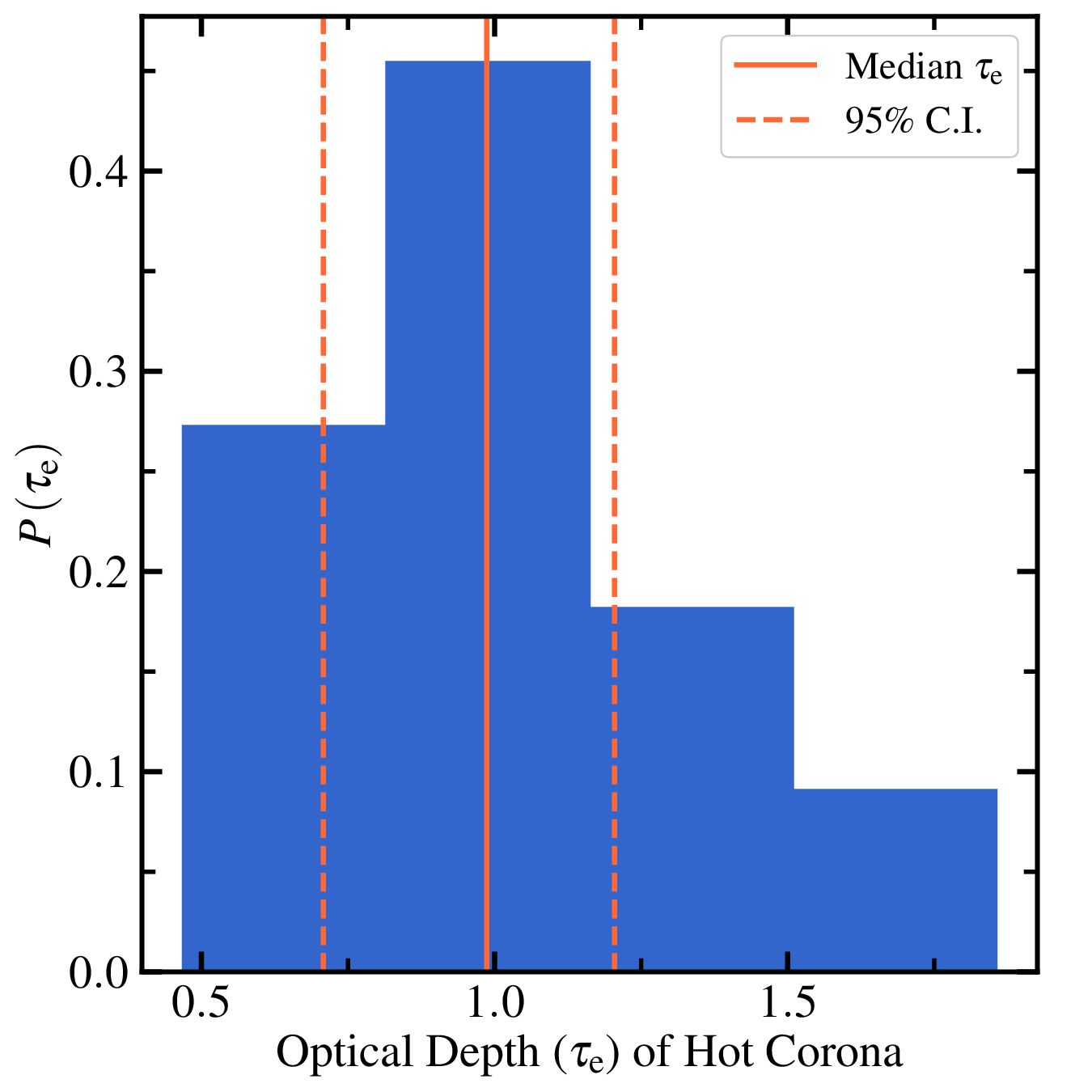}
\caption{Probability distribution of measured temperature ($kT_{\rm e}$) and inferred optical depth ($\tau_{\rm e}$) of the hot corona for 11 AGN from this work. The median values of hot coronal temperature and optical depth for the sample are $kT_{\rm e}=54_{-12}^{+11}$\keV{} and $\tau_{\rm e}=0.98_{-0.28}^{+0.22}$, respectively.}
\end{center}
\label{kTe_tau}
\end{figure*}

\begin{figure*}
\centering
\begin{center}
\includegraphics[scale=0.38,angle=-0]{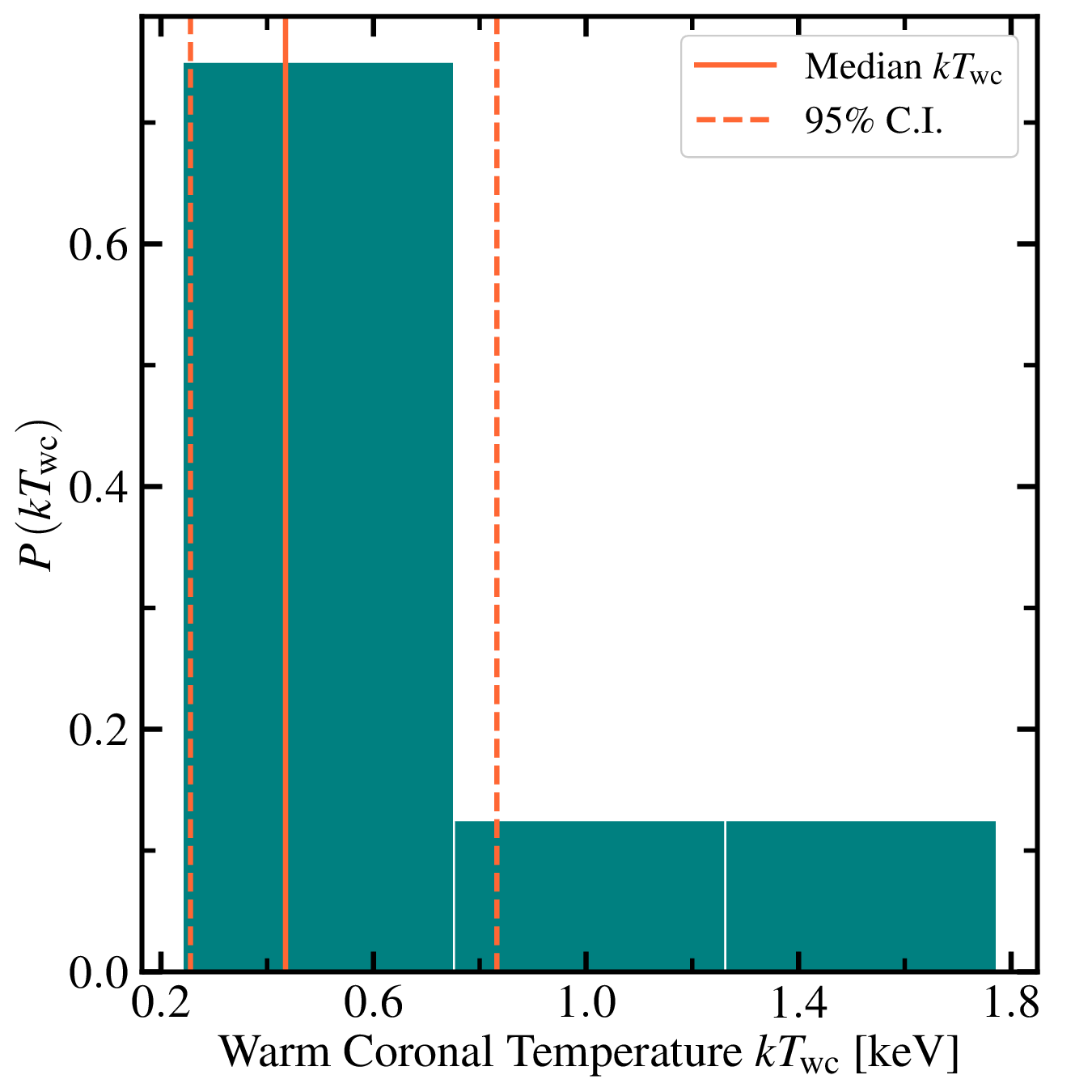}
\includegraphics[scale=0.38,angle=-0]{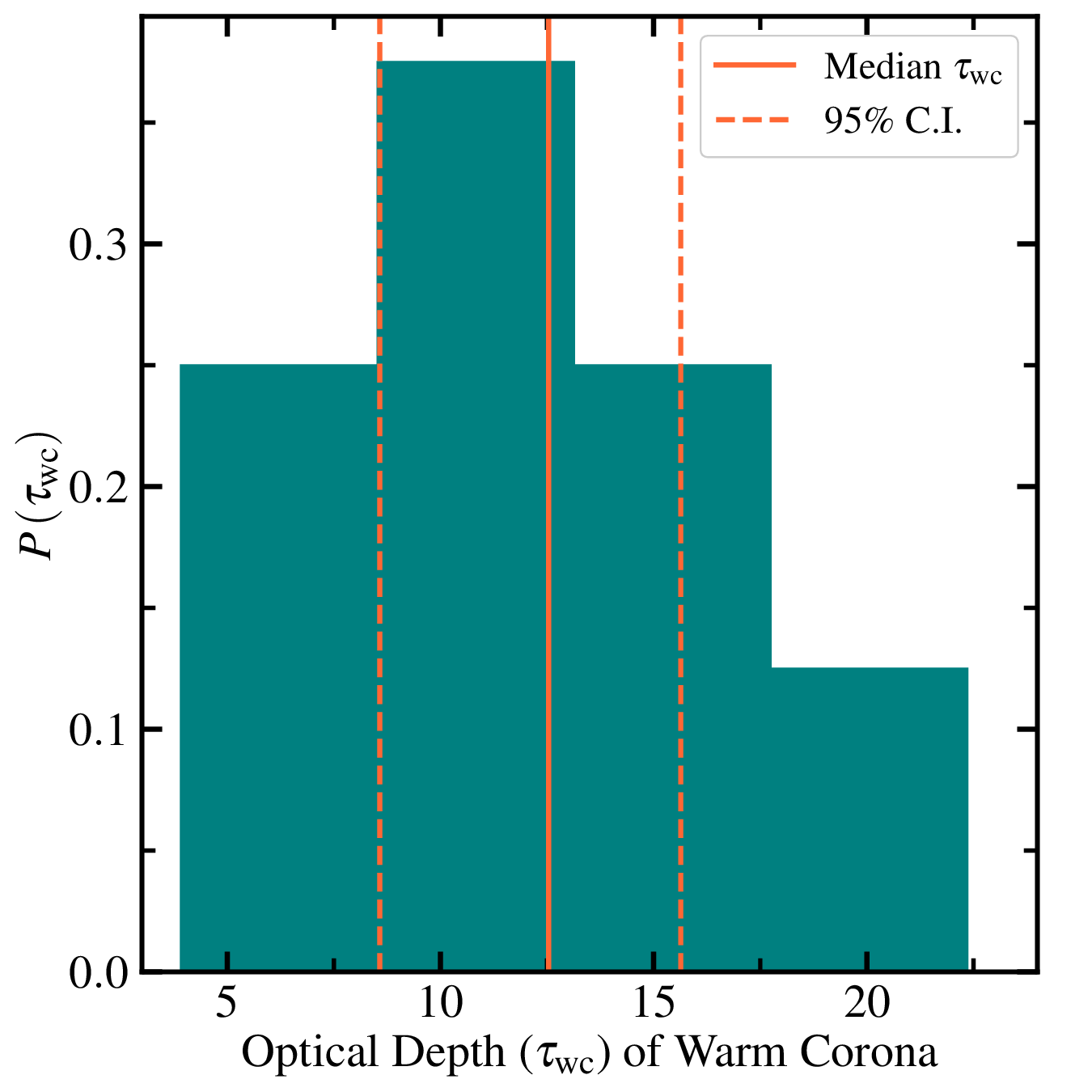}
\caption{Probability distribution of measured temperature and optical depth of the warm corona for 8 AGN in our sample that require an additional warm Comptonization to describe the observed soft X-ray excess. The sample median values of the warm coronal temperature and optical depth are $kT_{\rm wc}=0.43_{-0.18}^{+0.40}$\keV{} and $\tau_{\rm wc}=12.5_{-3.9}^{+3.1}$, respectively.} 
\end{center}
\label{kTs_tau_s}
\end{figure*}

Fig.~9 (right) shows the distribution of the $f$-parameter for the sample, with a median value of $f=0.68_{-0.25}^{+0.25}$ estimated at the $2\sigma$ confidence level. According to the density solution of SZ94 in equation~(\ref{eq1}), if all other input parameters are held constant, the black hole mass spanning $\log M_{\rm BH} \sim 5.5-9.0$ for the sample can provide approximately $2$ orders of magnitude variation in disk density. The dimensionless mass accretion rate ($\dot{m}$), ranging from $\sim 0.1-2.5$ for the sample, can alone offer about $3$ orders of magnitude variation in the measured density. However, the $f$-parameter varying in the range of $\sim 0.01-0.99$ can cause up to $6$ orders of magnitude variation in the density parameter, thereby justifying the wide range of fitted disk densities for the sample.

\subsection{Hot and Warm Coronal Properties}
The spectrum of the primary X-ray source or hot corona, which illuminates the accretion disk and produces reflection features in the {\tt relxillCp} and {\tt xillverCp} models, is described by the thermally Comptonized continuum model {\tt nthComp}. The key parameters of the {\tt nthComp} model are the photon index ($\Gamma$) of the primary continuum and the electron temperature ($kT_{\rm e}$) of the hot corona, both of which are presented in Table~\ref{table_output_parameters}. We first compare the hot coronal temperatures measured through our \xmm{}+\nustar{} broadband spectral modeling of the sample with previous measurements reported in the literature \citep{Ricci_2017,Akylas_Georg_2021,Kang_Wang_2022}, as shown in Figure~10. For AGN when only a cut-off energy ($E_{\rm c}$) estimate is available, we convert it to electron temperature using $kT_{\rm e}=E_{\rm c}/2.5$, since the cut-off energy of the primary continuum is typically estimated to be $\sim$2--3 times the electron temperature \citep{Petrucci_2001}. No temperature comparison is possible for PG~0844+349 and PG~1426+015, as their coronal temperature or cut-off energy measurements are not available in the literature. Our broadband spectroscopy provides the first measurement of the coronal temperature for these two AGN. As evident from Fig.~10, the electron temperatures measured in this work are consistent with those reported in the literature within their $2\sigma$ confidence intervals. The left panel of Fig.~11 presents our measured temperature distribution of the hot corona for the sample, which has a median value of $54_{-12}^{+11}$\keV{} at the $2\sigma$ confidence level.

Another physical parameter that characterizes the corona is its optical depth. However, optical depth is not a free parameter in the {\tt nthComp} model. Therefore, we need to estimate the optical depth ($\tau_{\rm e}$) of the hot corona, which is related to the electron temperature ($kT_{\rm e}$) and photon index ($\Gamma$) of the {\tt nthComp} model via the formula \citep{Shakura_Titarchuk_1980}:
\begin{equation}
\tau_{\rm e}=\sqrt{2.25+\frac{3}{\frac{kT_{{\rm e}}}{m_{{\rm e}}c^{2}}\times\left[(\Gamma+0.5)^{2}-2.25\right]}}-1.5
\end{equation}
An anti-correlation between $kT_{\rm e}$ and $\tau_{\rm e}$ is expected from equation~(2) itself. Therefore, we did not conduct a correlation analysis between these two parameters. Such an analysis is only meaningful when $\tau_{\rm e}$ is independently measured via models, as demonstrated by \cite{Tortosa_2018}. The calculated optical depth ($\tau_{\rm e}$) of the hot corona for our sample lies in the range of $\sim 0.15-1.86$, with a mean value of $\sim 1.0$. This range aligns well with the optical depth range of $\sim 0.19-1.78$ inferred for the AGN sample in \cite{Fabian_2015}. The distribution of hot coronal optical depths for our AGN sample is presented in Fig.~11 (right), showing a median value of $0.98_{-0.28}^{+0.22}$ at $2\sigma$ confidence. The optical depth range inferred for the hot corona in our study is reasonable, as demonstrated by the numerical simulations of \cite{Haardt_Maraschi_1993}. Furthermore, to clearly observe the relativistic reflection features from the inner disk, the coronal optical depth is required to be less than or close to unity, as argued by \cite{Fabian_1994}.

\begin{table*}
\centering
\caption{The best-fit source spectral models and spin measurements obtained from both hard X-ray (3--78\keV{}) and broadband (0.3--78\keV{}) relativistic reflection spectroscopy of the sample. Columns~(2) and (3) list the best-fit hard X-ray source spectral model and the corresponding spin parameter obtained from the hard X-ray spectral modeling. Columns~(4) and (5) display the broadband best-fit source spectral model and the spin parameter determined from the broadband spectral modeling for each source. } 
\begin{center}
\scalebox{0.9}{%
\begin{tabular}{cccccccccccc}
\hline 
Source  & 3--78\keV{} Best-fit Model & $a^{\ast}_{\rm [3-78~keV]}$  & 0.3--78\keV{} Best-fit Model &  $a^{\ast}_{\rm [0.3-78~keV]}$   \\ [0.1cm]
(1)    &   (2)   &   (3)   &   (4)  &  (5)  \\     [0.1cm]                                                  
\hline 

UGC 6728 &  {\tt relxillCp+nthComp}  & $0.658_{-0.518}^{+0.249}$ & {\tt relxillCp+nthComp}  & $0.730_{-0.610}^{+0.114}$ \\ [0.15cm]
         
Mrk 1310 &  {\tt relxillCp+nthComp}  & $0.736_{-0.606}^{+0.219}$ & {\tt compTT+relxillCp+nthComp} & $0.752_{-0.605}^{+0.190}$ \\ [0.15cm]
         
NGC 4748 & {\tt relxillCp+xillverCp+nthComp} & $0.705_{-0.573}^{+0.240}$ & {\tt compTT+relxillCp+xillverCp+nthComp} & $0.796_{-0.663}^{+0.123}$ \\ [0.15cm]
         
Mrk 110 & {\tt relxillCp+xillverCp+nthComp} & $\ge 0.970$ & {\tt compTT+relxillCp+xillverCp+nthComp} & $0.978_{-0.024}^{+0.016}$ \\ [0.15cm]
         
Mrk 279 & {\tt relxillCp+xillverCp+nthComp} & $\ge 0.968$ & {\tt compTT+relxillCp+xillverCp+nthComp} & $\ge 0.991$ \\ [0.15cm]
         
Mrk 590 & ({\tt relxillCp+zGauss\_N+zGauss2\_N+nthComp} & $0.901_{-0.130}^{+0.048}$ & ({\tt relxillCp+zGauss\_N+zGauss2\_N+nthComp} & $0.890_{-0.093}^{+0.041}$ \\ [0.15cm]
                 
Mrk 79 & {\tt relxillCp+xillverCp+nthComp} & $0.944_{-0.762}^{+0.038}$ & {\tt compTT+relxillCp+xillverCp+nthComp} & $0.880_{-0.039}^{+0.048}$ \\ [0.15cm]
         
PG 1229+204 &  {\tt relxillCp+nthComp}  & $0.676_{-0.553}^{+0.226}$ &  {\tt relxillCp+nthComp}  & $0.974_{-0.775}^{+0.024}$ \\ [0.15cm]
         
PG 0844+349 &  {\tt relxillCp+nthComp}  & $\le 0.968$ &  {\tt compTT+relxillCp+nthComp} & $0.292_{-0.157}^{+0.555}$ \\ [0.15cm]
         
PG 0804+761 & {\tt relxillCp+zGauss\_N+nthComp}  & $\le 0.931$ & {\tt compTT+relxillCp+zGauss\_N+nthComp} & $\le 0.706$ \\ [0.15cm]
         
PG 1426+015 & {\tt relxillCp+xillverCp+nthComp} & $0.460_{-0.297}^{+0.463}$ & {\tt compTT+relxillCp+xillverCp+nthComp} & $0.441_{-0.122}^{+0.488}$ \\ [0.15cm]

\hline

\end{tabular}}
\end{center} 
\label{spin_hard_full_bands}           
\end{table*}

\begin{figure*}
\centering
\begin{center}
\includegraphics[scale=0.28,angle=-0]{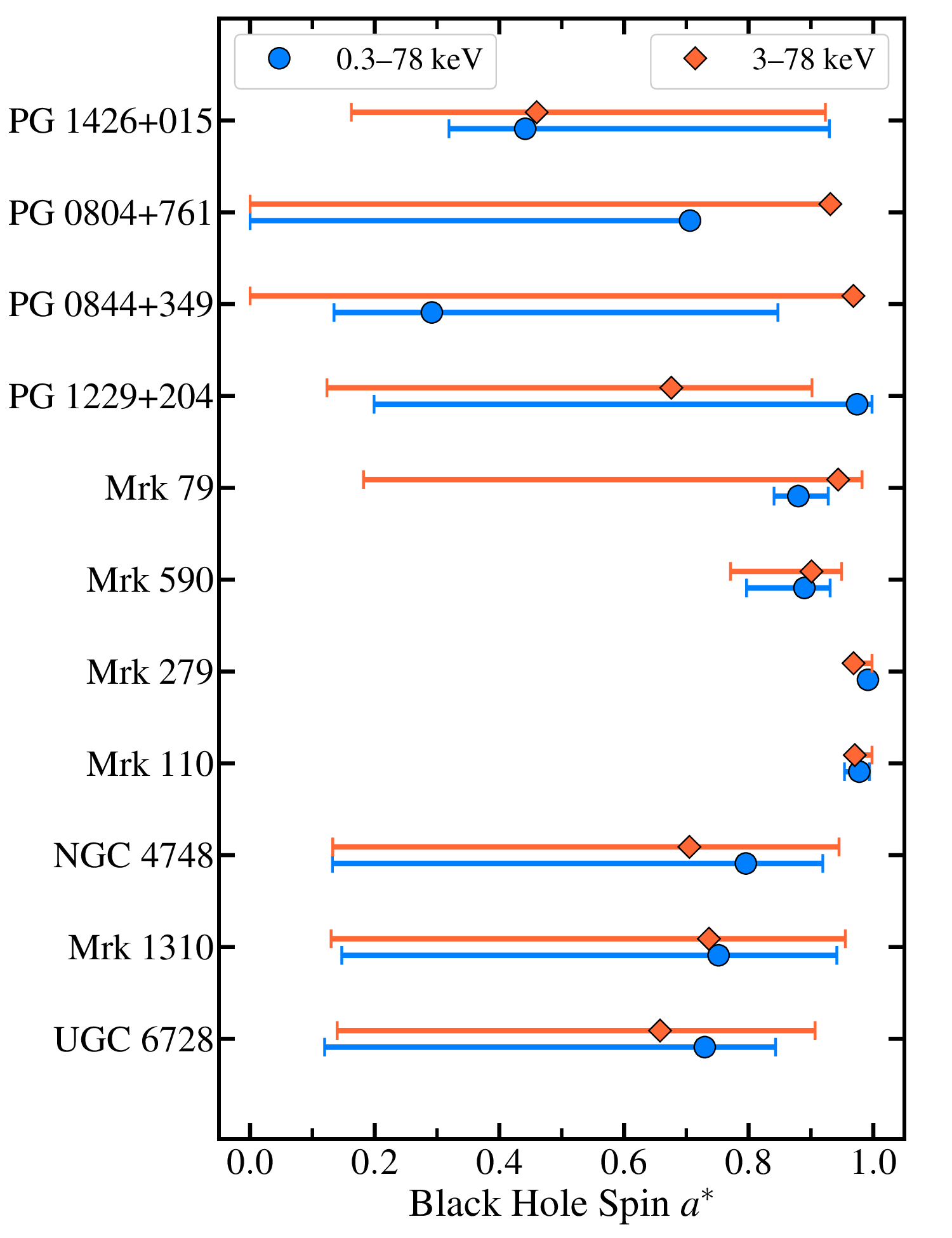}
\includegraphics[scale=0.28,angle=-0]{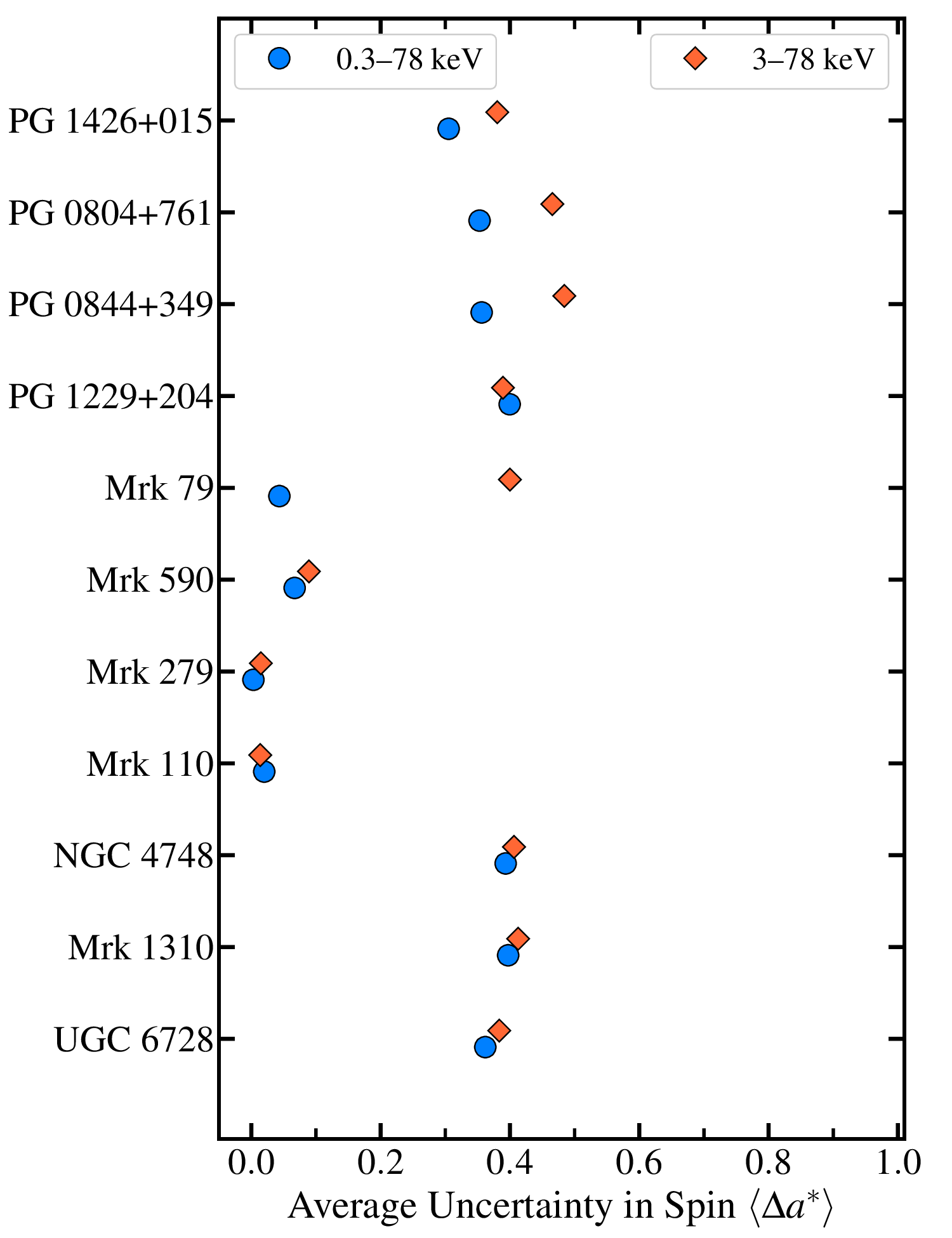}
\caption{The left panel presents black hole spins measured using hard X-ray (3--78\keV{}; orange diamonds) and broadband (0.3--78\keV{}; blue circles) relativistic reflection spectroscopy. All errors are measured through MCMC at the 90\% confidence level. Details of the best-fit spectral models for both energy ranges are provided in Table~III. The right panel shows the average uncertainties for spin measurements in each case.}
\end{center}
\label{compare_spin}
\end{figure*}

The physical origin of the warm corona remains an open question. It may correspond to an anomalously heated surface layer of the accretion disk, as suggested by \cite{Kawanaka_Mineshige_2024}. In the warm Comptonization model ({\tt CompTT}), optical/UV seed photons from the accretion disk are Compton up-scattered in an optically thick, low-temperature corona, producing excess emission in the soft X-ray band of AGN. The key parameters of the {\tt CompTT} model are the temperature ($kT_{\rm wc}$) and optical depth ($\tau_{\rm wc}$) of the warm corona, which are listed in Table~\ref{table_output_parameters} for the 8 AGN in our sample that require an additional warm Comptonization component to model the observed soft X-ray excess. The distribution of the measured temperature of the warm corona for these 8 AGN is presented in Fig.~12 (left), revealing a median value of $kT_{\rm wc}=0.43_{-0.18}^{+0.40}$\keV{} at the $2\sigma$ confidence level. The right panel of Fig.~12 shows the distribution of the measured optical depth for the warm corona in these 8 AGN, which has a median value of $\tau_{\rm wc}=12.5_{-3.9}^{+3.1}$ at $2\sigma$ confidence. Our sample spans almost the full range of temperature and optical depth allowed for the warm corona.

\begin{figure*}
\centering
\begin{center}
\includegraphics[scale=0.38,angle=-0]{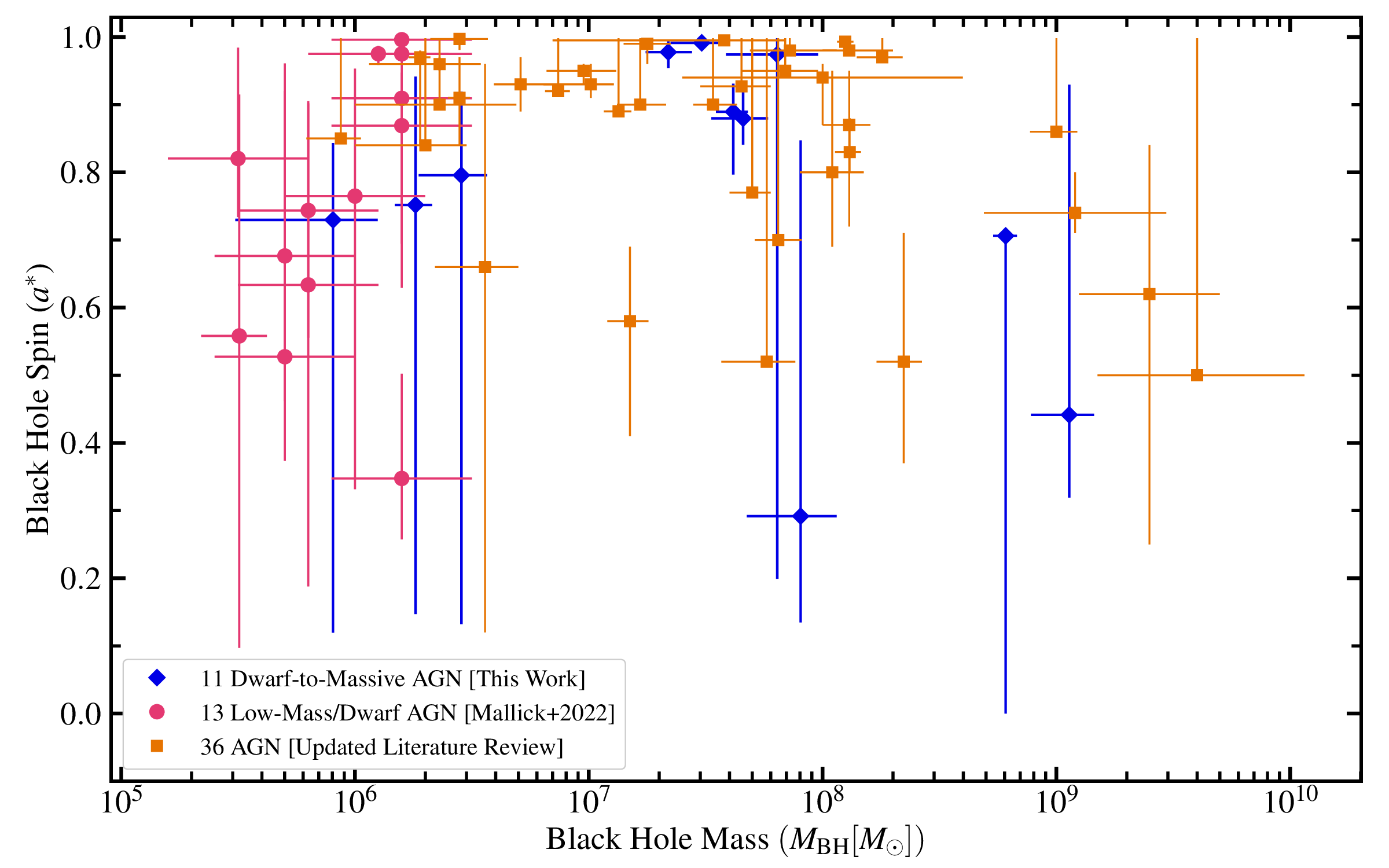}
\caption{Evolution of dimensionless black hole spin as a function of black hole mass constructed using the most updated spin and mass measurements. The blue diamonds show the spin measurements of the 11 AGN from this work. The pink circles represent spin measurements of low-mass AGN (including dwarf AGN) from \cite{Mallick_2022}, while the orange squares denote spin measurements compiled from the updated literature review. The error bars on the spin values indicate the 90\% confidence intervals.}
\end{center}
\label{spin}
\end{figure*}

\begin{figure*}
\centering
\begin{center}
\includegraphics[scale=0.38,angle=-0]{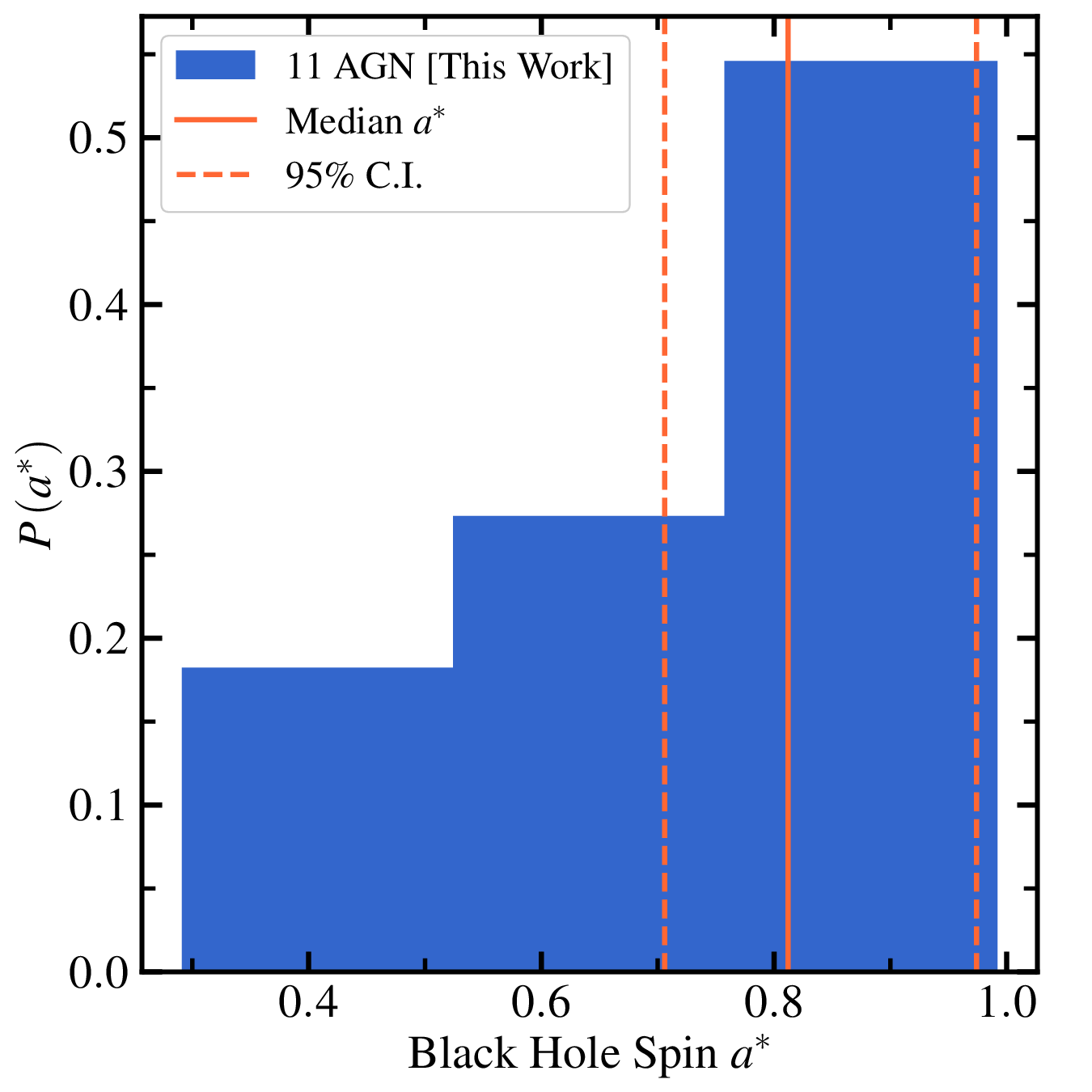}
\includegraphics[scale=0.38,angle=-0]{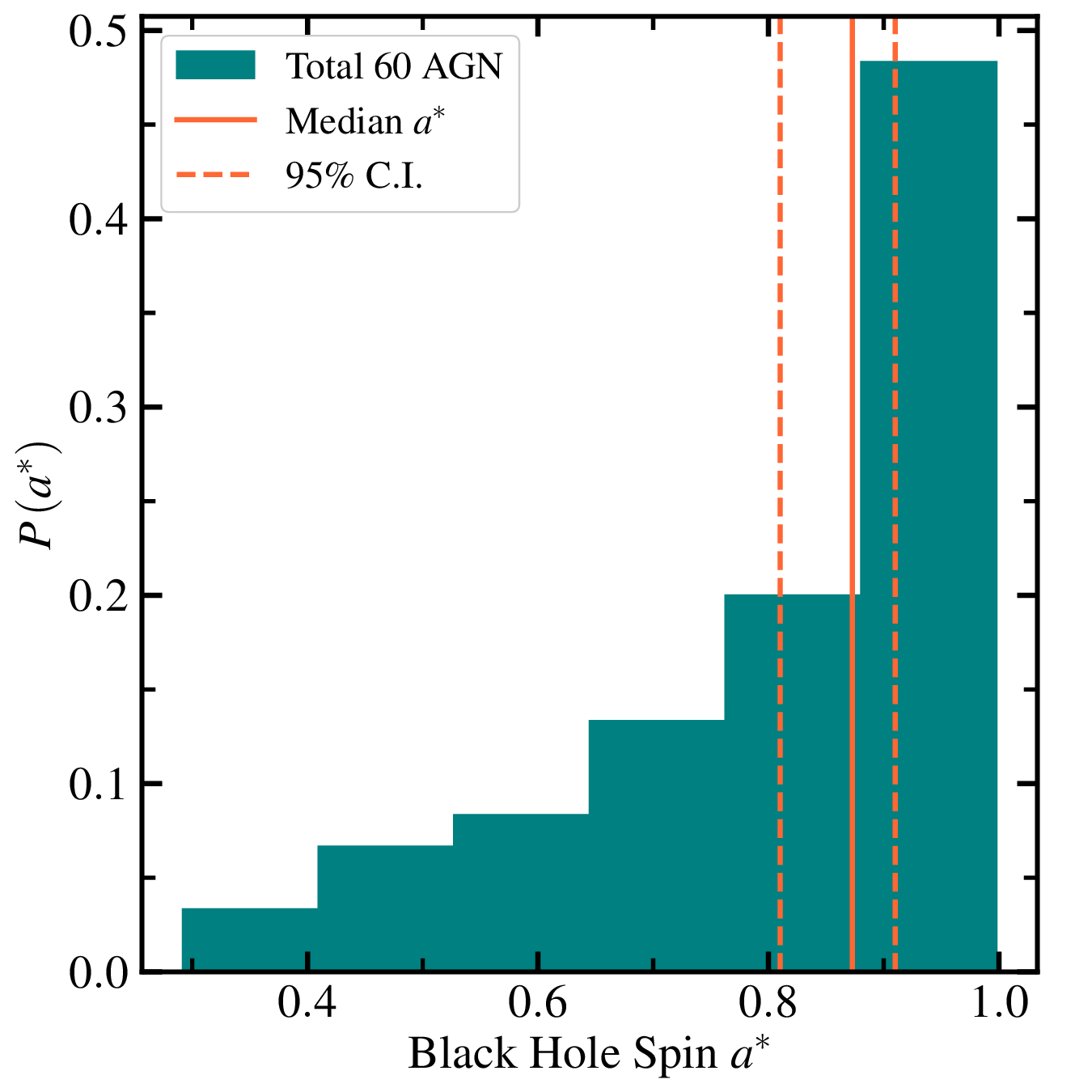}
\caption{Probability distribution of black hole spin for 11 AGN from this work and all 60 AGN with updated spin measurements, including those 11 AGN. Our spin measurements are increasing or refining the available spin population of AGN by $\sim 20$\%. The sample medians of the spin parameter for the 11 AGN in our sample and the total 60 AGN are $0.81_{-0.11}^{+0.16}$ and $0.87_{-0.07}^{+0.04}$, respectively, at the 95\% confidence level.}
\end{center}
\label{distribution_spin}
\end{figure*}

\begin{figure*}
\centering
\begin{center}
\includegraphics[scale=0.255,angle=-0]{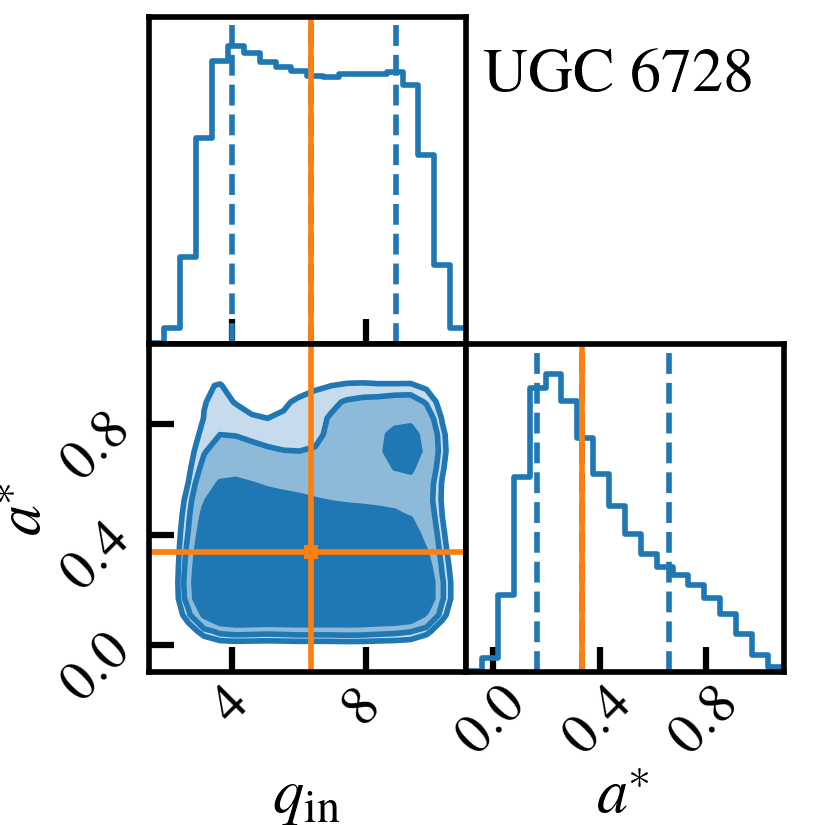}
\includegraphics[scale=0.255,angle=-0]{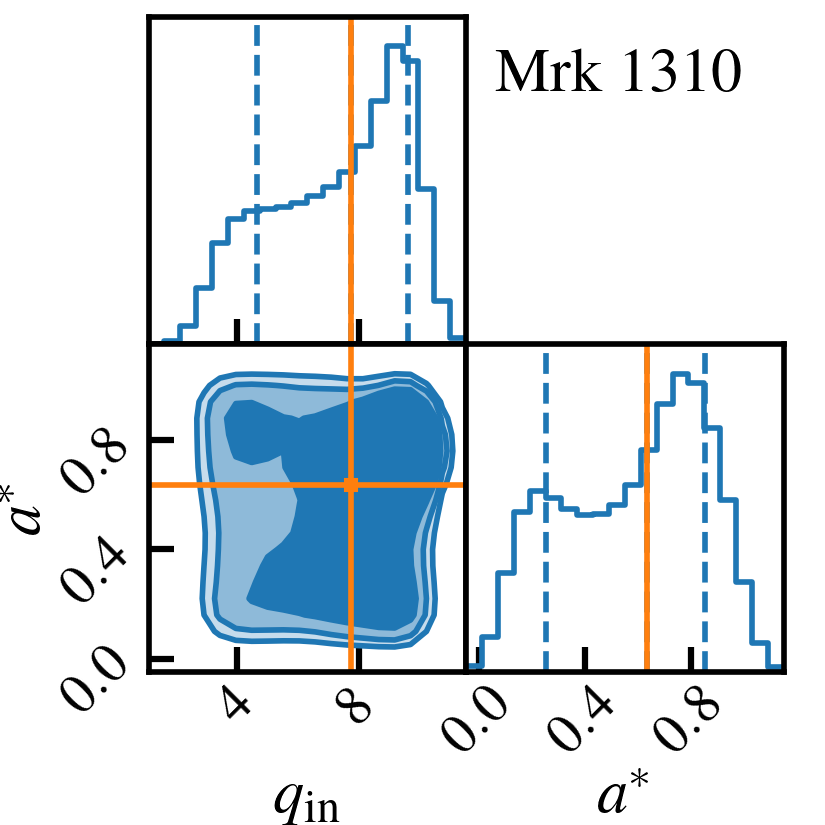}
\includegraphics[scale=0.255,angle=-0]{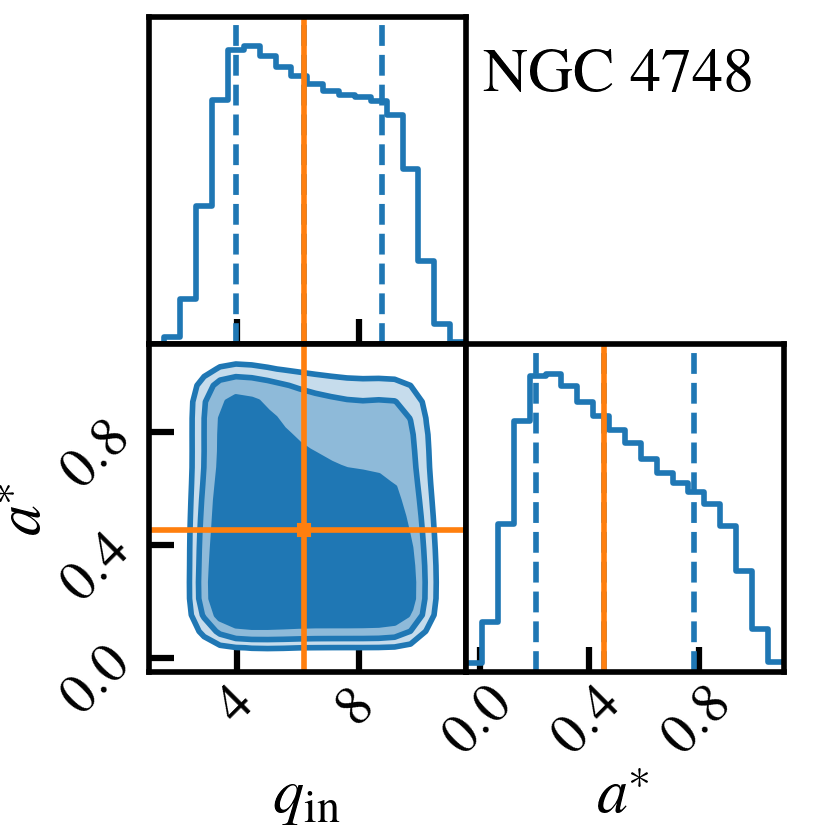}
\includegraphics[scale=0.255,angle=-0]{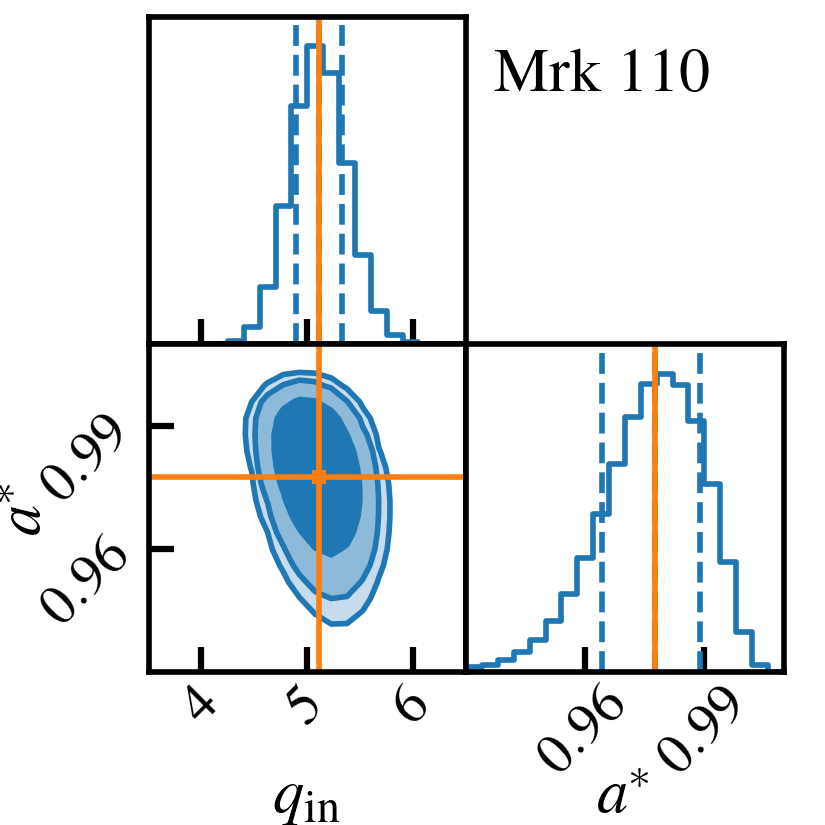}
\includegraphics[scale=0.255,angle=-0]{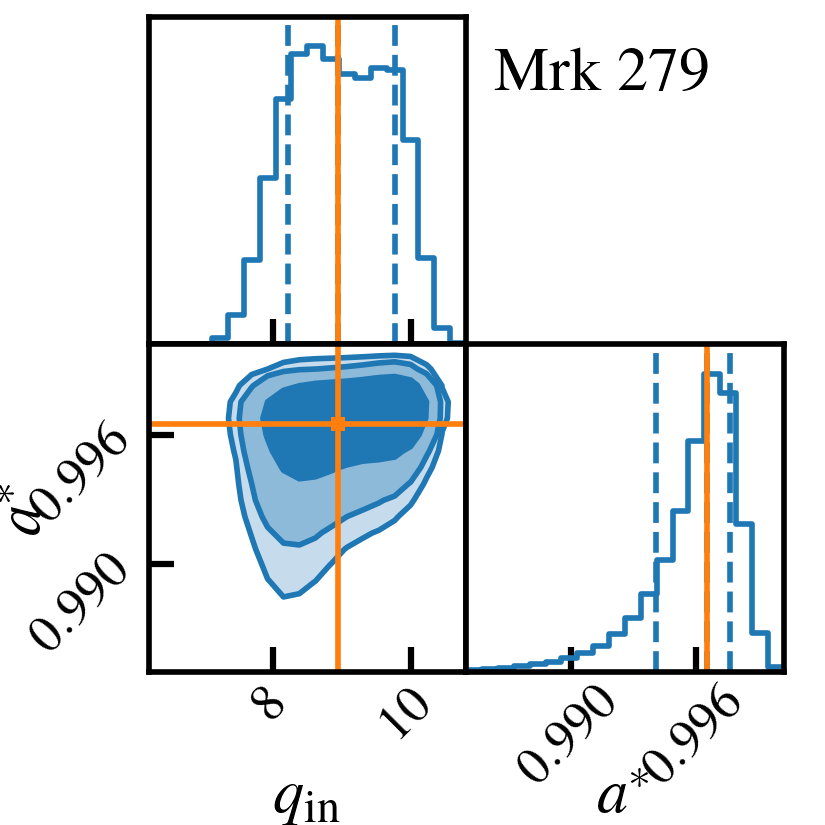}
\includegraphics[scale=0.255,angle=-0]{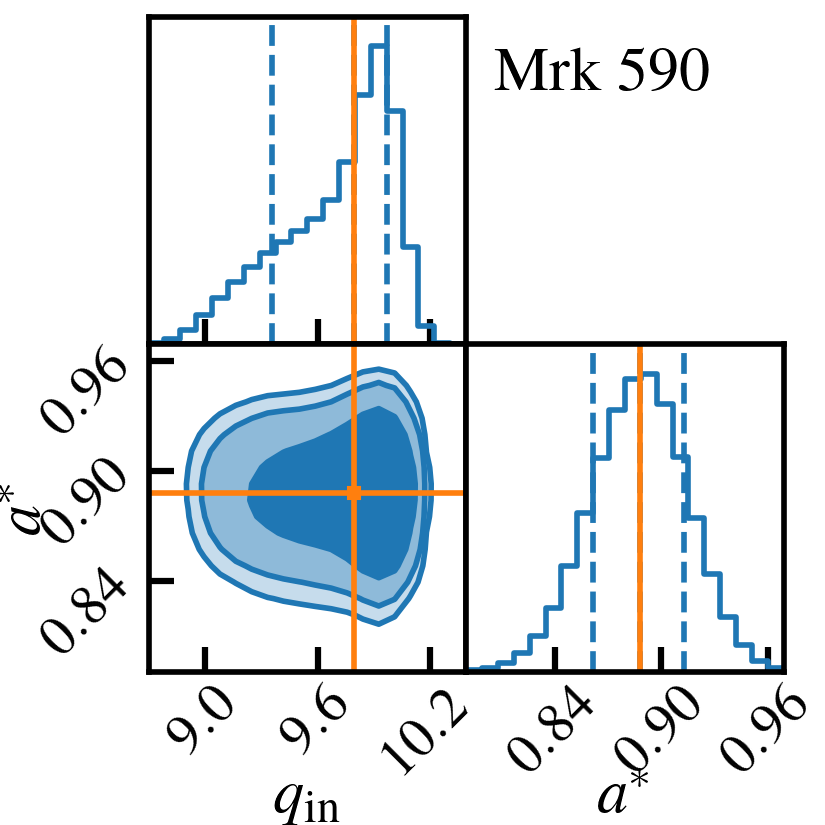}
\includegraphics[scale=0.255,angle=-0]{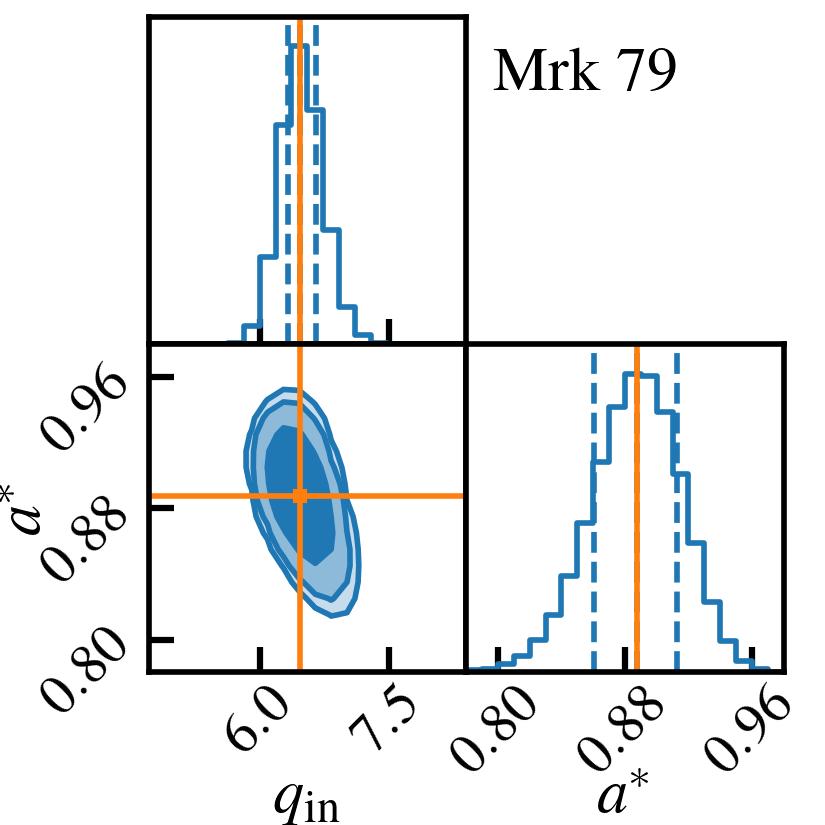}
\includegraphics[scale=0.255,angle=-0]{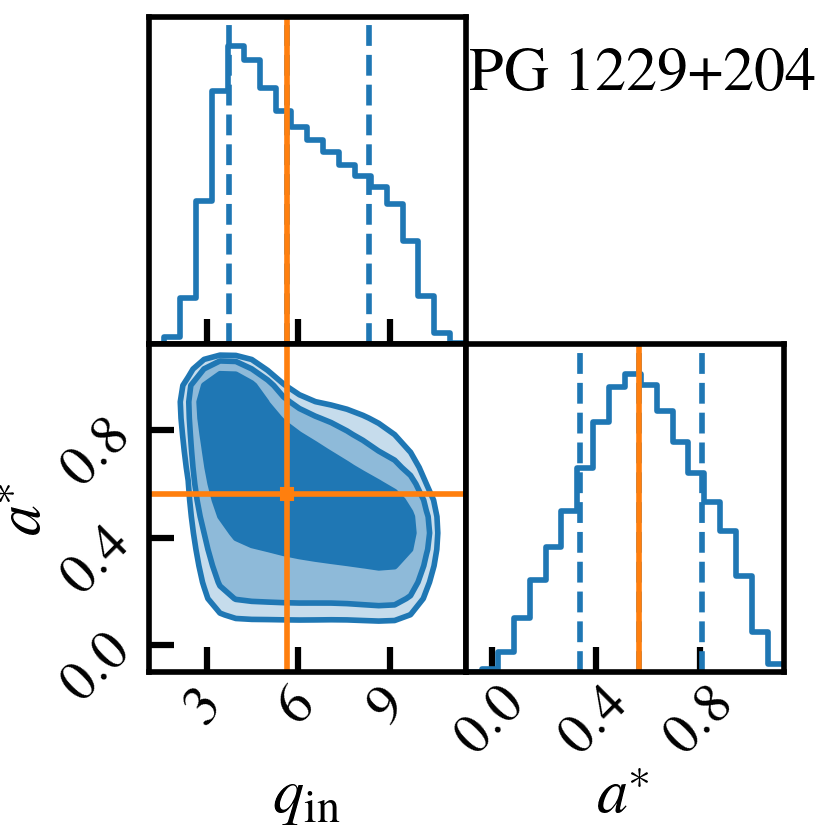}
\includegraphics[scale=0.255,angle=-0]{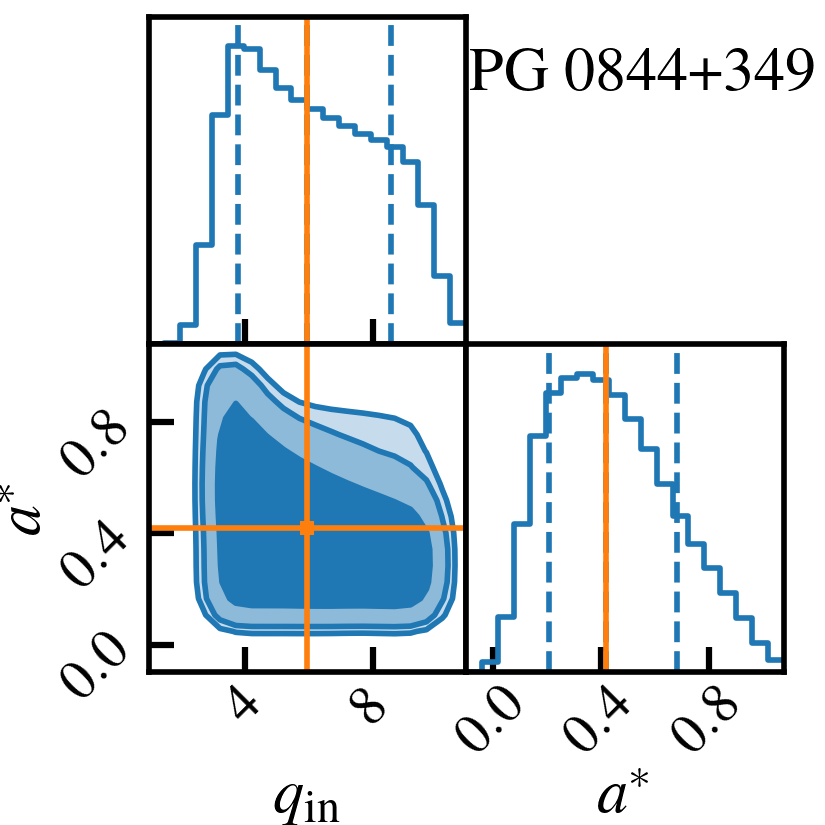}
\includegraphics[scale=0.255,angle=-0]{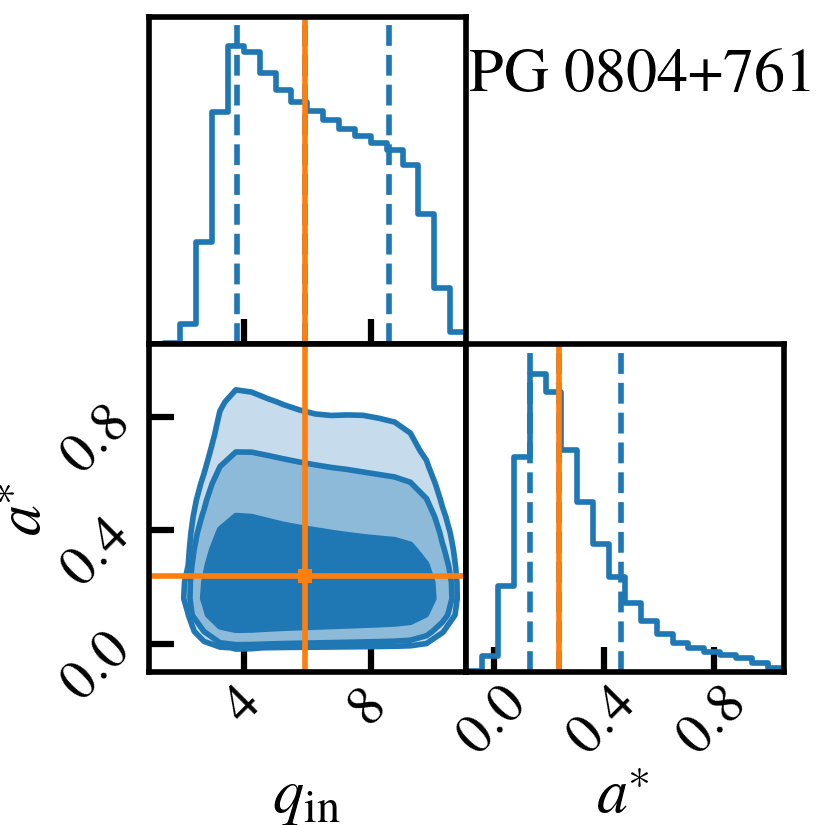}
\includegraphics[scale=0.255,angle=-0]{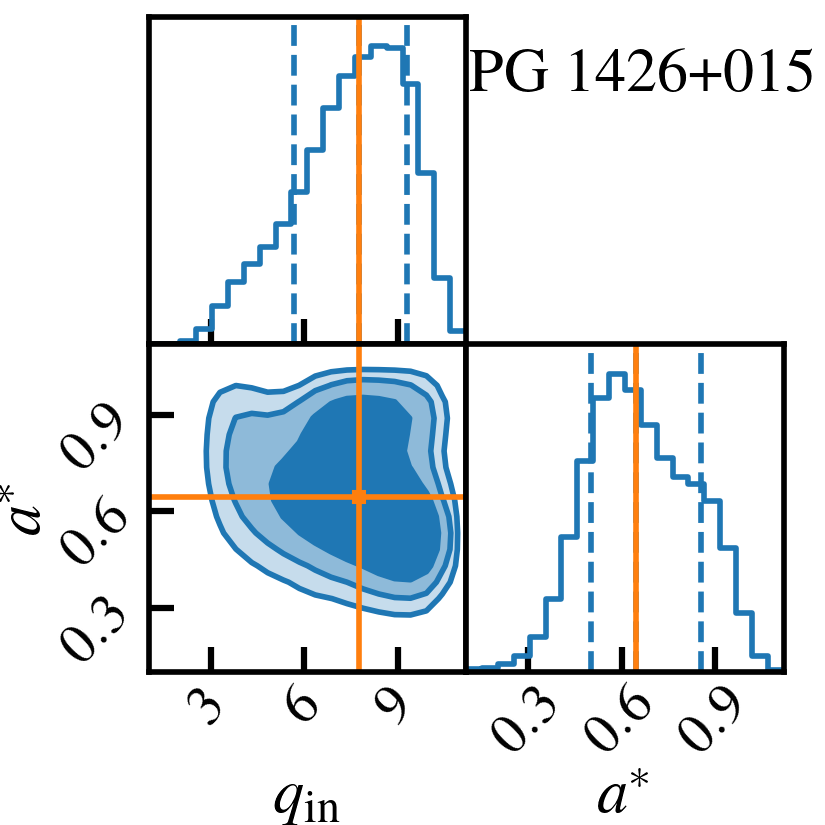}
\caption{Corner plots between dimensionless black hole spin ($a^{\ast}$) and inner emissivity index ($q_{\rm in}$) of the sample. Vertical lines show the median and 68.3\% confidence intervals. The dark, medium, and light blue areas represent 68.3\%, 90\%, and 95\% confidence levels, respectively, with square symbols indicating the parameter medians.}
\end{center}
\label{mcmc_plot_qin_spin}
\end{figure*}

\subsection{Black Hole Spin Measurements from the Relativistic Reflection Features}
\label{sec:spin}
We measure black hole spin by modeling relativistic disk reflection features, primarily the broad Fe~K emission line and soft X-ray excess, using the {\tt relxillCp} model under the assumption that emission originates from the innermost stable circular orbit. The accuracy of spin measurements mainly depends on the red wing of the broad Fe~K emission line. Since the soft X-ray band is contaminated by warm coronal emission in most sources, we first estimate the spin parameter by modeling the hard X-ray (3--78\keV{}) spectra, which contain the broad Fe~K emission feature, for all sources in the sample, as quantitatively demonstrated in Table~\ref{test_broad_FeK}. Additionally, with the use of \nustar{} data above 10\keV{}, we obtain better constraints on the hard X-ray continuum and detect Compton reflection hump, which can provide a more accurate determination of the red wing of the broad Fe~K emission line. The black hole spin values obtained from the best-fit hard X-ray (3--78\keV{}) spectra and their 90\% credible intervals determined through MCMC are presented in Table~III and shown as orange diamonds in the left panel of Figure~13.

Although warm Comptonization ({\tt compTT}) dominates the soft X-ray (0.3--2\keV{}) band for some sources (e.g. PG~1426+015), relativistic disk reflection ({\tt relxillCp}) also contributes to this excess emission (see Fig.~4). The best-fit spin values measured through broadband (0.3--78\keV{}) spectral modeling and corresponding 90\% credible intervals obtained from MCMC analyses of the sample are presented in Table~III and depicted by blue circles in the left panel of Fig.~13. The average uncertainties in the black hole spin parameter, as measured from the hard X-ray and broadband spectral modeling of the sample, are shown in the right panel of Fig.~13. Overall, the inclusion of soft X-ray spectra provides tighter constraints on the spin parameter for the sample, notably for the 4 AGN: Mrk~79, PG~0844+349, PG~0804+761, and PG~1426+015. In these four AGN, warm coronal emission was present in the soft X-ray band in addition to the relativistic disk reflection. However, Mrk~79 exhibits relativistic disk reflection-dominated soft X-ray spectra (Fig.~4), which substantially reduces spin uncertainties and significantly refines the measured spin through broadband spectroscopy (Fig.~13).

In Fig.~14, we present the dimensionless black hole spin parameter as a function of black hole mass for 11 AGN measured using our broadband spectral modeling, together with the 13 low-mass dwarf AGN from \cite{Mallick_2022} and 36 AGN with the most recent reliable spin-mass measurements. With the addition of 11 new, refined spin measurements through broadband \xmm{}+\nustar{} relativistic reflection spectroscopy, the total spin sample size has reached $N=60$. Thus, this work is increasing or refining the available spin measurements in the AGN population by approximately 20\%. The distribution of black hole spins for our sample is shown in Fig.~15 (left panel), which has a median of $0.81_{-0.11}^{+0.16}$ at the $2\sigma$ confidence level. The right panel of Fig.~15 shows the spin distribution for all $60$ AGN, where the median is $0.87_{-0.07}^{+0.04}$ at $2\sigma$ confidence.

The measurement of high spins is not the shortcoming of the relativistic reflection model. We have previously measured the black hole spin parameter in faint dwarf AGN using the same relativistic reflection spectroscopy and found a range of spin values, from low and moderate to high \citep{Mallick_2022}. A fundamental parameter governed by black hole spin is the radiative efficiency ($\eta$) of the accretion flow, given by
\begin{equation}
\eta=1-\left(1-\frac{2r_{\rm g}}{3r_{\rm isco}}\right)^{1/2}.
\end{equation}
For prograde orbits restricted to $\theta=\pi/2$ plane, the formula for $r_{\rm isco}$ as derived by \cite{Bardeen_1972} is: 
\begin{equation}
r_{{\rm isco}}=\left(3+Z_{2}-\left[(3-Z_{1})(3+Z_{1}+2Z_{2})\right]^{1/2}\right)r_{{\rm g}},
\end{equation}
where,  
\begin{equation}
Z_{1}=1+(1-{a^{\ast}}^{2})^{1/3}\left[(1+a^{\ast})^{1/3}+(1-a^{\ast})^{1/3}\right],
\end{equation}

\begin{equation}
Z_{2}=\left(3{a^{\ast}}^{2}+Z_{1}^{2}\right)^{1/2}.
\end{equation}
As evident from the above equations, radiative efficiency is purely a function of black hole spin. AGN with high spins ($a^{\ast}>0.9$) exhibit high radiative efficiency ($\eta>0.2$). In addition to spin, radiative efficiency directly influences the luminosity of an accreting object. For a steady-state accretion flow, the total radiated luminosity is $L=\eta \dot{M}c^{2}$, where $\dot{M}$ is the accretion rate. Highly spinning black holes have high radiative efficiency, which makes them more luminous even if they have a comparable accretion rate. Therefore, accreting black holes with high spins are more likely to dominate flux-limited samples (e.g. \cite{Vasudevan_2016}).

As a final note, we emphasize that in the relativistic reflection method for spin estimation, high spin measurements are likely to be more reliable than low spin measurements. This is because the method requires blurred reflection features, ideally the broadened Fe~K line, which can only be produced when the inner edge of the disk is very close to the black hole and the corona is compact or located near the black hole. If the Fe~K line is not very broadened, the black hole may not have a high spin even if the corona is compact. Alternatively, the corona may not be compact or located near the black hole, even if the black hole has a high spin value. In essence, spin measurements are influenced by the inner edge of the illuminated part of the disk, while intrinsic spin governs the true inner edge of the disk. Therefore, a correlation between measured spin and coronal compactness may indicate that we are actually sampling the illuminated inner edges of the disks rather than their true ISCOs. To evaluate the reliability of our spin measurements, we search for any correlations between black hole spin and the inner emissivity index, which serves as a proxy for the corona's compactness or proximity to the black hole. In Figure 16, we present contour plots of spin versus the inner emissivity index for our sample, which show no significant correlation or degeneracy between the two parameters. This finding demonstrates that our spin measurements accurately probe the true ISCOs of the central SMBHs in these 11 AGN.

\section{Summary \& Conclusions}
\label{sec:conclusion}
In this paper, we apply the updated higher-density disk reflection model {\tt relxillCp} to joint \xmm{}+\nustar{} broadband spectra of a sample of Type-1 AGN spanning nearly the complete range of central black hole masses from $M_{\rm BH}\sim10^{5.5}M_{\odot}$ to $10^{9}M_{\odot}$. We systematically investigate the origin of both hard and soft X-ray excess in all sources of the sample, and assess the relevance of both high-density disk reflection and warm Comptonization for the observed soft X-ray excess using a Bayesian framework. Additionally, we compute the disk-to-corona power transfer fraction for the first time in any accreting system. Finally, we constrain the SMBH spin parameter across mass scales using both hard X-ray (3--78\keV{}) and broadband (0.3--78\keV{}) relativistic reflection spectroscopy. The main results and conclusions of our work are summarized below:

\begin{enumerate}
\item
The relativistic reflection model, incorporating a variable-density accretion disk with a broken power-law emissivity profile, can describe the soft X-ray excess, together with the broad Fe~K emission line and Compton hump in 3 out of 11 AGN, namely UGC~6728, Mrk~590, and PG~1229+204. For the remaining 8 sources, an additional low-temperature or warm Comptonization component is necessary to accurately model the soft X-ray excess, which suggests its hybrid origin.

\item 
The measured temperature and optical depth of the warm corona for these 8 AGN in our sample are $kT_{\rm wc}\sim 0.2-2$\keV{} and $\tau_{\rm wc}\sim 4-22$, respectively, spanning nearly the entire theoretically allowed range for the warm corona. The median values are $kT_{\rm wc}=0.43_{-0.18}^{+0.40}$\keV{} and $\tau_{\rm wc}=12.5_{-3.9}^{+3.1}$, respectively. 
Among these 8 AGN, the measured disk density is significantly higher than the canonical value of $n_{\rm e}=10^{15}$~cm$^{-3}$, even with the presence of a warm corona in 3 AGN (Mrk~110, Mrk~279, and Mrk~79). For the remaining 5 AGN (Mrk~1310, NGC~4748, PG~0844$+$349, PG~0804$+$761, and PG~1426$+$015), the lower limit of the disk density is consistent with the canonical value.

\item 
The inner accretion disk of the AGN sample is found to be ionized and dense with a median ionization of $\xi \sim 10^{2}$~erg~cm$^{-2}$~s$^{-1}$ and a median density of $n_{\rm e} \sim 10^{16.5}\rm{cm}^{-3}$ without requiring a very high super-solar iron abundance. The iron abundance of the accretion disk is near-solar for the sample, with a median value of $\sim 2$ relative to the solar abundance.

\item
We did not find any anti-correlation between disk density and either black hole mass times the accretion rate squared or black hole mass alone. This indicates that the intrinsic scatter in the fraction of power transferred from the accretion disk to the corona is substantial, and our sample spans a broad range of disk-to-corona power transfer fractions.

\item
For the first time, we calculate the disk-to-corona power transfer fraction for each source, finding a sample median of $f=0.68_{-0.25}^{+0.25}$ at $2\sigma$ confidence. Importantly, we observe a strong positive correlation between the $f$-parameter and black hole mass times the accretion rate squared, $\log(M_{\rm BH}\dot{m}^{2})$, which is consistent with the theoretical prediction of a radiation pressure-supported accretion disk.

\item
The coupling between the accretion disk and hot corona is directly evident from the fraction of the disk power transferred into the hot corona. The measured electron temperature and optical depth of the hot corona have medians of $kT_{\rm e}=54_{-12}^{+11}$\keV{} and $\tau_{\rm e}=0.98_{-0.28}^{+0.22}$, respectively, for the sample. The hot corona has a reasonably low optical depth to make the relativistic reflection from the inner accretion disk observable.

\item

By utilizing joint \xmm{}+\nustar{} relativistic reflection spectroscopy, we are increasing or refining the AGN population for which a spin measurement is available by around 20\% across the mass range of $\log M_{\rm BH} \sim 5.5-9.0$. This allows us to achieve a total spin sample of 60 AGN. The median spins for 11 AGN in our sample and all 60 AGN are $0.81_{-0.11}^{+0.16}$ and $0.87_{-0.07}^{+0.04}$, respectively, at $2\sigma$ confidence.

\end{enumerate}

\section{Future Prospects}
\label{sec:future_prospect}
The RELXILL family of reflection models requires further development to provide a more robust determination of the parameter space and the origin of the soft X-ray excess. Currently, the {\tt relxillCp} model assumes a broken power-law emissivity profile for the coronal illumination of the accretion disk. However, this assumption has not yet been fully validated due to uncertainties regarding the coronal geometry. Future observations with \ixpe{} \citep{Weisskopf_2022} will be crucial for constraining the geometry of the AGN corona in this context.

The other noteworthy simplification in current reflection models is the assumption that the density and ionization structure remain uniform across the entire disk. In reality, accretion disks are likely to exhibit variations in density and ionization structure, both radially and vertically. Thus, determining the origin of the soft X-ray excess will require the development of a disk reflection model that accounts for radial and vertical variations in disk density and ionization state. In addition to refining these models, we must design advanced X-ray telescopes capable of detecting the features in X-ray spectra predicted by these more realistic models.

Next-generation X-ray observatories, such as \athena{} \citep{Cruise_2025}, \colibri{} \citep{Heyl_2020}, and \heroix{} (Thomas et al., in prep) will significantly advance the scope of this study in multiple ways. Overall, these future missions will allow similar analyses to be performed at much higher redshifts, with better accuracy in parameter estimation and faster processing.

\athena{} with its advanced sensitivity in soft X-rays and high spatial resolution will extend the results of this study to fainter AGN populations, thereby probing the disk-to-corona power transfer in lower luminosity regimes, even for non-central, gas-starved supermassive black holes (see e.g. \cite{DiMatteo_2023}). \athena{} and \heroix{} will jointly provide tighter constraints on the temperatures of warm and hot coronae by observing a larger and more diverse sample of AGN across a wide range of redshifts and mass scales. They will play a pivotal role in refining spin measurements for a significantly larger number of supermassive black holes (e.g. \cite{Cappelluti_2024}), especially for high(er)-redshift AGN, which remain underrepresented in current spin demographic studies. Those constraints will also be crucial to improving our knowledge of the population of black hole seeds (e.g. \cite{Pacucci_Loeb_2022}), typically formed at redshifts $z=20-30$ \citep{Barkana_Loeb_2001}.

Both \athena{} and \colibri{} will be instrumental in testing competing models for the soft X-ray excess (high-density disk reflection versus warm Comptonization) due to their superior capability to resolve these components spectroscopically. The determination of the X-ray spectral energy distribution of the faint AGN population recently discovered by \jwst{} (see e.g. \cite{Harikane_2023}) will also open up new possibilities to investigate peculiar spectral shapes and assess the spectral impact of super-Eddington accretors as recently investigated by several studies (see e.g. \cite{Pacucci_Narayan_2024}).

In summary, several future X-ray facilities will offer immense possibilities for enhancing and expanding the work performed in this paper.

\section*{Acknowledgements}
L.M. acknowledges support from her CITA National Fellowship (reference \#DIS-2022-568580) administered by the University of Toronto. L.M. also acknowledges unlimited High-Performance Computing (HPC) support from CITA. 

C.P. acknowledges support from PRIN MUR SEAWIND funded by European Union - NextGenerationEU and INAF Grant BLOSSOM. 

A.G.M. acknowledges partial support from Narodowe Centrum Nauki (NCN) grants 2018/31/G/ST9/03224 and 2019/35/B/ST9/03944. 

S.S.H. acknowledges support from NSERC's CRC and Discovery Grant Programs. 

F.P. acknowledges support from a Clay Fellowship administered by the Smithsonian Astrophysical Observatory. F.P. also acknowledges support from the BHI Fellowship administered by the Black Hole Initiative at Harvard University.

L.M. thanks Javier Garcia, Jiachen Jiang, Dan Wilkins, and Luigi Gallo for their insightful suggestions and discussions. We gratefully acknowledge the AGN community for their valuable feedback on the preliminary draft of this manuscript posted on arXiv.

This research has made use of archival data from the \xmm{} and \nustar{} observatories provided by the High Energy Astrophysics Science Archive Research Center (HEASARC), which is a service of the Astrophysics Science Division at NASA/GSFC. This research has used \nustar{} Data Analysis Software (NuSTARDAS) jointly developed by the ASI Science Data Center (ASDC, Italy) and the California Institute of Technology (USA).

\section*{Data Availability}
All data used in this study are publicly available from the NASA High Energy Astrophysics Science Archive Research Center ({\tt HEASARC} \cite{heasarc_2026}).

\bibliography{ms_prd}

\let\restartappendixnumbering\apptablenumbers

\def\resetapptablenumbers{\global\c@table=0
\global\c@figure=0
\global\c@equation=0
\def\thetable{\thesection\the\c@table}
\def\fnum@table{{\bf\tablename~\thetable}}%
\def\thefigure{\thesection\the\c@figure}
\def\fnum@figure{{\bf\figurename~\thefigure}}%
}

\appendix
\section{Modeling Details of Individual AGN}
\label{sec:individual}
Here, we discuss the hard-to-soft X-ray spectral fitting details for individual AGN in the sample.

\subsubsection{UGC~6728}
The source exhibits soft X-ray excess below $\sim$2\keV{} in the \xmm{} spectra and a broad Fe~K emission line with a width of $\sim$0.13\keV{} in the 6$-$7\keV{} range of the \nustar{} spectra (Fig.~A1). We noticed only a weak Compton hump in the 15$-$30\keV{} range. The confidence level of relativistic disk reflection ({\tt relxillCp}) for fitting the broad Fe~K emission line is 99.95\% with $\Delta {\rm DIC}=12.6>10$. The fitting of the Compton reflection hump with the distant reflection ({\tt xillverCp}) model is found to be insignificant, with a confidence level of 86.4\% and $\Delta {\rm DIC}=-4.3<0$. However, the Compton hump is well modeled by the relativistic disk reflection ({\tt relxillCp}) model with 95.72\% confidence and $\Delta {\rm DIC}=14.2>10$. Therefore, the hard X-ray (3$-$78\keV{}) spectra require mainly the relativistic disk reflection model {\tt relxillCp} to fit the broad Fe~K emission together with the Compton hump, and the model {\tt TBabs*(relxillCp+nthComp)} best fits the hard X-ray spectra with $\chi^{2}/\rm{dof}=375.8/360$. Once we extrapolate the hard X-ray best-fit model down to 0.3\keV{}, we find that the {\tt relxillCp} model can self-consistently fit the soft X-ray excess emission, yielding a very good fit with $\chi^{2}/\rm{dof} = 514.7/460$. No structural residuals are seen in the entire energy band. To test the relevance of the warm coronal emission for the origin of soft X-ray excess, we added the warm Comptonization model {\tt compTT}, which provided ${\rm DIC}_{\rm with~WC}=545.2$ through Bayesian analysis. The confidence level of the warm coronal emission as evaluated by the MLR test is only 77.18\%. The difference between the Deviance Information Criteria without and with the warm Comptonization component is $\Delta {\rm DIC}={\rm DIC}_{\rm without~WC}-{\rm DIC}_{\rm with~WC}= 543.9-545.2 =-1.3<0$, implying that an extra warm Comptonization is not required to fit the observed soft X-ray excess in UGC~6728. Considering the LOS Galactic absorption ({\tt TBabs}), the broadband (0.3$-$78\keV{}) best-fit model expression is 
\begin{linenomath}
\[
{\tt TBabs*(relxillCp+nthComp)}.
\]
\end{linenomath}
Fig.~A2 shows the broadband \xmm{}/\nustar{} spectra, the best-fit count spectral model with components, and the corresponding residuals. We plot the best-fit averaged spectral energy flux model with {\tt relxillCp} and {\tt nthComp} components in Fig.~A3. The best-fit spectral model parameters and their 90\% confidence intervals determined through MCMC parameter exploration are presented in Table~\ref{table_output_parameters}. The best-fit values for the disk density, iron abundance, black hole spin, and disk inclination angle are $\log [n_{\rm e}/\rm{cm}^{-3}] \le 17.7$, $A_{\rm Fe} \le 4.3$, $a^{\ast}=0.73_{-0.61}^{+0.114}$, and $\theta=50_{-21}^{+4}$~degree, respectively. The temperature of the hot corona is measured to be $kT_{\rm e}\ge 42$\keV{}. Bayesian analysis suggested no difference between the high-density and canonical disk reflection models, which is supported by the disk density parameter reaching its lower limit of $\log [n_{\rm e}/\rm{cm}^{-3}]=15$.

\subsubsection{Mrk~1310}
The source revealed soft X-ray excess below $\sim$1.5\keV{} in the \xmm{} spectra and a broad Fe~K emission line of width $\sim$0.14\keV{} in the 6$-$7\keV{} band of the \nustar{} spectra (Fig.~A1). The Compton hump was not detected in the \nustar{} spectra. The presence of a Compton hump, as modeled by either distant reflection ({\tt xillverCp}) or relativistic reflection ({\tt relxillCp}), is found to be insignificant with confidence levels of only 35.54\% and 10.94\%, and $\Delta {\rm DIC}=-1.1$ and $-1.8$, respectively. The confidence level of the detected broad Fe~K emission feature fitted by the relativistic disk reflection ({\tt relxillCp}) model is 99.92\% with $\Delta {\rm DIC}=6.1>2$. Therefore, the hard X-ray spectra of Mrk~1310 are best described by the model, {\tt TBabs*({relxillCp$+$nthComp})}, with $\chi^{2}/\rm{dof}=193.2/158$. The extrapolation of the hard X-ray best-fit model down to 0.3\keV{} can describe the broadband spectra well with $\chi^{2}/\rm{dof}=277.1/228$. No significant features are seen in the residual plot. However, we still test for the presence of a warm corona by adding the warm Comptonization model ({\tt compTT}), which provides $\chi^{2}/\rm{dof}=263.4/223$ and ${\rm DIC}_{\rm with~WC}=297.9$. The warm coronal emission has a confidence level of 98.24\% as calculated by the MLR test. The measured difference between the Deviance Information Criteria without and with the warm Comptonization component is $\Delta {\rm DIC}={\rm DIC}_{\rm without~WC}-{\rm DIC}_{\rm with~WC}=308.9-297.9=11.0 >10$. Both the MLR test and Bayesian model selection metric imply that the warm corona contributes to the observed soft X-ray excess in Mrk~1310. Therefore, the broadband best-fit model, including the LOS Galactic absorption ({\tt TBabs}), is
\begin{linenomath}
\[
{\tt TBabs*(compTT+relxillCp+nthComp)}.
\]
\end{linenomath}
In Fig.~A2, we show the broadband \xmm{}/\nustar{} spectra, the best-fit count spectral model with components, and the corresponding residual plot. Fig.~A3 shows the best-fit averaged spectral energy flux model with {\tt compTT}, {\tt relxillCp}, and {\tt nthComp} components. In Table~\ref{table_output_parameters}, we present the best-fit spectral model parameters and their 90\% confidence intervals obtained from the MCMC calculation. The best-fit values of disk density, iron abundance, black hole spin, and disk inclination angle are $\log [n_{\rm e}/\rm{cm}^{-3}] \le 17.3$, $A_{\rm Fe}=1.4_{-0.7}^{+2.3}$, $a^{\ast}=0.752_{-0.605}^{+0.19}$, and $\theta=49_{-31}^{+7}$~degree, respectively. The lower limit of the hot coronal temperature is estimated to be 26\keV{}. The Bayesian analysis did not find any difference between the high-density and canonical disk reflection models since the density parameter is pegged at the lower bound of $\log [n_{\rm e}/\rm{cm}^{-3}]=15$.

\subsubsection{NGC~4748}
The \xmm{}/\nustar{} spectra of NGC~4748 reveal a narrow Fe~K$_{\alpha}$ core at $\sim$6.4\keV{} with a broad Fe~K emission feature of width $\sim$0.35\keV{} in the 6$-$7\keV{} band, a Compton hump above 15\keV{}, and soft X-ray excess below $\sim$2\keV{} (Fig.~A1). The broad Fe~K emission line is modeled by relativistic disk reflection ({\tt relxillCp}) and has a confidence level of 98.8\% with $\Delta {\rm DIC}=9.7>2$. The confidence levels of the Compton hump fitted by distant reflection ({\tt xillverCp}) or relativistic reflection ({\tt relxillCp}) are 99.11\% and 97.19\%, with $\Delta {\rm DIC}=5.3$ and $4.2$, respectively. Therefore, the model, {\tt TBabs*(relxillCp$+$xillverCp$+$nthComp)}, best explains the hard X-ray (3$-$78\keV{}) spectra with $\chi^{2}/\rm{dof}=207.2/197$, where the narrow 6.4\keV{} Fe~K$_{\alpha}$ line and part of the Compton hump are modeled by distant reflection ({\tt xillverCp}), while the broad Fe~K emission line and most of the Compton hump are modeled by relativistic reflection ({\tt relxillCp}). The extrapolation of the hard X-ray best-fit model can fit the broadband (0.3$-$78\keV{}) X-ray spectra with $\chi^{2}/\rm{dof}=281.0/242$, where the soft X-ray excess is modeled by {\tt relxillCp}. However, we notice some excess emission in the hard X-ray band, which means {\tt relxillCp} cannot explain both the soft and hard X-ray excess emission self-consistently. We then add the warm Comptonization model {\tt compTT} and find that the spectral model, {\tt TBabs*(compTT+relxillCp+xillverCp+nthComp)}, explains the broadband spectra the best with $\chi^{2}/\rm{dof}=258.3/238$. There are no structural residuals in the entire energy band. The Bayesian model selection metric very strongly prefers warm Comptonization over high-density disk reflection for fitting of soft X-ray excess with $\Delta {\rm DIC}={\rm DIC}_{\rm without~WC}-{\rm DIC}_{\rm with~WC}= 306.7-290.6 =16.1>10$. The MLR test finds that the confidence level of the warm coronal emission is 99.99\%. Fig.~A2 shows the broadband \xmm{}/\nustar{} spectra, the best-fit count spectral model with components, and the corresponding residual plot. We plot the best-fit averaged spectral energy flux model with {\tt compTT}, {\tt relxillCp}, {\tt xillverCp}, and {\tt nthComp} components in Fig.~A3. The expression for the broadband best-fit model considering the LOS Galactic absorption ({\tt TBabs}), 
\begin{linenomath}
\[
{\tt TBabs*(compTT+relxillCp+xillverCp+nthComp)}.
\]
\end{linenomath}
The best-fit source spectral model parameters and their 90\% confidence intervals derived through MCMC computation, are presented in Table~\ref{table_output_parameters}. The best-fit values of disk density, iron abundance, black hole spin, and disk inclination angle are estimated to be $\log [n_{\rm e}/\rm{cm}^{-3}] \le 17.1$, $A_{\rm Fe} \le 4.9$, $a^{\ast}=0.796_{-0.663}^{+0.123}$, and $\theta=51_{-33}^{+5}$~degree, respectively. We find the lower limit for the hot coronal temperature to be 63\keV{}. Our joint \xmm{}/\nustar{} spectral modeling obtained stringent constraints on the spin parameter, which was previously unconstrained by JJ19's \xmm{} spectroscopy alone.

\subsubsection{Mrk~110}
The source exhibits soft X-ray excess below $\sim$2\keV{} and a narrow Fe~K$_{\alpha}$ core at $\sim$6.37\keV{} along with a broad Fe~K emission feature of width $\sim$0.15\keV{} in the 6$-$7\keV{} band, and a Compton hump like structure in the 15$-$30\keV{} band (Fig.~A1). The modeling of the Compton hump with distant reflection ({\tt xillverCp}) or relativistic reflection ({\tt relxillCp}) is found to be significant with $>$99.99\% confidence and $\Delta {\rm DIC}=28.0$ and $20.1$, respectively. The broad Fe~K emission line fitted by the relativistic disk reflection ({\tt relxillCp}) model has a significance of $>$99.99\% with $\Delta {\rm DIC}=38.4>10$. Therefore, the model {\tt TBabs*({relxillCp$+$xillverCp$+$nthComp})} best represents the hard X-ray (3$-$78\keV{}) spectra of Mrk~110 with $\chi^{2}/\rm{dof}=1233.7/1155$, where the distant reflection ({\tt xillverCp}) component fits the narrow Fe~K$_{\alpha}$ core and most of the Compton hump, while the relativistic disk reflection ({\tt relxillCp}) component fits the broad Fe~K emission line and part of the Compton hump. By extrapolating the hard X-ray best-fit model down to 0.3\keV{}, we find that the model is unable to fit the broadband (0.3$-$78\keV{}) spectra satisfactorily with significant residuals observed in the hard X-ray band, providing $\chi^{2}/\rm{dof}=1904.4/1434$. This suggests that the relativistic reflection model {\tt relxillCp} cannot self-consistently fit both the soft and hard X-ray excess emission, and an extra warm Comptonization for soft X-ray excess is perhaps needed, as found by \cite{Porquet_2024}. Hence, we add the warm Comptonization model {\tt compTT} and find that the model {\tt TBabs*({compTT$+$relxillCp$+$xillverCp$+$nthComp})} represents the broadband spectra well with $\chi^{2}/\rm{dof}=1647/1425$, where the soft X-ray excess is fitted by {\tt compTT}, and {\tt relxillCp} describes the broad Fe~K emission feature and part of the Compton hump. The Bayesian analysis strongly prefers the warm coronal origin of soft X-ray excess over high-density disk reflection with $\Delta {\rm DIC}={\rm DIC}_{\rm without~WC}-{\rm DIC}_{\rm with~WC}= 1996.5-1752.8 = 243.7>>10$. The significance of the warm coronal emission as computed by the MLR test is $>$99.99\%. In Fig.~A2, we show the broadband \xmm{}/\nustar{} spectra, the best-fit count spectral model with components, and the corresponding residuals. Fig.~A3 presents the best-fit averaged spectral energy flux model with {\tt compTT}, {\tt relxillCp}, {\tt xillverCp}, and {\tt nthComp} components. With the LOS Galactic absorption ({\tt TBabs}), we can write the broadband best-fit model as 
\begin{linenomath}
\[
{\tt TBabs*(compTT+relxillCp+xillverCp+nthComp)}.
\]
\end{linenomath}
Table~\ref{table_output_parameters} presents the best-fit spectral model parameters and their 90\% confidence intervals, obtained by exploring the complete parameter space through MCMC. The best-fit values of disk density, iron abundance, black hole spin, and disk inclination angle are $\log [n_{\rm e}/\rm{cm}^{-3}] =19.0_{-0.2}^{+0.2}$, $A_{\rm Fe}=1.5_{-0.4}^{+0.7}$, $a^{\ast}=0.978_{-0.024}^{+0.016}$, and $\theta\le 24$~degree, respectively. The lower limit of the hot coronal temperature is measured to be 118\keV{}. Through joint \xmm{}/\nustar{} spectroscopy, we can constrain the disk density for which only an upper limit was calculated by JJ19's \xmm{} spectral modeling alone. The Bayesian analysis strongly supports the higher-density disk against the canonical disk reflection model with $\Delta {\rm DIC}>10$.

\subsubsection{Mrk~279} 
The \xmm{}/\nustar{} spectra of Mrk~279 show two narrow lines at $\sim$6.38\keV{} and $\sim$6.98\keV{} corresponding to Fe~K$_{\alpha}$ and Fe~K$_{\beta}$ emission cores, respectively, one broad Fe~K emission line of width $\sim$0.24\keV{} in the 6$-$7\keV{} band, a Compton hump above 15\keV{}, and soft X-ray excess below $\sim$2\keV{} (Fig.~A1). The broad Fe~K emission line is modeled by relativistic disk reflection ({\tt relxillCp}), which has a significance of $>$99.99\% with $\Delta {\rm DIC}=44.0>10$. The modeling of the Compton hump by distant reflection ({\tt xillverCp}) or relativistic reflection ({\tt relxillCp}) is found to be significant with $\Delta {\rm DIC}= 70.6$ and 68.4, respectively, and has a confidence level of $>$99.99\%. Therefore, to fit the hard X-ray (3$-$78\keV{}) spectra of the source, we employ the model {\tt TBabs*(relxillCp$+$xillverCp$+$nthComp)}, which provided the best-fit with $\chi^{2}/\rm{dof}=1031.3/956$. Here, the two narrow Fe~K emission line cores and most of the Compton hump are modeled by the distant reflection ({\tt xillverCp}) model, and the relativistic disk reflection ({\tt relxillCp}) model fits the broad Fe~K emission and part of the Compton hump. The extrapolation of the hard X-ray best-fit model can fit the broadband (0.3$-$78\keV{}) X-ray spectra well with $\chi^{2}/\rm{dof}=1376.9/1181$. However, there are some excess residuals in the soft X-ray band, which may indicate the need for an extra soft X-ray component. We then add the warm Comptonization ({\tt compTT}) model and find that the model {\tt TBabs*(compTT+relxillCp+xillverCp+nthComp)} best fits the broadband spectra of the source with $\chi^{2}/\rm{dof}=1307.6/1170$ and ${\rm DIC}_{\rm with~WC}=1413.0$, where both {\tt compTT} and {\tt relxillCp} contribute to the soft X-ray excess. The presence of warm Comptonization is significant with a confidence level of $>$99.99\%, which is further supported by our Bayesian analysis with the difference between the Deviance Information Criteria without and with warm Comptonization is $\Delta {\rm DIC}={\rm DIC}_{\rm without~WC}-{\rm DIC}_{\rm with~WC}=1463.2-1413.0 =50.2 > 10$. The broadband \xmm{}/\nustar{} spectra, the best-fit count spectral model along with all components, and the residual plot are shown in Fig.~A2. We plot the best-fit averaged spectral energy flux model with {\tt compTT}, {\tt relxillCp}, {\tt xillverCp}, and {\tt nthComp} components in Fig.~A3. Including the LOS Galactic absorption ({\tt TBabs}), the broadband best-fit model expression is
\begin{linenomath}
\[
{\tt TBabs*(compTT+relxillCp+xillverCp+nthComp)}.
\]
\end{linenomath}
We explore the complete parameter space through the MCMC method and list the best-fit source spectral model parameters and their 90\% confidence intervals in Table~\ref{table_output_parameters}. The best-fit values of disk density, iron abundance, black hole spin, and disk inclination angle are $\log [n_{\rm e}/\rm{cm}^{-3}]\ge 19.3$, $A_{\rm Fe}=4.0_{-1.5}^{+0.8}$, $a^{\ast}\ge 0.991$, and $\theta=47_{-11}^{+2}$~degree, respectively. The temperature of the hot corona is measured to be $kT_{\rm e}=40_{-14}^{+24}$\keV{}. The Bayesian analysis strongly preferred the higher-density disk over the canonical disk reflection model with $\Delta {\rm DIC}>10$.

\subsubsection{Mrk~590}
In the \xmm{}/\nustar{} spectra of the source, we find narrow Fe~K$_{\alpha}$ and Fe~K$_{\beta}$ emission line cores at $\sim$6.38\keV{} and $\sim$7.0\keV{}, respectively along with a broad Fe~K emission feature of width $\sim$0.3\keV{} in the 6$-$7\keV{} band, and soft X-ray excess below $\sim$1\keV{} (Fig.~A1). There is no strong Compton hump observed in the \nustar{} spectra. The confidence levels of the Compton hump as modeled by either distant reflection ({\tt xillverCp}) or relativistic reflection ({\tt relxillCp}) are only 89.06\% with $\Delta {\rm DIC}= 0.6$, and 87.3\% with $\Delta {\rm DIC}= -8$, respectively. Since the Compton hump is not significant, to fit the Fe~K$_{\alpha}$ and Fe~K$_{\beta}$ emission cores, we used two narrow ($\sigma=10$\ev{}) Gaussian emission lines, {\tt zGauss\_N} and {\tt zGauss2\_N}, respectively. The broad Fe~K emission feature is modeled by relativistic disk reflection ({\tt relxillCp}) and has a significance of 97.27\% with $\Delta {\rm DIC}=5.4>2$. Therefore, the hard X-ray (3$-$78\keV{}) spectra of the source are best described by the model {\tt TBabs*(relxillCp$+$zGauss\_N$+$zGauss2\_N$+$nthComp)} with $\chi^{2}/\rm{dof}=1640.5/1623$. Once we extrapolate the hard X-ray best-fit model down to 0.3\keV{}, the same model can fit the broadband (0.3$-$78\keV{}) spectra well, providing a reasonable fit with $\chi^{2}/\rm{dof}=2030.2/2028$. Self-consistently, the relativistic disk reflection ({\tt relxillCp}) model explains the observed soft X-ray excess. We still incorporate the warm Comptonization ({\tt compTT}) model and test its relevance for the soft X-ray excess using both the MLR test and Bayesian analysis. The confidence level of warm Comptonization as evaluated by the MLR test is 89.53\%. The estimated difference between the Deviance Information Criteria without and with warm Comptonization is $\Delta {\rm DIC}={\rm DIC}_{\rm without~WC}-{\rm DIC}_{\rm with~WC}= 2229.5-2236.9 =-7.4<0$, confirming that high-density disk reflection is sufficient enough to fit the observed soft X-ray excess and an extra warm Comptonization is not required. Considering the LOS Galactic absorption ({\tt TBabs}), the broadband best-fit model expression can be written as
\begin{linenomath}
\[
{\tt TBabs*(relxillCp+zGauss\_N+zGauss2\_N+nthComp)}.
\]
\end{linenomath}
We show the broadband \xmm{}/\nustar{} spectra, the best-fit count spectral model with components, and the corresponding residuals in Fig.~A2. The best-fit averaged spectral energy flux model with {\tt relxillCp}, {\tt zGauss\_N}, {\tt zGauss2\_N}, and {\tt nthComp} components are presented in Fig.~A3. Table~\ref{table_output_parameters} shows the best-fit spectral model parameters and their 90\% confidence intervals, where the complete parameter space is explored through MCMC. The best-fit values of disk density, iron abundance, black hole spin, and disk inclination angle are $\log [n_{\rm e}/\rm{cm}^{-3}] \ge 19.1$, $A_{\rm Fe}=2.6_{-0.3}^{+0.5}$, $a^{\ast}=0.89_{-0.093}^{+0.041}$, and $\theta=41_{-4}^{+5}$~degree, respectively. The measured electron temperature of the hot corona is $kT_{\rm e}=49_{-19}^{+37}$\keV{}. The joint \xmm{}+\nustar{} relativistic reflection spectroscopy provided more stringent constraints on the black hole spin parameter, which was previously unconstrained by JJ19's \xmm{} spectral fitting alone. Through Bayesian analysis, we verify that the higher-density disk is preferred over the canonical disk reflection model with $\Delta {\rm DIC}=8.7>2$.

\subsubsection{Mrk~79}
The source Mrk~79 revealed narrow Fe~K$_{\alpha}$ and Fe~K$_{\beta}$ emission line cores at $\sim$6.38\keV{} and $\sim$6.95\keV{}, respectively, together with a broad Fe~K emission feature of width $\sim$0.3\keV{} in the 6$-$7\keV{} band, a Compton hump in the 15$-$40\keV{} range, and variable soft X-ray excess emission below $\sim$2\keV{} (Fig.~A1). The broad Fe~K emission feature is modeled by relativistic disk reflection ({\tt relxillCp}), which has a confidence level of 99.9\% and $\Delta {\rm DIC}=13.6>10$. The Compton hump modeled by distant reflection ({\tt xillverCp}) or relativistic reflection ({\tt relxillCp}) has confidence levels of 90.72\% and 91.43\%, and $\Delta {\rm DIC}=2.3$ and $5.3>2$, respectively. Therefore, the model {\tt TBabs*(relxillCp$+$xillverCp$+$nthComp)} best describes the hard X-ray (3$-$78\keV{}) spectra of the source, providing $\chi^{2}/\rm{dof}=716/718$, where {\tt xillverCp} models Fe~K$_{\alpha}$/K$_{\beta}$ emission line cores and part of the Compton hump, while the broad Fe~K emission and most of the Compton hump are modeled by {\tt relxillCp}. When extrapolated down to 0.3\keV{}, the hard X-ray best-fit model can fit the broadband spectra well with $\chi^{2}/\rm{dof}=1138.6/972$. However, we notice some positive residuals in the soft X-ray band, which could indicate the presence of an additional soft X-ray component. Therefore, we add the warm Comptonization model {\tt compTT} to test for the presence of a warm corona and find that the broadband spectra of the source are best described by the model {\tt TBabs*(compTT+relxillCp+xillverCp+nthComp)} with $\chi^{2}/\rm{dof}=1075.0/963$ and ${\rm DIC}_{\rm with~WC}=1196.3$, where soft X-ray excess is modeled by both {\tt compTT} and {\tt relxillCp}. However, relativistic disk reflection contributes more to the soft X-ray excess than warm Comptonization. The MLR test finds that warm Comptonization has a confidence level of 99\%, which is further supported by our Bayesian analysis, revealing a difference of $\Delta {\rm DIC}={\rm DIC}_{\rm without~WC}-{\rm DIC}_{\rm with~WC}= 1211.3-1196.3 =15.0>10$ in the Deviance Information Criteria between models without and with warm Comptonization. Therefore, we can write the broadband best-fit model expression as 
\begin{linenomath} 
\[
{\tt TBabs*(compTT+relxillCp+xillverCp+nthComp)}.
\]
\end{linenomath}
Fig.~A2 shows the broadband \xmm{}/\nustar{} spectra, the best-fit count spectral model with components, and the corresponding residuals. We present the best-fit averaged spectral energy flux model with {\tt compTT}, {\tt relxillCp}, {\tt xillverCp}, and {\tt nthComp} components in Fig.~A4. The best-fit spectral model parameters and their 90\% confidence intervals derived through MCMC are presented in Table~\ref{table_output_parameters}. Our MCMC analysis confirms that all the reflection model parameters are constrained with the best-fit values of disk density, iron abundance, black hole spin, and disk inclination angle are $\log [n_{\rm e}/\rm{cm}^{-3}]=18.0_{-0.2}^{+0.2}$, $A_{\rm Fe}=0.7_{-0.1}^{+0.1}$, $a^{\ast}=0.88_{-0.039}^{+0.048}$, and $\theta=35_{-3}^{+4}$~degree, respectively. We obtained a more precise measurement of the black hole spin parameter, which was not constrained by the relativistic reflection spectroscopy of only \xmm{} spectra (see Fig.~7). The Bayesian model selection metric strongly supports the higher-density disk reflection model over the canonical one with $\Delta {\rm DIC}>10$.

\subsubsection{PG~1229+204}
The \xmm{}/\nustar{} spectra of PG~1229+204 unveiled a strong soft X-ray excess below $\sim$2\keV{}, a broad Fe~K emission line of width $\sim$0.2\keV{} in the 6$-$7\keV{} band, and an excess emission above $\sim$10\keV{} indicating a Compton hump (Fig.~A1). However, the detection significance of the Compton hump modeled by either distant reflection ({\tt xillverCp}) or relativistic reflection ({\tt relxillCp}) is less than 1$\sigma$ with $\Delta {\rm DIC}=-6.0$ and $-1.8 < 0$, respectively. The broad Fe~K emission line is modeled by relativistic disk reflection ({\tt relxillCp}) and has a confidence level of 94.99\% with $\Delta {\rm DIC}= 9.7 > 2$. Therefore, the hard X-ray (3$-$78\keV{}) spectra of the source are best modeled by {\tt TBabs*(relxillCp$+$nthComp)} with $\chi^{2}/\rm{dof} = 123.1/113$. Once we extrapolate the hard X-ray best-fit model down to 0.3\keV{}, we can fit the complete 0.3$-$78\keV{} energy band by the same model, providing a reasonable fit with $\chi^{2}/\rm{dof} = 162.0/154$, where the relativistic disk reflection ({\tt relxillCp}) model explains the soft X-ray excess emission self-consistently. However, we still tested the relevance of warm corona for soft X-ray excess and added the warm Comptonization model {\tt compTT}. The significance of warm Comptonization as evaluated by the MLR test is only 56.63\%. Additionally, our Bayesian analysis finds that the difference between the Deviance Information Criteria without and with warm Comptonization is $\Delta {\rm DIC}={\rm DIC}_{\rm without~WC}-{\rm DIC}_{\rm with~WC}= 188.2-201.9 =-13.7<0$, confirming that the relativistic disk reflection is sufficient for the modeling of soft X-ray excess and an additional warm Comptonization component is not required. Considering the LOS Galactic absorption ({\tt TBabs}), the broadband best-fit model can be written as 
\begin{linenomath}
\[
{\tt TBabs*(relxillCp+nthComp)}.
\]
\end{linenomath}
The broadband \xmm{}/\nustar{} spectra, the best-fit count spectral model with components, and the corresponding residual plot are shown in Fig.~A2. We plot the best-fit averaged spectral energy flux model with {\tt relxillCp} and {\tt nthComp} components in Fig.~A4. The complete parameter space of the best-fit model was explored using MCMC computation. In Table~\ref{table_output_parameters}, we present the best-fit spectral model parameters and their 90\% confidence intervals. The best-fit values of disk density, iron abundance, black hole spin, and disk inclination angle are $\log [n_{\rm e}/\rm{cm}^{-3}] \le 16.6$, $A_{\rm Fe} =1.6_{-0.8}^{+1.6}$, $a^{\ast}=0.974_{-0.775}^{+0.024}$, and $\theta=37_{-8}^{+8}$~degree, respectively. The temperature of the hot corona is measured to be $kT_{\rm e}\ge 64$\keV{}.

\subsubsection{PG~0844+349}
The \xmm{}/\nustar{} spectra of PG~0844+349 show a variable soft X-ray excess below around 1$-$2\keV{}, a broad Fe~K emission line of width $\sim$0.6\keV{} in the 5$-$7\keV{} band, and a possible presence of Compton hump above 10\keV{} (Fig.~A1). The broad Fe~K emission line modeled by relativistic disk reflection ({\tt relxillCp}) has a confidence level of $>$99.99\% with $\Delta {\rm DIC}=30.2>10$. However, the Compton hump modeled by either distant reflection ({\tt xillverCp}) or relativistic reflection ({\tt relxillCp}) has significance levels below 1$\sigma$ with $\Delta {\rm DIC}=-4.1$ and $-7.7 < 0$, respectively. Therefore, we model the hard X-ray (3$-$78\keV{}) spectra using {\tt TBabs*(relxillCp$+$nthComp)}, which provided the best-fit with $\chi^{2}/\rm{dof}=88/86$. When extrapolated down to 0.3\keV{}, the hard X-ray best-fit model cannot fit the broadband spectra well, resulting in significant residuals in the hard X-ray band, with $\chi^{2}/\rm{dof}=192.0/155$. This implies that the relativistic disk reflection ({\tt relxillCp}) model is incapable of explaining both soft and hard X-ray excess emission self-consistently. By incorporating warm Comptonization ({\tt compTT}), we find that the model {\tt TBabs*(compTT$+$relxillCp$+$nthComp)} successfully fits the broadband spectra with $\chi^{2}/\rm{dof}=158.7/150$ and ${\rm DIC}_{\rm with~WC}=191.4$, where both {\tt compTT} and {\tt relxillCp} contribute to the observed soft X-ray excess. The MLR test indicates that the warm corona has a significance level of $>$99.99\%. This is further supported by our Bayesian analysis, which finds that the difference between the Deviance Information Criteria without and with warm Comptonization is $\Delta {\rm DIC}={\rm DIC}_{\rm without~WC}-{\rm DIC}_{\rm with~WC} = 236.7-191.4 = 45.3>10$. Therefore, the broadband best-fit model containing the LOS Galactic absorption ({\tt TBabs}) can be expressed as 
\begin{linenomath}
\[
{\tt TBabs*(compTT+relxillCp+nthComp)}.
\]
\end{linenomath}
Fig.~A2 shows the broadband \xmm{}/\nustar{} spectra, the best-fit count spectral model with components, and the corresponding residuals. In Fig.~A4, we present the best-fit averaged spectral energy flux model with {\tt compTT}, {\tt relxillCp}, and {\tt nthComp} components. The best-fit spectral model parameters and their 90\% confidence intervals are presented in Table~\ref{table_output_parameters}, where we explore the complete parameter space and calculate parameter uncertainties using the MCMC method. The best-fit values of disk density, iron abundance, black hole spin, and disk inclination angle are $\log [n_{\rm e}/\rm{cm}^{-3}] \le 18.5$, $A_{\rm Fe}=4.5_{-2.8}^{+0.4}$, $a^{\ast}=0.292_{-0.157}^{+0.555}$, and $\theta=34_{-7}^{+11}$~degree, respectively. The lower limit of the hot coronal temperature is found to be 54\keV{}. Our broadband relativistic reflection spectroscopy constrains the black hole spin parameter, which was previously unconstrained through \xmm{} reflection spectroscopy alone, as shown in Fig.~7.

\subsubsection{PG~0804+761}
The source revealed a soft X-ray excess below $\sim$2\keV{}, a narrow Fe~K$_{\alpha}$ emission line core at $\sim$6.4\keV{} along with a broad Fe~K emission feature of width $\sim$0.3\keV{} in the 6$-$7\keV{} band, and a possible Compton hump above 10\keV{}, as shown in Fig.~A1. However, the significance levels of the Compton hump as modeled by distant reflection ({\tt xillverCp}) or relativistic reflection ({\tt relxillCp}) are only 81.53\% and 78.1\%, with $\Delta {\rm DIC}=-1.8$, and $-4.0 <0$, respectively. Since the Compton hump is not statistically significant, we fit the Fe~K$_{\alpha}$ emission core using a narrow ($\sigma=10$\ev{}) Gaussian emission ({\tt zGauss\_N}) line. The relativistic disk reflection ({\tt relxillCp}) model fits the broad Fe~K emission line with a confidence level of 97.72\% and $\Delta {\rm DIC} =5.1>2$. Thus, the hard X-ray (3$-$78\keV{}) spectra of the source are best represented by the model {\tt TBabs*(relxillCp$+$zGauss\_N$+$nthComp)} with $\chi^{2}/\rm{dof}=159.5/191$. We then extrapolated the best-fit hard X-ray model down to 0.3\keV{} to fit the broadband (0.3$-$78\keV{}) spectra, which resulted in $\chi^{2}/\rm{dof}=301.6/277$. However, we noticed excess residuals in the hard X-ray band, suggesting that the {\tt relxillCp} model cannot self-consistently explain both the soft and hard X-ray excess in the source. Therefore, we include the warm Comptonization model {\tt compTT} for the soft X-ray excess, and the model {\tt TBabs*(compTT+relxillCp+zGauss\_N+nthComp)} provides the best fit to the broadband spectra with $\chi^{2}/\rm{dof}=252.2/272$ and ${\rm DIC}_{\rm with~WC}=310.1$, where both {\tt compTT} and {\tt relxillCp} contribute to the observed soft X-ray excess. The MLR test shows that the warm Comptonization component is significant at the $>$99.99\% level, which is further supported by our Bayesian analysis, where the difference between the Deviance Information Criteria without and with warm Comptonization is $\Delta {\rm DIC}={\rm DIC}_{\rm without~WC}-{\rm DIC}_{\rm with~WC}=348.5-310.1=38.4 >10$. The broadband best-fit model expression incorporating the LOS Galactic absorption ({\tt TBabs}) is
\begin{linenomath}
\[
{\tt TBabs*(compTT+relxillCp+zGauss\_N+nthComp)}.
\]
\end{linenomath}
In Fig.~A2, we show the broadband \xmm{}/\nustar{} spectra, the best-fit count spectral model with components, and the corresponding residual plot. The best-fit averaged spectral energy flux model with components is plotted in Fig.~A4. In Table~\ref{table_output_parameters}, we present the best-fit spectral model parameters and their 90\% confidence intervals obtained through MCMC parameter space exploration of the best-fit model. The best-fit values of disk density, iron abundance, black hole spin, and disk inclination angle are estimated to be $\log [n_{\rm e}/\rm{cm}^{-3}]\le 17.6$, $A_{\rm Fe}=2.1_{-1.0}^{+2.6}$, $a^{\ast}\le 0.706$, and $\theta=44_{-7}^{+4}$~degree, respectively. The lower limit measured for the electron temperature of the hot corona is 52\keV{}.

\subsubsection{PG~1426+015}
The \xmm{}/\nustar{} spectra of the source revealed a soft X-ray excess below $\sim$2\keV{}, a narrow Fe~K$_{\alpha}$ emission line core at $\sim$6.38\keV{} together with a broad Fe~K emission feature of width $\sim$0.5\keV{} in the 6$-$7\keV{} band, and a strong Compton hump in the 15$-$40\keV{} range (Fig.~A1). The Compton hump modeled using either distant reflection ({\tt xillverCp}) or relativistic reflection ({\tt relxillCp}) has significance levels of 96.49\% and 96.60\%, with $\Delta {\rm DIC}=8.1 (>6)$ and $13.7(>10)$, respectively. The broad Fe~K emission line is modeled by relativistic disk reflection (relxillCp) and is significant at the 99.95\% level with $\Delta {\rm DIC}=11.4>10$. Therefore, the hard X-ray (3$-$78\keV{}) spectra of the source are best modeled by {\tt TBabs*(relxillCp$+$xillverCp$+$nthComp)} with $\chi^{2}/\rm{dof}=373.8/350$, where the distant reflection ({\tt xillverCp}) component fits the narrow Fe~K$_{\alpha}$ emission line core and most of the Compton hump, while the broad Fe~K emission line and part of the Compton hump are explained by relativistic disk reflection ({\tt relxillCp}). However, the extrapolation of the hard X-ray best-fit model is unable to explain the broadband (0.3$-$78\keV{}) X-ray spectra well, yielding $\chi^{2}/\rm{dof}=524.9/420$ with significant residuals in the hard X-ray band, which implies that the {\tt relxillCp} model cannot fit both soft and hard X-ray excess self-consistently. Therefore, we add the warm Comptonization model ({\tt compTT}) for soft X-ray excess and find that the model {\tt TBabs*(compTT+relxillCp+xillverCp+nthComp)} fits the broadband spectra well with $\chi^{2}/\rm{dof}=477.8/415$. No structures or features are seen in the residual spectra. The significance of the warm Comptonization component is $>$99.99\% as determined by the MLR test. The Bayesian model selection metric confirms the relevance of warm Comptonization over high-density disk reflection for soft X-ray excess, since the difference between the Deviance Information Criteria without and with the warm Comptonization component is $\Delta {\rm DIC}={\rm DIC}_{\rm without~WC}-{\rm DIC}_{\rm with~WC}= 575.8-534.6 =41.2>10$. We can write the broadband best-fit model considering the LOS Galactic absorption ({\tt TBabs}) as
\begin{linenomath}
\[
{\tt TBabs*(compTT+relxillCp+xillverCp+nthComp)}.
\]
\end{linenomath}
Fig.~A2 shows the broadband \xmm{}/\nustar{} spectra, the best-fit count spectral model with components, and the corresponding residuals. We plot the best-fit averaged spectral energy flux model with all four ({\tt compTT}, {\tt relxillCp}, {\tt xillverCp}, and {\tt nthComp}) components in Fig.~A4. Our MCMC analysis explored the complete parameter space of the best-fit model. The resulting best-fit spectral model parameters and their 90\% confidence intervals are presented in Table~\ref{table_output_parameters}. The best-fit values of disk density, iron abundance, black hole spin, and disk inclination angle are $\log [n_{\rm e}/\rm{cm}^{-3}] \le 16.2$, $A_{\rm Fe}=1.6_{-0.7}^{+0.9}$, $a^{\ast}=0.442_{-0.122}^{+0.488}$, and $\theta=22_{-6}^{+27}$~degree, respectively. We measured the electron temperature of the hot corona to be $kT_{\rm e}\ge 52$\keV{}. While performing the \xmm{} spectral modeling of the source, JJ19 fixed the spin parameter at 0.998. Through joint \xmm{}/\nustar{} relativistic reflection spectroscopy, we provide the first measurement of the black hole spin parameter for PG~1426+015.

\pagebreak

\setcounter{table}{0}
\renewcommand{\thetable}{A\arabic{table}}

\onecolumngrid

\begin{longtable}{ccccccccccc}

\caption{Observing Log of the AGN sample. Observations marked with `$p$' are affected by pile-up. To mitigate pile-up effects, the inner radii of the annuli used for the affected observations are as follows: 8~arcsec for Obs.~ID 0723100401 (Large Window mode) of NGC~4748; 12~arcsec for Obs.~ID 0083960101 (Full Window mode) of Mrk~279; 8~arcsec for Obs.~ID 0103660201 (Full Window mode) of PG~0844+349; 5~arcsec for Obs.~ID 0852210101 (Small Window mode) of PG~1426+015 and 8~arcsec for Obs.~ID 0102040501 (Full Window mode) of PG~1426+015.}  
\label{table_obs_log} \\

\hline

\hline \multicolumn{8}{c}{}\\  
\endfirsthead

\multicolumn{8}{c}%
{{ \tablename\ \thetable{}}} \\
\hline

\endhead
			
\hline \multicolumn{8}{r}{{\it Continued on next page}} \\
\endfoot
			
\hline 
\endlastfoot
Source  & Observatory & Camera & Obs. ID & Start Time  & Elapsed Time  & Net Exposure & Net Count Rate & Net Counts  \\
   &   &   &  & [MJD] & [ks] & [ks] &  [ct/s] &    \\
(1)    &   (2)   &   (3)   &   (4)  &  (5)   & (6) &   (7)  &    (8)  & (9)   \\                                                         
\hline 

UGC 6728 & \xmm{} & EPIC-MOS1 & 0312191601  & 53789.8  & 12.15 & 8.22 & 0.87 & 7.16E+03 \\ [0.1cm] 
         &        & EPIC-MOS2 &  &  & 11.58 & 8.74 & 0.87 & 7.60E+03 \\ [0.1cm] 

  & \nustar{} & FPMA & 60376007002 & 58039.8 & 68.39 & 51.80 & 0.40 & 2.07E+04 \\ [0.1cm] 
  &   & FPMB &   &   &   & 50.73 & 0.37 & 1.86E+04 \\ [0.1cm] 
  &   & FPMA & 60160450002 & 57579.5 & 27.59 & 21.20 & 0.19 & 3.95E+03 \\ [0.1cm] 
  &   & FPMB &   &   &   & 21.36 & 0.17 & 3.72E+03 \\ [0.1cm] 
\hline 
Mrk 1310 & \xmm{} & EPIC-pn & 0831790501 & 58487.3 & 24.16 & 16.32 & 1.76 & 2.88E+04 \\ [0.1cm] 
  &   & EPIC-pn & 0723100301 & 56635.4 & 55.18 & 33.74 & 0.15 & 4.97E+03 \\ [0.1cm] 
  & \nustar{} & FPMA & 60160465002 & 57556.5 & 38.57 & 21.13 & 0.23 & 4.86E+03 \\ [0.1cm] 
  &   & FPMB &   &   &   & 21.08 & 0.21 & 4.49E+03 \\ [0.1cm] 
\hline 
NGC 4748 & \xmm{} & EPIC-pn & 0723100401$^{p}$ & 56671.3 & 65.38 & 26.42 & 4.21 & 1.11E+05 \\ [0.1cm] 
  & \nustar{} & FPMA & 60663002002 & 59238.4 & 154.82 & 79.91 & 0.18 & 1.41E+04 \\ [0.1cm] 
  &   & FPMB &   &   &   & 79.07 & 0.16 & 1.27E+04 \\ [0.1cm] 
\hline 
Mrk 110 & \xmm{} & EPIC-pn & 0201130501 & 53324.2 & 46.97 & 32.30 & 21.18 & 6.84E+05 \\ [0.1cm] 
  &   & EPIC-pn & 0852590101 & 58804.4 & 42.64 & 29.03 & 19.48 & 5.65E+05 \\ [0.1cm] 
  &   & EPIC-pn & 0852590201 & 58946.0 & 46.64 & 31.69 & 15.23 & 4.83E+05 \\ [0.1cm] 
  &   & EPIC-pn & 0840220801 & 58792.4 & 41.14 & 28.37 & 14.53 & 4.12E+05 \\ [0.1cm] 
  &   & EPIC-pn & 0840220901 & 58794.4 & 38.74 & 26.35 & 15.48 & 4.08E+05 \\ [0.1cm] 
  &   & EPIC-pn & 0840220701 & 58790.4 & 41.74 & 28.41 & 10.07 & 2.86E+05 \\ [0.1cm] 
  & \nustar{} & FPMA & 60201025002 & 57776.8 & 386.30 & 177.33 & 1.09 & 1.93E+05 \\ [0.1cm] 
  &   & FPMB &   &   &   & 176.53 & 1.03 & 1.82E+05 \\ [0.1cm] 
  &   & FPMA & 60502022002 & 58803.2 & 163.20 & 86.77 & 0.75 & 6.50E+04 \\ [0.1cm] 
  &   & FPMB &   &   &   & 85.97 & 0.69 & 5.93E+04 \\ [0.1cm] 
  &   & FPMA & 60502022004 & 58944.6 & 168.44 & 88.67 & 0.64 & 5.66E+04 \\ [0.1cm] 
  &   & FPMB &   &   &   & 87.75 & 0.59 & 5.15E+04 \\ [0.1cm] 
\hline 
Mrk 279 & \xmm{} & EPIC-pn & 0302480401 & 53689.7 & 59.37 & 40.58 & 19.33 & 7.85E+05 \\ [0.1cm] 
  &   & EPIC-pn & 0302480501 & 53691.7 & 59.37 & 37.76 & 16.56 & 6.25E+05 \\ [0.1cm] 
  &   & EPIC-pn & 0302480601 & 53694.0 & 37.77 & 21.76 & 17.50 & 3.81E+05 \\ [0.1cm] 
  &   & EPIC-pn & 0872391301 & 59203.3 & 28.64 & 19.60 & 11.47 & 2.25E+05 \\ [0.1cm] 
  &   & EPIC-pn & 0083960101$^{p}$ & 52401.5 & 30.07 & 13.00 & 3.22 & 4.19E+04 \\ [0.1cm] 
  & \nustar{} & FPMA & 60601011004 & 59066.5 & 317.27 & 199.41 & 0.16 & 3.21E+04 \\ [0.1cm] 
  &   & FPMB &   &   &   & 197.85 & 0.14 & 2.77E+04 \\ [0.1cm] 
  &   & FPMA & 60160562002 & 58785.4 & 37.75 & 27.28 & 0.66 & 1.81E+04 \\ [0.1cm] 
  &   & FPMB &   &   &   & 27.07 & 0.63 & 1.71E+04 \\ [0.1cm] 
  &   & FPMA & 60601011006 & 59073.0 & 79.61 & 52.38 & 0.21 & 1.09E+04 \\ [0.1cm] 
  &   & FPMB &   &   &   & 51.97 & 0.19 & 9.94E+03 \\ [0.1cm] 
  &   & FPMA & 60601011002 & 59064.9 & 97.04 & 58.24 & 0.17 & 9.74E+03 \\ [0.1cm] 
  &   & FPMB &   &   &   & 58.05 & 0.15 & 8.58E+03 \\ [0.1cm] 
\hline 
Mrk 590 & \xmm{} & EPIC-pn & 0201020201 & 53190.9 & 61.45 & 41.85 & 3.37 & 1.41E+05 \\ [0.1cm] 
  &   & EPIC-pn &   & 53190.4 & 45.31 & 29.23 & 3.17 & 9.27E+04 \\ [0.1cm] 
  &   & EPIC-pn & 0865470301 & 59217.5 & 25.14 & 17.10 & 3.84 & 6.56E+04 \\ [0.1cm] 
  &   & EPIC-pn & 0865470201 & 59034.6 & 25.14 & 17.25 & 3.23 & 5.58E+04 \\ [0.1cm] 
  &   & EPIC-pn & 0870840301 & 59981.2 & 36.14 & 24.73 & 2.01 & 4.97E+04 \\ [0.1cm] 
  &   & EPIC-pn & 0912400101 & 59788.4 & 25.14 & 17.26 & 2.75 & 4.75E+04 \\ [0.1cm] 
  &   & EPIC-pn & 0870840101 & 59437.1 & 8.94 & 6.04 & 5.65 & 3.41E+04 \\ [0.1cm] 
  &   & EPIC-pn & 0870840401 & 59439.1 & 5.34 & 3.64 & 6.69 & 2.43E+04 \\ [0.1cm] 
  &   & EPIC-pn & 0870840201 & 59603.1 & 23.94 & 16.34 & 1.12 & 1.83E+04 \\ [0.1cm] 
  &   & EPIC-pn & 0109130301 & 52275.5 & 10.00 & 6.80 & 2.43 & 1.65E+04 \\ [0.1cm] 
  & \nustar{} & FPMA & 80502630002 & 58726.4 & 136.75 & 65.02 & 0.35 & 2.29E+04 \\ [0.1cm] 
  &   & FPMB &   &   &   & 64.05 & 0.31 & 2.00E+04 \\ [0.1cm] 
  &   & FPMA & 80502630004 & 58869.5 & 97.75 & 50.03 & 0.36 & 1.79E+04 \\ [0.1cm] 
  &   & FPMB &   &   &   & 49.64 & 0.33 & 1.66E+04 \\ [0.1cm] 
  &   & FPMA & 80502630006 & 59224.5 & 84.21 & 41.26 & 0.17 & 7.13E+03 \\ [0.1cm] 
  &   & FPMB &   &   &   & 41.00 & 0.16 & 6.41E+03 \\ [0.1cm] 
  &   & FPMA & 80602604002 & 59571.0 & 114.55 & 53.05 & 0.13 & 7.05E+03 \\ [0.1cm] 
  &   & FPMB &   &   &   & 52.42 & 0.12 & 6.41E+03 \\ [0.1cm] 
  &   & FPMA & 60761012002 & 59444.2 & 39.76 & 18.65 & 0.34 & 6.40E+03 \\ [0.1cm] 
  &   & FPMB &   &   &   & 16.96 & 0.32 & 5.39E+03 \\ [0.1cm] 
  &   & FPMA & 80402610002 & 58418.4 & 38.50 & 21.07 & 0.25 & 5.26E+03 \\ [0.1cm] 
  &   & FPMB &   &   &   & 20.96 & 0.23 & 4.78E+03 \\ [0.1cm] 
  &   & FPMA & 80602604004 & 59981.1 & 84.75 & 40.58 & 0.12 & 4.69E+03 \\ [0.1cm] 
  &   & FPMB &   &   &   & 39.88 & 0.11 & 4.31E+03 \\ [0.1cm] 
  &   & FPMA & 90201043002 & 57724.5 & 95.57 & 50.62 & 0.08 & 4.20E+03 \\ [0.1cm] 
  &   & FPMB &   &   &   & 50.45 & 0.07 & 3.46E+03 \\ [0.1cm] 
  &   & FPMA & 60160095002 & 57423.7 & 39.03 & 21.21 & 0.08 & 1.60E+03 \\ [0.1cm] 
  &   & FPMB &   &   &   & 21.16 & 0.07 & 1.39E+03 \\ [0.1cm] 
\hline 
Mrk 79 & \xmm{} & EPIC-pn & 0870880101 & 59117.6 & 29.64 & 20.11 & 12.72 & 2.56E+05 \\ [0.1cm] 
  &   & EPIC-pn & 0400070201 & 54008.5 & 20.67 & 13.84 & 15.11 & 2.09E+05 \\ [0.1cm] 
  &   & EPIC-pn & 0400070401 & 54178.1 & 20.31 & 13.57 & 10.24 & 1.39E+05 \\ [0.1cm] 
  &   & EPIC-pn & 0400070301 & 54040.6 & 19.97 & 13.92 & 9.95 & 1.39E+05 \\ [0.1cm] 
  &   & EPIC-pn & 0103860801 & 51826.2 & 2.41 & 1.61 & 11.46 & 1.84E+04 \\ [0.1cm] 
  &   & EPIC-pn & 0103862101 & 52025.8 & 5.12 & 3.52 & 5.21 & 1.83E+04 \\ [0.1cm] 
  & \nustar{} & FPMA & 60601010002 & 59115.5 & 125.42 & 64.44 & 0.63 & 4.08E+04 \\ [0.1cm] 
  &   & FPMB &   &   &   & 64.45 & 0.58 & 3.72E+04 \\ [0.1cm] 
  &   & FPMA & 60601010004 & 59117.2 & 73.09 & 38.19 & 0.65 & 2.47E+04 \\ [0.1cm] 
  &   & FPMB &   &   &   & 38.10 & 0.59 & 2.25E+04 \\ [0.1cm] 
\hline  
PG 1229+204 & \xmm{} & EPIC-pn & 0301450201 & 53560.5 & 25.07 & 17.17 & 2.74 & 4.71E+04 \\ [0.1cm] 
  & \nustar{} & FPMA & 60061229002 & 57597.4 & 37.52 & 18.61 & 0.09 & 1.75E+03 \\ [0.1cm] 
  &   & FPMB &   &   &   & 18.39 & 0.09 & 1.58E+03 \\ [0.1cm] 
\hline
PG 0844+349 & \xmm{} & EPIC-pn & 0103660201$^{p}$ & 51853.0 & 21.24 & 5.98 & 2.38 & 1.43E+04 \\ [0.1cm] 
  &   & EPIC-pn & 0554710101 & 54954.2 & 12.67 & 11.33 & 0.42 & 4.72E+03 \\ [0.1cm] 
  & \nustar{} & FPMA & 60463024002 & 59224.0 & 41.43 & 17.70 & 0.05 & 8.14E+02 \\ [0.1cm] 
  &   & FPMB &   &   &   & 17.59 & 0.04 & 7.46E+02 \\ [0.1cm] 
\hline 
PG 0804+761 & \xmm{} & EPIC-pn & 0605110101 & 55265.5 & 44.50 & 16.18 & 10.36 & 1.68E+05 \\ [0.1cm] 
  &   & EPIC-pn & 0605110201 & 55267.5 & 36.25 & 15.56 & 7.88 & 1.23E+05 \\ [0.1cm] 
  & \nustar{} & FPMA & 60160322002 & 57480.1 & 26.30 & 16.94 & 0.18 & 3.13E+03 \\ [0.1cm] 
  &   & FPMB &   &   &   & 16.87 & 0.18 & 2.96E+03 \\ [0.1cm] 
\hline 
PG 1426+015 & \xmm{} & EPIC-pn & 0852210101$^{p}$ & 58872.5 & 106.14 & 70.48 & 3.05 & 2.15E+05 \\ [0.1cm] 
  &   & EPIC-pn & 0102040501$^{p}$ & 51753.5 & 6.06 & 0.68 & 4.16 & 2.85E+03 \\ [0.1cm] 
  & \nustar{} & FPMA & 60501049002 & 58871.9 & 202.55 & 94.12 & 0.17 & 1.56E+04 \\ [0.1cm] 
  &   & FPMB &   &   &   & 93.82 & 0.16 & 1.48E+04 \\ [0.1cm] 
  &   & FPMA & 60061254002 & 58340.4 & 62.21 & 32.48 & 0.20 & 6.62E+03 \\ [0.1cm] 
  &   & FPMB &   &   &   & 32.32 & 0.19 & 6.05E+03 \\ [0.1cm] 
\hline  

\end{longtable}


\twocolumngrid

\setcounter{figure}{0}
\renewcommand{\thefigure}{A\arabic{figure}}

\begin{figure*}
\centering
\begin{center}
\includegraphics[scale=0.29,angle=-0]{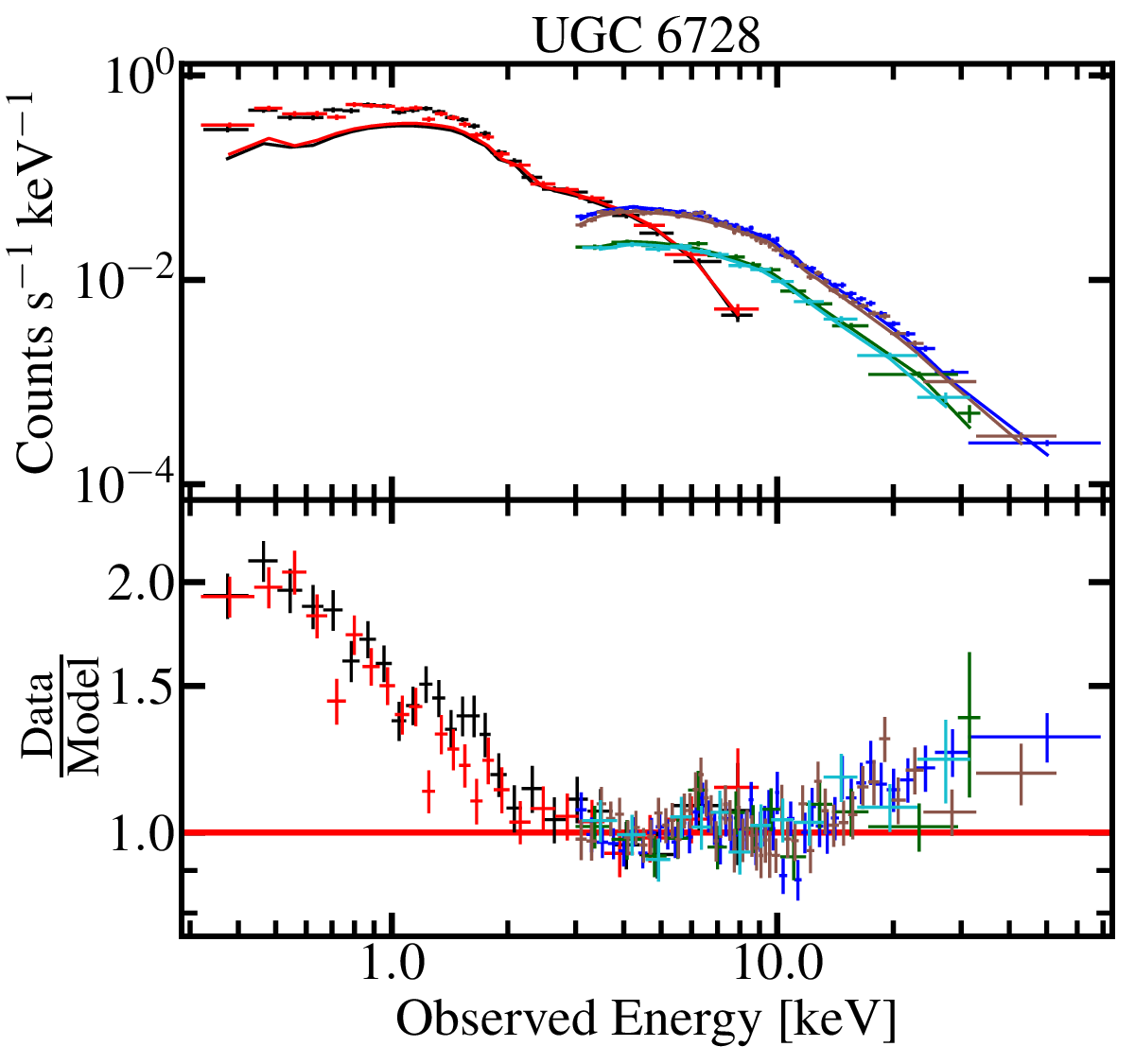}
\includegraphics[scale=0.29,angle=-0]{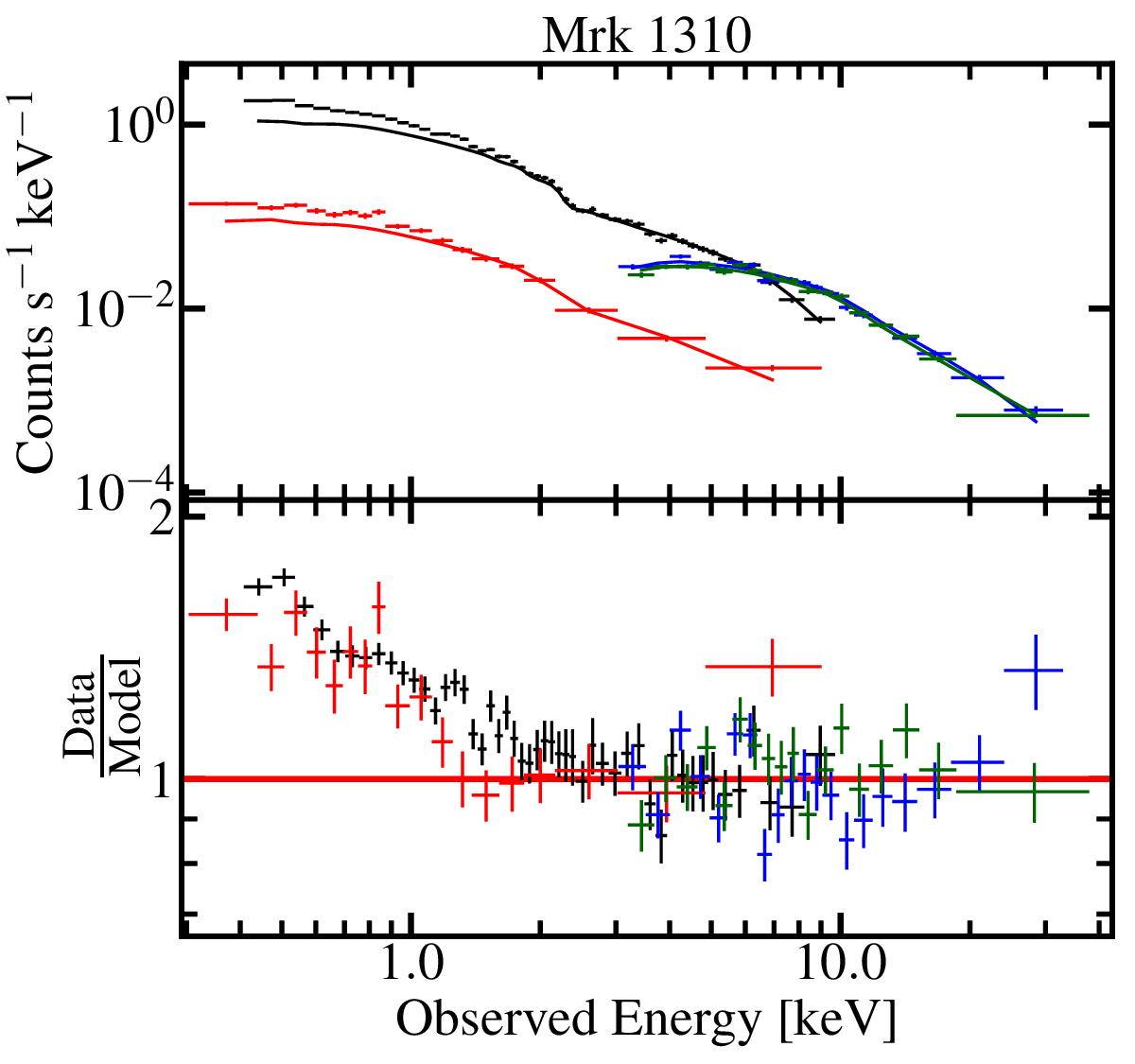}
\includegraphics[scale=0.29,angle=-0]{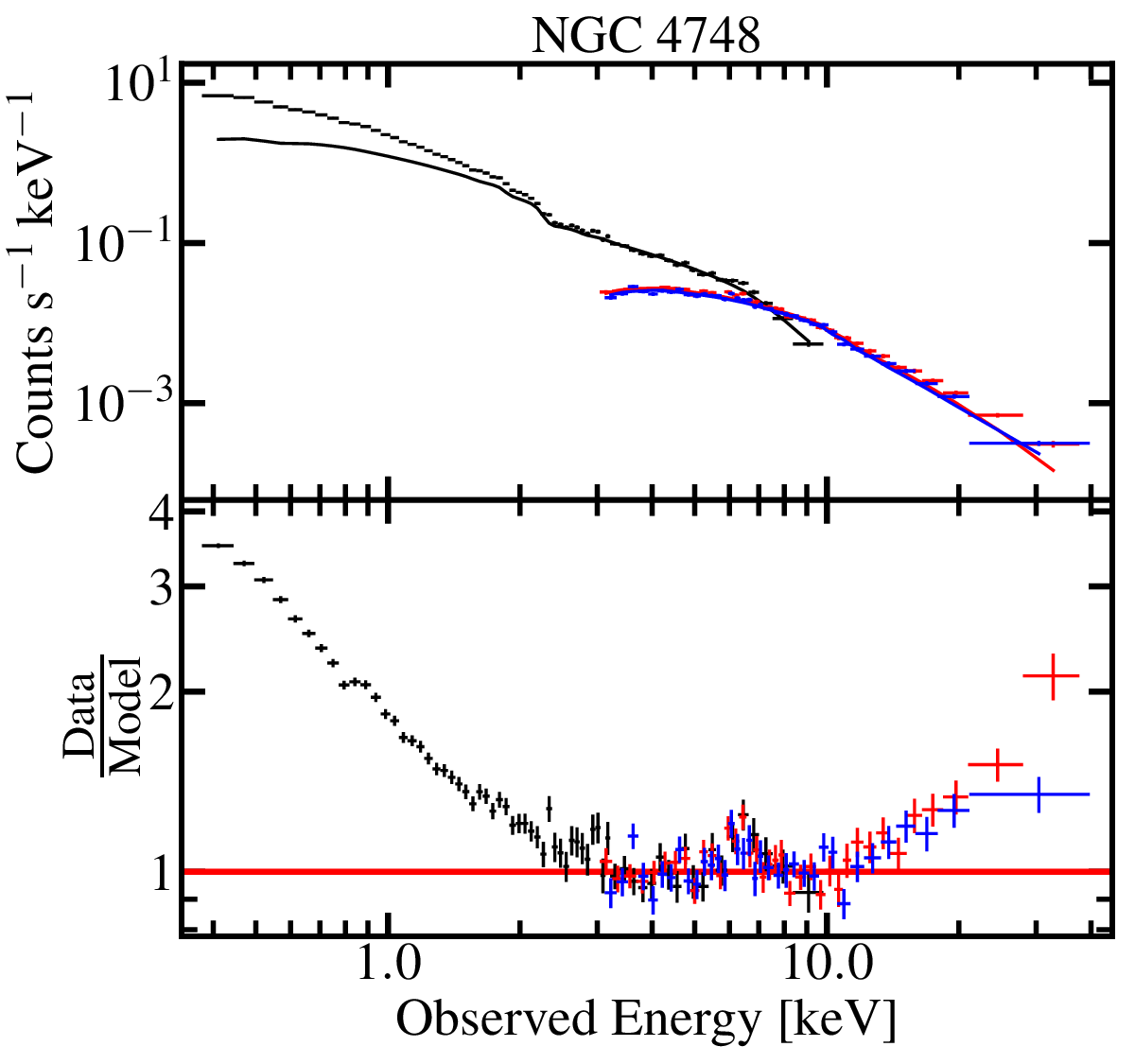}
\includegraphics[scale=0.29,angle=-0]{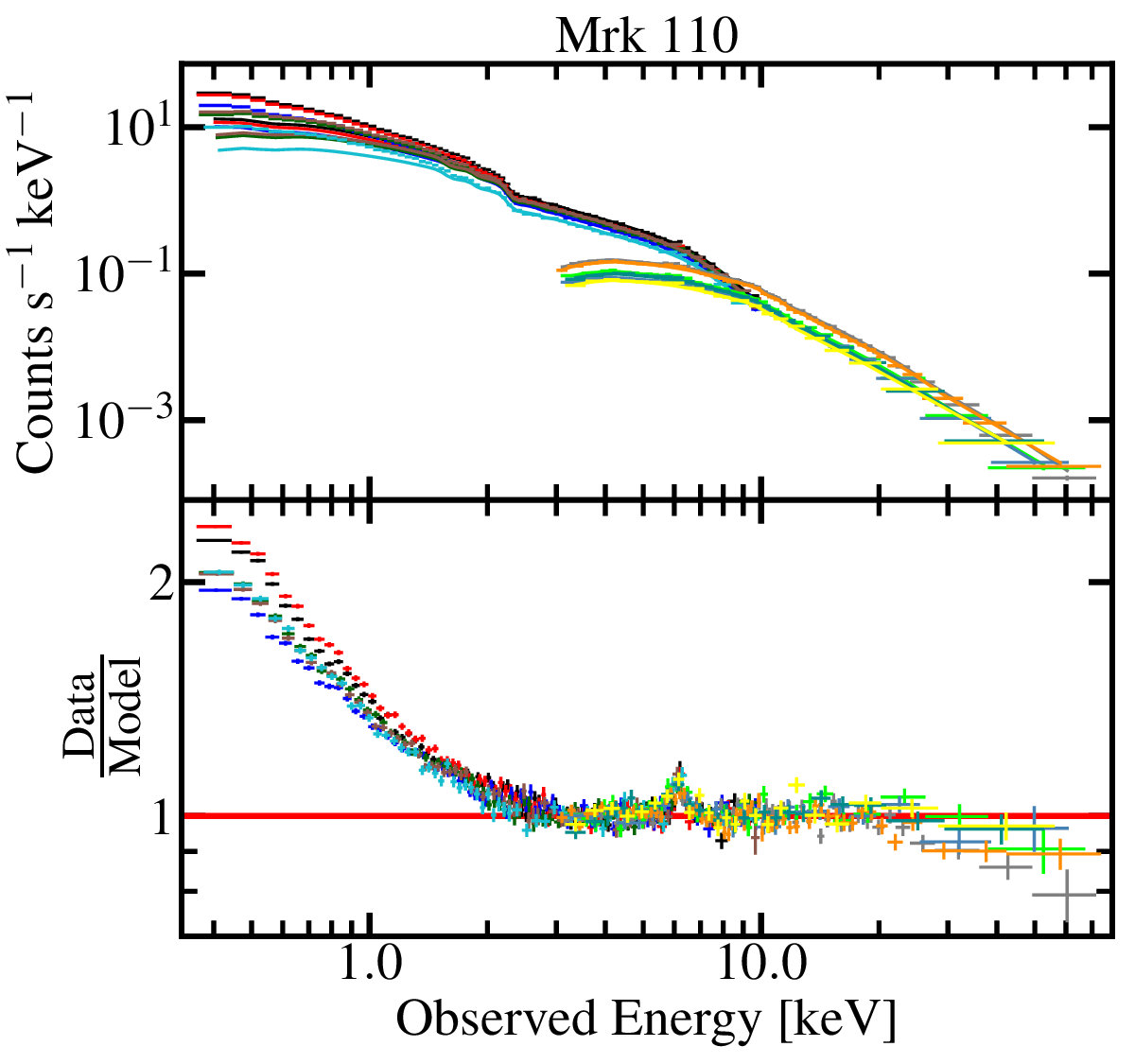}
\includegraphics[scale=0.29,angle=-0]{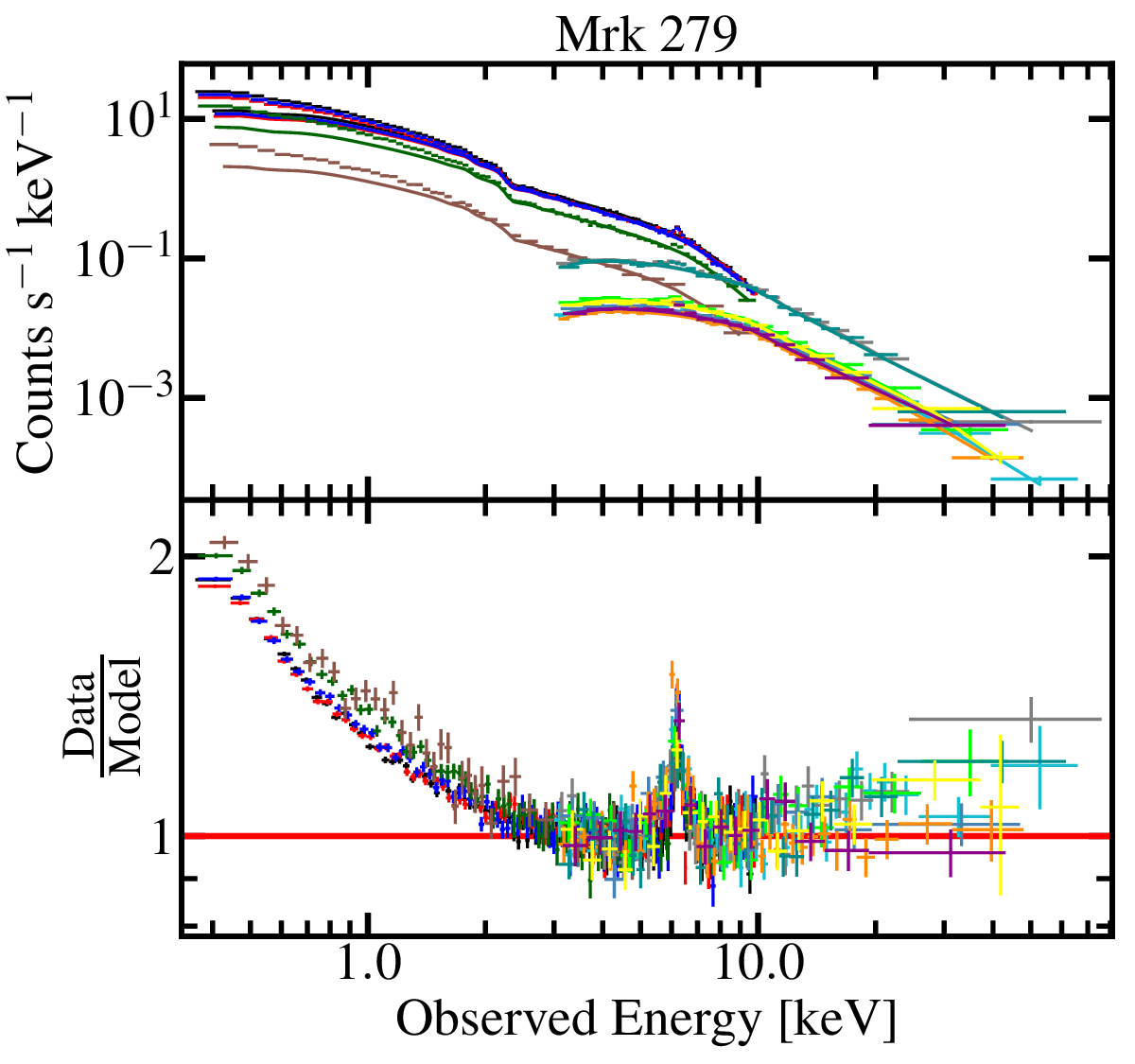}
\includegraphics[scale=0.29,angle=-0]{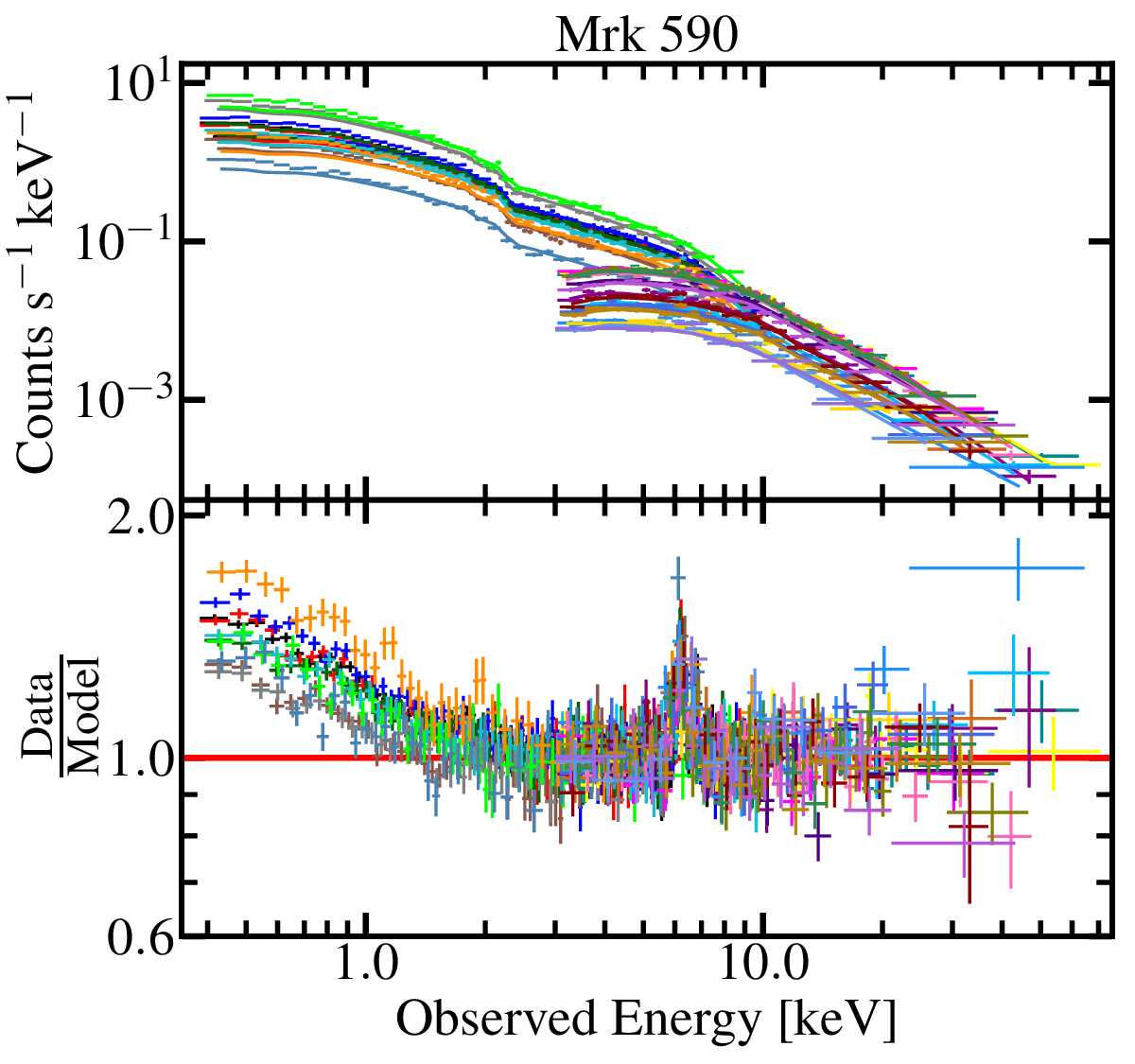}
\includegraphics[scale=0.29,angle=-0]{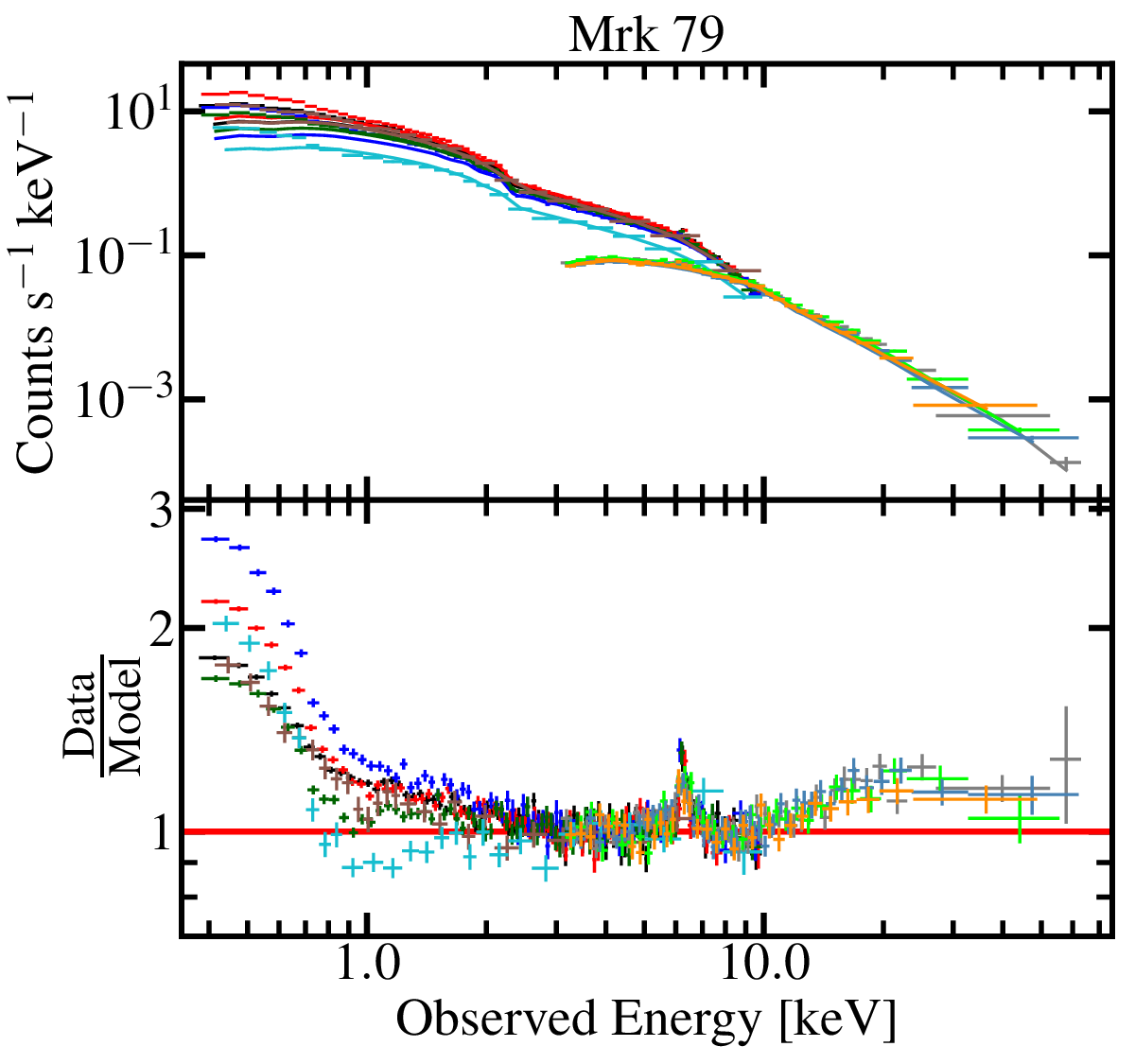}
\includegraphics[scale=0.29,angle=-0]{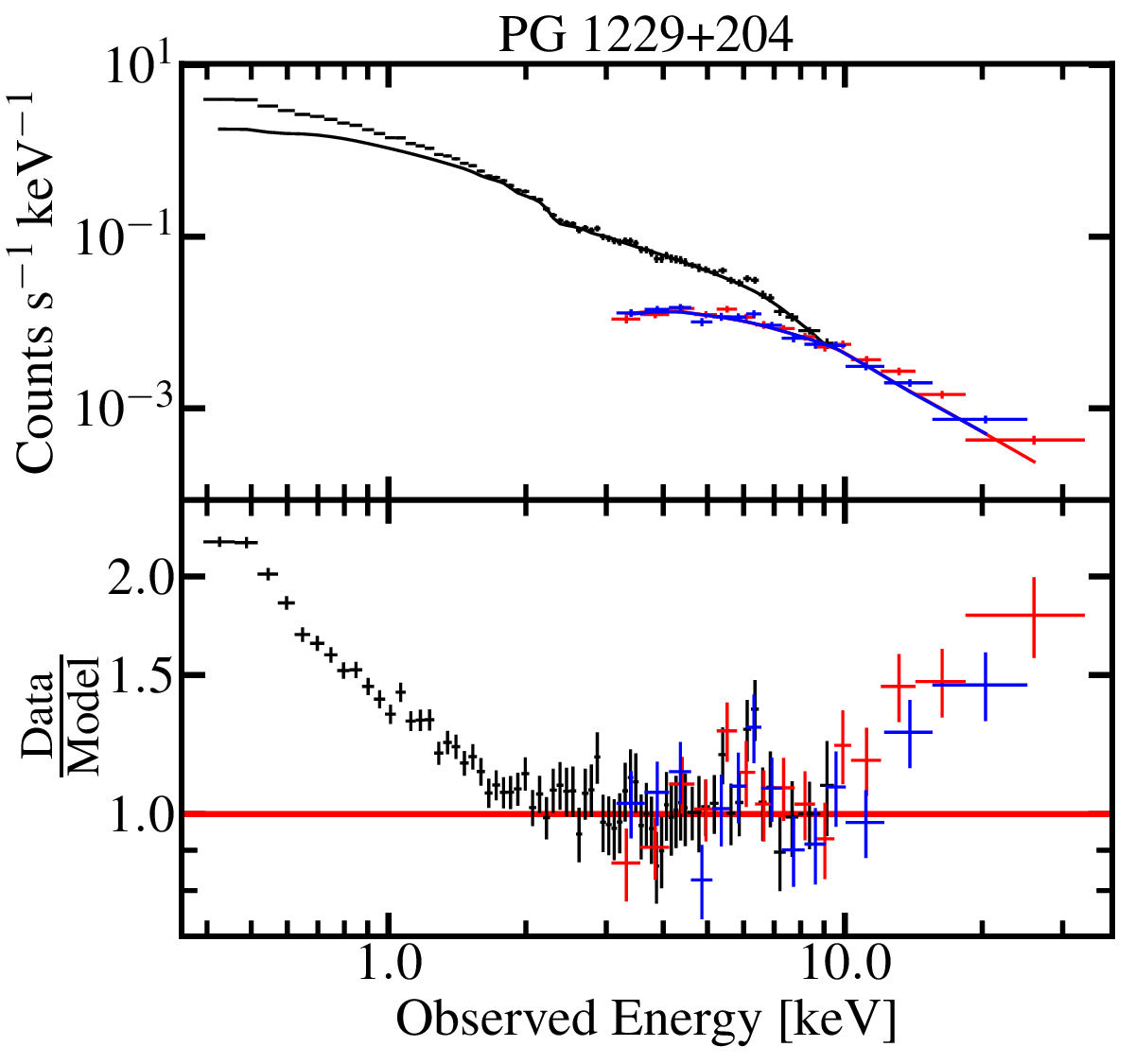}
\includegraphics[scale=0.29,angle=-0]{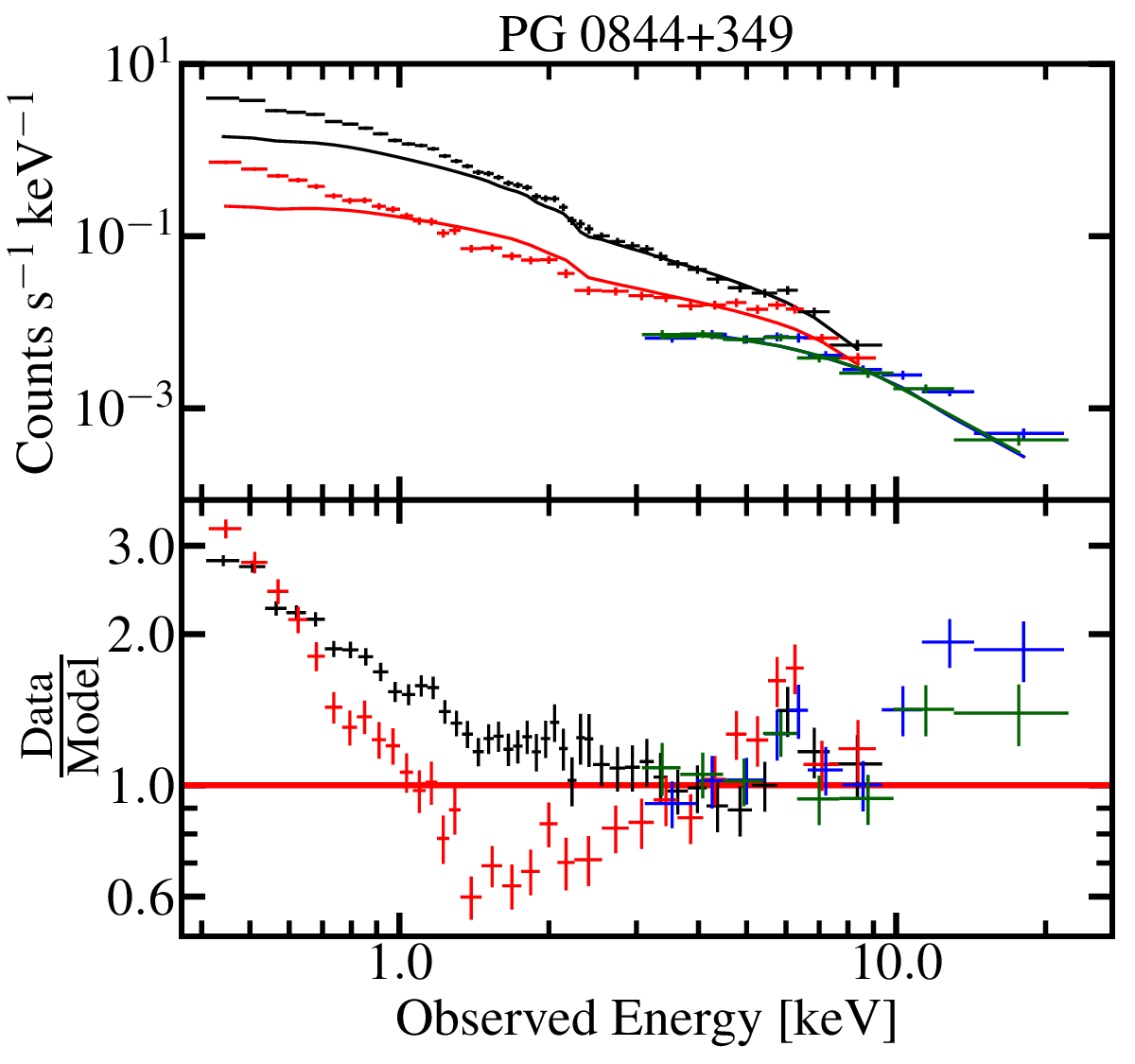}
\includegraphics[scale=0.29,angle=-0]{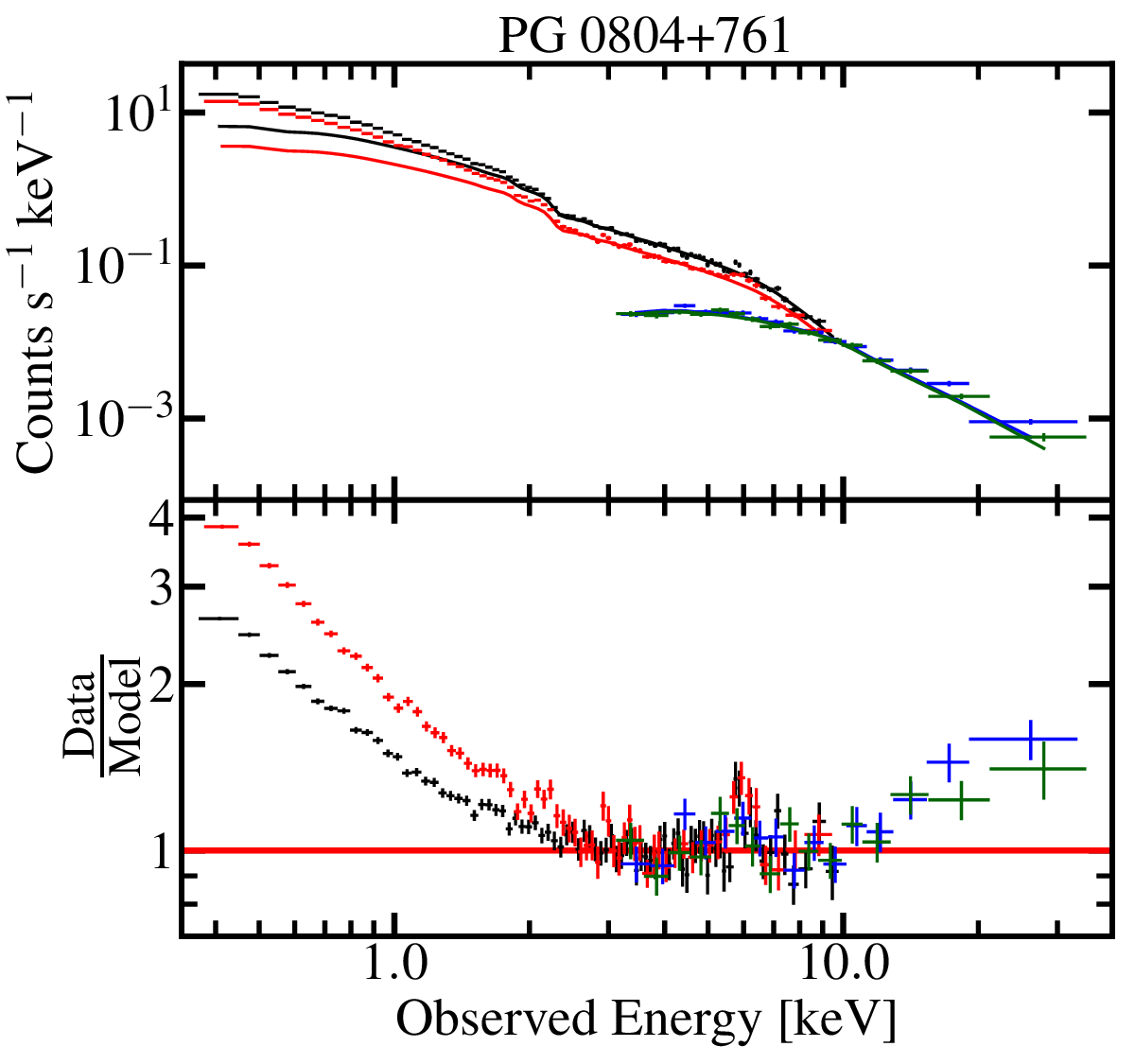}
\includegraphics[scale=0.29,angle=-0]{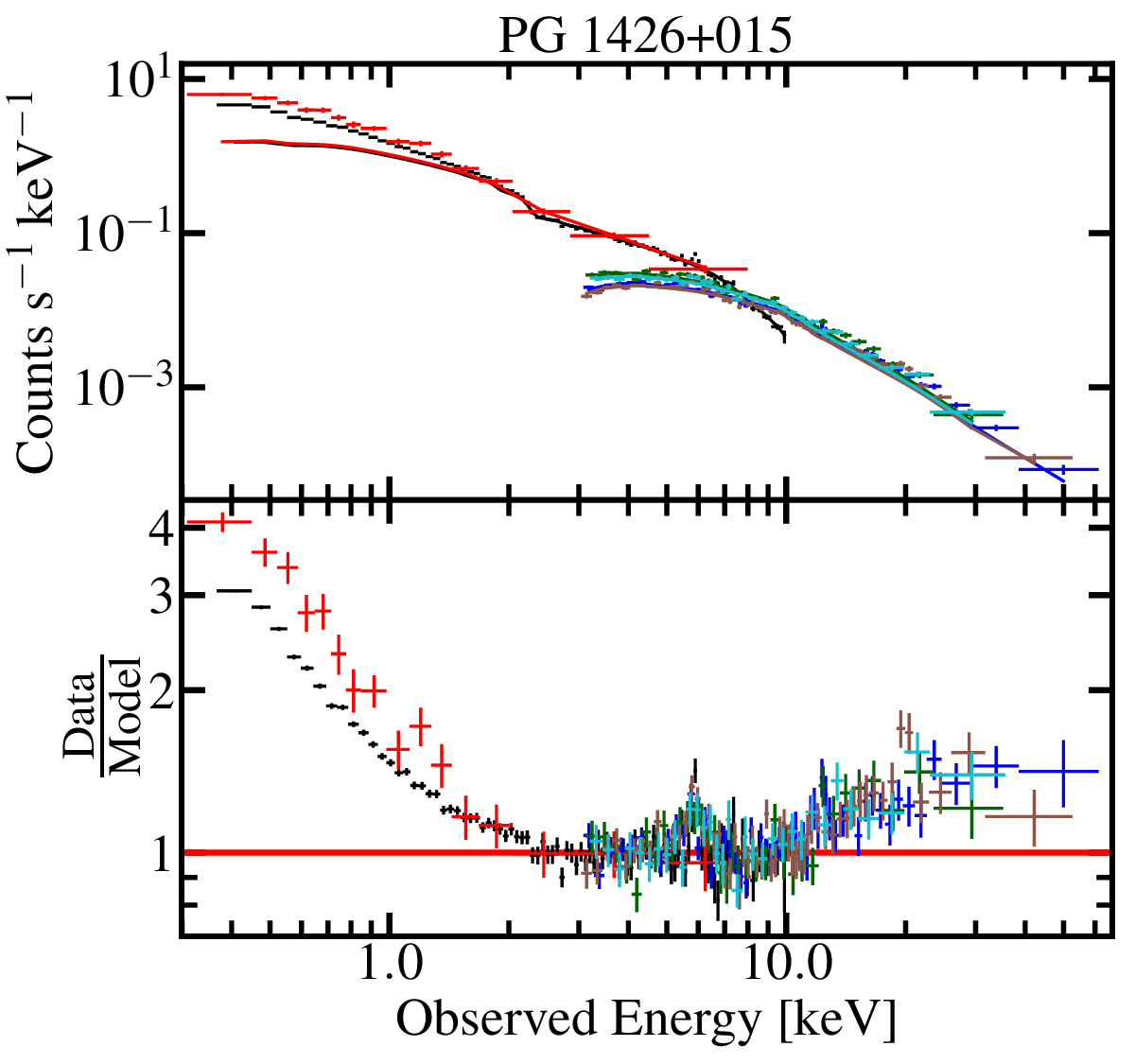}
\caption{The top panels show the \xmm{} EPIC and \nustar{} FPM spectral data fitted by the Galactic absorption corrected power-law model ({\tt{TBabs$*$zpowerlw}}) in the range 3$-$5\keV{} and 7$-$10\keV{} and then extrapolated over the complete energy range for each source. The bottom panel is the corresponding data-to-model ratio. The ratio plot shows soft X-ray excess below around 1$-$2\keV{} and Fe~K emission in the 6$-$7\keV{} range for each source, along with a hard X-ray excess in the 15$-$30\keV{} range for some sources. The spectra are re-binned to visualize and demonstrate the shape of the residuals more clearly.}
\label{ratio_plot}
\end{center}
\end{figure*}

\begin{table*}
\centering
\caption{Details of the maximum likelihood ratio (MLR) and Bayesian statistical tests for quantitative assessment of the broad Fe~K emission feature in the 5--7\keV{} band. Model~1 and Model~2 in columns~(2) and (3) show the source models used to fit the 3--10\keV{} spectra before and after adding the relativistic reflection ({\tt relxillCp}) component for the broad Fe~K emission. Columns~(4) and (5) report the model statistics obtained from fitting the 3--10\keV{} spectra with Model~1 and Model~2, respectively. Column~(6) reports the confidence level of the detected broad Fe~K emission feature, as computed by the MLR test. Columns~(7) and (8) present the Deviance Information Criteria, ${\rm DIC}_{1}$ and ${\rm DIC}_{2}$, without and with the relativistic reflection ({\tt relxillCp}) component, respectively. Column~(9) shows the difference between ${\rm DIC}_{1}$ and ${\rm DIC}_{2}$, affirming the significance of the broad Fe~K emission feature.} 
\begin{center}
\begin{tabular}{cccccccccccc}
\hline 
Source & Model~1 & Model~2 & $\frac{\chi_{1}^{2}}{{\rm d.o.f}_{1}}$ & $\frac{\chi_{2}^{2}}{{\rm d.o.f}_{2}}$  & C.I.-MLR & ${\rm DIC}_{\rm 1}$  &  ${\rm DIC}_{\rm 2}$  & $\Delta {\rm DIC_{12}}$    \\
                 & (without {\tt relxillCp}) & (with {\tt relxillCp}) & & & $(\%)$ &  &  &  ${\rm DIC}_{\rm 1}-{\rm DIC}_{\rm 2}$   \\ [0.1cm]
  
(1)    &   (2)   &   (3)   &   (4)  &  (5)   & (6)  & (7) & (8) & (9)  \\  [0.1cm]
                                                     
\hline 

UGC 6728 & {\tt nthComp} & {\tt nthComp+relxillCp} & $\frac{183.5}{174}$ & $\frac{153.6}{165}$ & $99.95$ & $200.2$ & $187.6$ & $12.6$ [Very Strong] \\ [0.35cm]
         
Mrk 1310 & {\tt nthComp} & {\tt nthComp+relxillCp}  &  $\frac{144.0}{104}$ & $\frac{115.5}{95}$ & $99.92$ & $151.5$ & $145.4$ & $6.1$ [Strong] \\ [0.35cm]
         
NGC 4748 & {\tt nthComp+zGauss\_N}  & ({\tt nthComp+zGauss\_N}  &  $\frac{123.0}{108}$ & $\frac{103.4}{100}$ & $98.8$ & $139.1$ & $129.4$ & $9.7$ [Strong]   \\ [0.1cm]

         &                          & {\tt +relxillCp})  &  &  &  &  &  &    \\ [0.35cm]
         
Mrk 110 &  {\tt nthComp+zGauss\_N}  & ({\tt nthComp+zGauss\_N} & $\frac{653.0}{558}$ & $\frac{591.6}{545}$ & $>99.99$ & $707.6$ & $669.2$ & $38.4$ [Very Strong]  \\ [0.1cm]

       &                          & {\tt +relxillCp})  &  &  &  &  &  &    \\ [0.35cm]

Mrk 279 & ({\tt nthComp+zGauss\_N} & ({\tt nthComp+zGauss\_N} & $\frac{618.0}{513}$ & $\frac{552.2}{498}$ & $>99.99$ & $694.8$ & $650.8$ & $44.0$ [Very Strong]   \\ [0.1cm]
        &   {\tt +zGauss2\_N})             & {\tt zGauss2\_N+relxillCp})  &  &  &  &  &  &    \\ [0.35cm]

Mrk 590 &  ({\tt nthComp+zGauss\_N} & ({\tt nthComp+zGauss\_N+} &  $\frac{969.0}{945}$ & $\frac{930.0}{921}$ & $97.27$ & $1157.2$ & $1151.8$ & $5.4$  [Positive]  \\ [0.1cm]
        &   {\tt +zGauss2\_N})             & {\tt zGauss2\_N+relxillCp})  &  &  &  &  &  &    \\ [0.35cm]

Mrk 79 & ({\tt nthComp+zGauss\_N} & ({\tt nthComp+zGauss\_N+}  &  $\frac{446.2}{434}$ & $\frac{411.8}{421}$ & $99.9$ & $511.5$ & $497.9$ & $13.6$ [Very Strong]  \\ [0.1cm]      
        &   {\tt +zGauss2\_N})             & {\tt zGauss2\_N+relxillCp})  &  &  &  &  &  &    \\ [0.35cm]
         
PG 1229+204 &  {\tt nthComp}  & {\tt nthComp+relxillCp}  &  $\frac{111.1}{91}$ & $\frac{95.6}{83}$ & $94.99$ & $119.7$ & $110.0$ & $9.7$ [Strong]  \\ [0.35cm]

PG 0844+349 & {\tt nthComp}  & {\tt nthComp+relxillCp} &  $\frac{106.3}{83}$ & $\frac{69.7}{74}$ & $>99.99$ & $133.2$ & $103.0$ & $30.2$ [Very Strong]  \\ [0.35cm]

PG 0804+761 & {\tt nthComp+zGauss\_N}  & ({\tt nthComp+zGauss\_N} &  $\frac{140.3}{148}$ & $\frac{121.0}{139}$ & $97.72$ & $164.2$ & $159.1$ & $5.1$ [Positive] \\ [0.1cm]
        &                 & {\tt +relxillCp})  &  &  &  &  &  &    \\ [0.35cm]

PG 1426+015 & {\tt nthComp+zGauss\_N}  & ({\tt nthComp+zGauss\_N} & $\frac{214.0}{168}$ & $\frac{184.1}{159}$ & $99.95$ & $237.1$ & $225.7$ & $11.4$ [Very Strong] \\ [0.1cm]
        &                 & {\tt +relxillCp})  &  &  &  &  &  &    \\ [0.35cm]

\hline

\end{tabular}
\end{center} 
\label{test_broad_FeK}           
\end{table*}


\begin{table*}
\centering
\caption{Details of the maximum likelihood ratio (MLR) and Bayesian statistical tests for quantitative assessment of the Compton hump. Model~1 and Model~2 in columns~(2) and (3), respectively, represent the primary power-law continuum ({\tt nthComp}) and Compton reflection plus the primary continuum ({\tt{xillverCp$+$nthComp}}). Columns~(4) and (5) show the model statistics obtained from fitting the 7--78\keV{} spectra with Model~1 and Model~2, respectively. Column~(6) reports the confidence level of the Compton hump over the primary power-law continuum and is evaluated by the MLR test. Columns~(7) and (8) show the Deviance Information Criteria, ${\rm DIC}_{1}$ and ${\rm DIC}_{2}$, for Model~1 and Model~2, respectively. Column~(9) presents the difference between ${\rm DIC}_{1}$ and ${\rm DIC}_{2}$, verifying the relevance of Compton reflection hump over the primary continuum.} 
\begin{center}
\scalebox{0.9}{%
\begin{tabular}{cccccccccccc}
\hline 
Source  & Model~1 & Model~2 & $\frac{\chi_{1}^{2}}{{\rm d.o.f}_{1}}$ & $\frac{\chi_{2}^{2}}{{\rm d.o.f}_{2}}$  & C.I.-MLR & ${\rm DIC}_{\rm 1}$  &  ${\rm DIC}_{\rm 2}$  & $\Delta {\rm DIC_{12}}$   \\
            &   &  &  &  & $(\%)$ &   & &  ${\rm DIC}_{\rm 1}-{\rm DIC}_{\rm 2}$    \\ [0.1cm]
  
(1)    &   (2)   &   (3)   &   (4)  &  (5)   & (6)  & (7) & (8)  & (9) \\   [0.1cm]                                                    
\hline 

UGC 6728 & {\tt nthComp} & {\tt nthComp+xillverCp} & $\frac{277.6}{236}$ & $\frac{270.6}{232}$ & $86.41$ & $300.0$ & $304.3$ & $-4.3$  [Negative]  \\ [0.15cm]
         
Mrk 1310 & {\tt nthComp} & {\tt nthComp+xillverCp} & $\frac{104.5}{93}$ & $\frac{102.0}{89}$ & $35.54$ & $111.4$ & $112.5$ & $-1.1$ [Negative]  \\ [0.15cm]
         
NGC 4748 & {\tt nthComp} & {\tt nthComp+xillverCp} & $\frac{144.7}{132}$ & $\frac{133.1}{129}$ & $99.11$ & $150.7$ & $145.4$ & $5.3$ [Positive] \\ [0.15cm]
         
Mrk 110 & {\tt nthComp} & {\tt nthComp+xillverCp} & $\frac{898.0}{801}$ & $\frac{861.8}{793}$ & $>99.99$ & $931.8$ & $903.8$ & $28.0$ [Very Strong]  \\ [0.15cm]
         
Mrk 279 & {\tt nthComp} & {\tt nthComp+xillverCp} & $\frac{717.3}{630}$ & $\frac{632.9}{620}$ & $>99.99$ & $745.9$ & $675.3$ & $70.6$ [Very Strong]  \\ [0.15cm]
         
Mrk 590 & {\tt nthComp} & {\tt nthComp+xillverCp} & $\frac{1037.4}{1026}$ & $\frac{1010.6}{1007}$ & $89.06$ & $1095.5$ & $1094.9$ & $0.6$ [Neutral]  \\ [0.15cm]
         
Mrk 79 & {\tt nthComp} & {\tt nthComp+xillverCp} & $\frac{427.0}{432}$ & $\frac{413.4}{424}$ & $90.72$ & $451.7$ & $449.4$ & $2.3$ [Positive]  \\ [0.15cm]

PG 1229+204 & {\tt nthComp} & {\tt nthComp+xillverCp} & $\frac{42.7}{54}$ & $\frac{42.5}{51}$ & $2.24$ & $50.3$ & $56.3$ & $-6.0$ [Negative] \\ [0.15cm]
         
PG 0844+349 & {\tt nthComp} & {\tt nthComp+xillverCp} & $\frac{24.4}{20}$ & $\frac{22.5}{16}$ & $24.59$ & $33.6$ & $37.7$ & $-4.1$ [Negative]  \\ [0.15cm]
         
PG 0804+761 & {\tt nthComp} & {\tt nthComp+xillverCp} & $\frac{87.6}{98}$ & $\frac{81.4}{94}$ & $81.53$ & $97.6$ & $99.4$ & $-1.8$ [Negative]  \\ [0.15cm]
         
PG 1426+015 & {\tt nthComp} & {\tt nthComp+xillverCp} & $\frac{245.9}{248}$ & $\frac{237.3}{245}$ & $96.49$ & $272.6$ & $264.5$ & $8.1$ [Strong]  \\ [0.15cm]

\hline

\end{tabular}}
\end{center} 
\label{test_hump_xil}           
\end{table*}


\begin{table*}
\centering
\caption{Details of the maximum likelihood ratio (MLR) and Bayesian statistical tests to quantitatively assess the relativistic nature of the Compton reflection hump. Model~1 and Model~2 in columns~(2) and (3) show the primary power-law continuum ({\tt nthComp}) and relativistic disk reflection plus the primary continuum ({\tt{rellxillCp$+$nthComp}}), respectively. Columns~(4) and (5) report the model statistics obtained by fitting the 7--78\keV{} spectra with Model~1 and Model~2, respectively. Column (6) shows the confidence level of the relativistic Compton reflection hump, evaluated by the MLR test. Columns~(7) and (8) present the Deviance Information Criteria, ${\rm DIC}_{1}$ and ${\rm DIC}_{2}$, for Model~1 and Model~2, respectively. Column~(9) shows the difference between ${\rm DIC}_{1}$ and ${\rm DIC}_{2}$, revealing the relativistic nature of the Compton reflection hump.} 
\begin{center}
\scalebox{0.9}{%
\begin{tabular}{cccccccccccc}
\hline 
Source  & Model~1 & Model~2 & $\frac{\chi_{1}^{2}}{{\rm d.o.f}_{1}}$ & $\frac{\chi_{2}^{2}}{{\rm d.o.f}_{2}}$  & C.I.-MLR & ${\rm DIC}_{\rm 1}$  &  ${\rm DIC}_{\rm 2}$  & $\Delta {\rm DIC_{12}}$   \\ 
            &   &  &  &  & $(\%)$ &   & &  ${\rm DIC}_{\rm 1}-{\rm DIC}_{\rm 2}$    \\  [0.1cm]
  
(1)    &   (2)   &   (3)   &   (4)  &  (5)   & (6)  & (7) & (8)  & (9) \\    [0.1cm]                                                   
\hline 

UGC 6728 & {\tt nthComp} & {\tt nthComp+rellxillCp}  & $\frac{277.6}{236}$ & $\frac{260.2}{227}$ & $95.72$ & $300.0$ & $285.8$ & $14.2$ [Very Strong] \\ [0.15cm]
         
Mrk 1310 & {\tt nthComp} & {\tt nthComp+rellxillCp}   & $\frac{104.5}{93}$ & $\frac{100.2}{84}$ & $10.94$ & $111.4$ & $113.2$ & $-1.8$ [Negative] \\ [0.15cm]
         
NGC 4748 & {\tt nthComp} & {\tt nthComp+rellxillCp}   & $\frac{144.7}{132}$ & $\frac{127.5}{124}$ & $97.19$ & $150.7$ & $146.5$ & $4.2$ [Positive] \\ [0.15cm]
         
Mrk 110 & {\tt nthComp} & {\tt nthComp+rellxillCp}  & $\frac{898.0}{801}$ & $\frac{848.8}{788}$ & $>99.99$ & $931.8$ & $911.7$ & $20.1$ [Very Strong]  \\ [0.15cm]
         
Mrk 279 & {\tt nthComp} & {\tt nthComp+rellxillCp}  & $\frac{717.3}{630}$ & $\frac{627.8}{615}$ & $>99.99$ & $745.9$ & $677.5$ & $68.4$  [Very Strong] \\ [0.15cm]
         
Mrk 590 & {\tt nthComp} & {\tt nthComp+rellxillCp} & $\frac{1037.4}{1026}$ & $\frac{1005.4}{1002}$ & $87.3$ & $1095.5$ & $1103.5$ & $-8.0$ [Negative] \\ [0.15cm]
         
Mrk 79 & {\tt nthComp} & {\tt nthComp+rellxillCp} & $\frac{427.0}{432}$ & $\frac{406.6}{419}$ & $91.43$ & $451.7$ & $446.4$ & $5.3$ [Positive] \\ [0.15cm]
         
PG 1229+204 & {\tt nthComp} & {\tt nthComp+rellxillCp} & $\frac{42.7}{54}$ & $\frac{40.8}{46}$ & $1.61$ & $50.3$ & $52.1$ & $-1.8$ [Negative] \\ [0.15cm]
         
PG 0844+349 & {\tt nthComp} & {\tt nthComp+rellxillCp} & $\frac{24.4}{20}$ & $\frac{21.0}{11}$ & $5.37$ & $33.6$ & $41.3$ & $-7.7$ [Negative] \\ [0.15cm]
         
PG 0804+761 & {\tt nthComp} & {\tt nthComp+rellxillCp} & $\frac{87.6}{98}$ & $\frac{75.7}{89}$ & $78.1$ & $97.6$ & $101.6$ & $-4.0$ [Negative] \\ [0.15cm]
         
PG 1426+015 & {\tt nthComp} & {\tt nthComp+rellxillCp} & $\frac{245.9}{248}$ & $\frac{227.8}{239}$ & $96.6$ & $272.6$ & $258.9$ & $13.7$ [Very Strong]  \\ [0.15cm]

\hline

\end{tabular}}
\end{center} 
\label{test_hump_rel}           
\end{table*}

\begin{table*}
\centering
\caption{Details of the maximum likelihood ratio (MLR) and Bayesian statistical tests performed to examine the relevance of an extra warm Comptonization ({\tt compTT}) component in addition to the variable-density relativistic disk reflection ({\tt relxillCp}) model for the observed soft X-ray excess. Model~1 and Model~2 in columns~(2) and (3), respectively, illustrate the source models used to fit the 0.3--78\keV{} spectra before and after the addition of an extra warm Comptonization ({\tt compTT}) component for the soft X-ray excess. Columns~(4) and (5) report the model statistics obtained from fitting the 0.3--78\keV{} spectra with Model~1 and Model~2, respectively. Column~(6) presents the confidence level of the warm coronal emission evaluated by the MLR test. Columns~(7) and (8), respectively, show the Deviance Information Criteria, ${\rm DIC}_{1}$ and ${\rm DIC}_{2}$, without and with warm Comptonization ({\tt compTT}). Column~(9) reports the difference between ${\rm DIC}_{1}$ and ${\rm DIC}_{2}$, verifying the contribution of warm coronal emission to the observed soft X-ray excess.}
\begin{center}
\begin{tabular}{cccccccccccc}
\hline 
Source & Model~1 & Model~2 & $\frac{\chi_{1}^{2}}{{\rm d.o.f}_{1}}$ & $\frac{\chi_{2}^{2}}{{\rm d.o.f}_{2}}$  & C.I.-MLR & ${\rm DIC}_{\rm 1}$  &  ${\rm DIC}_{\rm 2}$  & $\Delta {\rm DIC_{12}}$     \\

                 & (without {\tt compTT}) & (with {\tt compTT}) & & & $(\%)$ &  &  &  ${\rm DIC}_{\rm 1}-{\rm DIC}_{\rm 2}$   \\ [0.1cm]
  
(1)    &   (2)   &   (3)   &   (4)  &  (5)   & (6)  & (7) & (8) & (9)  \\  [0.1cm]
\hline 

UGC 6728 & {\tt nthComp+relxillCp} & ({\tt nthComp+relxillCp} & $\frac{514.7}{460}$ & $\frac{507.8}{455}$ & $77.18$ & $543.9$ & $545.2$ & $-1.3$ [Negative] \\ [0.1cm]
         &                          & {\tt +compTT})  &  &  &  &  &  &    \\ [0.45cm]

Mrk 1310 & {\tt nthComp+relxillCp} & ({\tt nthComp+relxillCp} & $\frac{277.1}{228}$ & $\frac{263.4}{223}$ & $98.24$ & $308.9$ & $297.9$ & $11.0$ [Very Strong] \\ [0.1cm]
         &                          & {\tt +compTT})  &  &  &  &  &  &    \\ [0.45cm]

NGC 4748 & ({\tt nthComp+xillverCp} & ({\tt nthComp+xillverCp} & $\frac{281.0}{242}$ & $\frac{258.3}{238}$ & $99.99$ & $306.7$ & $290.6$ & $16.1$ [Very Strong] \\ [0.1cm]
         &  {\tt +relxillCp})        & {\tt +relxillCp+compTT})  &  &  &  &  &  &    \\ [0.45cm]              
         
Mrk 110 &  ({\tt nthComp+xillverCp} & ({\tt nthComp+xillverCp} & $\frac{1904.4}{1434}$ & $\frac{1647.0}{1425}$ & $>99.99$ & $1996.5$ & $1752.8$ & $243.7$ [Very Strong] \\ [0.1cm]
           &  {\tt +relxillCp})        & {\tt +relxillCp+compTT})  &  &  &  &  &  &    \\ [0.45cm]                    
         
Mrk 279 &  ({\tt nthComp+xillverCp} & ({\tt nthComp+xillverCp} & $\frac{1376.9}{1181}$ & $\frac{1307.6}{1170}$ & $>99.99$ & $1463.2$ & $1413.0$ & $50.2$ [Very Strong] \\ [0.1cm]
         &  {\tt +relxillCp})        & {\tt +relxillCp+compTT})  &  &  &  &  &  &    \\ [0.45cm]

Mrk 590 & ({\tt nthComp+zGauss\_N}  & ({\tt nthComp+zGauss\_N} & $\frac{2030.2}{2028}$ & $\frac{2002.0}{2008}$ & $89.53$ & $2229.5$ & $2236.9$ & $-7.4$ [Negative] \\ [0.1cm]
        & {\tt +zGauss2\_N}         & {\tt +zGauss2\_N}  &  &  &  &  &  &    \\ [0.1cm] 
        &  {\tt +relxillCp})          & {\tt +relxillCp+compTT})  &  &  &  &  &  &    \\ [0.45cm]

Mrk 79 &  ({\tt nthComp+xillverCp} & ({\tt nthComp+xillverCp} & $\frac{1138.6}{972}$ & $\frac{1075.0}{963}$ & $>99.99$ & $1211.3$ & $1196.3$ & $15.0$ [Very Strong] \\ [0.1cm]
       &  {\tt +relxillCp})        & {\tt +relxillCp+compTT})  &  &  &  &  &  &    \\ [0.45cm]

PG 1229+204 & {\tt nthComp+relxillCp} & ({\tt nthComp+relxillCp} & $\frac{162.0}{154}$ & $\frac{158.2}{150}$ & $56.63$ & $188.2$ & $201.9$ & $-13.7$ [Negative] \\ [0.1cm]
            &                         & {\tt +compTT})  &  &  &  &  &  &    \\ [0.45cm]

PG 0844+349 &  {\tt nthComp+relxillCp} & ({\tt nthComp+relxillCp} & $\frac{192.0}{155}$ & $\frac{158.7}{150}$ & $>99.99$ & $236.7$ & $191.4$ & $45.3$ [Very Strong] \\ [0.1cm]
            &                           & {\tt +compTT})  &  &  &  &  &  &    \\ [0.45cm]               
                           
PG 0804+761 &  ({\tt nthComp+zGauss\_N} &  ({\tt nthComp+zGauss\_N} & $\frac{301.6}{277}$ & $\frac{252.2}{272}$ & $>99.99$ & $348.5$ & $310.1$ & $38.4$ [Very Strong] \\ [0.1cm]
            &  {\tt +relxillCp})       & {\tt +relxillCp+compTT})  &  &  &  &  &  &    \\ [0.45cm]

PG 1426+015 &  ({\tt nthComp+xillverCp} & ({\tt nthComp+xillverCp} & $\frac{524.9}{420}$ & $\frac{477.8}{415}$ & $>99.99$ & $575.8$ & $534.6$ & $41.2$ [Very Strong] \\ [0.1cm]
            &  {\tt +relxillCp})        & {\tt +relxillCp+compTT})  &  &  &  &  &  &    \\ [0.45cm]

\hline

\end{tabular}
\end{center} 
\label{test_wc}           
\end{table*}

\begin{figure*}
\centering
\begin{center}
\includegraphics[scale=0.25,angle=-0]{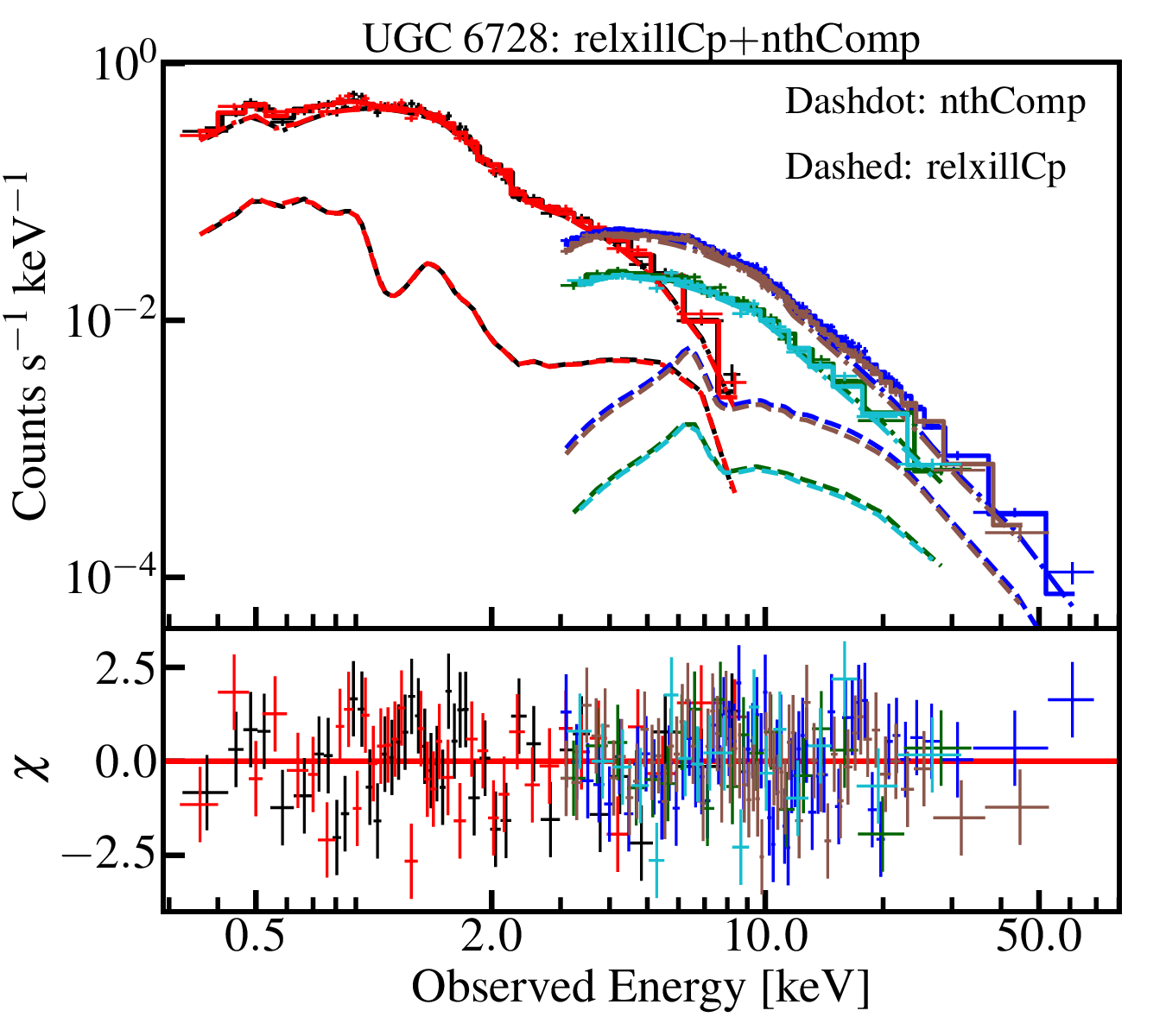}
\includegraphics[scale=0.25,angle=-0]{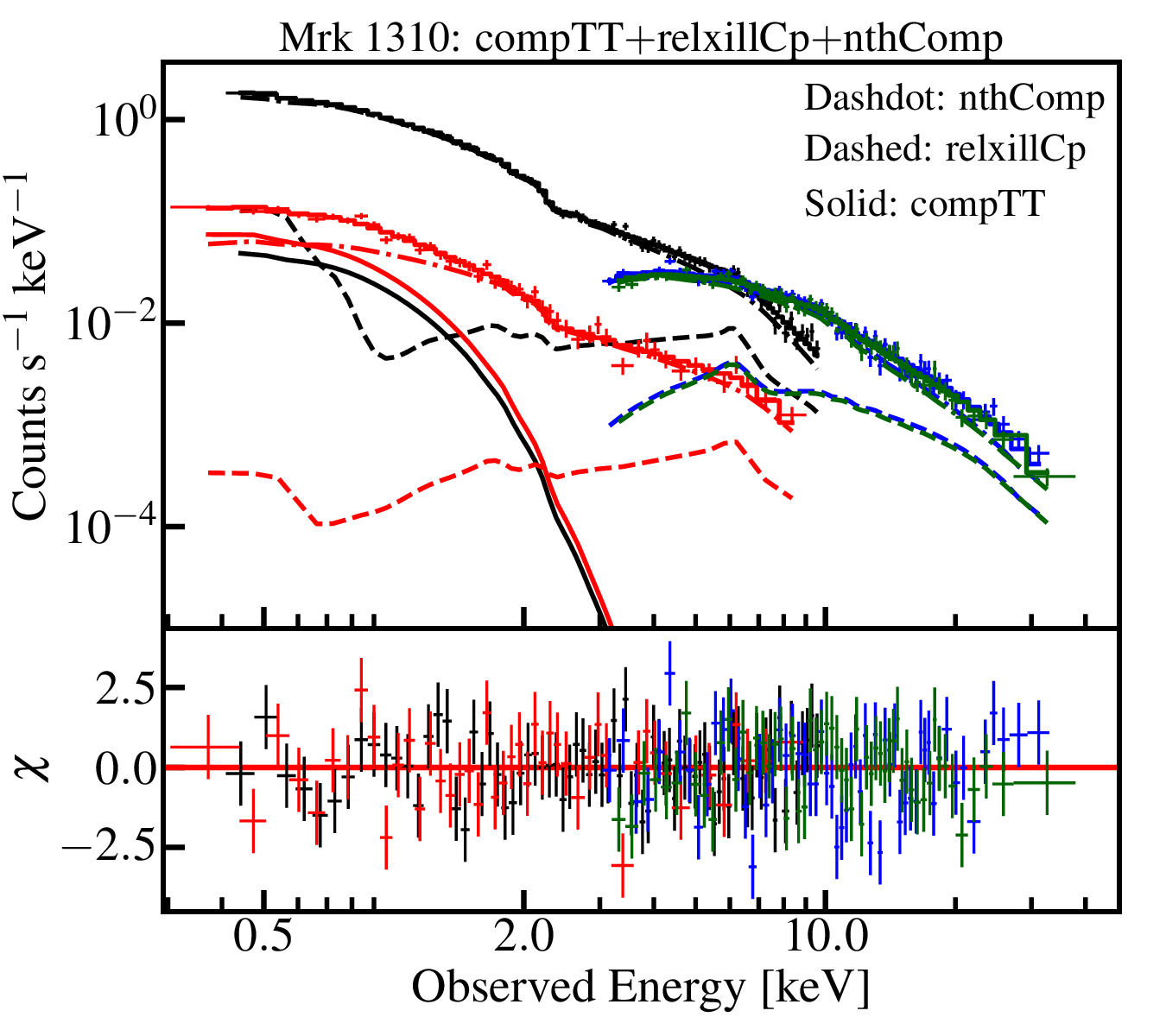}
\includegraphics[scale=0.25,angle=-0]{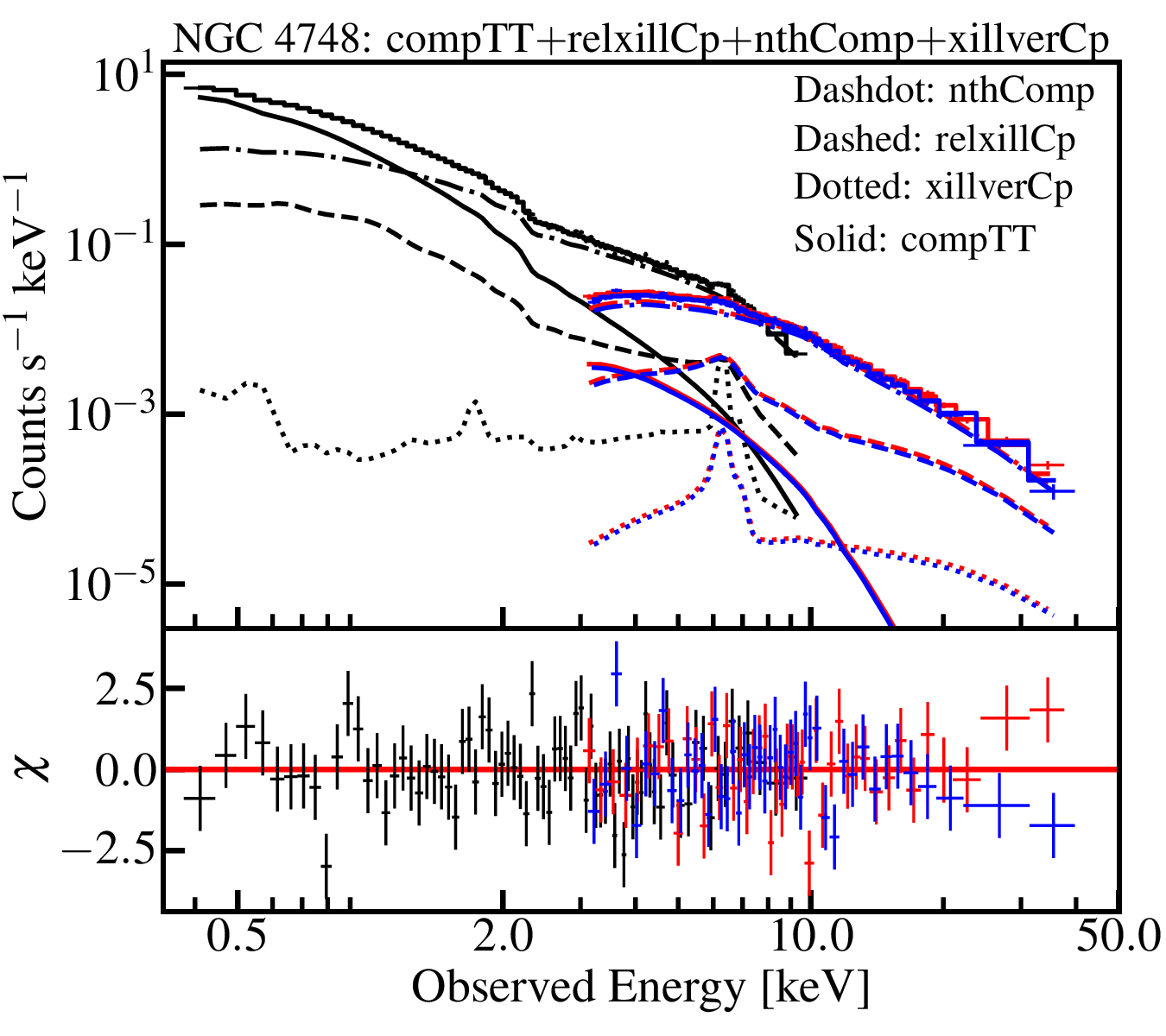}
\includegraphics[scale=0.25,angle=-0]{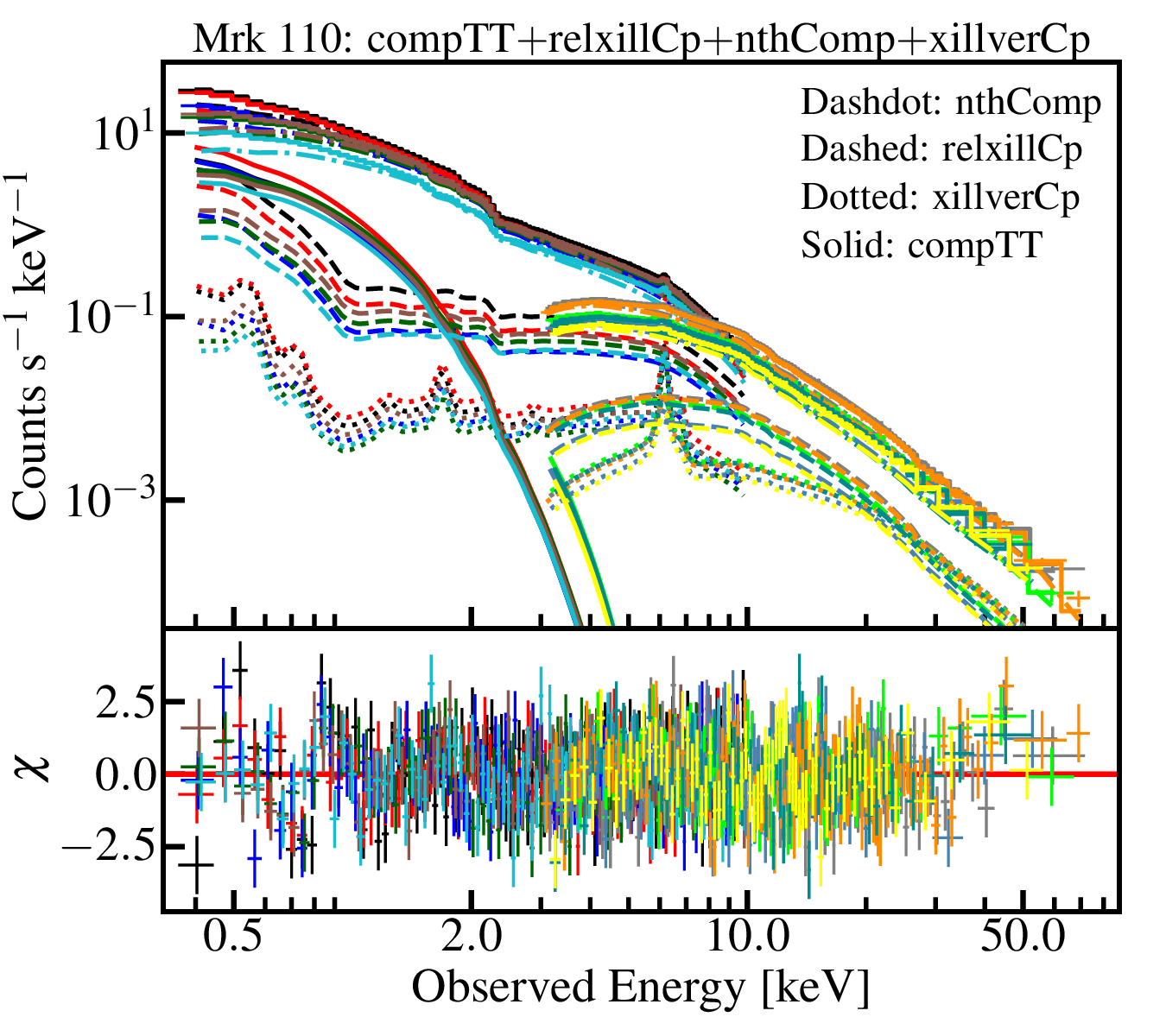}
\includegraphics[scale=0.25,angle=-0]{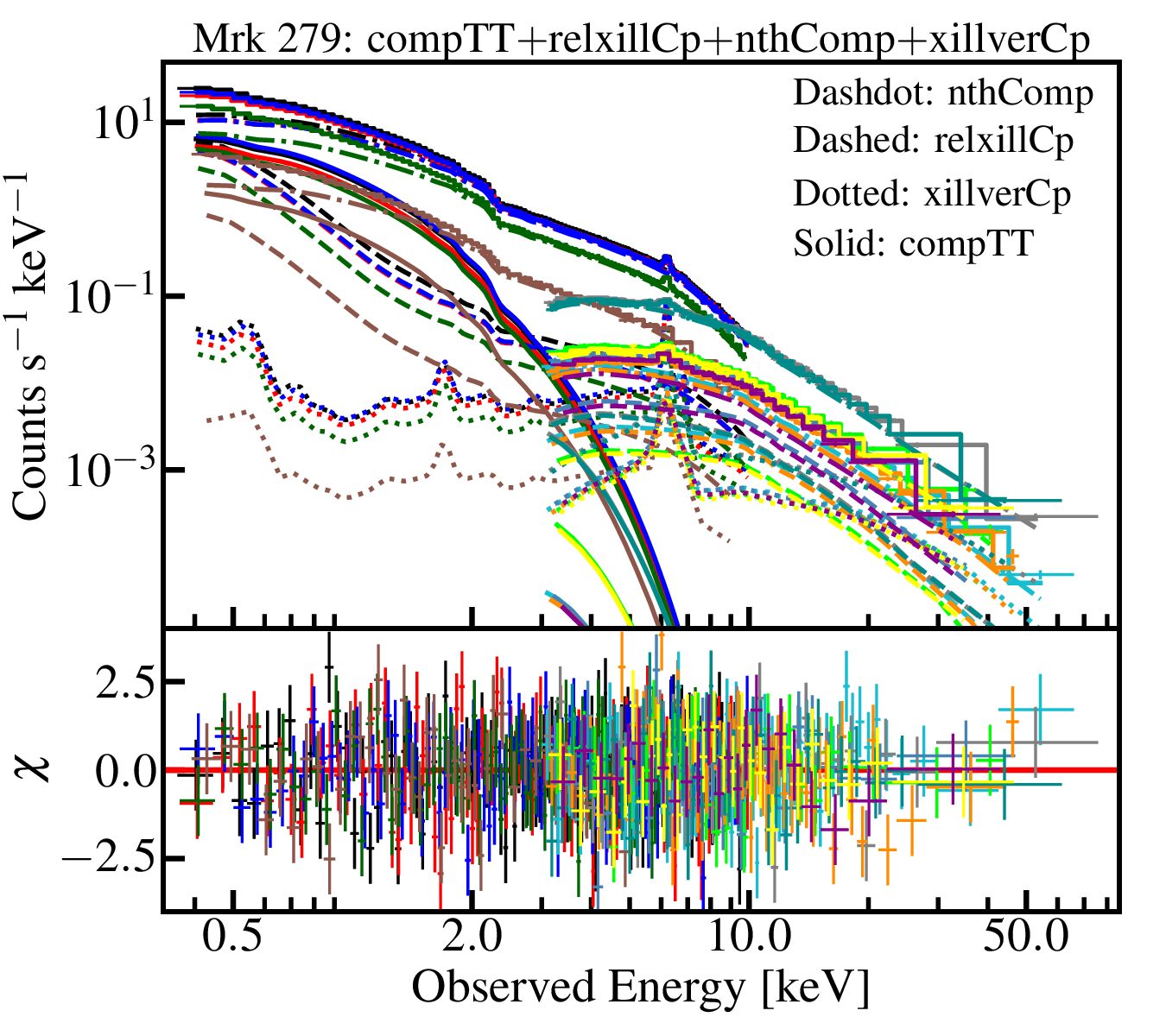}
\includegraphics[scale=0.25,angle=-0]{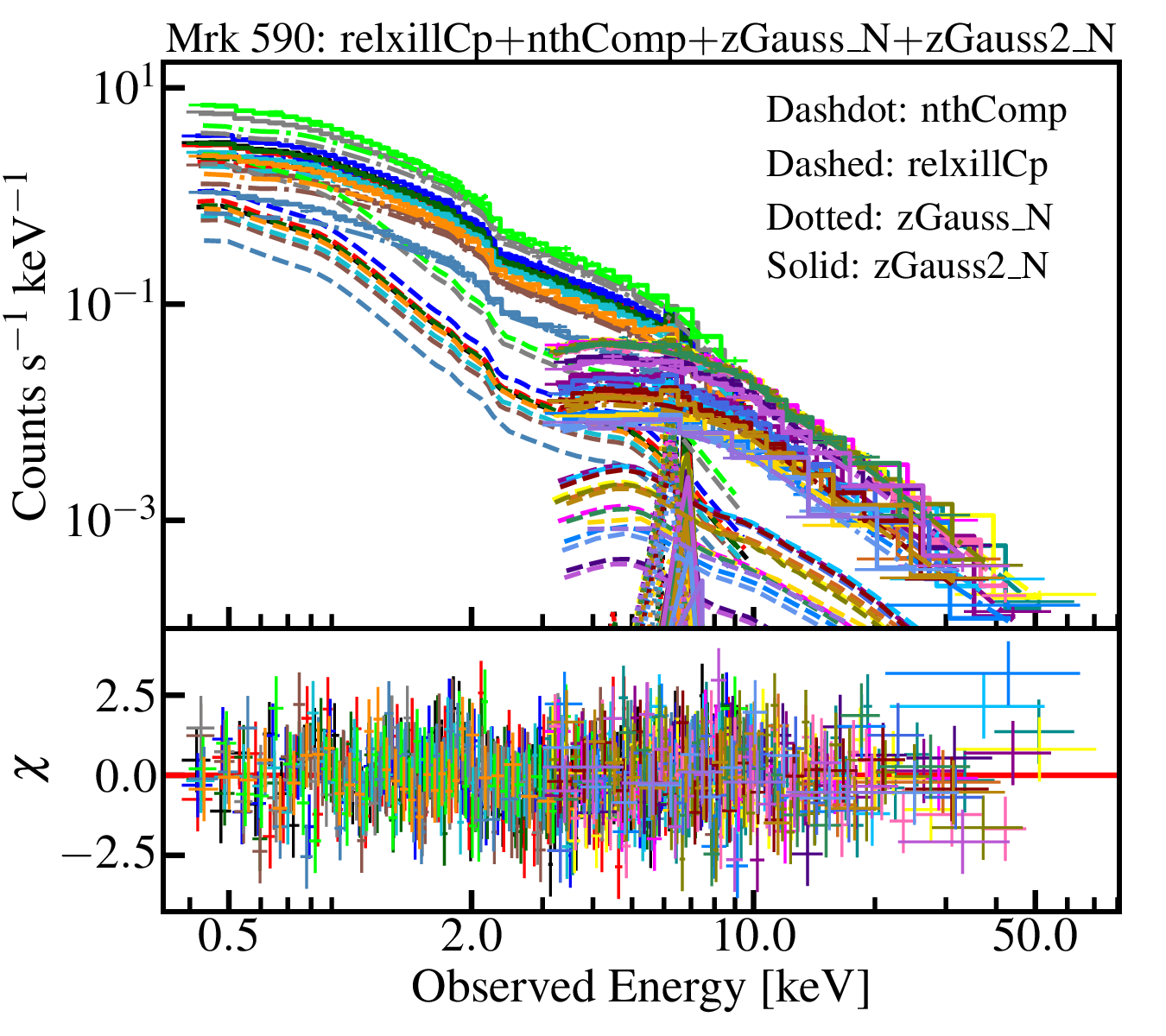}
\includegraphics[scale=0.25,angle=-0]{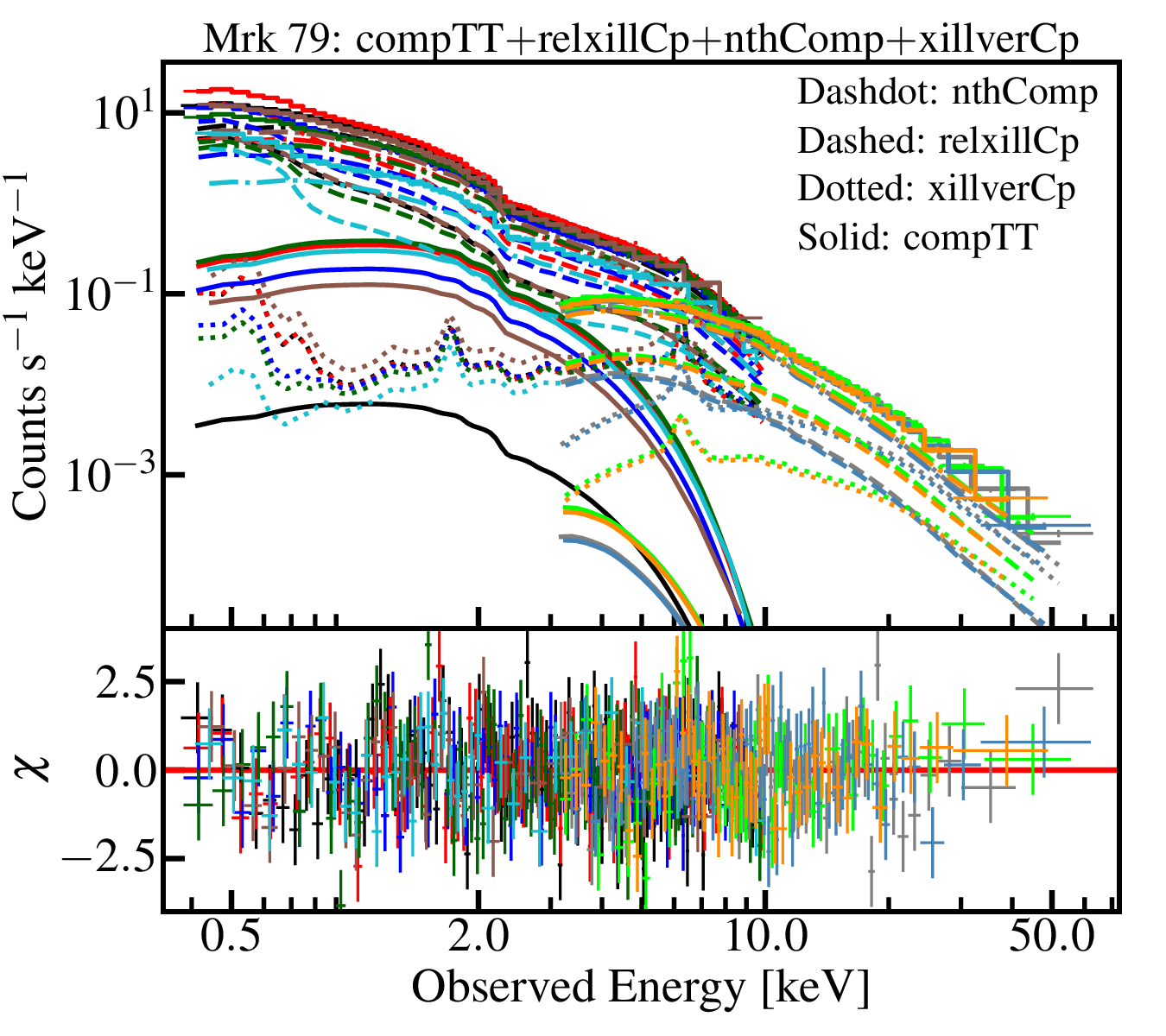}
\includegraphics[scale=0.25,angle=-0]{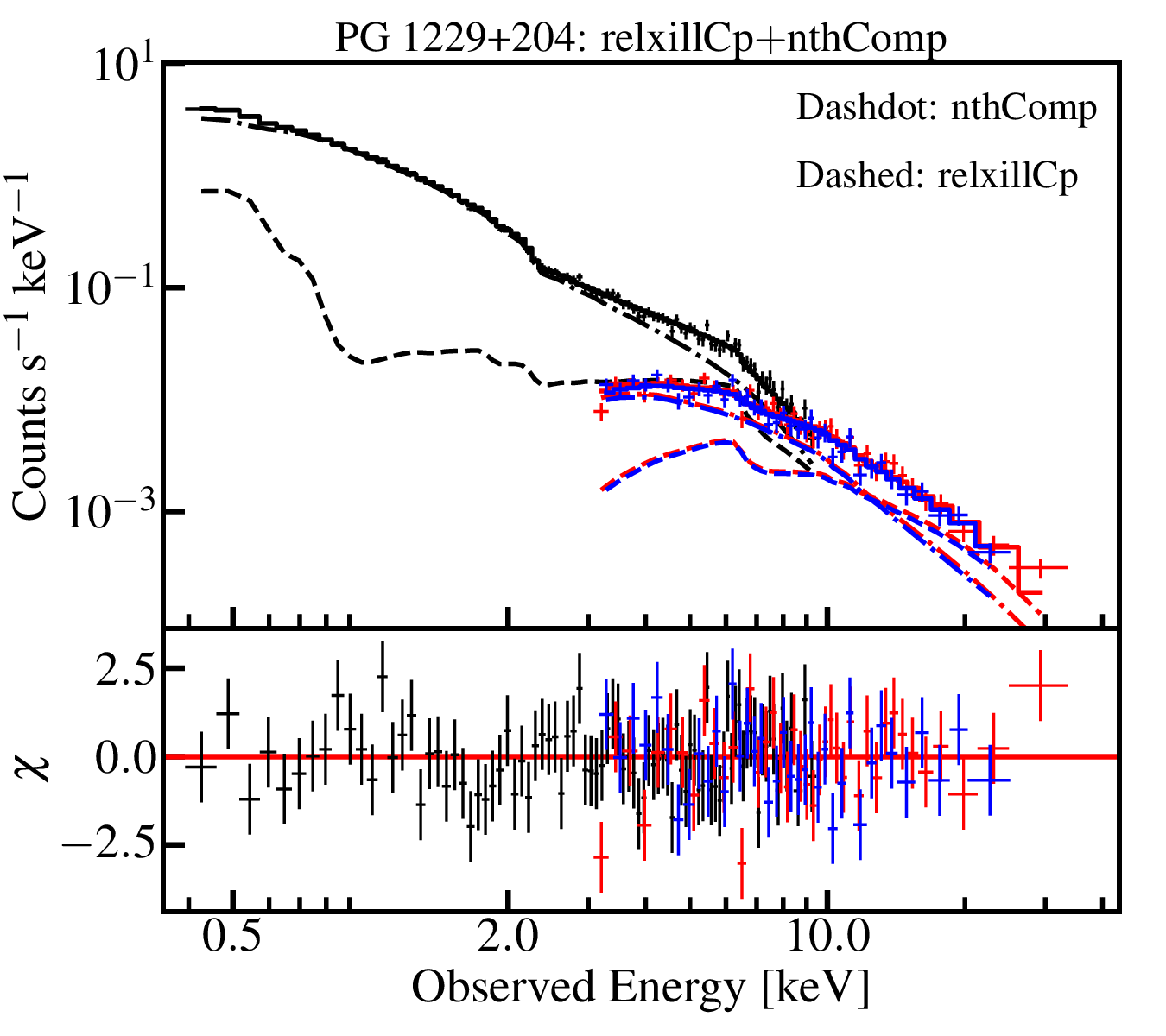}
\includegraphics[scale=0.25,angle=-0]{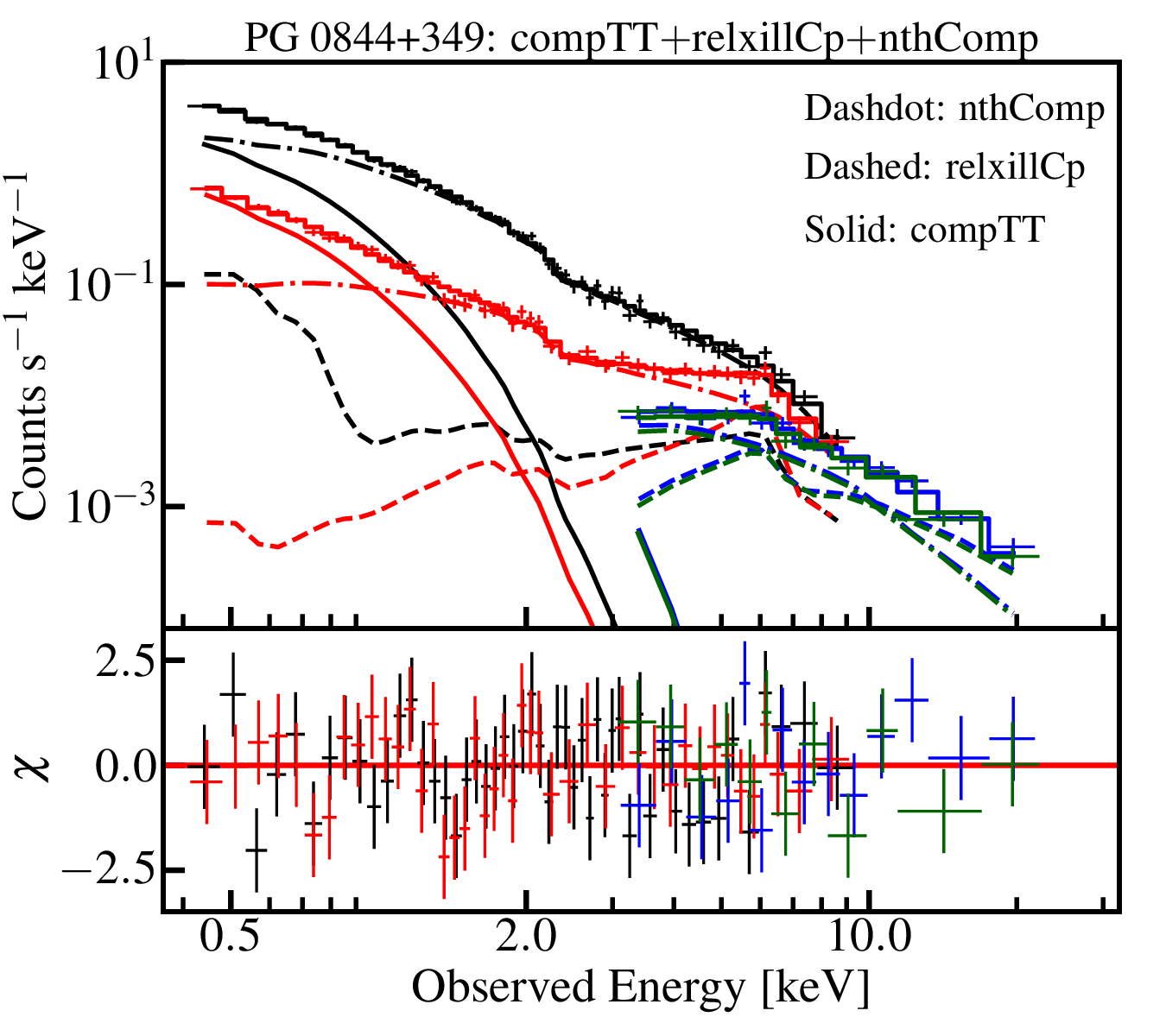}
\includegraphics[scale=0.25,angle=-0]{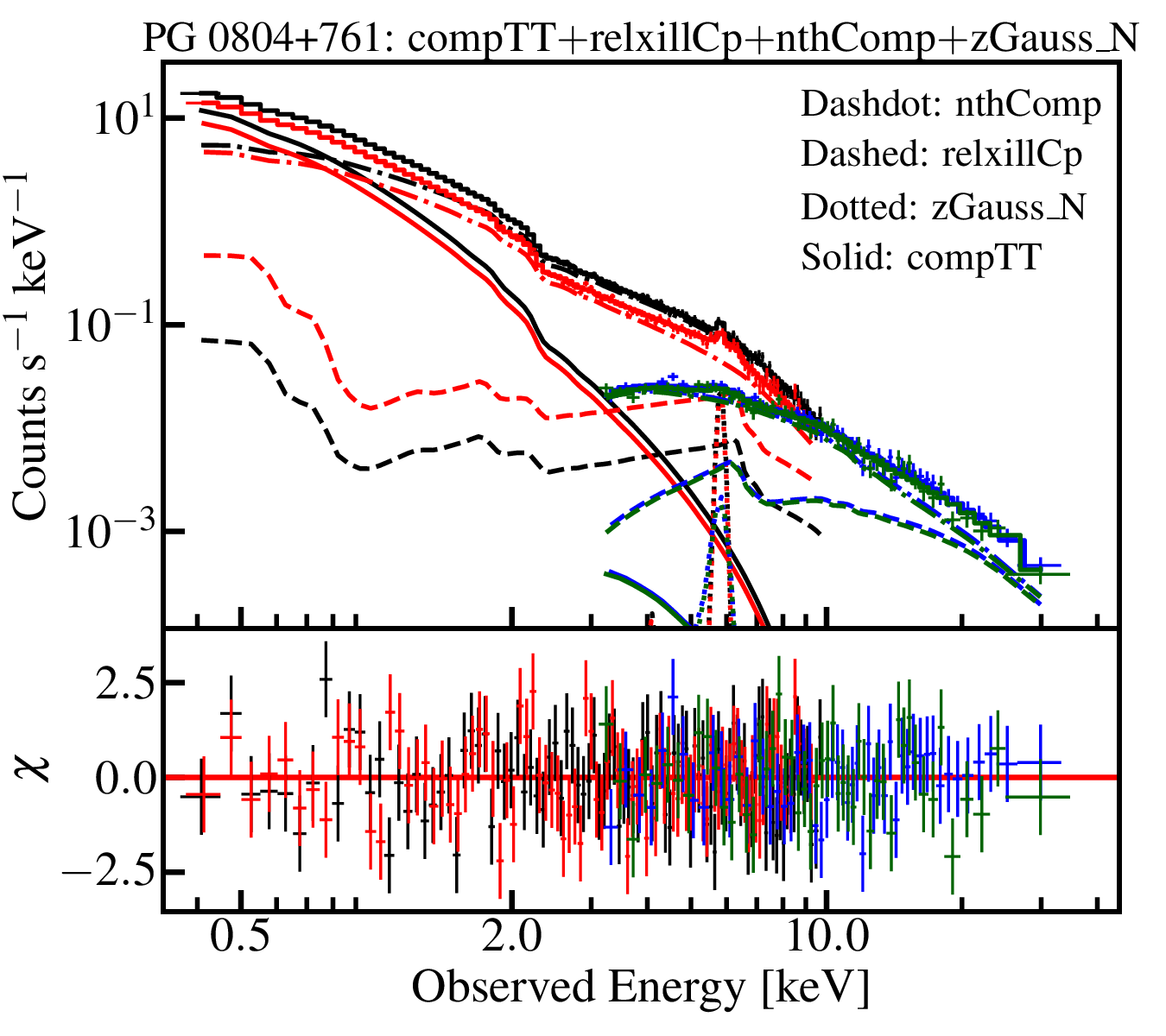}
\includegraphics[scale=0.25,angle=-0]{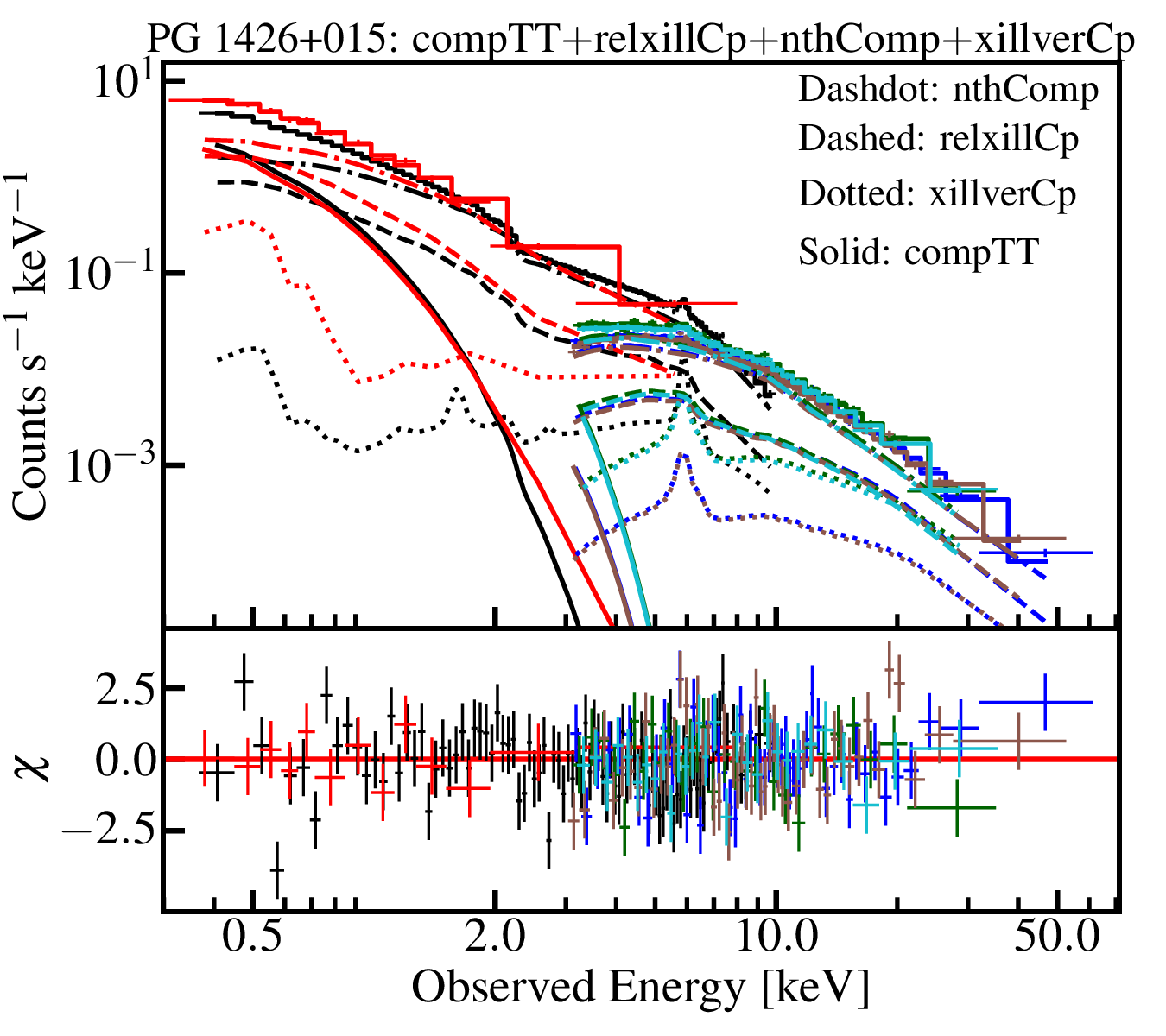}
\caption{The \xmm{}/EPIC, \nustar{}/FPMA, and FPMB count spectra with the best-fit models and residuals as a function of energy. Five sets of best-fit models have been obtained for the sample. The spectra are binned up only for plotting purposes.}
\label{best_fit_plot}
\end{center}
\end{figure*}

\begin{figure*}
\centering
\begin{center}
\includegraphics[scale=0.21,angle=-0]{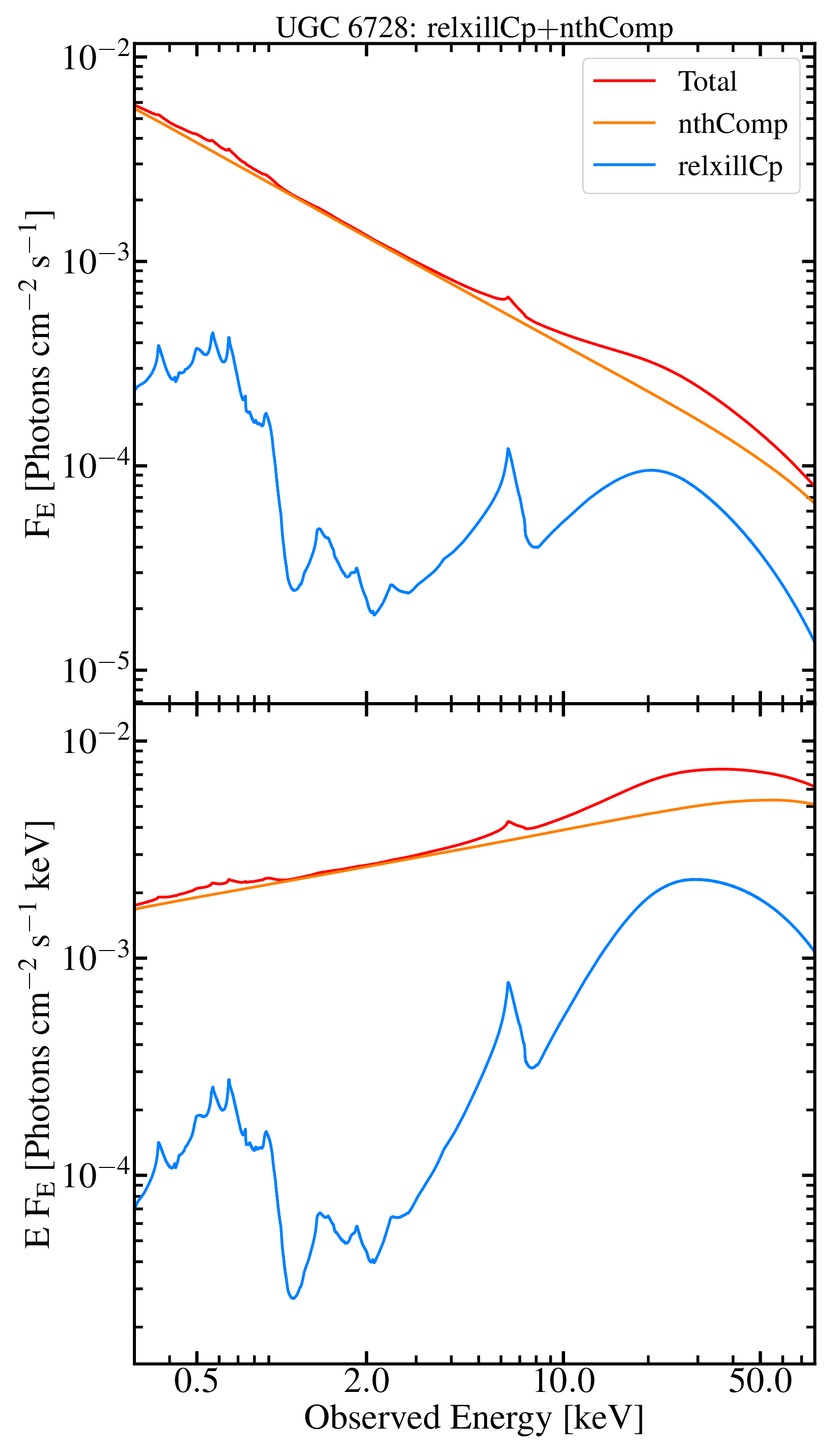}
\includegraphics[scale=0.21,angle=-0]{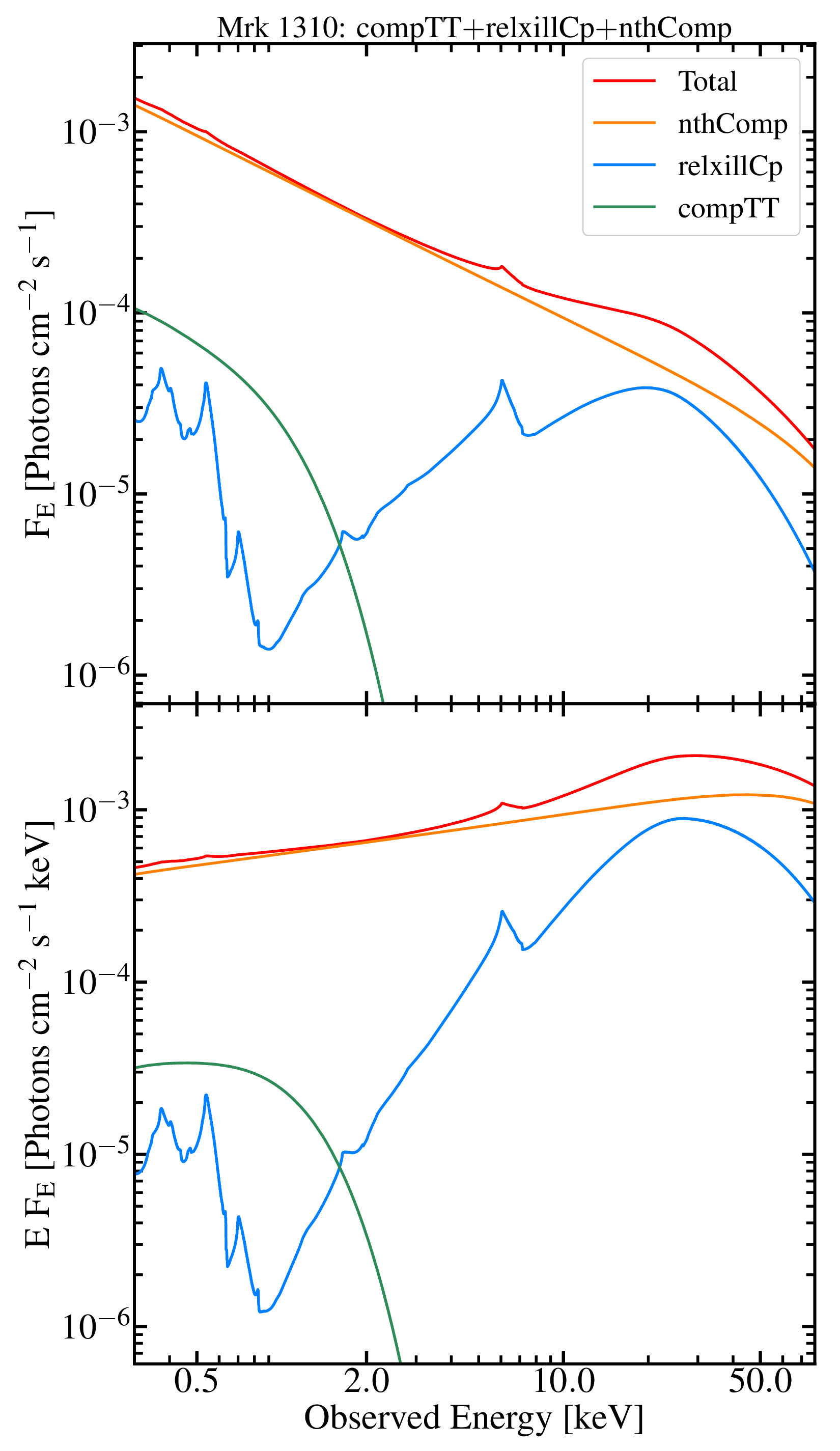}
\includegraphics[scale=0.21,angle=-0]{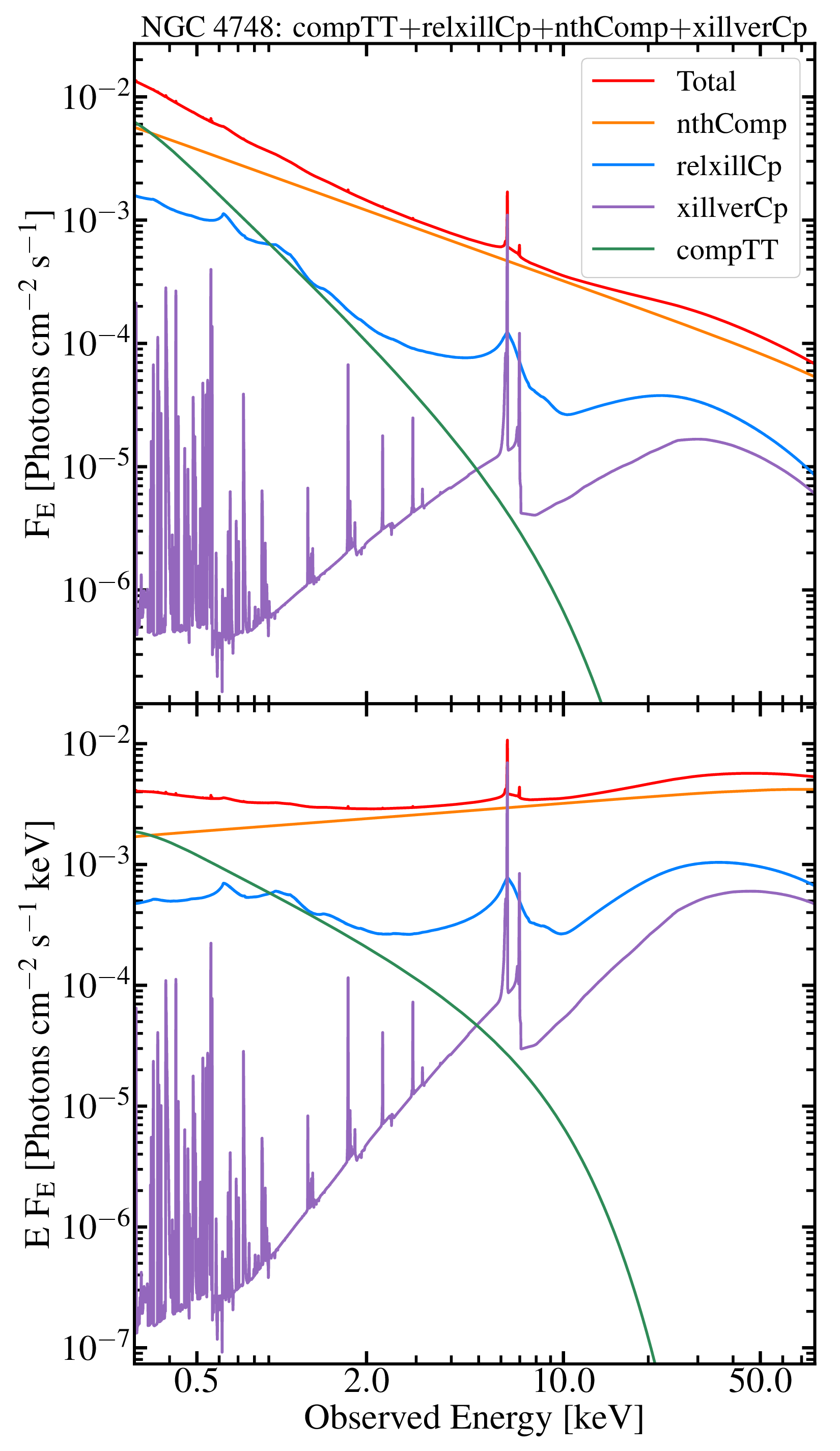}
\includegraphics[scale=0.21,angle=-0]{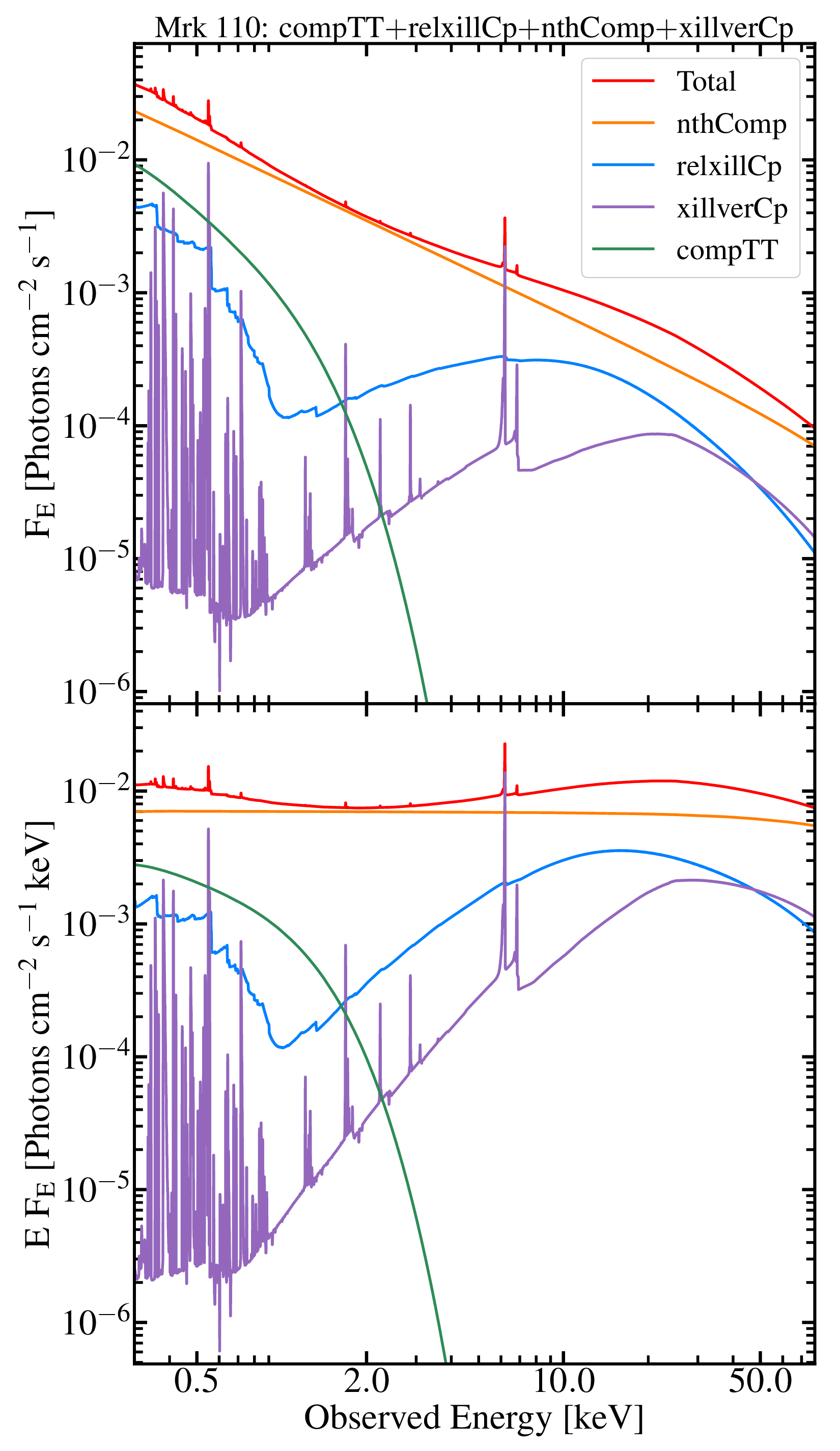}
\includegraphics[scale=0.21,angle=-0]{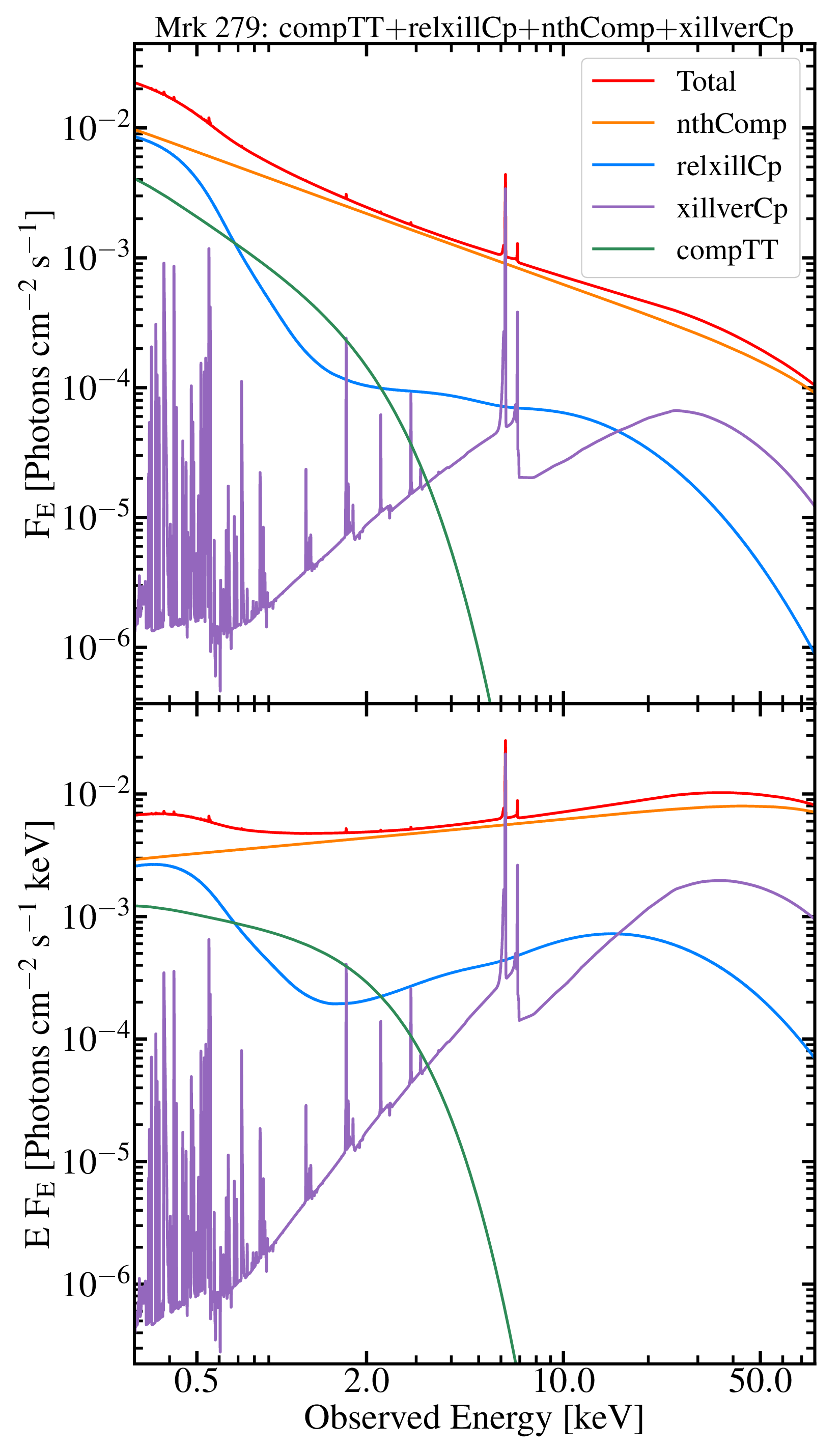}
\includegraphics[scale=0.21,angle=-0]{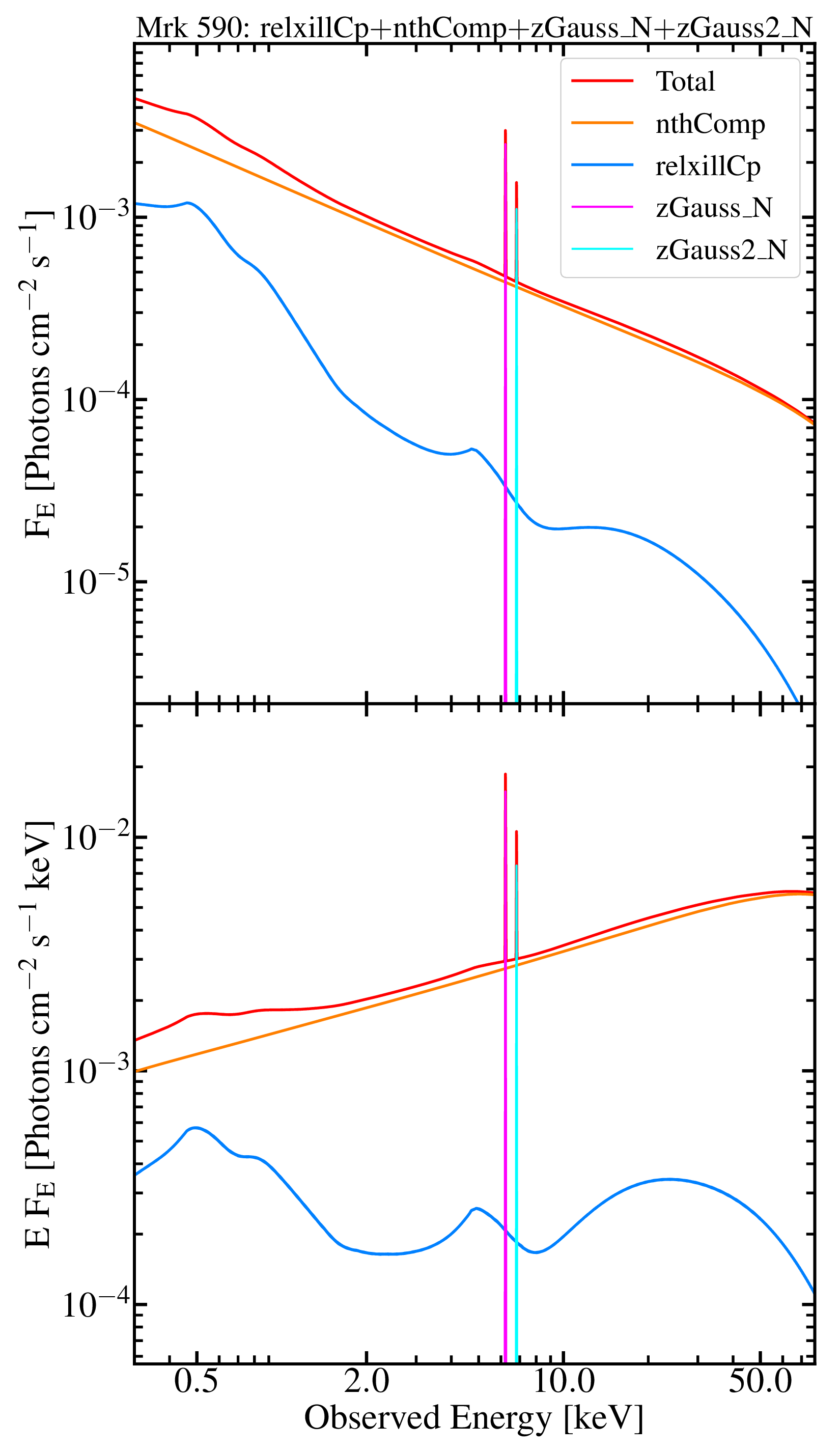}
\caption{The best-fit averaged spectral energy flux model (in red) with various model components for each source in the sample. The orange, blue, purple, green, and magenta/cyan solid lines represent the primary coronal emission ({\tt nthComp}), relativistic disk reflection ({\tt rellxillCp}), distant reflection ({\tt xillverCp}), warm coronal emission ({\tt compTT}), and narrow Fe~K$_\alpha$/K$_\beta$ emission cores ({\tt zGauss\_N}, {\tt zGauss2\_N}), respectively.} 
\label{best_emo_plot1}
\end{center}
\end{figure*}

\begin{figure*}
\centering
\begin{center}
\includegraphics[scale=0.21,angle=-0]{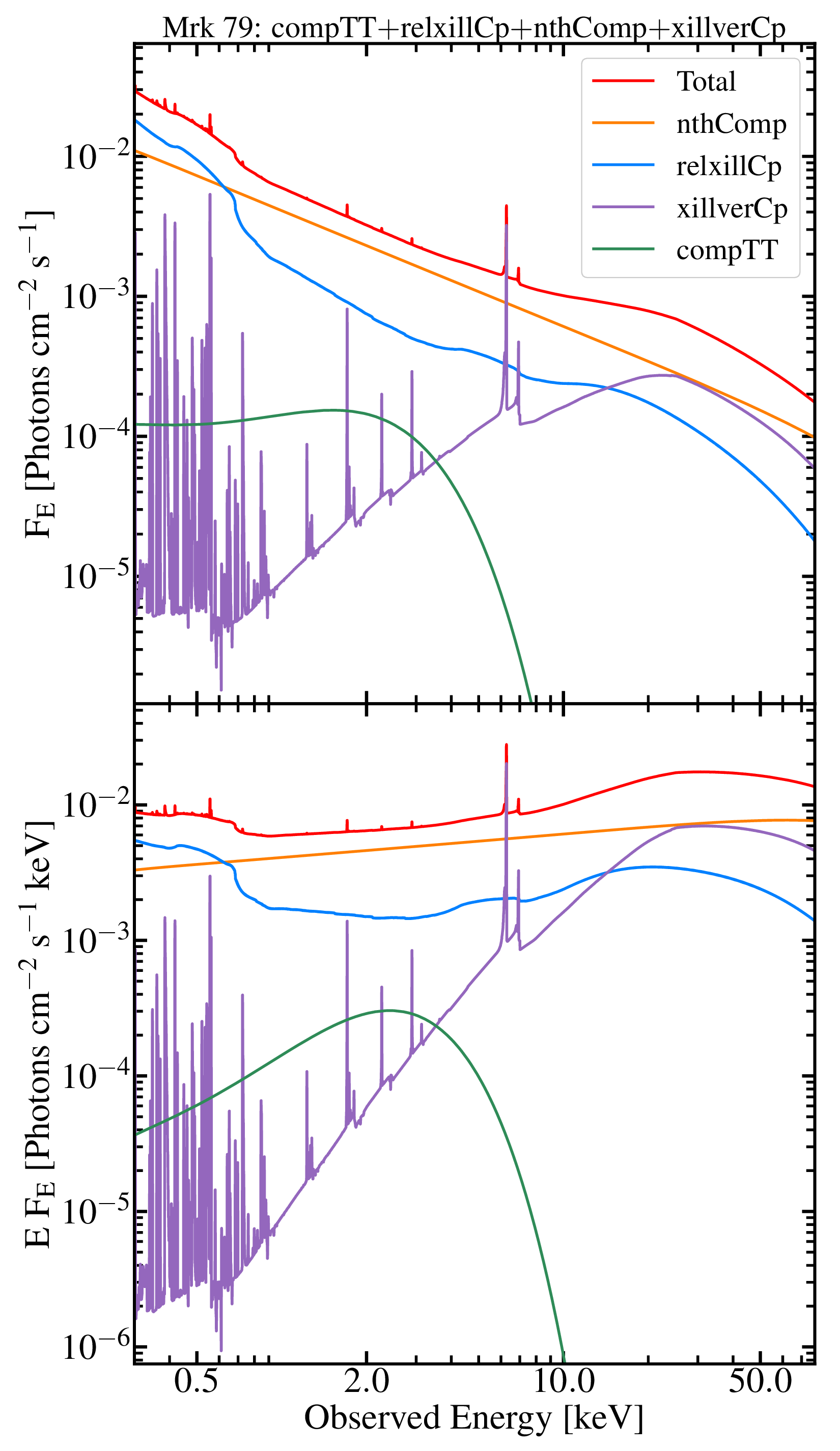}
\includegraphics[scale=0.21,angle=-0]{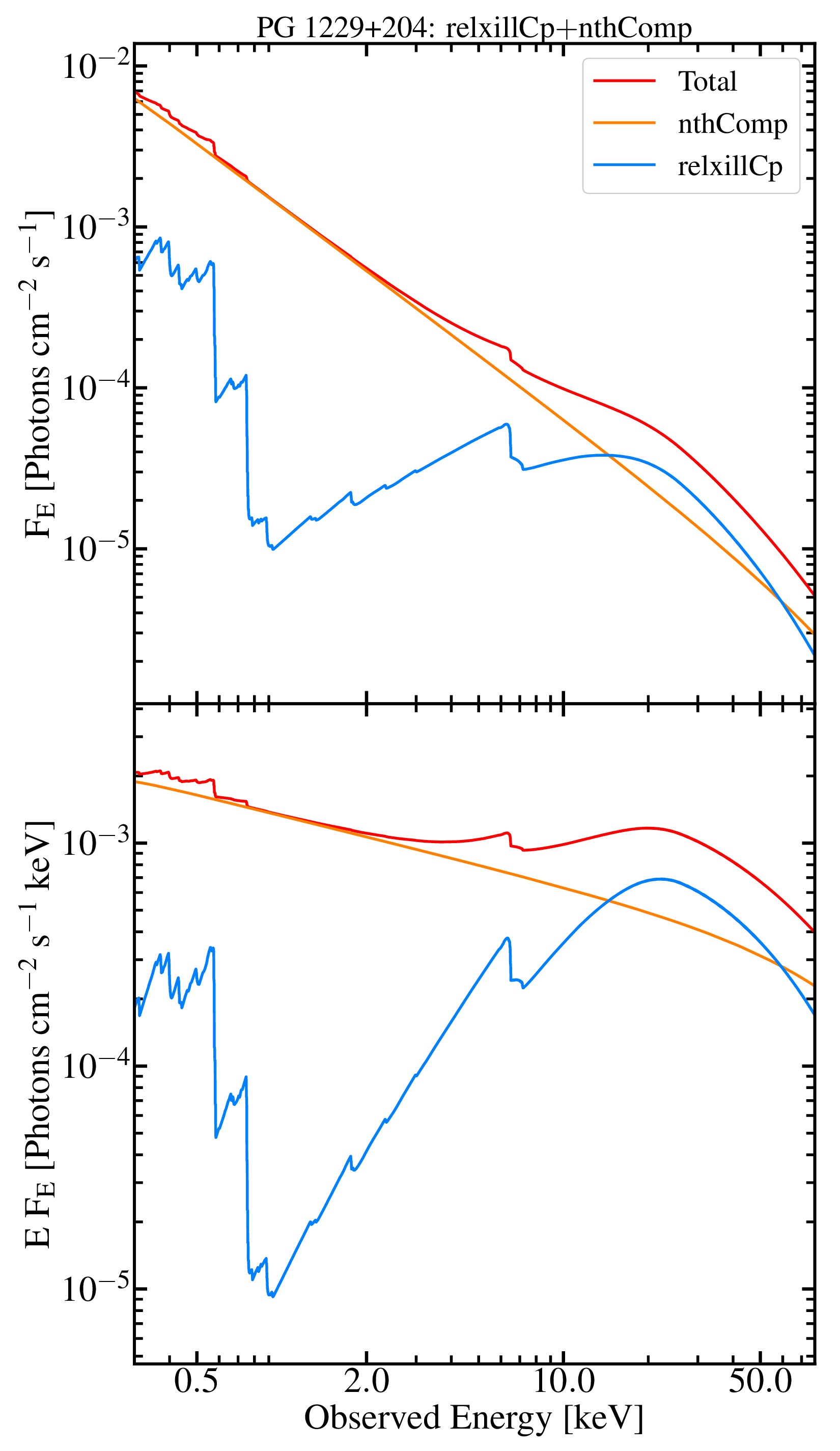}
\includegraphics[scale=0.21,angle=-0]{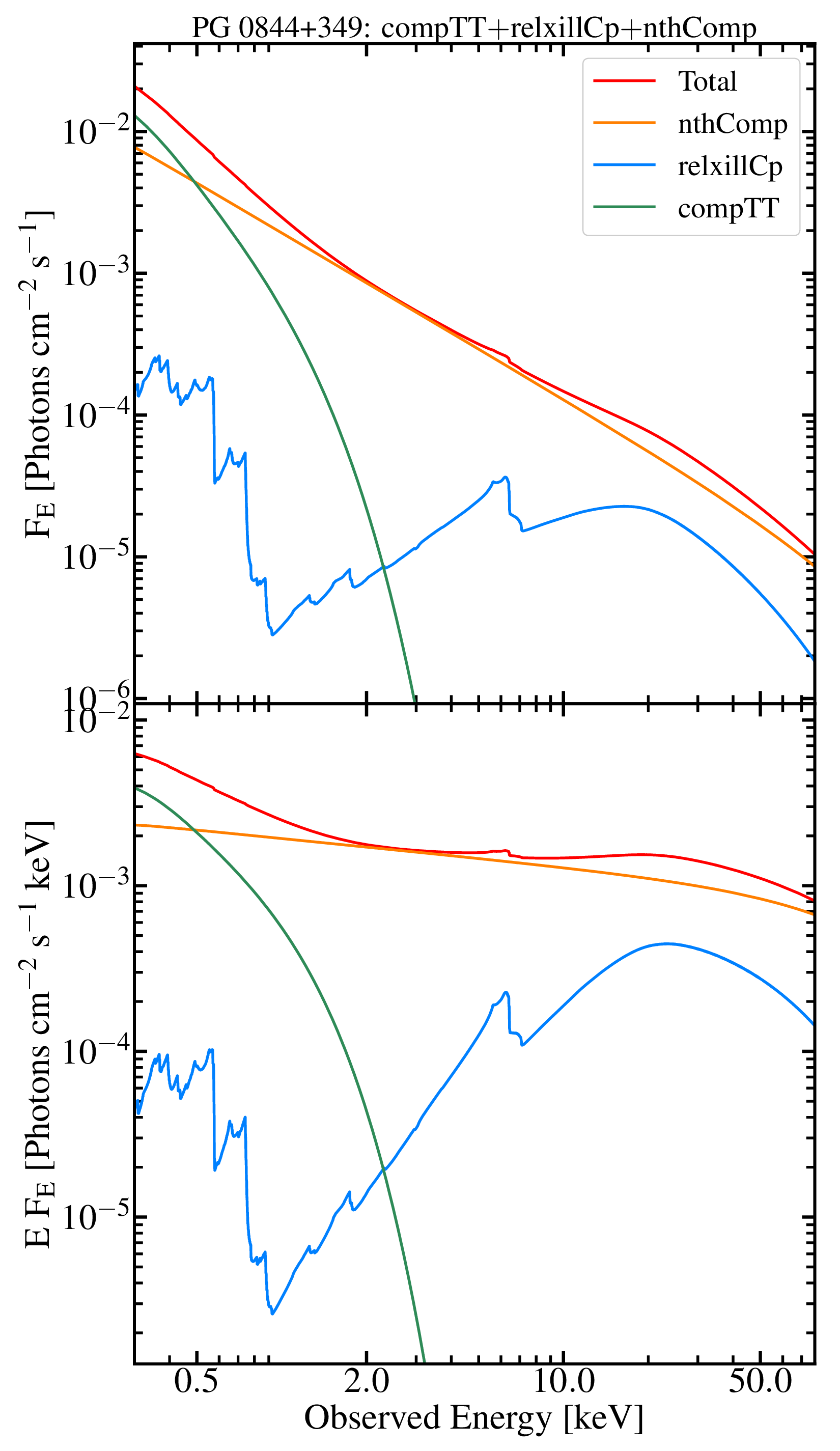}
\includegraphics[scale=0.21,angle=-0]{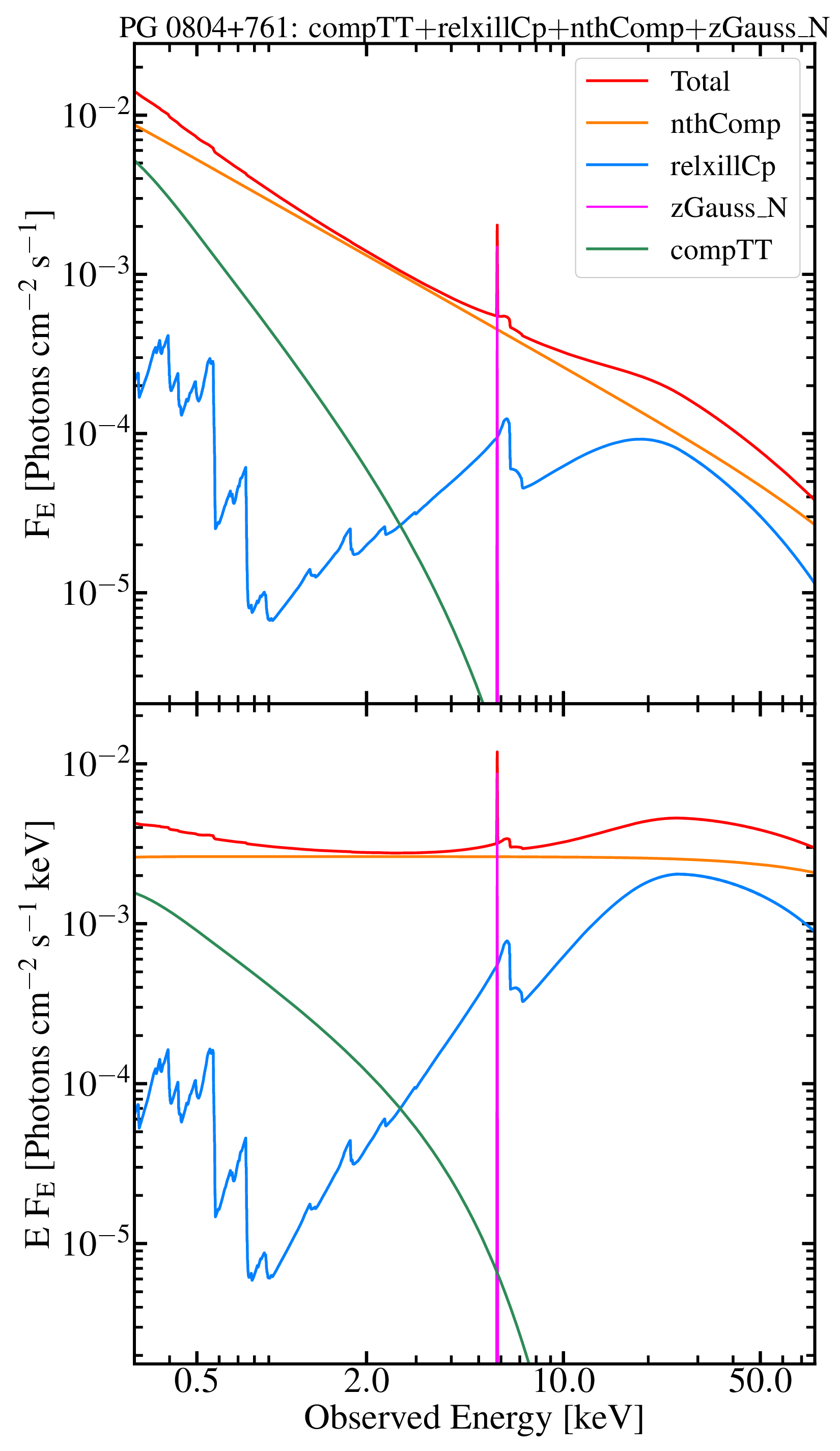}
\includegraphics[scale=0.21,angle=-0]{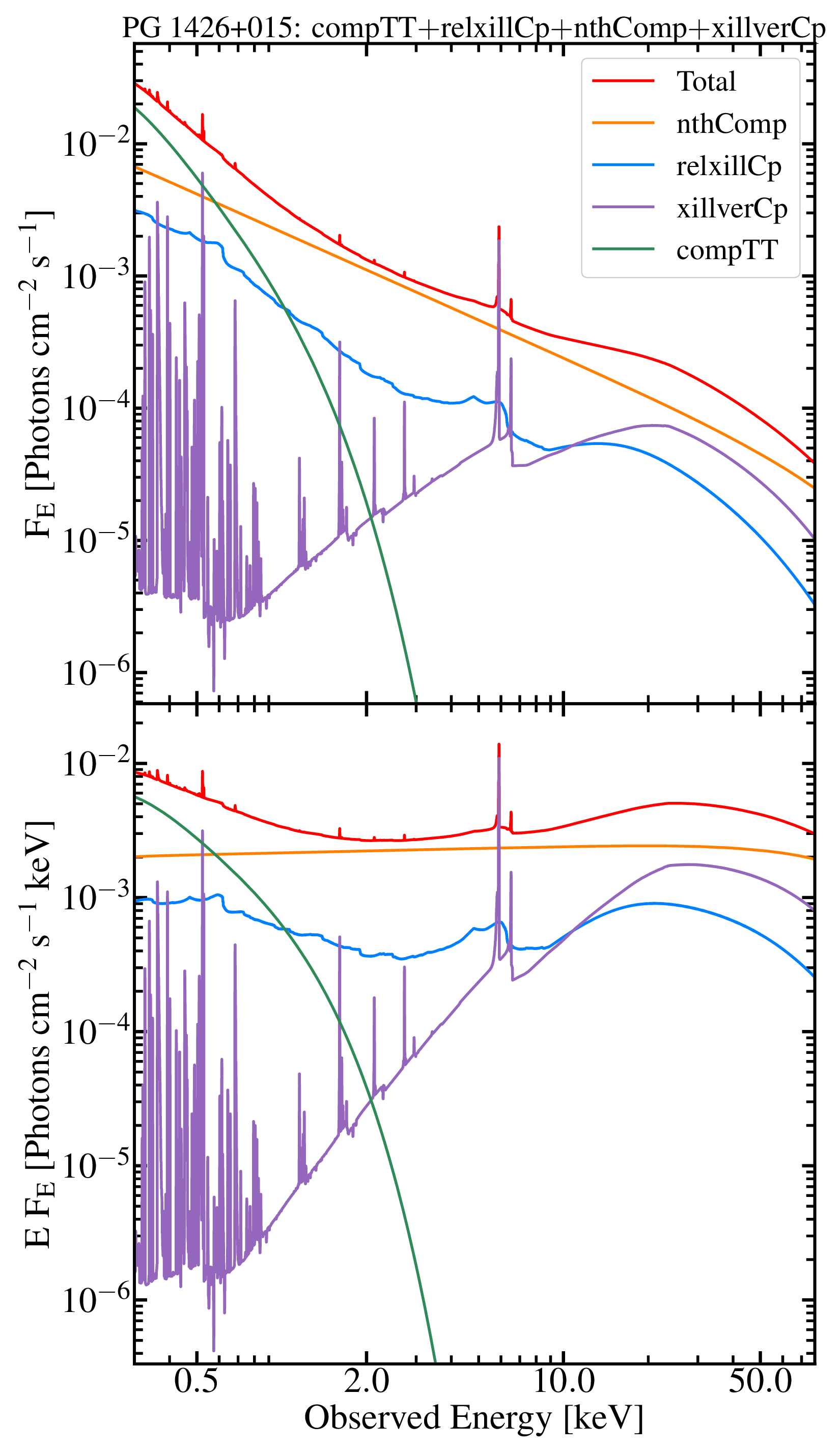}
\caption{Continued from previous page.}
\label{best_emo_plot2}
\end{center}
\end{figure*}

\clearpage
\begin{turnpage}
\begin{table*}[t!] 
\vspace*{-2.0cm}   

\caption{The best-fit model parameters obtained from the joint fitting of the broadband (0.3$-$78~keV) {\textit{XMM-Newton}}$+${\textit{NuSTAR}} spectra for each source. The source name and the corresponding best-fit spectral model are shown in bold font. The errors represent 90\% confidence intervals calculated through MCMC analyses. The `${\dagger}$' and `${\ddagger}$' symbols mark simultaneous or quasi-simultaneous {\textit{XMM-Newton}}/{\textit{NuSTAR}} observations. The `$\ast$' symbol denotes the parameters tied between observations. $F_{\rm nth}$, $F_{\rm rel}$, and $F_{\rm xil}$, respectively, indicate flux of the primary continuum ({\tt nthComp}), relativistic reflection ({\tt relxillCp}) and distant reflection ({\tt xillverCp}) model components in units of $10^{-12}$~erg~cm$^{-2}$~s$^{-1}$, and are measured in the 0.3$-$50~keV range. The observation start time, $T_{\rm start}$, is in MJD. The disk inclination angle ($\theta$), density ($n_{\rm e}$), ionization ($\xi$), and hot coronal temperature ($kT_{\rm e}$) are in units of degree, ${\rm cm^{-3}}$, ${\rm erg~cm~s^{-1}}$, and keV, respectively. The inner emissivity index ($q_{\rm in}$) and black hole spin ($a^{\ast}$) are dimensionless. The iron abundance, $A_{\rm Fe}$, is calculated relative to the solar abundance. $E_{\rm c}$ and $E_{\rm c2}$ are the centroid energies of the narrow ($\sigma=10$~eV) Fe~K$_\alpha$ and Fe~K$_\beta$ emission lines in keV modeled using {\tt zGauss\_N} and {\tt zGauss2\_N}, respectively. $F_{\rm NGa}$ and $F_{\rm NGa2}$ indicate flux values of the narrow Fe~K$_\alpha$ and Fe~K$_\beta$ emission lines, respectively, in units of $10^{-14}$~erg~cm$^{-2}$~s$^{-1}$ and are measured in the 5$-$7~keV range. $F_{\rm wc}$ is in the unit of $10^{-12}$~erg~cm$^{-2}$~s$^{-1}$ and denotes the warm corona flux measured in the range of 0.3--2~keV. $kT_{\rm wc}$ and $\tau_{\rm wc}$ represent the warm coronal temperature and optical depth, respectively. $\frac{\chi^{2}}{d.o.f}$ shows the best-fit model statistic.}
\label{table_output_parameters}

\begin{ruledtabular}
\begin{tabular}{lcccccccccccccccccccc}

\multicolumn{19}{c}{\textbf{UGC~6728: Model $\equiv$ {relxillCp [Relativistic Reflection] $+$ nthComp [Hot Corona]}} } \\ [0.15cm] 

  & $T_{\rm start}$  & $q_{\rm in}$ & $\theta$ & $a^{\ast}$ & $\log n_{\rm e}$ & $A_{\rm Fe}$ & $\log\xi$ & $F_{\rm rel}$ & $\Gamma$ &  $kT_{\rm e}$ &  $F_{\rm nth}$  &  $\frac{\chi^{2}}{d.o.f}$ \\ [0.25cm] 
\hline

EPIC & $53789.8$ & $\ge 3.3$ & $50_{-21}^{+4}$ & $0.73_{-0.61}^{+0.114}$ & $\le 17.7$ & $\le 4.3$ & $1.8_{-1.2}^{+0.7}$ & $8.4_{-1.7}^{+1.7}$ & $1.82_{-0.09}^{+0.04}$ & $\ge 42$ & $18.1_{-0.5}^{+0.5}$ & $\frac{514.7}{460}$ &  &  &  &  &  \\ [0.2cm]
FPM & $58039.8$ & $-$ & $50^{\ast}$ & $0.73^{\ast}$ & $-$ &  $-$ & $1.8^{\ast}$ & $7.5_{-0.9}^{+0.9}$ & $1.78_{-0.05}^{+0.07}$ &  $-$ & $45.1_{-0.9}^{+0.8}$ &  &  &  &  &  &  \\ [0.2cm]
FPM & $57579.5$ & $-$ & $50^{\ast}$ & $0.73^{\ast}$ & $-$ &  $-$ & $1.8^{\ast}$ & $2.5_{-1.1}^{+1.1}$ & $1.71_{-0.12}^{+0.08}$ &  $-$ & $23.9_{-1.0}^{+1.0}$ &  &  &  &  &  &  \\ [0.25cm]

\hline 
\hline 
\multicolumn{19}{c}{\textbf{Mrk~1310: Model $\equiv$ {relxillCp [Relativistic Reflection] $+$ nthComp [Hot Corona] $+$ CompTT [Warm Corona]}} } \\ [0.15cm]  

  & $T_{\rm start}$  & $q_{\rm in}$ & $\theta$ & $a^{\ast}$ & $\log n_{\rm e}$ & $A_{\rm Fe}$ & $\log\xi$ & $F_{\rm rel}$ & $\Gamma$ &  $kT_{\rm e}$ &  $F_{\rm nth}$  &   &   & $kT_{\rm wc}$ & $\tau_{\rm wc}$ & $F_{\rm wc}$ & $\frac{\chi^{2}}{d.o.f}$   \\[0.25cm]
\hline

EPIC & $58487.3$ & $\ge 3.5$ & $49_{-31}^{+7}$ & $0.752_{-0.605}^{+0.19}$ & $\le 17.3$ & $1.4_{-0.7}^{+2.3}$ & $0.3_{-0.2}^{+0.7}$ & $3.2_{-0.6}^{+0.6}$ & $1.93_{-0.11}^{+0.05}$ & $\ge 26$ & $8.6_{-0.2}^{+0.2}$ &  &  & $0.24_{-0.05}^{+0.11}$ & $22.4_{-7.4}^{+6.7}$ & $\le 0.3$ & $\frac{263.4}{223}$ \\ [0.2cm]
EPIC & $56635.4$ & $-$ & $49^{\ast}$ & $0.752^{\ast}$ & $-$ & $1.4^{\ast}$ & $0.3^{\ast}$ & $0.6_{-0.4}^{+0.4}$ & $\le 1.86$ & $-$ & $1.2_{-0.1}^{+0.1}$ &  &  & $0.24^{\ast}$ & $22.4^{\ast}$ & $\le 0.1$ &  \\ [0.2cm]
FPM & $57556.5$ & $-$ & $49^{\ast}$ & $0.752^{\ast}$ & $-$ & $1.4^{\ast}$ & $0.3^{\ast}$ & $4.7_{-1.0}^{+1.0}$ & $1.91_{-0.15}^{+0.26}$ & $-$ & $27.3_{-1.1}^{+1.2}$ &  &  & $0.24^{\ast}$ & $22.4^{\ast}$ & $\le 0.01$ &  \\ [0.25cm]

\hline 
\hline 
\multicolumn{19}{c}{\textbf{NGC~4748: Model $\equiv$ {relxillCp [Relativistic Reflection] $+$ nthComp [Hot Corona] $+$ xillverCp [Distant Reflection] $+$ CompTT [Warm Corona]}} } \\ [0.15cm]  
  & $T_{\rm start}$  & $q_{\rm in}$ & $\theta$ & $a^{\ast}$ & $\log n_{\rm e}$ & $A_{\rm Fe}$ & $\log\xi$ & $F_{\rm rel}$ & $\Gamma$ &  $kT_{\rm e}$ &  $F_{\rm nth}$  &  $F_{\rm xil}$ &  & $kT_{\rm wc}$ & $\tau_{\rm wc}$ & $F_{\rm wc}$ & $\frac{\chi^{2}}{d.o.f}$   \\[0.25cm]
\hline 

EPIC & $56671.3$ & $\ge 3.3$ & $51_{-33}^{+5}$ & $0.796_{-0.663}^{+0.123}$ & $\le 17.1$ & $\le 4.9$ & $3.1_{-2.0}^{+0.8}$ & $3.5_{-1.5}^{+1.5}$ & $1.77_{-0.21}^{+0.11}$ & $\ge 63$ & $23.2_{-1.0}^{+1.0}$ & $1.0_{-0.6}^{+0.6}$ &  & $1.8_{-1.0}^{+0.2}$ & $4.0_{-0.1}^{+2.5}$ & $12.9_{-1.1}^{+2.5}$ & $\frac{258.3}{238}$ \\ [0.2cm]
FPM & $59238.4$ & $-$ & $51^{\ast}$ & $0.796^{\ast}$ & $-$ & $-$ & $3.1^{\ast}$ & $5.5_{-2.4}^{+1.7}$ & $1.88_{-0.13}^{+0.17}$ & $-$ & $22.5_{-1.1}^{+1.3}$ & $\le 0.9$ &  & $1.8^{\ast}$ & $4.0^{\ast}$ & $\le 0.6$ &  \\ [0.25cm]

\end{tabular}
\end{ruledtabular}
\end{table*}
\end{turnpage}

\addtocounter{table}{-1} 
\begin{turnpage}
\begin{table*}[t]
\caption{Continued from previous page.}

\begin{ruledtabular}
\setlength{\tabcolsep}{1.2pt}
\begin{tabular}{lcccccccccccccccccccc}

\multicolumn{19}{c}{\textbf{Mrk~110: Model $\equiv$ {relxillCp [Relativistic Reflection] $+$ nthComp [Hot Corona] $+$ xillverCp [Distant Reflection] $+$ CompTT [Warm Corona]}} } \\ [0.15cm]  
  & $T_{\rm start}$  & $q_{\rm in}$ & $\theta$ & $a^{\ast}$ & $\log n_{\rm e}$ & $A_{\rm Fe}$ & $\log\xi$ & $F_{\rm rel}$ & $\Gamma$ &  $kT_{\rm e}$ &  $F_{\rm nth}$  &  $F_{\rm xil}$ &  & $kT_{\rm wc}$ & $\tau_{\rm wc}$ & $F_{\rm wc}$ & $\frac{\chi^{2}}{d.o.f}$   \\[0.25cm]
\hline

EPIC & $53324.2$ & $5.2_{-0.4}^{+0.3}$ & $\le 24$ & $0.978_{-0.024}^{+0.016}$ & $19.0_{-0.2}^{+0.2}$ & $1.5_{-0.4}^{+0.7}$ & $0.6_{-0.2}^{+0.1}$ & $21.6_{-1.0}^{+1.0}$ & $2.01_{-0.03}^{+0.02}$ & $\ge 118$ & $71.4_{-0.3}^{+0.3}$ & $5.0_{-1.4}^{+1.4}$ &  & $0.25_{-0.02}^{+0.02}$ & $15.6_{-0.9}^{+1.2}$ & $4.1_{-0.6}^{+1.2}$ & $\frac{1647.0}{1425}$ \\ [0.2cm]
EPIC & $58804.4^{\dagger}$ & $5.2^{\ast}$ & $-$ & $0.978^{\ast}$ & $19.0^{\ast}$ & $1.5^{\ast}$ & $0.6^{\ast}$ & $14.3_{-0.8}^{+0.8}$ & $1.99_{-0.03}^{+0.02}$ &  $-$ & $67.1_{-0.3}^{+0.3}$ & $5.5_{-0.7}^{+0.7}$ &  & $0.25^{\ast}$ & $15.6^{\ast}$ & $5.8_{-0.7}^{+0.8}$ &  \\ [0.2cm]
EPIC & $58946.0^{\ddagger}$ & $5.2^{\ast}$ & $-$ & $0.978^{\ast}$ & $19.0^{\ast}$ & $1.5^{\ast}$ & $0.6^{\ast}$ & $9.5_{-0.7}^{+0.7}$ & $1.90_{-0.03}^{+0.02}$ &  $-$ & $62.9_{-0.3}^{+0.3}$ & $4.6_{-0.7}^{+0.7}$ &  & $0.25^{\ast}$ & $15.6^{\ast}$ & $4.0_{-0.5}^{+0.7}$ &  \\ [0.2cm]
EPIC & $58792.4$ & $5.2^{\ast}$ & $-$ & $0.978^{\ast}$ & $19.0^{\ast}$ & $1.5^{\ast}$ & $0.6^{\ast}$ & $13.7_{-1.2}^{+1.2}$ & $1.94_{-0.05}^{+0.04}$ &  $-$ & $69.9_{-0.4}^{+0.4}$ & $4.2_{-1.6}^{+1.6}$ &  & $0.25^{\ast}$ & $15.6^{\ast}$ & $5.8_{-1.0}^{+1.5}$ &  \\ [0.2cm]
EPIC & $58794.4$ & $5.2^{\ast}$ & $-$ & $0.978^{\ast}$ & $19.0^{\ast}$ & $1.5^{\ast}$ & $0.6^{\ast}$ & $15.1_{-1.2}^{+1.2}$ & $1.99_{-0.05}^{+0.03}$ &  $-$ & $70.2_{-0.4}^{+0.4}$ & $5.4_{-1.6}^{+1.6}$ &  & $0.25^{\ast}$ & $15.6^{\ast}$ & $5.1_{-0.9}^{+1.6}$ &  \\ [0.2cm]
EPIC & $58790.4$ & $5.2^{\ast}$ & $-$ & $0.978^{\ast}$ & $19.0^{\ast}$ & $1.5^{\ast}$ & $0.6^{\ast}$ & $12.2_{-1.1}^{+1.1}$ & $1.87_{-0.08}^{+0.03}$ &  $-$ & $53.3_{-0.4}^{+0.4}$ & $6.0_{-1.6}^{+1.6}$ &  & $0.25^{\ast}$ & $15.6^{\ast}$ & $4.2_{-0.8}^{+1.5}$ &  \\ [0.2cm]
FPM & $57776.8$ & $5.2^{\ast}$ & $-$ & $0.978^{\ast}$ & $19.0^{\ast}$ & $1.5^{\ast}$ & $0.6^{\ast}$ & $13.4_{-1.0}^{+1.0}$ & $1.94_{-0.02}^{+0.01}$ &  $-$ & $117.9_{-1.1}^{+1.1}$ & $7.1_{-0.8}^{+0.8}$ &  & $0.25^{\ast}$ & $15.6^{\ast}$ & $\le 0.01$ &  \\ [0.2cm]
FPM & $58803.2^{\dagger}$ & $5.2^{\ast}$ & $-$ & $0.978^{\ast}$ & $19.0^{\ast}$ & $1.5^{\ast}$ & $0.6^{\ast}$ & $14.3^{\ast}$ & $1.99^{\ast}$ &  $-$ & $67.1^{\ast}$ & $5.5^{\ast}$ &  & $0.25^{\ast}$ & $15.6^{\ast}$ & $5.8^{\ast}$ &  \\ [0.2cm]
FPM & $58944.6^{\ddagger}$ & $5.2^{\ast}$ & $-$ & $0.978^{\ast}$ &$19.0^{\ast}$ & $1.5^{\ast}$ & $0.6^{\ast}$ & $9.5^{\ast}$ & $1.90^{\ast}$ &  $-$ & $62.9^{\ast}$ & $4.6^{\ast}$ &  & $0.25^{\ast}$ & $15.6^{\ast}$ & $4.0^{\ast}$ &  \\ [0.25cm]

\hline 
\hline 
\multicolumn{19}{c}{\textbf{Mrk~279: Model $\equiv$ {relxillCp [Relativistic Reflection] $+$ nthComp [Hot Corona] $+$ xillverCp [Distant Reflection] $+$ CompTT [Warm corona]}} } \\ [0.15cm] 
  & $T_{\rm start}$  & $q_{\rm in}$ & $\theta$ & $a^{\ast}$ & $\log n_{\rm e}$ & $A_{\rm Fe}$ & $\log\xi$ & $F_{\rm rel}$ & $\Gamma$ &  $kT_{\rm e}$ &  $F_{\rm nth}$  &  $F_{\rm xil}$ &  & $kT_{\rm wc}$ & $\tau_{\rm wc}$ & $F_{\rm wc}$ & $\frac{\chi^{2}}{d.o.f}$   \\[0.25cm]
\hline

EPIC & $53689.7$ & $\ge 7.9$ & $47_{-11}^{+2}$ & $\ge 0.991$ & $\ge 19.3$ & $4.0_{-1.5}^{+0.8}$ & $2.3_{-0.4}^{+0.1}$ & $7.9_{-0.3}^{+0.3}$ & $1.81_{-0.03}^{+0.02}$ & $40_{-14}^{+24}$ & $80.0_{-0.4}^{+0.4}$ & $5.1_{-0.6}^{+0.6}$ &  & $0.45_{-0.04}^{+0.11}$ & $12.3_{-2.3}^{+1.3}$ & $7.6_{-1.2}^{+2.1}$ & $\frac{1307.6}{1170}$ \\ [0.2cm]
EPIC & $53691.7$ & $-$ & $47^{\ast}$ & $-$ & $-$ & $4.0^{\ast}$ & $2.3^{\ast}$ & $6.2_{-0.3}^{+0.3}$ & $1.78_{-0.03}^{+0.03}$ & $40^{\ast}$ & $75.7_{-0.4}^{+0.4}$ & $4.9_{-0.7}^{+0.7}$ &  & $0.45^{\ast}$ & $12.3^{\ast}$ & $6.2_{-1.0}^{+1.6}$ &  \\ [0.2cm]
EPIC & $53694.0$ & $-$ & $47^{\ast}$ & $-$ & $-$ & $4.0^{\ast}$ & $2.3^{\ast}$ & $6.3_{-0.4}^{+0.4}$ & $1.79_{-0.04}^{+0.03}$ & $40^{\ast}$ & $74.4_{-0.6}^{+0.6}$ & $5.8_{-0.9}^{+0.9}$ &  & $0.45^{\ast}$ & $12.3^{\ast}$ & $8.2_{-1.2}^{+2.0}$ &  \\ [0.2cm]
EPIC & $59203.3$ & $-$ & $47^{\ast}$ & $-$ & $-$ & $4.0^{\ast}$ & $2.3^{\ast}$ & $3.6_{-0.3}^{+0.3}$ & $1.78_{-0.04}^{+0.04}$ & $40^{\ast}$ & $48.2_{-0.5}^{+0.5}$ & $3.1_{-0.7}^{+0.8}$ &  & $0.45^{\ast}$ & $12.3^{\ast}$ & $5.2_{-1.0}^{+1.5}$ &  \\ [0.2cm]
EPIC & $52401.5$ & $-$ & $47^{\ast}$ & $-$ & $-$ & $4.0^{\ast}$ & $2.3^{\ast}$ & $4.4_{-0.8}^{+0.8}$ & $1.76_{-0.08}^{+0.08}$ & $40^{\ast}$ & $51.1_{-1.2}^{+1.2}$ & $2.7_{-1.8}^{+1.9}$ &  & $0.45^{\ast}$ & $12.3^{\ast}$ & $6.4_{-2.1}^{+2.1}$ &  \\ [0.2cm]
FPM & $59066.5$ & $-$ & $47^{\ast}$ & $-$ & $-$ & $4.0^{\ast}$ & $2.3^{\ast}$ & $6.2_{-1.4}^{+1.4}$ & $1.78_{-0.04}^{+0.04}$ & $40^{\ast}$ & $16.7_{-0.9}^{+0.9}$ & $3.2_{-0.2}^{+0.3}$ &  & $0.45^{\ast}$ & $12.3^{\ast}$ & $\le 0.01$ &  \\ [0.2cm]
FPM & $58785.4$ & $-$ & $47^{\ast}$ & $-$ & $-$ & $4.0^{\ast}$ & $2.3^{\ast}$ & $\le 20.8$ & $1.90_{-0.03}^{+0.03}$ & $40^{\ast}$ & $76.4_{-7.5}^{+6.1}$ & $5.8_{-1.1}^{+1.1}$ &  & $0.45^{\ast}$ & $12.3^{\ast}$ & $\le 1.1$ &  \\ [0.2cm]
FPM & $59073.0$ & $-$ & $47^{\ast}$ & $-$ & $-$ & $4.0^{\ast}$ & $2.3^{\ast}$ & $3.4_{-3.2}^{+3.7}$ & $1.74_{-0.04}^{+0.04}$ & $40^{\ast}$ & $25.9_{-2.7}^{+2.2}$ & $2.9_{-0.6}^{+0.6}$ &  & $0.45^{\ast}$ & $12.3^{\ast}$ & $\le 0.3$ &  \\ [0.2cm]
FPM & $59064.9$ & $-$ & $47^{\ast}$ & $-$ & $-$ & $4.0^{\ast}$ & $2.3^{\ast}$ & $12.2_{-3.1}^{+3.8}$ & $1.82_{-0.05}^{+0.07}$ & $40^{\ast}$ & $14.7_{-2.5}^{+2.0}$ & $3.0_{-0.5}^{+0.5}$ &  & $0.45^{\ast}$ & $12.3^{\ast}$ & $\le 0.01$ &  \\ [0.25cm]

\end{tabular}
\end{ruledtabular}
\end{table*}
\end{turnpage}

\addtocounter{table}{-1} 
\begin{turnpage}
\begin{table*}[t!]
\caption{Continued from previous page.}

\begin{ruledtabular}
\begin{tabular}{lcccccccccccccccccccc}

\multicolumn{19}{c}{\textbf{Mrk~590: Model $\equiv$ {relxillCp [Relativistic Reflection] $+$ nthComp [Hot Corona] $+$ zGauss\_N [Narrow Fe~K$_\alpha$ Emission Line]$+$ zGauss2\_N [Narrow Fe~K$_\beta$ Emission Line]}} } \\ [0.15cm] 
  & $T_{\rm start}$  & $q_{\rm in}$ & $\theta$ & $a^{\ast}$ & $\log n_{\rm e}$ & $A_{\rm Fe}$ & $\log\xi$ & $F_{\rm rel}$ & $\Gamma$ &  $kT_{\rm e}$ &  $F_{\rm nth}$  & $E_{\rm c}$ & $F_{\rm NGa}$ & $E_{\rm c2}$ & $F_{\rm NGa2}$ &  $\frac{\chi^{2}}{d.o.f}$ \\[0.25cm]
\hline 

EPIC & $53190.9$ & $\ge 9.0$ & $41_{-4}^{+5}$ & $0.89_{-0.093}^{+0.041}$ & $\ge 19.1$ & $2.6_{-0.3}^{+0.5}$ & $3.0_{-0.1}^{+0.1}$ & $1.9_{-0.1}^{+0.1}$ & $1.68_{-0.01}^{+0.01}$ & $49_{-19}^{+37}$ & $22.4_{-0.2}^{+0.2}$ & $6.39_{-0.01}^{+0.01}$ & $9.2_{-1.8}^{+1.8}$ & $6.99_{-0.02}^{+0.04}$ & $5.5_{-1.8}^{+1.9}$ &  $\frac{2030.2}{2028}$ \\ [0.2cm]
EPIC & $59217.5$ & $-$ & $41^{\ast}$ & $0.89^{\ast}$ & $-$ & $2.6^{\ast}$ & $3.0^{\ast}$ & $2.6_{-0.2}^{+0.2}$ & $1.66_{-0.02}^{+0.02}$ & $49^{\ast}$ & $25.2_{-0.4}^{+0.4}$ & $6.39^{\ast}$ & $10.9_{-3.0}^{+3.2}$ & $6.99^{\ast}$ & $\le 5.3$ &  &  \\ [0.2cm]
EPIC & $59034.6$ & $-$ & $41^{\ast}$ & $0.89^{\ast}$ & $-$ & $2.6^{\ast}$ & $3.0^{\ast}$ & $1.9_{-0.2}^{+0.2}$ & $1.69_{-0.02}^{+0.02}$ & $49^{\ast}$ & $21.0_{-0.3}^{+0.3}$ & $6.39^{\ast}$ & $10.8_{-2.8}^{+3.0}$ & $6.99^{\ast}$ & $4.5_{-2.7}^{+2.9}$ &  &  \\ [0.2cm]
EPIC & $59981.2^{\dagger}$ & $-$ & $41^{\ast}$ & $0.89^{\ast}$ & $-$ & $2.6^{\ast}$ & $3.0^{\ast}$ & $1.5_{-0.1}^{+0.1}$ & $1.64_{-0.01}^{+0.01}$ & $49^{\ast}$ & $14.0_{-0.2}^{+0.2}$ & $6.39^{\ast}$ & $7.5_{-1.3}^{+1.4}$ & $6.99^{\ast}$ & $\le 3.7$ &  &  \\ [0.2cm]
EPIC & $59788.4$ & $-$ & $41^{\ast}$ & $0.89^{\ast}$ & $-$ & $2.6^{\ast}$ & $3.0^{\ast}$ & $1.6_{-0.1}^{+0.1}$ & $1.66_{-0.02}^{+0.02}$ & $49^{\ast}$ & $19.3_{-0.3}^{+0.3}$ & $6.39^{\ast}$ & $8.2_{-2.5}^{+2.7}$ & $6.99^{\ast}$ & $\le 4.7$ &  &  \\ [0.2cm]
EPIC & $59437.1$ & $-$ & $41^{\ast}$ & $0.89^{\ast}$ & $-$ & $2.6^{\ast}$ & $3.0^{\ast}$ & $4.6_{-0.4}^{+0.4}$ & $1.74_{-0.02}^{+0.03}$ & $49^{\ast}$ & $30.4_{-0.7}^{+0.8}$ & $6.39^{\ast}$ & $9.1_{-5.6}^{+6.0}$ & $6.99^{\ast}$ & $\le 10.9$ &  &  \\ [0.2cm]
EPIC & $59439.1$ & $-$ & $41^{\ast}$ & $0.89^{\ast}$ & $-$ & $2.6^{\ast}$ & $3.0^{\ast}$ & $5.6_{-0.5}^{+0.5}$ & $1.73_{-0.03}^{+0.03}$ & $49^{\ast}$ & $37.1_{-1.0}^{+1.0}$ & $6.39^{\ast}$ & $\le 6.6$ & $6.99^{\ast}$ & $\le 9.6$  &  &  \\ [0.2cm]
EPIC & $59603.1$ & $-$ & $41^{\ast}$ & $0.89^{\ast}$ & $-$ & $2.6^{\ast}$ & $3.0^{\ast}$ & $0.9_{-0.1}^{+0.1}$ & $1.63_{-0.02}^{+0.03}$ & $49^{\ast}$ & $7.8_{-0.2}^{+0.2}$ & $6.39^{\ast}$ & $9.1_{-2.2}^{+2.3}$ & $6.99^{\ast}$ & $3.7_{-2.0}^{+2.2}$ &  &  \\ [0.2cm]
EPIC & $52275.5$ & $-$ & $41^{\ast}$ & $0.89^{\ast}$ & $-$ & $2.6^{\ast}$ & $3.0^{\ast}$ & $1.8_{-0.2}^{+0.2}$ & $1.67_{-0.03}^{+0.03}$ & $49^{\ast}$ & $15.8_{-0.5}^{+0.5}$ & $6.39^{\ast}$ & $13.3_{-4.1}^{+4.5}$ & $6.99^{\ast}$ & $\le 7.3$ &  &  \\ [0.2cm]
FPM & $58726.4$ & $-$ & $41^{\ast}$ & $0.89^{\ast}$ & $-$ & $2.6^{\ast}$ & $3.0^{\ast}$ & $\le 3.1$ & $1.70_{-0.01}^{+0.01}$ & $49^{\ast}$ & $47.8_{-2.2}^{+0.4}$ & $6.39^{\ast}$ & $13.8_{-3.5}^{+3.6}$ & $6.99^{\ast}$ & $\le 6.3$ &  &  \\ [0.2cm]
FPM & $58869.5$ & $-$ & $41^{\ast}$ & $0.89^{\ast}$ & $-$ & $2.6^{\ast}$ & $3.0^{\ast}$ & $\le 6.0$ & $1.71_{-0.01}^{+0.01}$ & $49^{\ast}$ & $44.4_{-2.4}^{+2.1}$ & $6.39^{\ast}$ & $18.8_{-3.8}^{+3.9}$ & $6.99^{\ast}$ & $8.2_{-3.6}^{+3.7}$ &  &  \\ [0.2cm]
FPM & $59224.5$ & $-$ & $41^{\ast}$ & $0.89^{\ast}$ & $-$ & $2.6^{\ast}$ & $3.0^{\ast}$ & $4.0_{-2.9}^{+2.9}$ & $1.69_{-0.02}^{+0.02}$ & $49^{\ast}$ & $21.1_{-2.0}^{+2.0}$ & $6.39^{\ast}$ & $12.9_{-3.3}^{+3.4}$ & $6.99^{\ast}$ &  $\le 5.4$ &  &  \\ [0.2cm]
FPM & $59571.0$ & $-$ & $41^{\ast}$ & $0.89^{\ast}$ & $-$ & $2.6^{\ast}$ & $3.0^{\ast}$ & $4.2_{-2.5}^{+2.5}$ & $1.66_{-0.02}^{+0.02}$ & $49^{\ast}$ & $16.4_{-1.7}^{+1.7}$ & $6.39^{\ast}$ & $9.5_{-2.6}^{+2.7}$ & $6.99^{\ast}$ & $3.9_{-2.4}^{+2.5}$ &  &  \\ [0.2cm]
FPM & $59444.2$ & $-$ & $41^{\ast}$ & $0.89^{\ast}$ & $-$ & $2.6^{\ast}$ & $3.0^{\ast}$ & $\le 7.4$ & $1.71_{-0.02}^{+0.02}$ & $49^{\ast}$ & $46.5_{-4.1}^{+1.9}$ & $6.39^{\ast}$ & $17.0_{-6.6}^{+6.9}$ & $6.99^{\ast}$ & $\le 11.1$ &  &  \\ [0.2cm]
FPM & $58418.4$ & $-$ & $41^{\ast}$ & $0.89^{\ast}$ & $-$ & $2.6^{\ast}$ & $3.0^{\ast}$ & $\le 5.2$ & $1.7_{-0.02}^{+0.02}$ & $49^{\ast}$ & $34.2_{-3.3}^{+1.0}$ & $6.39^{\ast}$ & $16.2_{-5.3}^{+5.5}$ & $6.99^{\ast}$ & $7.3_{-4.9}^{+5.2}$ &  &  \\ [0.2cm]
FPM & $59981.1^{\dagger}$ & $-$ & $41^{\ast}$ & $0.89^{\ast}$ & $-$ & $2.6^{\ast}$ & $3.0^{\ast}$ & $1.5^{\ast}$ & $1.64^{\ast}$ & $49^{\ast}$ & $14.0^{\ast}$ & $6.39^{\ast}$ & $7.5^{\ast}$ & $6.99^{\ast}$ & $-$ &  &  \\ [0.2cm]
FPM & $57724.5$ & $-$ & $41^{\ast}$ & $0.89^{\ast}$ & $-$ & $2.6^{\ast}$ & $3.0^{\ast}$ & $\le 3.3$ & $1.65_{-0.03}^{+0.03}$ & $49^{\ast}$ & $10.9_{-1.5}^{+1.2}$ & $6.39^{\ast}$ & $8.0_{-2.2}^{+2.3}$ & $6.99^{\ast}$ & $\le 4.5$ &  &  \\ [0.2cm]
FPM & $57423.7$ & $-$ & $41^{\ast}$ & $0.89^{\ast}$ & $-$ & $2.6^{\ast}$ & $3.0^{\ast}$ & $\le 5.0$ & $1.63_{-0.04}^{+0.04}$ & $49^{\ast}$ & $10.8_{-2.4}^{+1.6}$ & $6.39^{\ast}$ & $5.4_{-3.3}^{+3.5}$ & $6.99^{\ast}$ & $\le 6.4$ &  &  \\ [0.25cm]

\end{tabular}
\end{ruledtabular}
\end{table*}
\end{turnpage}

\addtocounter{table}{-1} 
\begin{turnpage}
\begin{table*}[t!]
\caption{Continued from previous page.}

\begin{ruledtabular}
\setlength{\tabcolsep}{1.3pt}
\begin{tabular}{lcccccccccccccccccccc}

\multicolumn{19}{c}{\textbf{Mrk~79: Model $\equiv$ {relxillCp [Relativistic Reflection] $+$ nthComp [Hot Corona] $+$ xillverCp [Distant Reflection] $+$ CompTT [Warm Corona]}} } \\ [0.15cm] 
 & $T_{\rm start}$  & $q_{\rm in}$ & $\theta$ & $a^{\ast}$ & $\log n_{\rm e}$ & $A_{\rm Fe}$ & $\log\xi$ & $F_{\rm rel}$ & $\Gamma$ &  $kT_{\rm e}$ &  $F_{\rm nth}$  &  $F_{\rm xil}$ &  & $kT_{\rm wc}$ & $\tau_{\rm wc}$ & $F_{\rm wc}$ & $\frac{\chi^{2}}{d.o.f}$   \\[0.25cm]  
\hline 

EPIC & $59117.6^{\dagger}$ & $6.5_{-0.3}^{+0.3}$ & $35_{-3}^{+4}$ & $0.88_{-0.039}^{+0.048}$ & $18.0_{-0.2}^{+0.2}$ & $0.7_{-0.1}^{+0.1}$ & $2.9_{-0.1}^{+0.1}$ & $17.8_{-0.6}^{+0.6}$ & $1.88_{-0.01}^{+0.01}$ & $\ge 93$ & $56.6_{-1.3}^{+1.2}$ & $8.8_{-1.0}^{+1.1}$ &  & $0.69_{-0.05}^{+0.20}$ & $\ge 15.1$ & $\le 0.1$ & $\frac{1075.2}{963}$ \\ [0.2cm]
EPIC & $54008.5$ & $6.5^{\ast}$ & $35^{\ast}$ & $0.88^{\ast}$ & $18.0^{\ast}$ & $0.7^{\ast}$ & $2.9^{\ast}$ & $41.1_{-0.9}^{+0.9}$ & $1.88_{-0.03}^{+0.02}$ & $-$ & $40.8_{-2.1}^{+2.0}$ & $13.4_{-2.5}^{+2.5}$ &  & $0.69^{\ast}$ & $-$ & $1.0_{-0.4}^{+0.9}$ &  \\ [0.2cm]
EPIC & $54178.1$ & $6.5^{\ast}$ & $35^{\ast}$ & $0.88^{\ast}$ & $18.0^{\ast}$ & $0.7^{\ast}$ & $2.9^{\ast}$ & $31.4_{-0.8}^{+0.8}$ & $1.80_{-0.03}^{+0.02}$ & $-$ & $32.8_{-1.9}^{+1.9}$ & $17.3_{-2.6}^{+2.7}$ &  & $0.69^{\ast}$ & $-$ & $0.5_{-0.4}^{+0.5}$ &  \\ [0.2cm]
EPIC & $54040.6$ & $6.5^{\ast}$ & $35^{\ast}$ & $0.88^{\ast}$ & $18.0^{\ast}$ & $0.7^{\ast}$ & $2.9^{\ast}$ & $19.4_{-0.7}^{+0.7}$ & $1.75_{-0.03}^{+0.03}$ & $-$ & $47.6_{-1.9}^{+1.9}$ & $17.9_{-3.0}^{+3.1}$ &  & $0.69^{\ast}$ & $-$ & $1.1_{-0.3}^{+1.0}$ &  \\ [0.2cm]
EPIC & $51826.2$ & $6.5^{\ast}$ & $35^{\ast}$ & $0.88^{\ast}$ & $18.0^{\ast}$ & $0.7^{\ast}$ & $2.9^{\ast}$ & $17.4_{-2.3}^{+2.5}$ & $1.91_{-0.05}^{+0.05}$ & $-$ & $46.6_{-5.4}^{+4.2}$ & $18.3_{-6.4}^{+6.6}$ &  & $0.69^{\ast}$ & $-$ & $\le 1.0$ &  \\ [0.2cm]
EPIC & $52025.8$ & $6.5^{\ast}$ & $35^{\ast}$ & $0.88^{\ast}$ & $18.0^{\ast}$ & $0.7^{\ast}$ & $2.9^{\ast}$ & $16.9_{-0.9}^{+1.0}$ & $1.62_{-0.05}^{+0.04}$ & $-$ & $26.7_{-3.1}^{+3.1}$ & $19.0_{-6.3}^{+6.5}$ &  & $0.69^{\ast}$ & $-$ & $0.9_{-0.4}^{+0.9}$ &  \\ [0.2cm]
FPM & $59115.5$ & $6.5^{\ast}$ & $35^{\ast}$ & $0.88^{\ast}$ & $18.0^{\ast}$ & $0.7^{\ast}$ & $2.9^{\ast}$ & $17.3_{-13.3}^{+13.3}$ & $1.98_{-0.03}^{+0.04}$ & $-$ & $66.7_{-8.7}^{+8.7}$ & $14.2_{-1.3}^{+1.2}$ &  & $0.69^{\ast}$ & $-$ & $\le 0.1$ &  \\ [0.2cm]
FPM & $59117.2^{\dagger}$ & $6.5^{\ast}$ & $35^{\ast}$ & $0.88^{\ast}$ & $18.0^{\ast}$ & $0.7^{\ast}$ & $2.9^{\ast}$ & $17.8^{\ast}$ & $1.88^{\ast}$ & $-$ & $56.6^{\ast}$ & $8.8^{\ast}$ &  & $0.69^{\ast}$ & $-$ & $-$ &  \\ [0.25cm]

\hline 
\hline 
\multicolumn{19}{c}{\textbf{PG~1229+204: Model $\equiv$ {relxillCp [Relativistic Reflection] $+$ nthComp [Hot Corona]}} } \\ [0.15cm] 
  & $T_{\rm start}$  & $q_{\rm in}$ & $\theta$ & $a^{\ast}$ & $\log n_{\rm e}$ & $A_{\rm Fe}$ & $\log\xi$ & $F_{\rm rel}$ & $\Gamma$ &  $kT_{\rm e}$ &  $F_{\rm nth}$  &  $\frac{\chi^{2}}{d.o.f}$ \\ [0.25cm]   
\hline 

EPIC & $53560.5$ & $\ge 3.2$ & $37_{-8}^{+8}$ & $0.974_{-0.775}^{+0.024}$ & $\le 16.6$ & $1.6_{-0.8}^{+1.6}$ & $\le 0.3$ & $4.2_{-0.3}^{+0.3}$ & $2.31_{-0.05}^{+0.03}$ & $\ge 64$ & $8.6_{-0.1}^{+0.1}$ & $\frac{162.0}{154}$ &  &  &  &  &  \\ [0.2cm]
FPM & $57597.4$ & $-$ & $37^{\ast}$ & $0.974^{\ast}$ & $-$ & $1.6^{\ast}$ & $-$ & $4.7_{-0.8}^{+0.8}$ & $2.26_{-0.30}^{+0.31}$ & $-$ & $12.3_{-1.1}^{+1.1}$ &  &  &  &  &  &  \\ [0.25cm]

\hline 
\hline 
\multicolumn{19}{c}{\textbf{PG~0844+349: Model $\equiv$ {relxillCp [Relativistic Reflection] $+$ nthComp [Hot Corona] $+$ CompTT [Warm Corona]}} } \\[0.15cm]
  & $T_{\rm start}$  & $q_{\rm in}$ & $\theta$ & $a^{\ast}$ & $\log n_{\rm e}$ & $A_{\rm Fe}$ & $\log\xi$ & $F_{\rm rel}$ & $\Gamma$ &  $kT_{\rm e}$ &  $F_{\rm nth}$  &   &   & $kT_{\rm wc}$ & $\tau_{\rm wc}$ & $F_{\rm wc}$ & $\frac{\chi^{2}}{d.o.f}$   \\[0.25cm]
\hline 

EPIC & $51853.0$ & $\ge 3.2$ & $34_{-7}^{+11}$ & $0.292_{-0.157}^{+0.555}$ & $\le 18.5$ & $4.5_{-2.8}^{+0.4}$ & $\le 1.8$ & $2.9_{-1.3}^{+1.4}$ & $2.17_{-0.20}^{+0.20}$ & $\ge 54$ & $14.9_{-0.4}^{+0.4}$ &  &  & $0.26_{-0.09}^{+0.07}$ & $12.0_{-1.8}^{+5.8}$ & $4.1_{-1.7}^{+1.6}$ & $\frac{158.7}{150}$ \\ [0.2cm]
EPIC & $54954.2$ & $-$ & $34^{\ast}$ & $0.292^{\ast}$ & $-$ & $4.5^{\ast}$ & $-$ & $10.0_{-2.3}^{+2.4}$ & $\le 1.87$ & $-$ & $4.4_{-0.3}^{+0.3}$ &  &  & $0.26^{\ast}$ & $12.0^{\ast}$ & $0.6_{-0.2}^{+0.1}$ &  \\ [0.2cm]
FPM & $59224.0$ & $-$ & $34^{\ast}$ & $0.292^{\ast}$ & $-$ & $4.5^{\ast}$ & $-$ & $5.1_{-1.2}^{+1.2}$ & $2.34_{-0.48}^{+0.20}$ & $-$ & $10.5_{-2.0}^{+2.0}$ &  &  & $0.26^{\ast}$ & $12.0^{\ast}$ & $\le 18.9$ &  \\ [0.25cm]

\end{tabular}
\end{ruledtabular}
\end{table*}
\end{turnpage}

\addtocounter{table}{-1} 
\begin{turnpage}
\begin{table*}[t!] 
\vspace*{-5.0cm}   

\caption{Continued from previous page.}

\begin{ruledtabular}
\setlength{\tabcolsep}{1.3pt}
\begin{tabular}{lcccccccccccccccccccc}

\multicolumn{19}{c}{\textbf{PG~0804+761: Model $\equiv$ {relxillCp [Relativistic Reflection] $+$ nthComp [Hot Corona] $+$ zGauss\_N [Narrow Fe~K$_\alpha$ Emission Line] $+$ CompTT [Warm Corona]}} }\\ [0.15cm] 
  & $T_{\rm start}$  & $q_{\rm in}$ & $\theta$ & $a^{\ast}$ & $\log n_{\rm e}$ & $A_{\rm Fe}$ & $\log\xi$ & $F_{\rm rel}$ & $\Gamma$ &  $kT_{\rm e}$ &  $F_{\rm nth}$ &  $E_{\rm c}$  &  $F_{\rm NGa}$ & $kT_{\rm wc}$ & $\tau_{\rm wc}$ & $F_{\rm wc}$ & $\frac{\chi^{2}}{d.o.f}$ \\[0.25cm]
\hline 

EPIC & $55265.5$ & $\ge 3.1$ & $44_{-7}^{+4}$ & $\le 0.706$ & $\le 17.6$ & $2.1_{-1.0}^{+2.6}$ & $\le 0.7$ & $3.0_{-1.3}^{+1.3}$ & $1.95_{-0.10}^{+0.08}$ & $\ge 52$ & $26.5_{-0.4}^{+0.4}$ & $6.4_{-0.04}^{+0.04}$ & $7.3_{-3.9}^{+3.1}$ & $\le 0.99$ & $\ge 5.5$ & $11.9_{-1.5}^{+1.2}$ & $\frac{252.2}{272}$ \\ [0.2cm]
EPIC & $55267.5$ & $-$ & $44^{\ast}$ & $-$ & $-$ & $2.1^{\ast}$ & $-$ & $7.3_{-0.9}^{+1.0}$ & $2.09_{-0.19}^{+0.07}$ & $-$ & $16.6_{-0.3}^{+0.3}$ & $6.4^{\ast}$ & $5.9_{-3.9}^{+3.2}$ & $-$ & $-$ & $8.9_{-1.4}^{+1.8}$ &  \\ [0.2cm]
FPM & $57480.1$ & $-$ & $44^{\ast}$ & $-$ & $-$ & $2.1^{\ast}$ & $-$ & $5.6_{-1.1}^{+1.3}$ & $1.96_{-0.20}^{+0.19}$ & $-$ & $21.9_{-1.6}^{+1.3}$ & $6.4^{\ast}$ & $\le 9.4$ & $-$ & $-$ & $\le 0.7$ &  \\ [0.25cm]

\hline 
\hline 
\multicolumn{19}{c}{\textbf{PG~1426+015: Model $\equiv$ {relxillCp [Relativistic Reflection] $+$ nthComp [Hot Corona] $+$ xillverCp [Distant Reflection] $+$ CompTT [Warm Corona]}} } \\ [0.15cm] 
 & $T_{\rm start}$  & $q_{\rm in}$ & $\theta$ & $a^{\ast}$ & $\log n_{\rm e}$ & $A_{\rm Fe}$ & $\log\xi$ & $F_{\rm rel}$ & $\Gamma$ &  $kT_{\rm e}$ &  $F_{\rm nth}$  &  $F_{\rm xil}$ &  & $kT_{\rm wc}$ & $\tau_{\rm wc}$ & $F_{\rm wc}$ & $\frac{\chi^{2}}{d.o.f}$   \\[0.25cm]
\hline 

EPIC & $58872.5^{\dagger}$ & $\ge 4.2$ & $22_{-6}^{+27}$ & $0.442_{-0.122}^{+0.488}$ & $\le 16.2$ & $1.6_{-0.7}^{+0.9}$ & $3.0_{-0.1}^{+0.2}$ & $3.9_{-0.9}^{+0.9}$ & $1.81_{-0.03}^{+0.04}$ & $\ge 52$ & $16.6_{-0.6}^{+0.6}$ & $2.4_{-0.4}^{+0.4}$ &  & $0.26_{-0.04}^{+0.03}$ & $11.6_{-1.1}^{+2.2}$ & $2.7_{-0.4}^{+0.3}$ & $\frac{477.8}{415}$ \\ [0.2cm]
EPIC & $51753.5$ & $-$ & $22^{\ast}$ & $0.442^{\ast}$ & $-$ & $1.6^{\ast}$ & $3.0^{\ast}$ & $\le 15.9$ & $2.08_{-0.25}^{+0.20}$ & $-$ & $17.8_{-5.7}^{+5.8}$ & $8.6_{-5.6}^{+6.1}$ &  & $0.26^{\ast}$ & $11.6^{\ast}$ & $3.3_{-3.1}^{+3.0}$ &  \\ [0.2cm]
FPM & $58871.9^{\dagger}$ & $-$ & $22^{\ast}$ & $0.442^{\ast}$ & $-$ & $1.6^{\ast}$ & $3.0^{\ast}$ & $3.9^{\ast}$ & $1.81^{\ast}$ & $-$ & $16.6^{\ast}$ & $2.4^{\ast}$ &  & $0.26^{\ast}$ & $11.6^{\ast}$ & $2.7^{\ast}$ &  \\ [0.2cm]
FPM & $58340.4$ & $-$ & $22^{\ast}$ & $0.442^{\ast}$ & $-$ & $1.6^{\ast}$ & $3.0^{\ast}$ & $7.1_{-4.8}^{+4.8}$ & $1.96_{-0.07}^{+0.15}$ & $-$ & $21.7_{-3.0}^{+3.0}$ & $3.4_{-1.4}^{+1.4}$ &  & $0.26^{\ast}$ & $11.6^{\ast}$ & $\le 39.0$ &  \\ [0.25cm]

\end{tabular}
\end{ruledtabular}
\end{table*}
\end{turnpage}

\begin{table*}
\centering
\caption{Details of the maximum likelihood ratio (MLR) and Bayesian statistical tests performed to assess the presence of relativistic disk reflection in the broadband spectral data. Model~1 and Model~2 in columns~(2) and (3) show the source models used to fit the 0.3--78\keV{} spectra without and with the relativistic disk reflection ({\tt relxillCp}) component, respectively. In Model~1, a warm Comptonization ({\tt compTT}) component was always included for all sources, whereas in Model~2, with {\tt relxillCp}, an extra warm Comptonization ({\tt compTT}) component was included only if it was statistically significant, as reported in Table~\ref{test_wc}. Columns~(4) and (5) show the model statistics obtained from fitting the 0.3--78\keV{} spectra with Model~1 and Model~2, respectively. Column~(6) reports the confidence level of relativistic disk reflection as computed by the MLR test. Columns~(7) and (8) show the Deviance Information Criteria, ${\rm DIC}_{1}$ and ${\rm DIC}_{2}$ for Model~1 and Model~2, respectively. Column~(9) reports the difference between ${\rm DIC}_{1}$ and ${\rm DIC}_{2}$, reconfirming the importance of relativistic disk reflection in the broadband spectral data for all sources in the sample.} 
\begin{center}
\begin{tabular}{cccccccccccc}
\hline 
Source &  Model~1 & Model~2 & $\frac{\chi_{1}^{2}}{{\rm d.o.f}_{1}}$ & $\frac{\chi_{2}^{2}}{{\rm d.o.f}_{2}}$  & C.I.-MLR & ${\rm DIC}_{\rm 1}$  &  ${\rm DIC}_{\rm 2}$  & $\Delta {\rm DIC_{12}}$   \\

  & (without {\tt relxillCp}  & (with {\tt relxillCp} & & & $(\%)$ &  &  &  ${\rm DIC}_{\rm 1}-{\rm DIC}_{\rm 2}$       \\
  
  & and with {\tt{compTT}}) &  and {\tt{compTT}} if needed)  &  &  & &   \\ [0.1cm]
   
(1)    &   (2)   &   (3)   &   (4)  &  (5)   & (6)  & (7) & (8) & (9) \\     [0.1cm]                                                  
\hline 

UGC 6728 & {\tt nthComp+compTT}  & {\tt nthComp+relxillCp} & $\frac{549.8}{464}$ & $\frac{514.7}{460}$ & $>99.99$ & $575.8$ & $543.9$ & $31.9$ [Very Strong] \\ [0.45cm]
         
Mrk 1310 & {\tt nthComp+compTT}  & ({\tt nthComp+compTT} & $\frac{283.4}{232}$ & $\frac{263.4}{223}$ & $98.21$ & $305.0$ & $297.9$ & $7.1$ [Strong] \\ [0.1cm]
         &                      & {\tt +relxillCp})  &  &  &  &  &  &    \\ [0.45cm]

NGC 4748 & ({\tt nthComp+xillverCp}  & ({\tt nthComp+xillverCp} & $\frac{275.2}{245}$ & $\frac{258.3}{238}$ & $98.19$ & $300.4$ & $290.6$ & $9.8$ [Strong] \\ [0.1cm]
         &  {\tt +compTT})           & {\tt +compTT+relxillCp})  &  &  &  &  &  &    \\ [0.45cm]

Mrk 110 & ({\tt nthComp+xillverCp}  & ({\tt nthComp+xillverCp} & $\frac{1954.0}{1437}$ & $\frac{1647.0}{1425}$ & $>99.99$ & $2027.4$ & $1752.8$ & $274.6$ [Very Strong] \\ [0.1cm]
         &  {\tt +compTT})           & {\tt +compTT+relxillCp})  &  &  &  &  &  &    \\ [0.45cm]         
         
Mrk 279 & ({\tt nthComp+xillverCp} & ({\tt nthComp+xillverCp} & $\frac{1426.5}{1184}$ & $\frac{1307.6}{1170}$ & $>99.99$ & $1504.2$ & $1413.0$ & $91.2$ [Very Strong] \\ [0.1cm]
         &  {\tt +compTT})           & {\tt +compTT+relxillCp})  &  &  &  &  &  &    \\ [0.45cm]         
         
Mrk 590 & ({\tt nthComp+zGauss\_N}   & ({\tt nthComp+zGauss\_N} & $\frac{2045.0}{2032}$ & $\frac{2030.2}{2028}$ & $99.49$ & $2319.9$ & $2229.5$ & $90.4$ [Very Strong]  \\ [0.1cm]
        & {\tt +zGauss2\_N+compTT})    & {\tt +zGauss2\_N+relxillCp})  &  &  &  &  &  &    \\ [0.45cm]            

Mrk 79 & ({\tt nthComp+xillverCp} & ({\tt nthComp+xillverCp} & $\frac{1685.1}{975}$ & $\frac{1075.0}{963}$ & $>99.99$ & $1746.3$ & $1196.3$ & $550.0$ [Very Strong] \\ [0.1cm]
       &  {\tt +compTT})           & {\tt +compTT+relxillCp})  &  &  &  &  &  &    \\ [0.45cm]         
         
PG 1229+204 & {\tt nthComp+compTT} & {\tt nthComp+relxillCp} & $\frac{188.7}{158}$ & $\frac{162.0}{154}$ & $>99.99$ & $203.8$ & $188.2$ & $15.6$ [Very Strong] \\ [0.45cm]
         
PG 0844+349 & {\tt nthComp+compTT} & ({\tt nthComp+compTT} & $\frac{190.4}{159}$ & $\frac{158.7}{150}$ & $99.98$ & $216.5$ & $191.4$ & $25.1$ [Very Strong] \\ [0.1cm]
            &                      & {\tt +relxillCp})  &  &  &  &  &  &    \\ [0.45cm]

PG 0804+761 & ({\tt nthComp+zGauss\_N} & ({\tt nthComp+zGauss\_N} & $\frac{287.8}{281}$ & $\frac{252.2}{272}$ & $>99.99$ & $320.5$ & $310.1$ & $10.4$ [Very Strong] \\ [0.1cm]
            &  {\tt +compTT})      & {\tt +compTT+relxillCp})  &  &  &  &  &  &    \\ [0.45cm]

PG 1426+015 & ({\tt nthComp+xillverCp} & ({\tt nthComp+xillverCp} & $\frac{527.4}{423}$ & $\frac{477.8}{415}$ & $>99.99$ & $567.7$ & $534.6$ & $33.1$ [Very Strong] \\ [0.1cm]
            &  {\tt +compTT})           & {\tt +compTT+relxillCp})  &  &  &  &  &  &    \\ [0.45cm]

\hline

\end{tabular}
\end{center} 
\label{test_rel}           
\end{table*}

\begin{figure*}
\centering
\begin{center}
\includegraphics[scale=0.3,angle=-0]{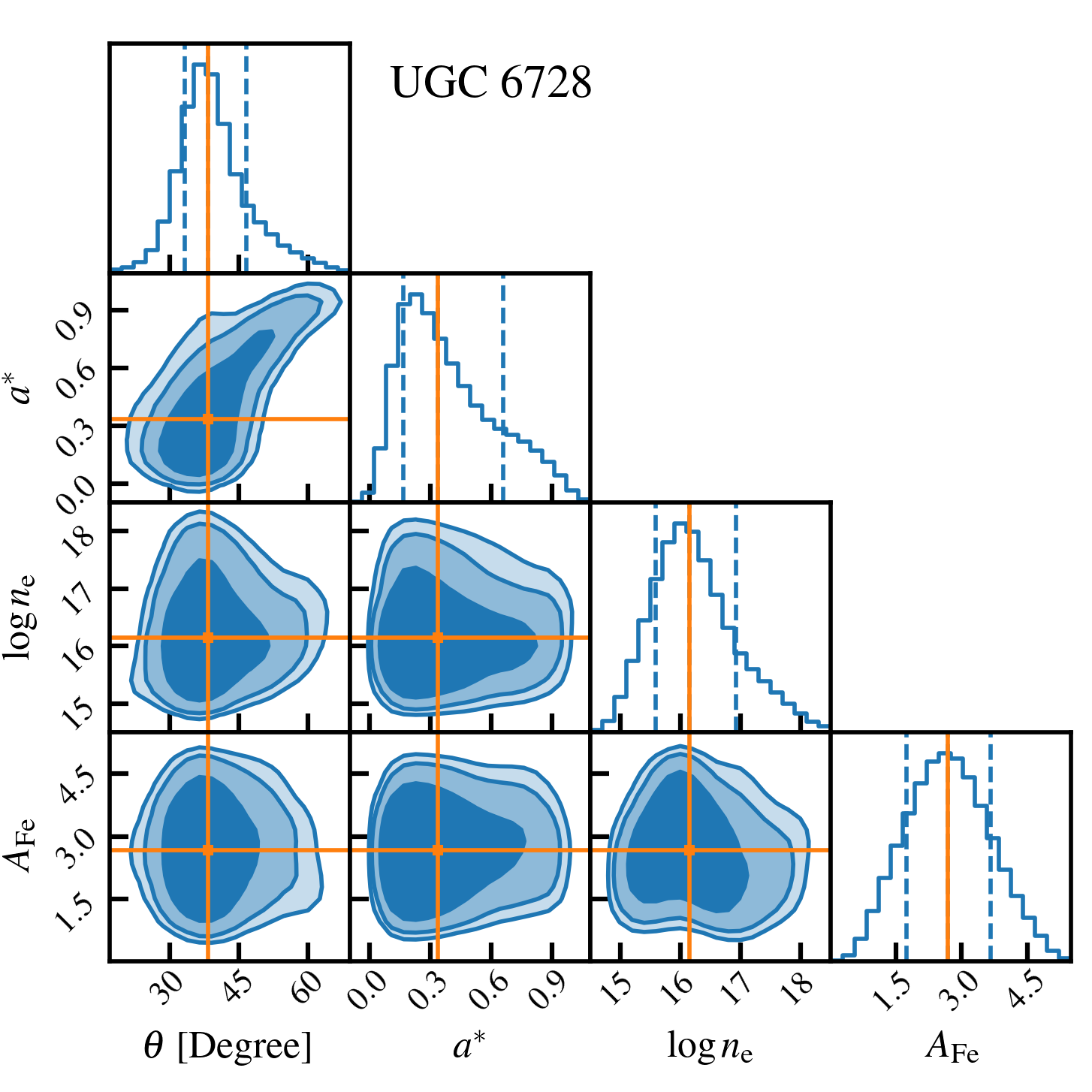}
\includegraphics[scale=0.3,angle=-0]{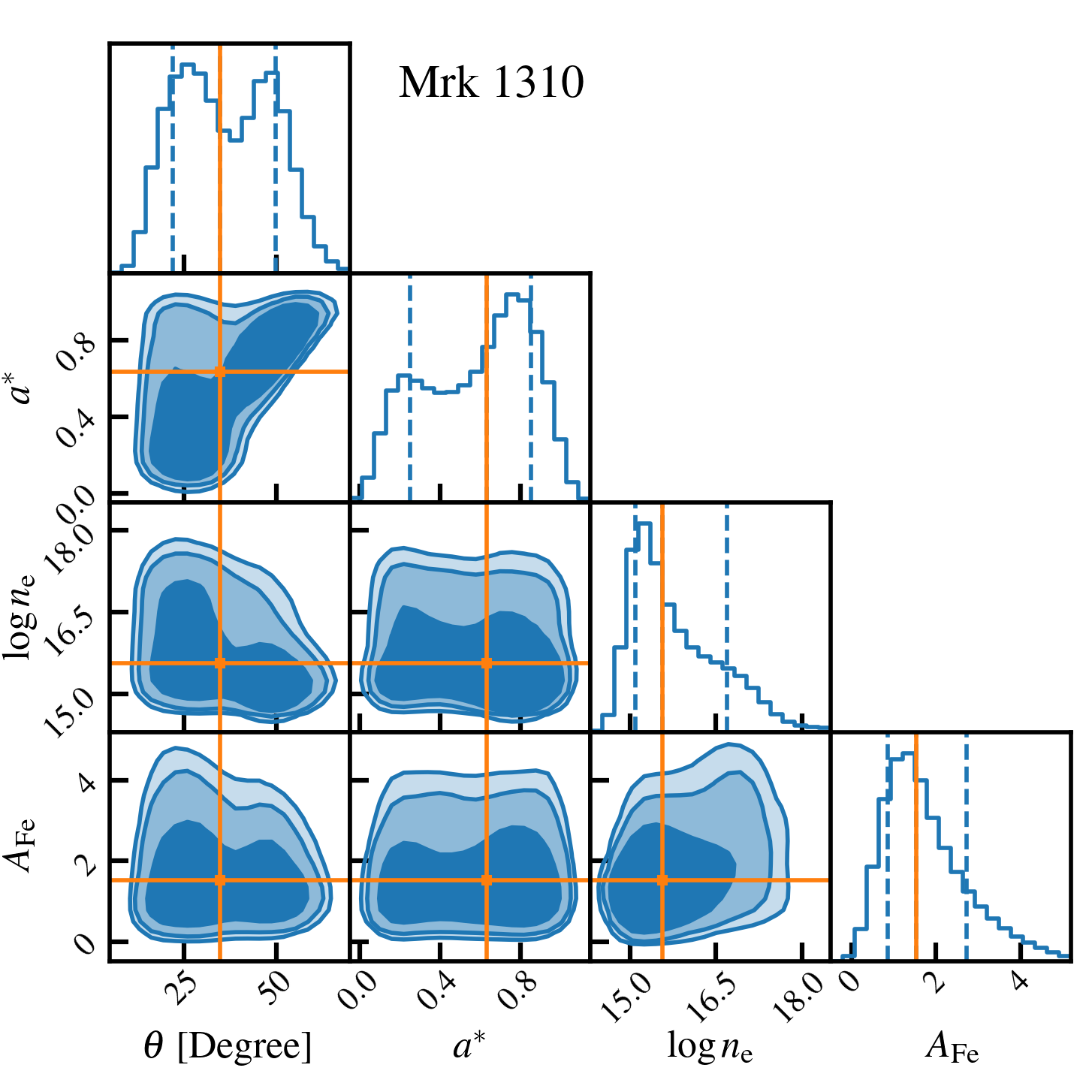}
\includegraphics[scale=0.3,angle=-0]{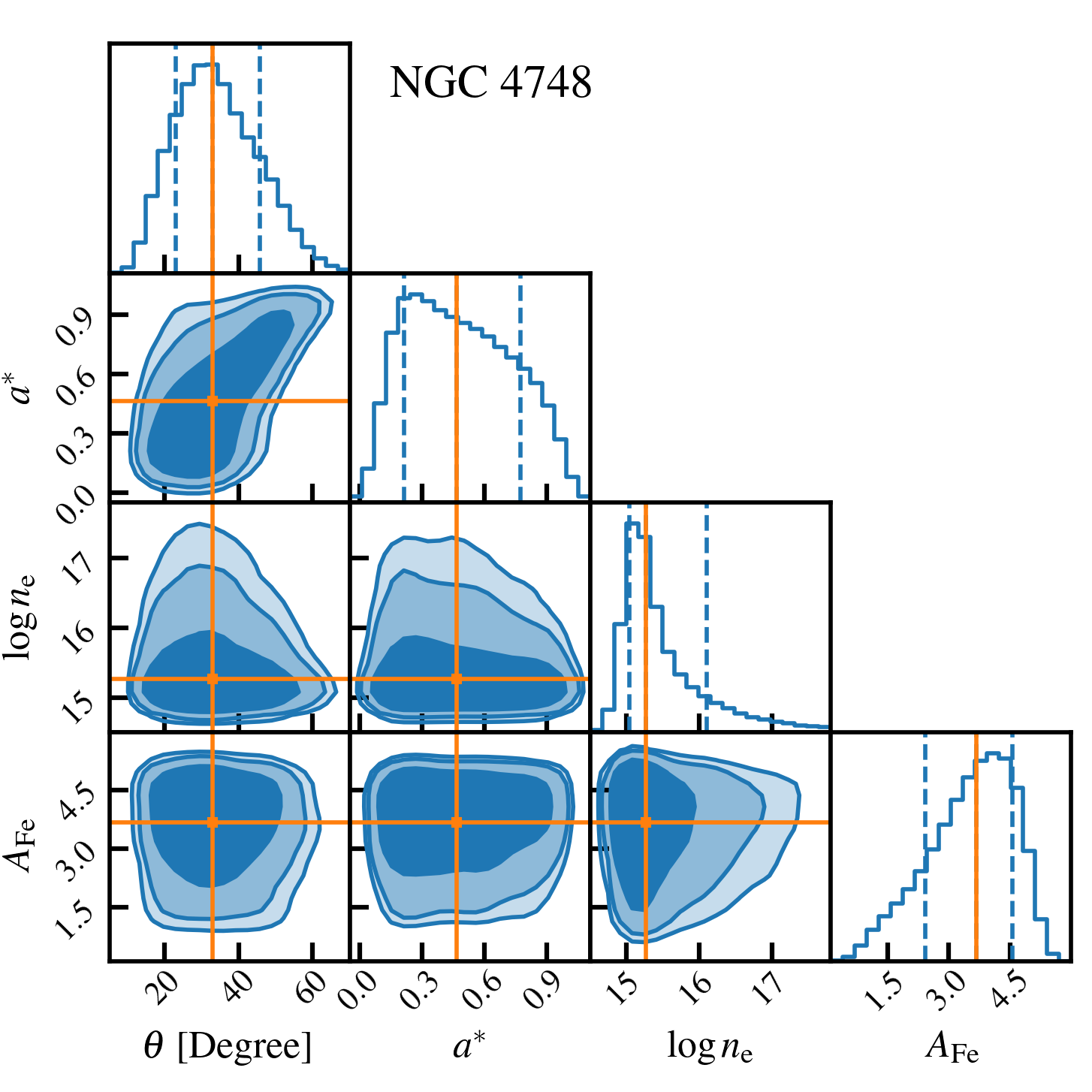}
\includegraphics[scale=0.3,angle=-0]{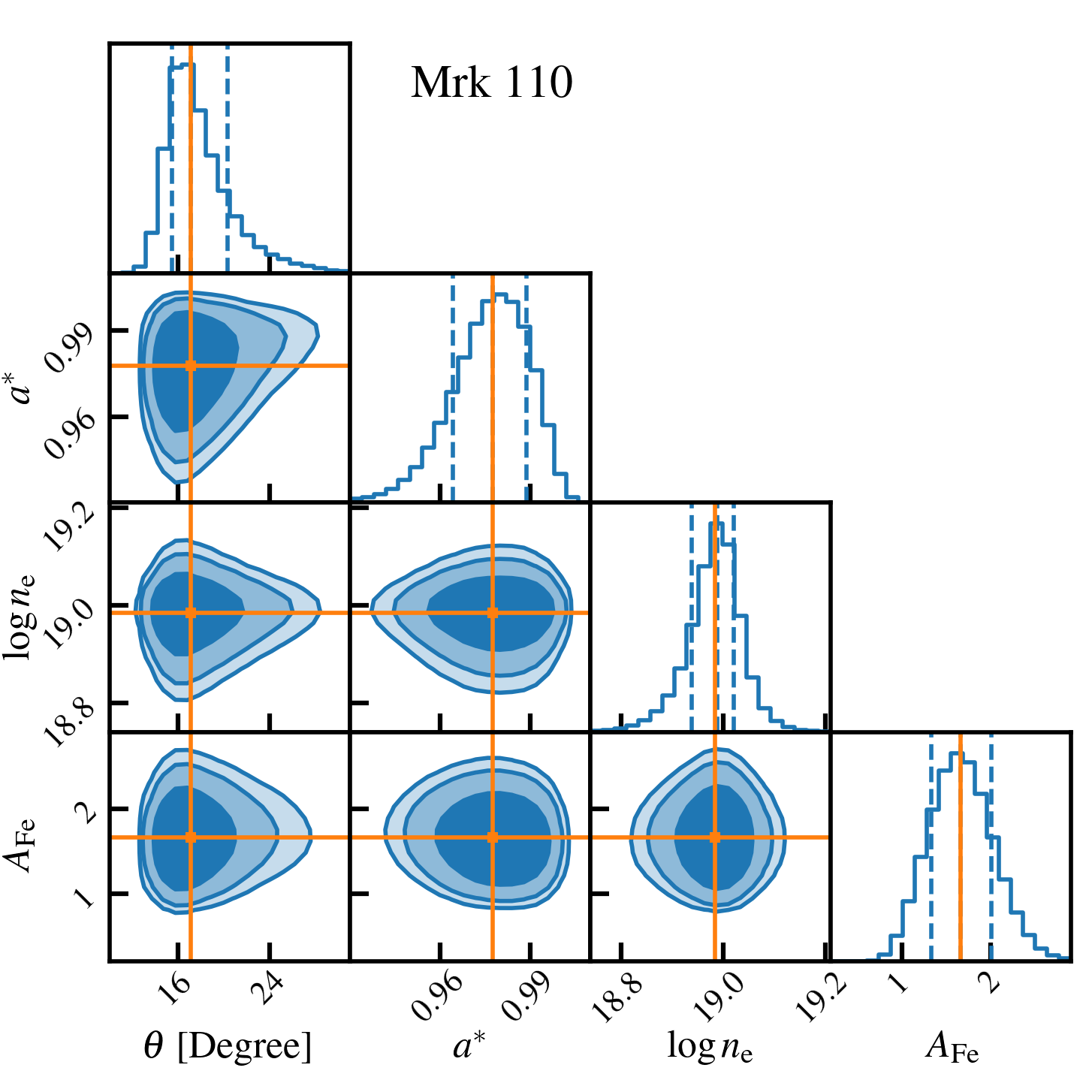}
\includegraphics[scale=0.3,angle=-0]{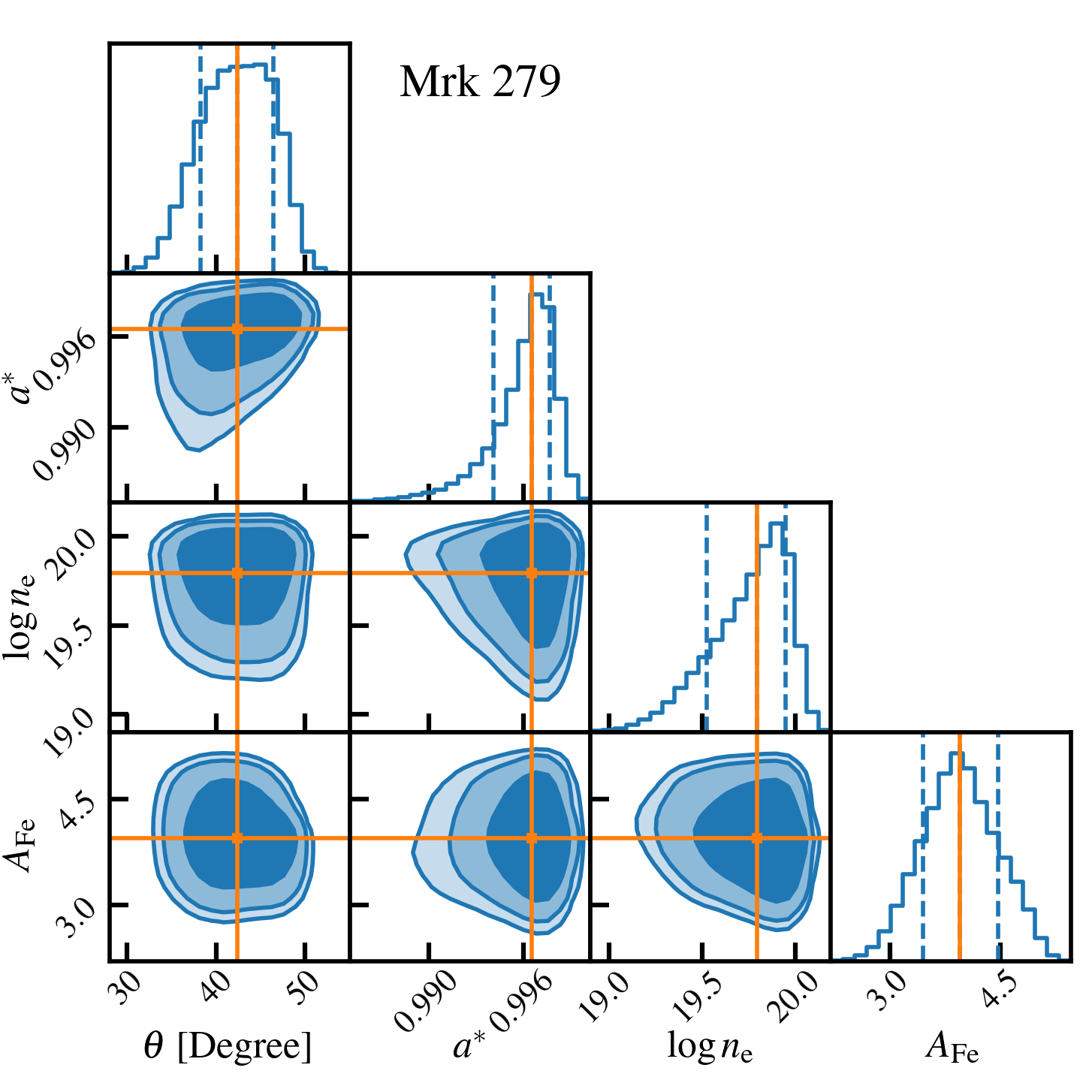}
\includegraphics[scale=0.3,angle=-0]{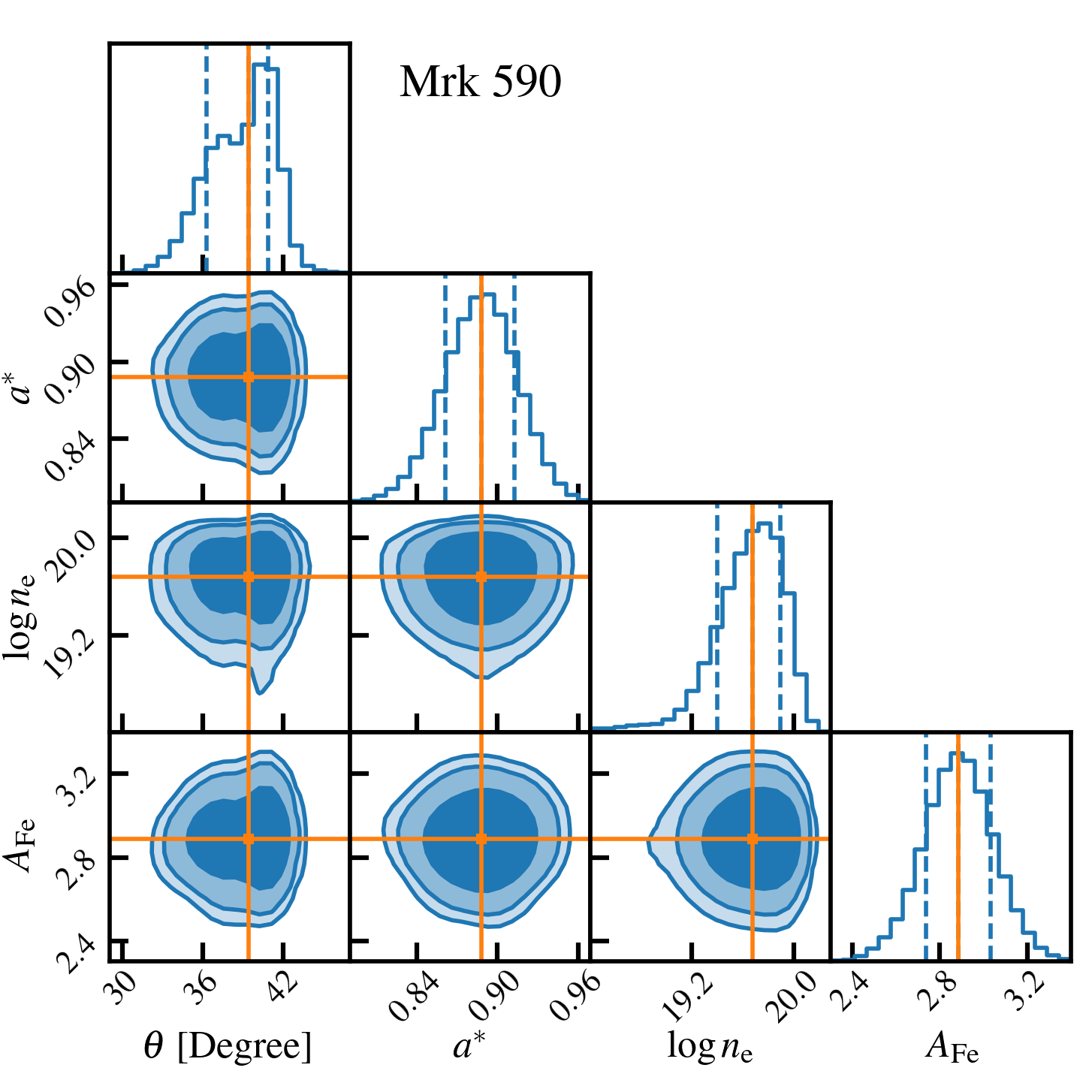}
\caption{Corner plots for black hole spin, disk density, iron abundance, and disk inclination angle. The on-diagonal histograms demonstrate the MCMC posterior parameter distributions. The median and 68.3\% confidence intervals are shown by vertical lines. The off-diagonal two-dimensional projections show MCMC contour plots for each pair of parameters. The dark, medium, and light blue areas represent 68.3\%, 90\%, and 95\% confidence levels, respectively, with square symbols indicating the parameter medians.}
\end{center}
\label{mcmc_plot_set1}
\end{figure*}

\begin{figure*}
\centering
\begin{center}
\includegraphics[scale=0.3,angle=-0]{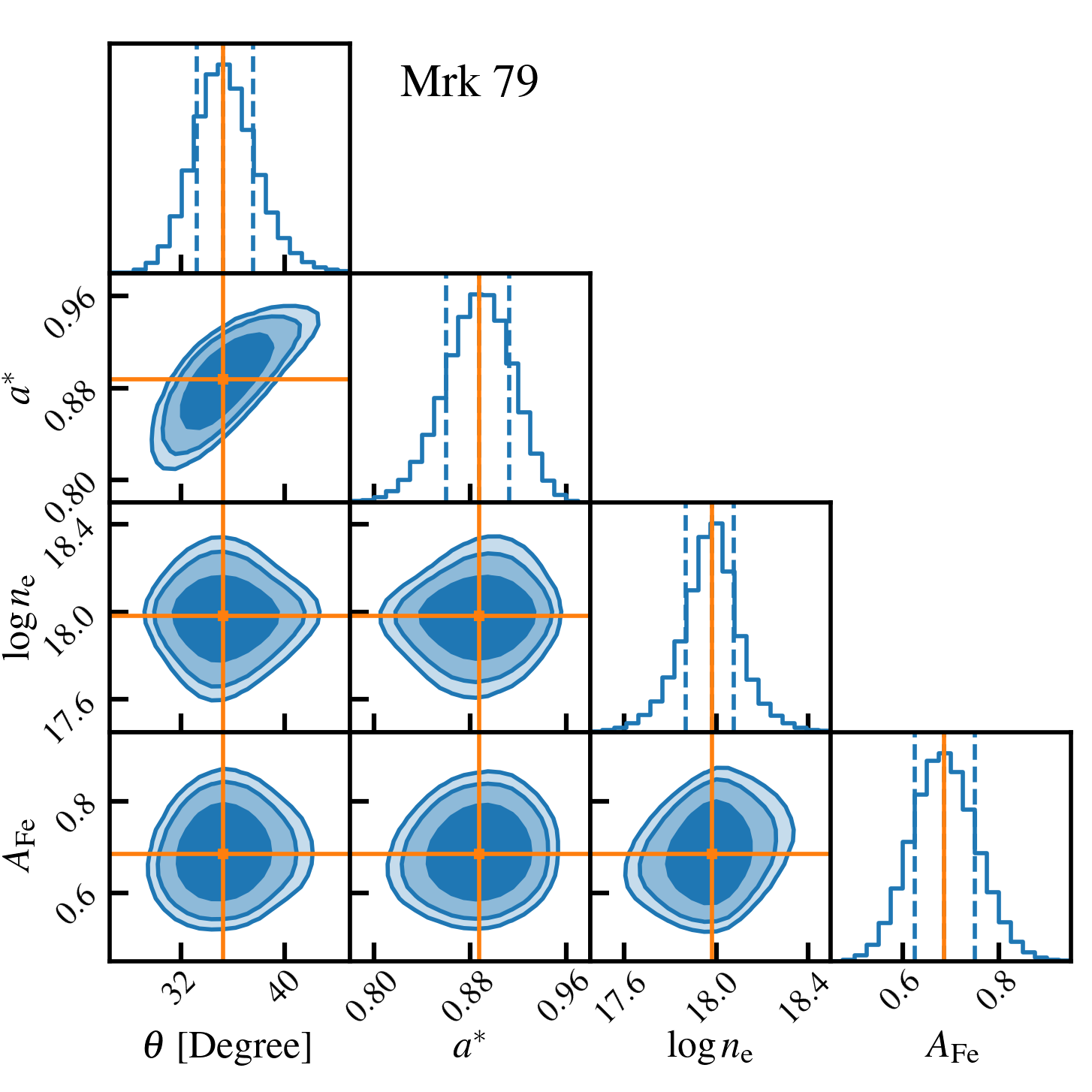}
\includegraphics[scale=0.3,angle=-0]{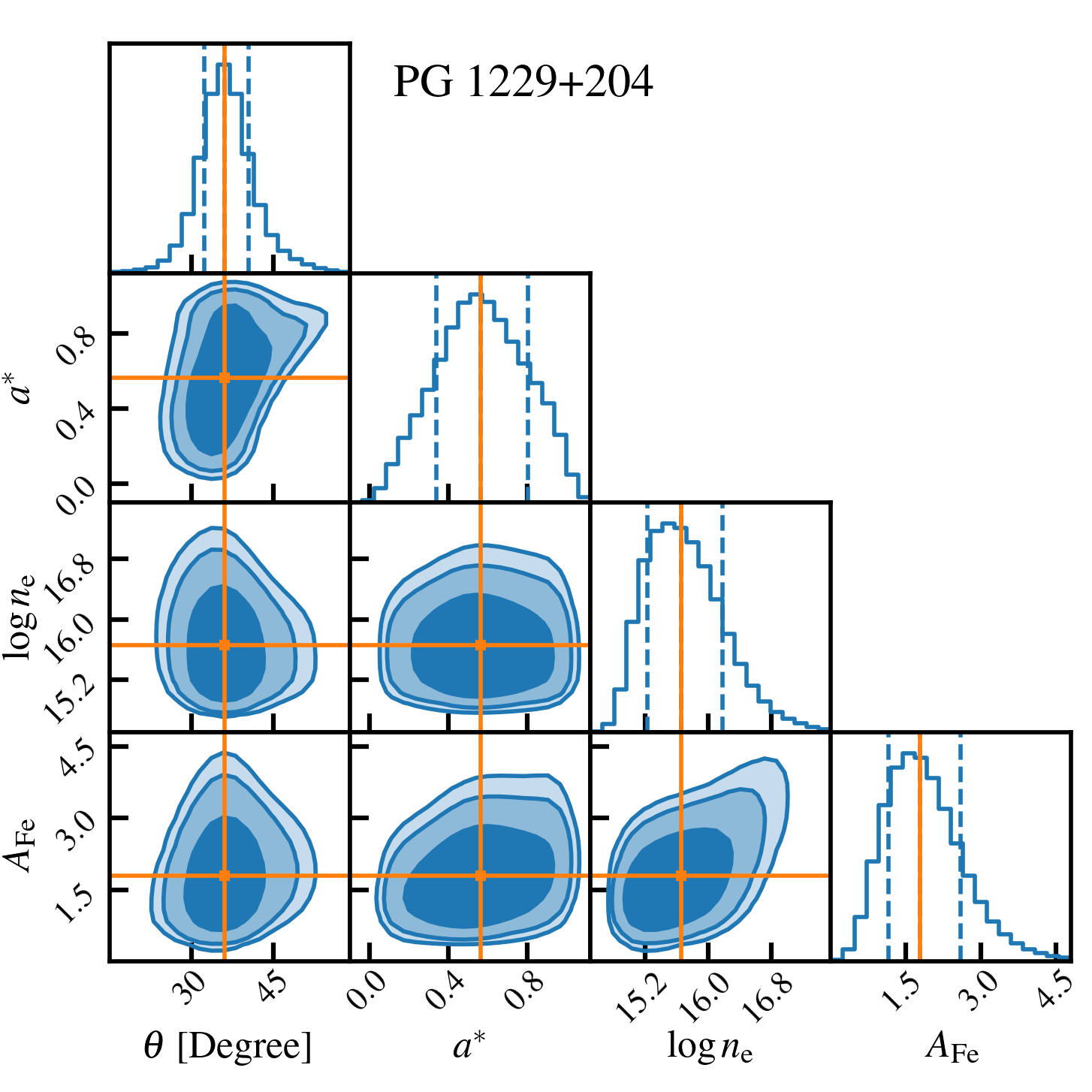}
\includegraphics[scale=0.3,angle=-0]{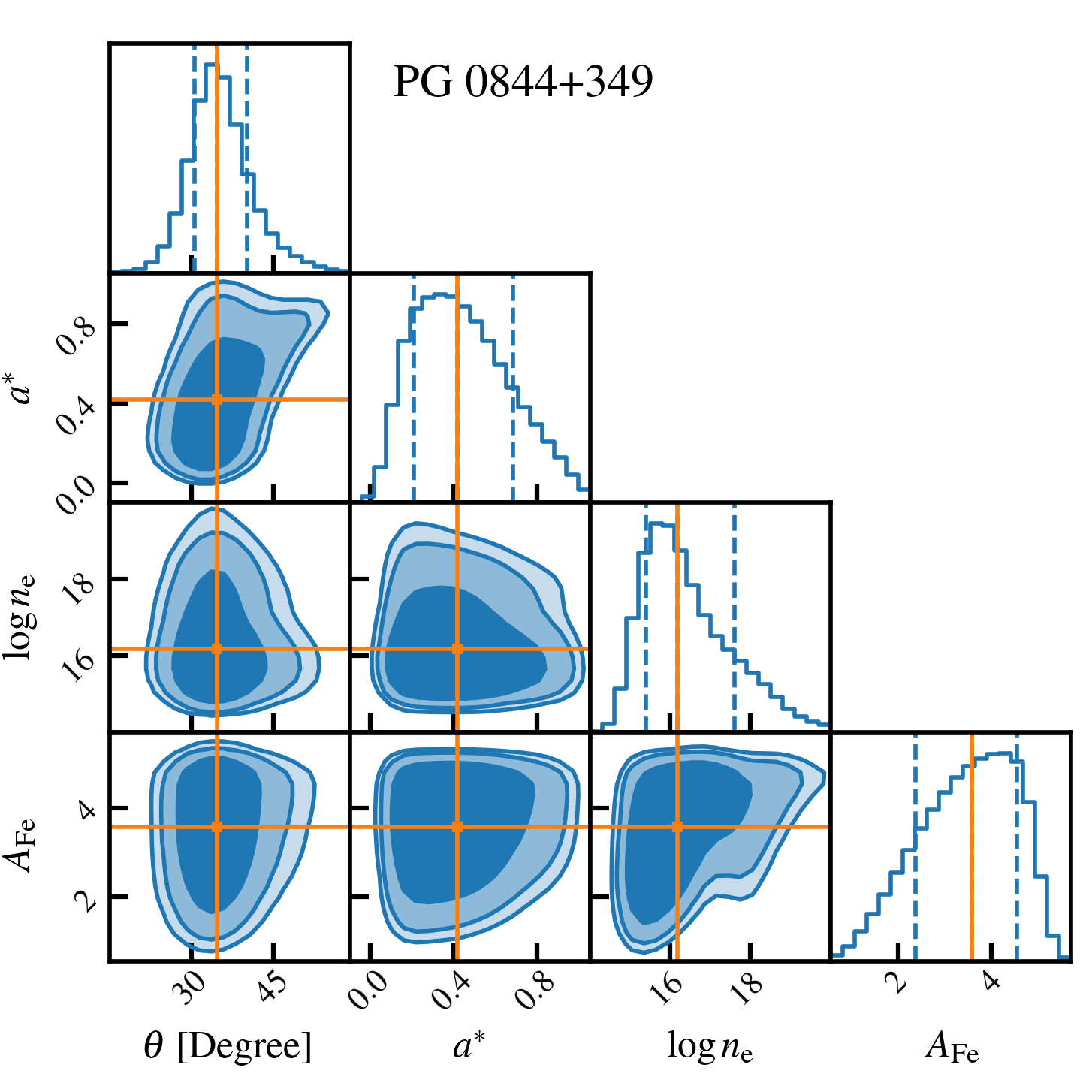}
\includegraphics[scale=0.3,angle=-0]{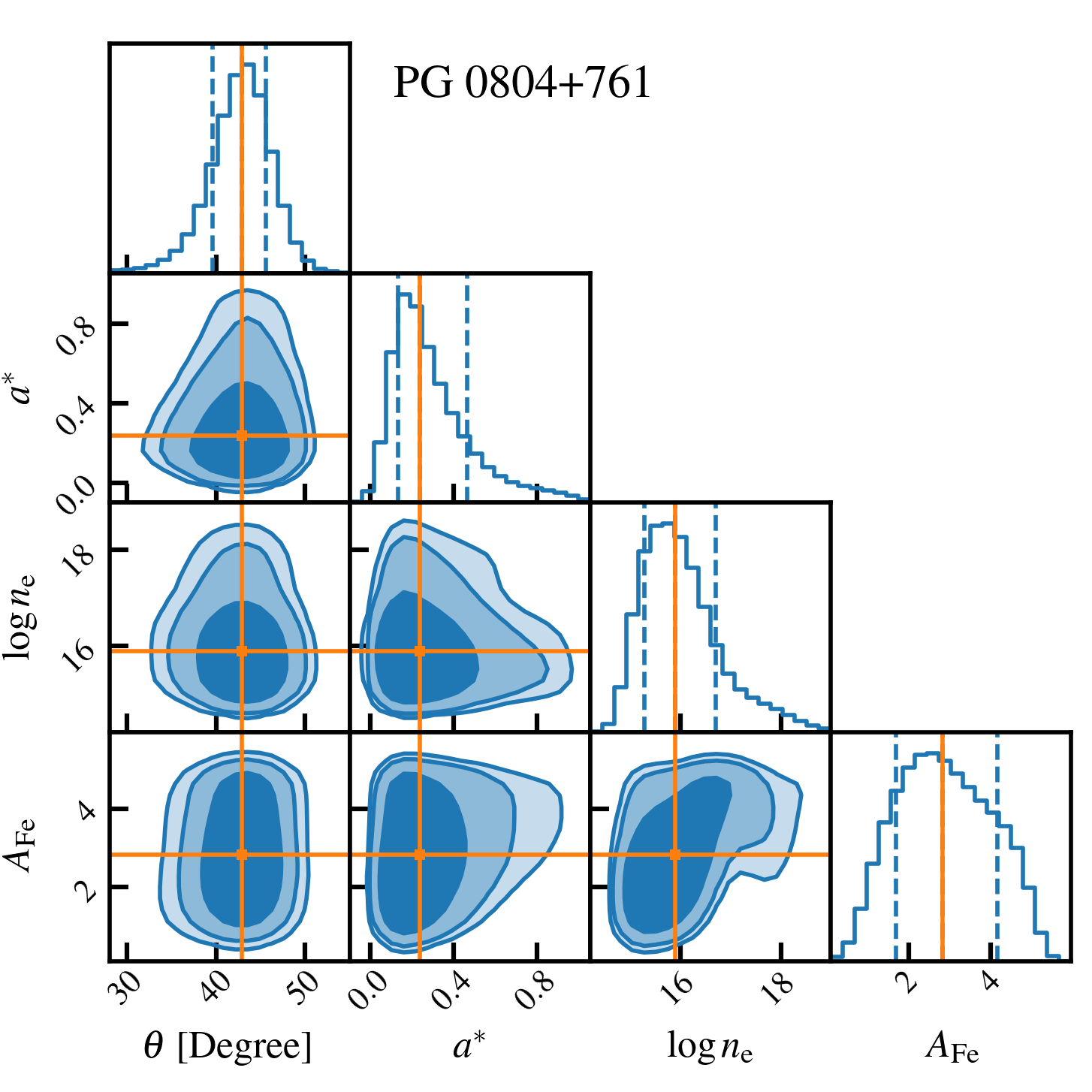}
\includegraphics[scale=0.3,angle=-0]{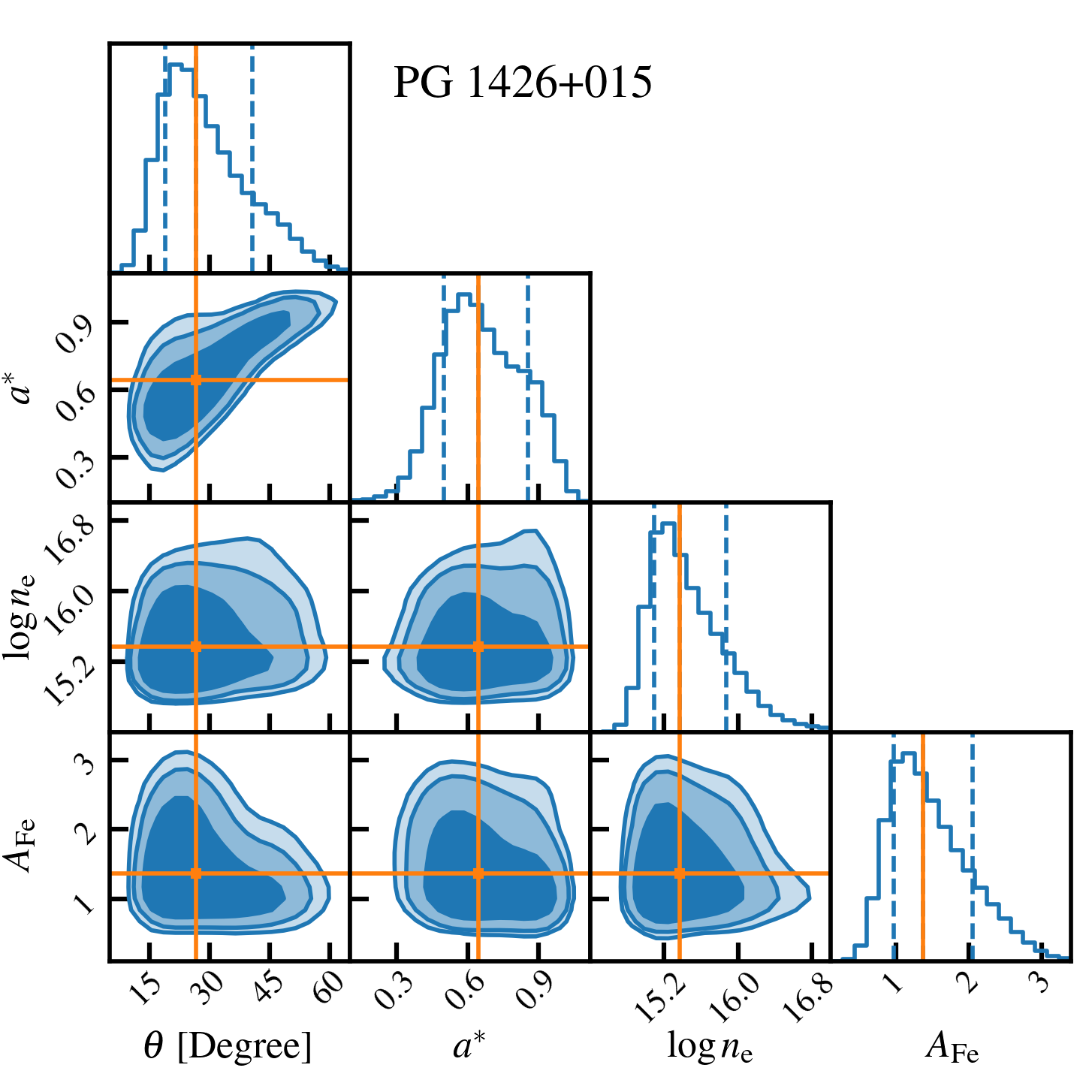}
\caption{Continued from previous page.}
\end{center}
\label{mcmc_plot_set2}
\end{figure*}

\begin{figure*}
\centering
\begin{center}
\includegraphics[scale=0.29,angle=-0]{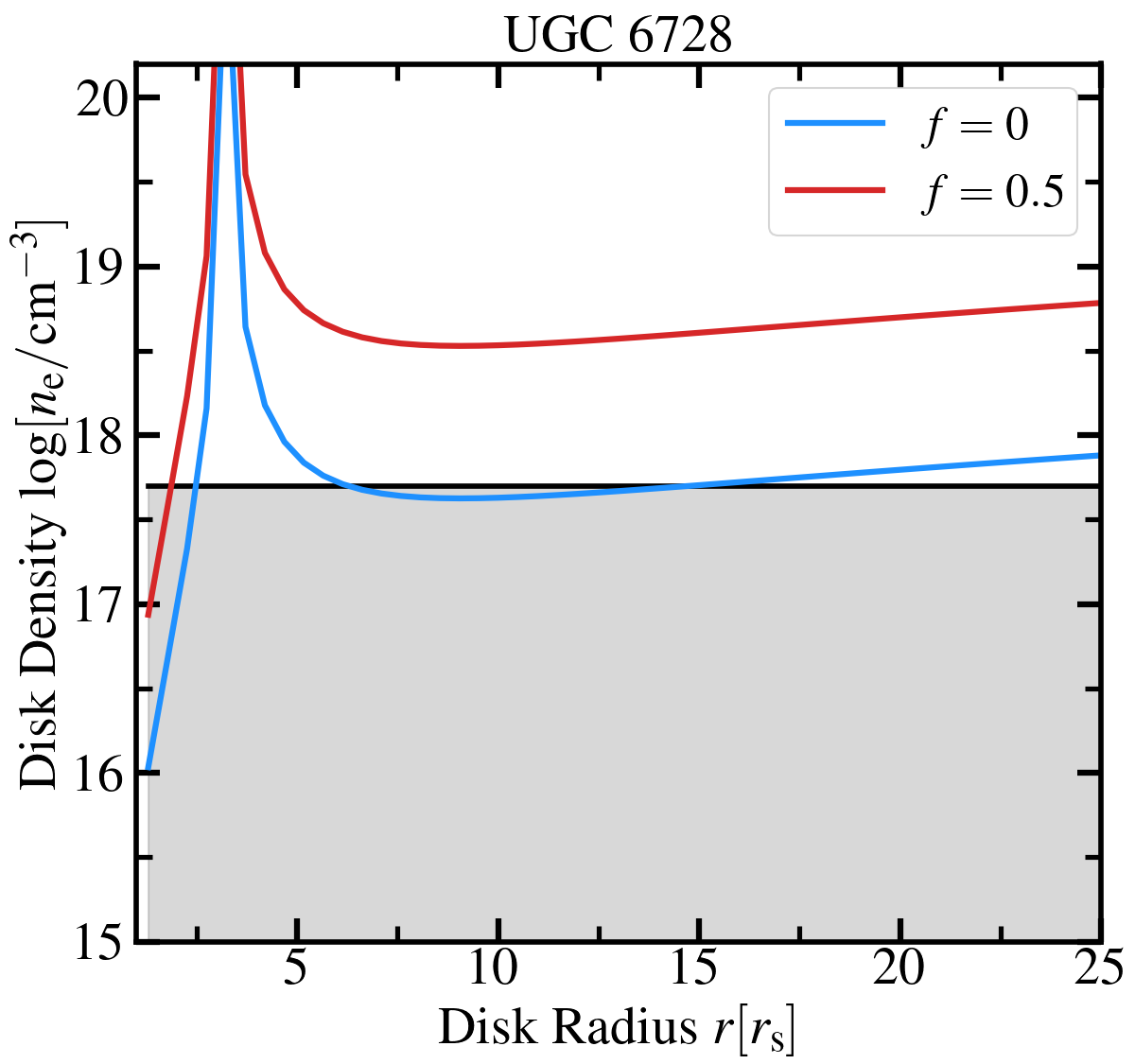}
\includegraphics[scale=0.29,angle=-0]{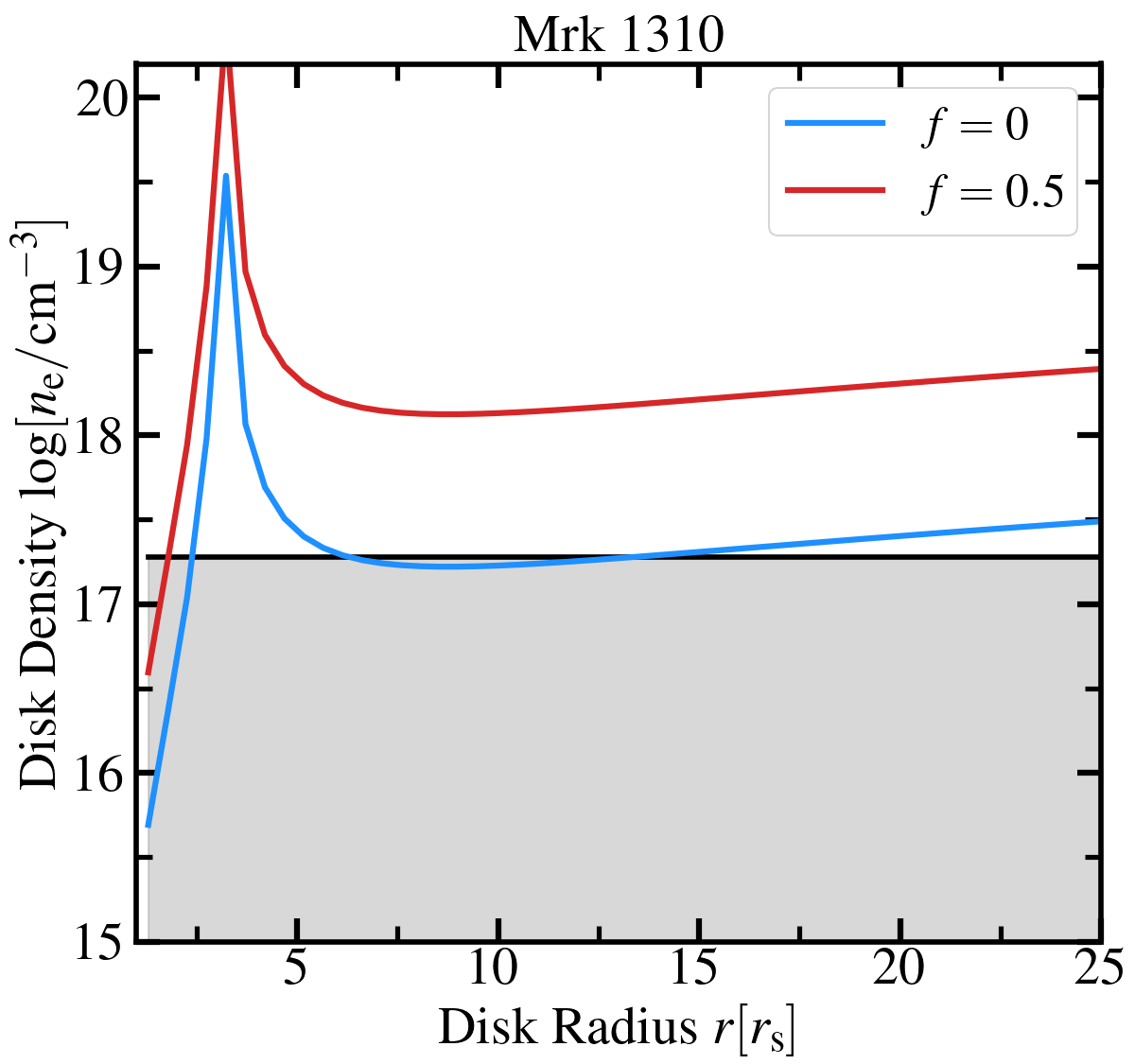}
\includegraphics[scale=0.29,angle=-0]{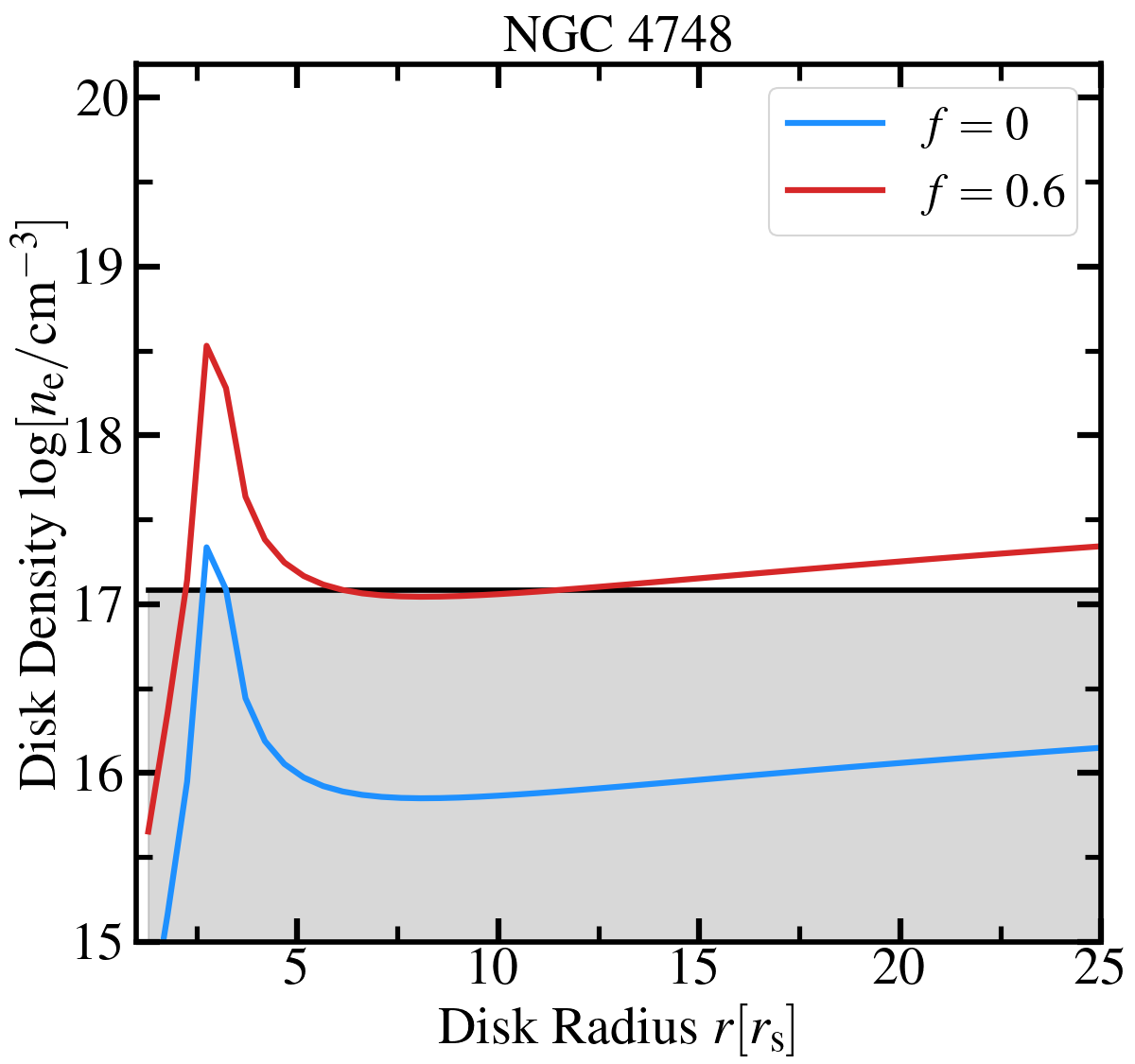}
\includegraphics[scale=0.29,angle=-0]{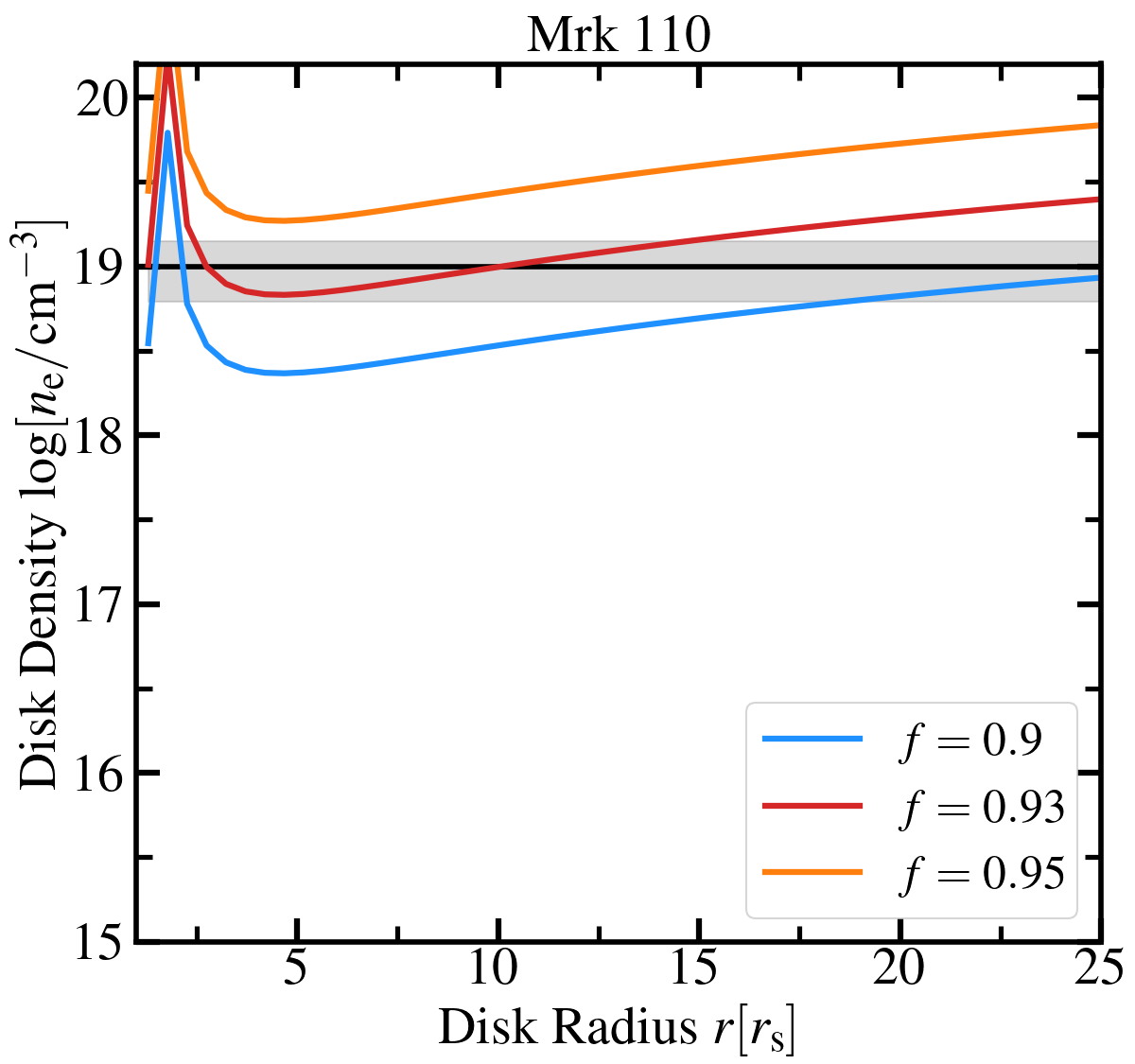}
\includegraphics[scale=0.29,angle=-0]{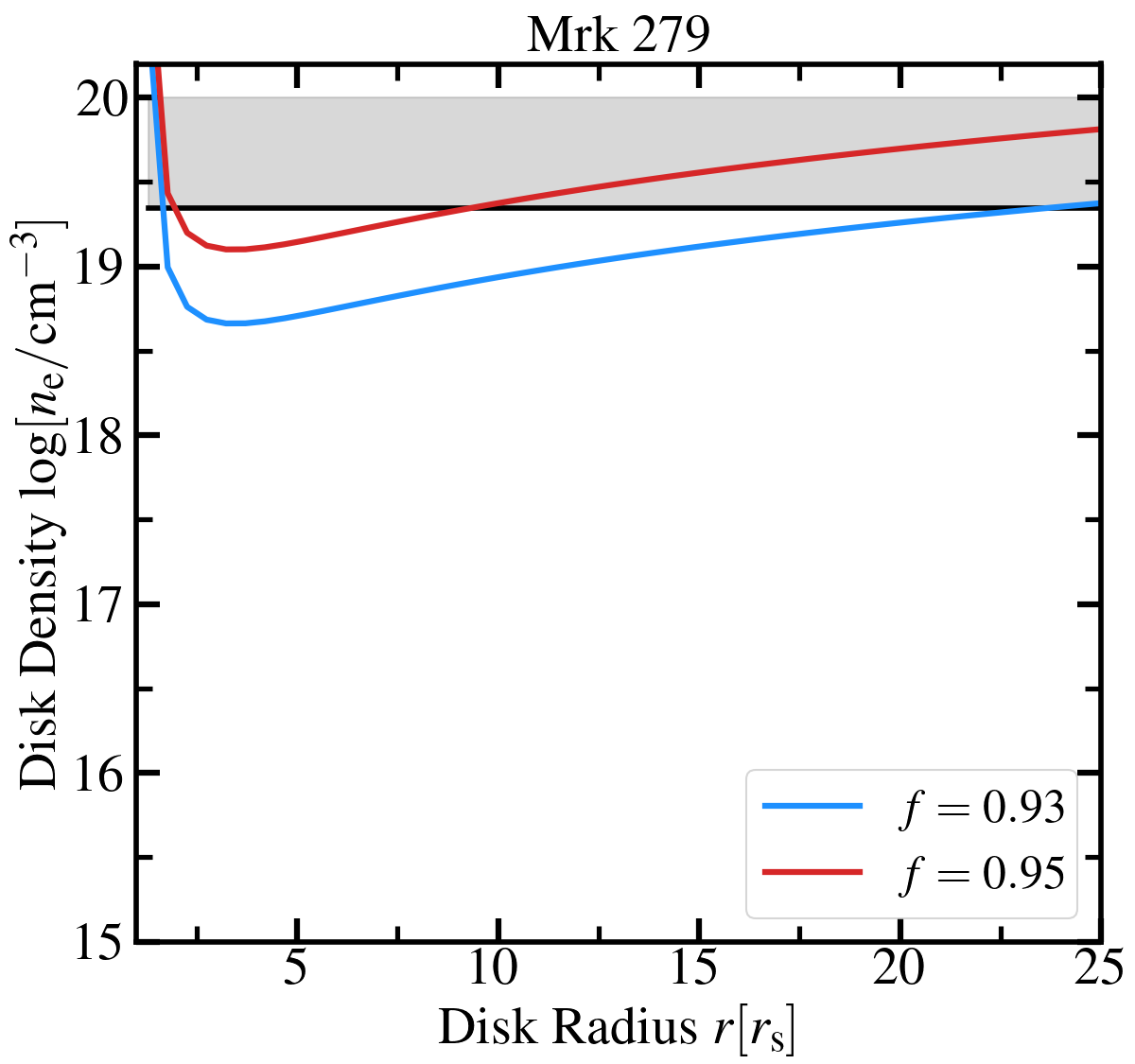}
\includegraphics[scale=0.29,angle=-0]{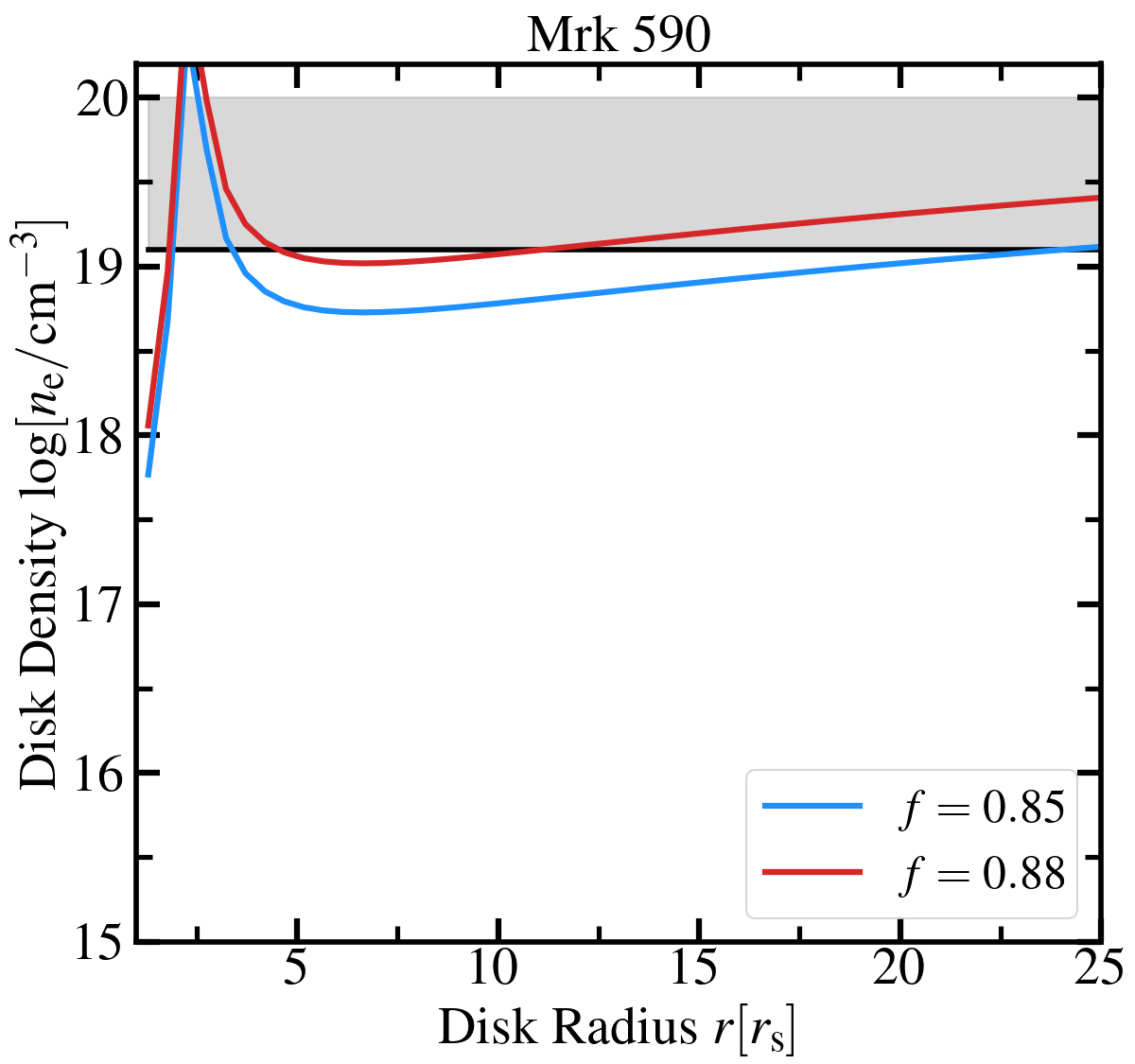}
\includegraphics[scale=0.29,angle=-0]{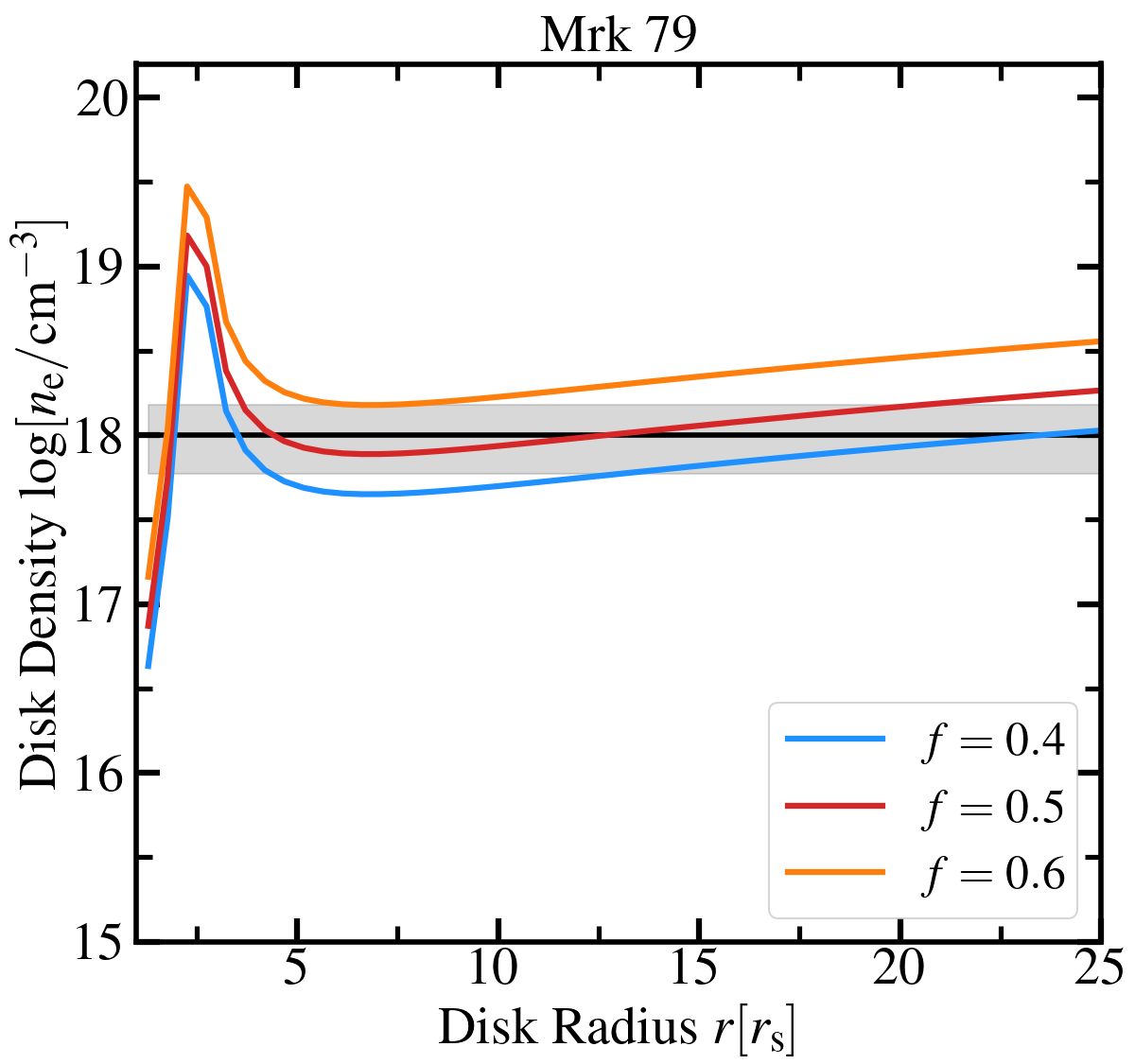}
\includegraphics[scale=0.29,angle=-0]{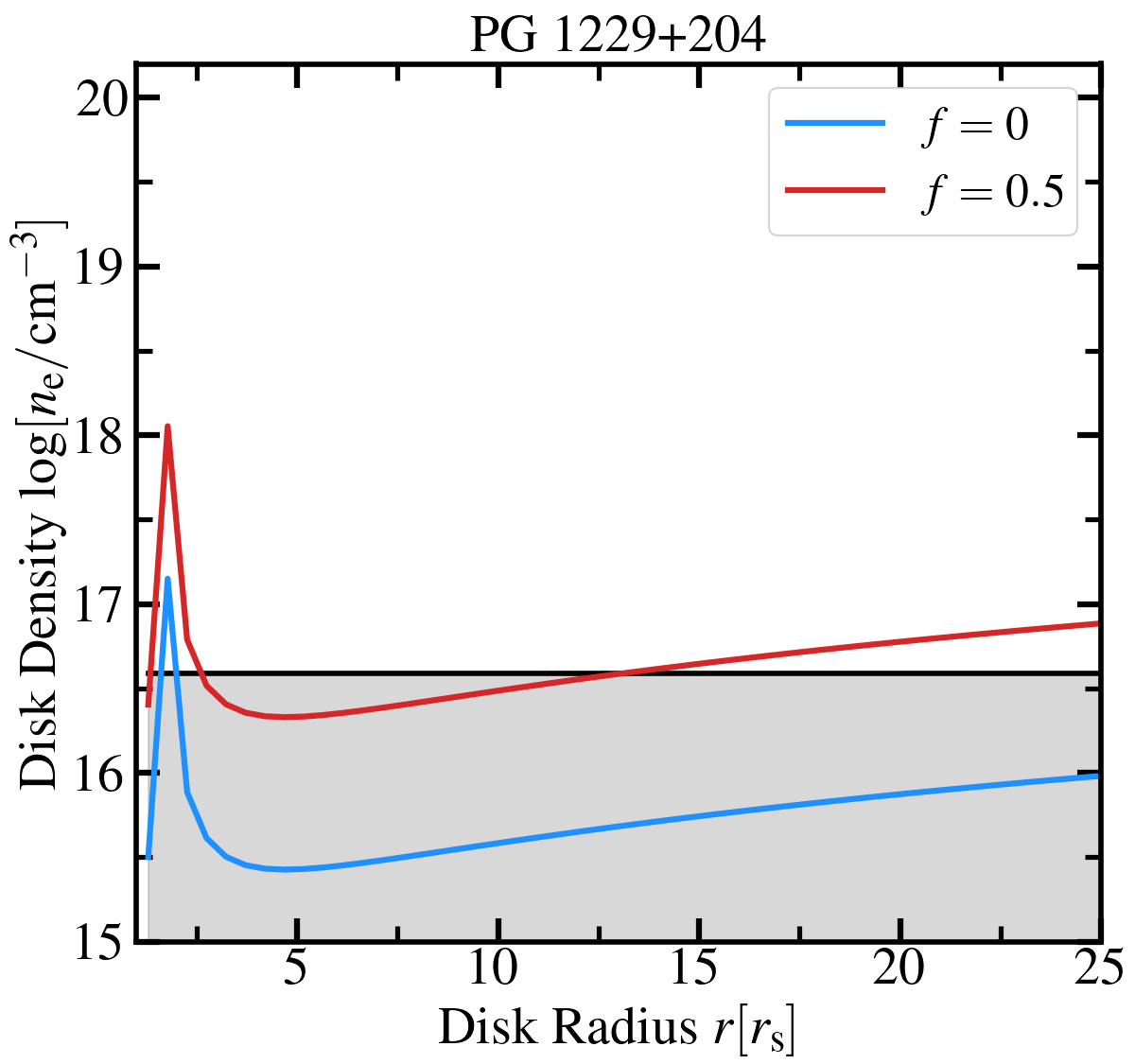}
\includegraphics[scale=0.29,angle=-0]{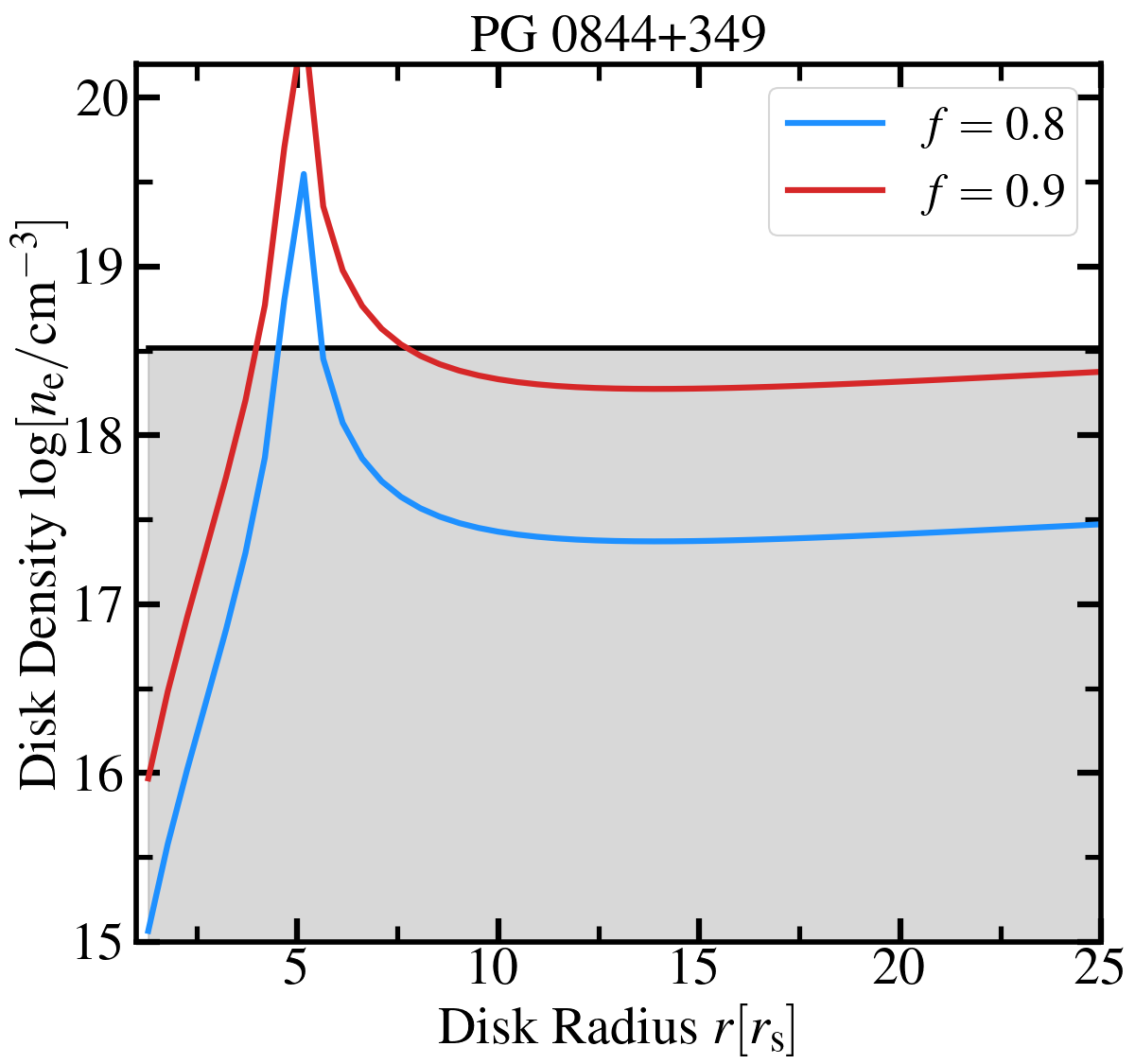}
\includegraphics[scale=0.29,angle=-0]{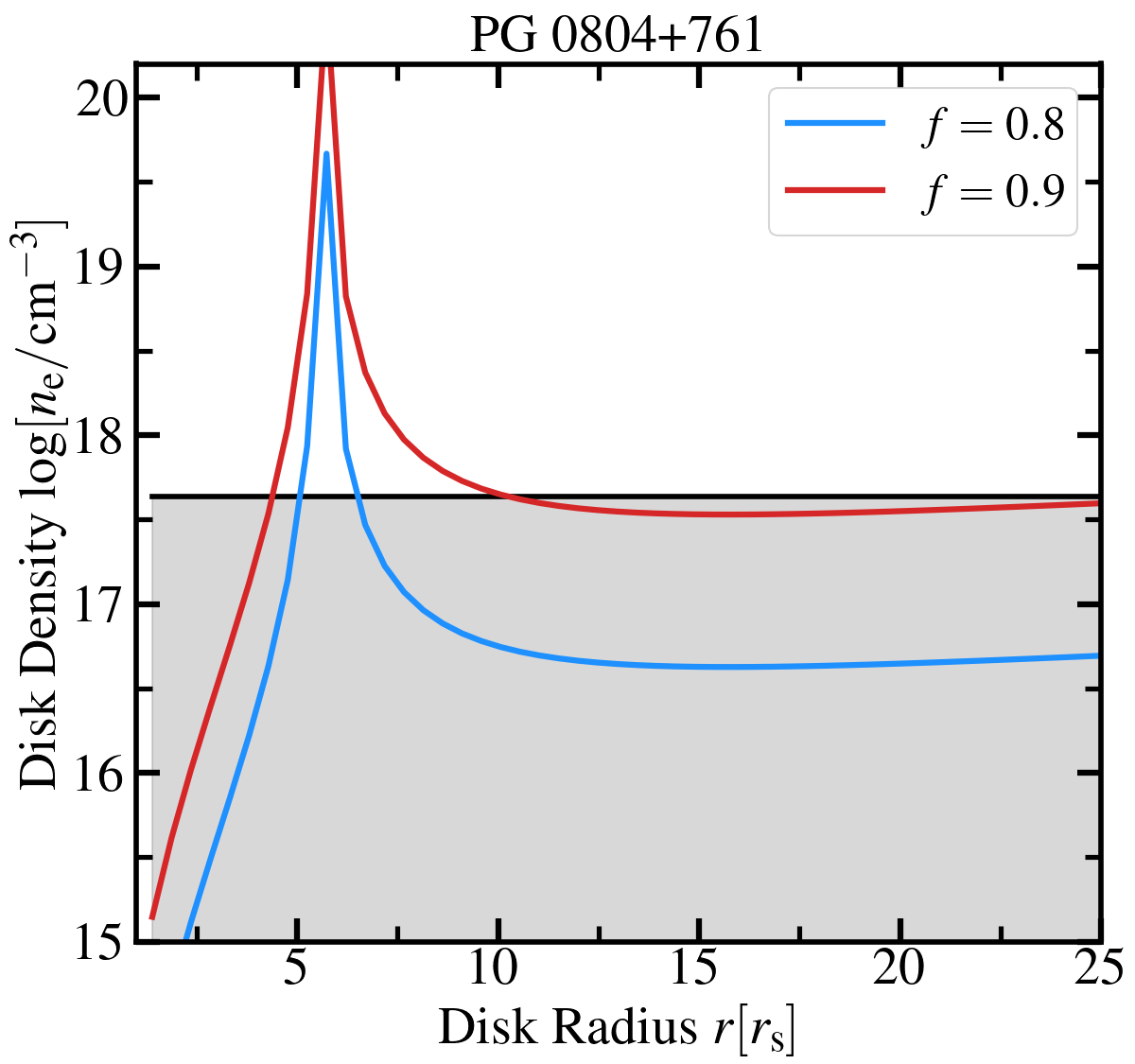}
\includegraphics[scale=0.29,angle=-0]{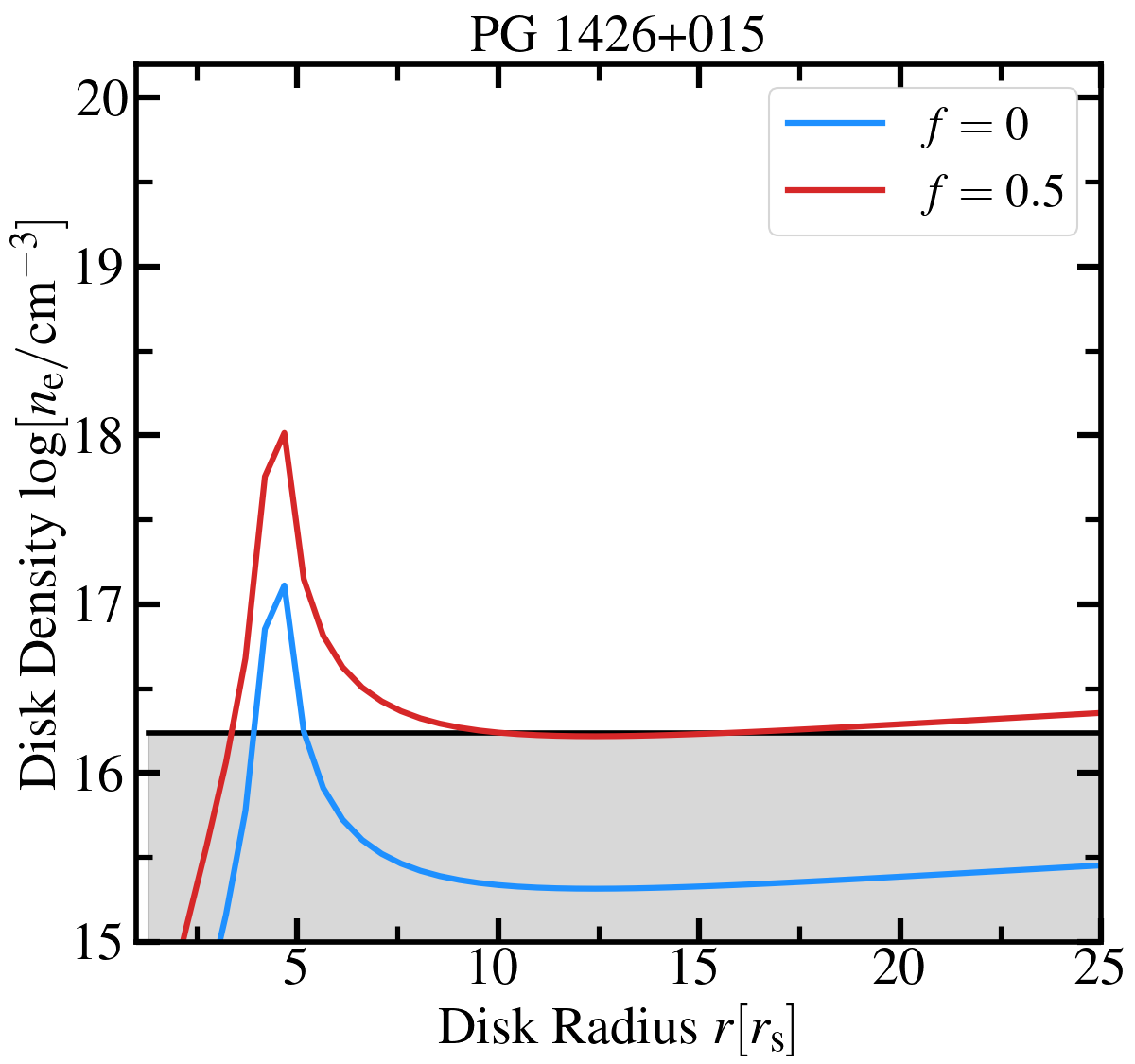}
\caption{The electron density $\left(\log[n_{\rm e}/{\rm cm^{-3}}] \right)$ of the accretion disk is shown as a function of disk radius ($r$) and calculated for different values of the disk-to-corona power transfer fraction ($f$) for each source. The black solid line and grey shaded area represent the best-fit $\log n_{e}$ and associated 90\% confidence intervals, respectively.}
\label{logne_r}
\end{center}
\end{figure*}

\end{document}